\documentclass[a4paper,oneside,11pt,titlepage]{book}

\newlength{\boxparindent}
\usepackage{graphics}
\usepackage{ifthen}
\usepackage{fancyheadings}
\usepackage{amsfonts}
\usepackage{amsmath}

\begin{document}

\newcommand{\color}[1]{}
\newcommand{\textcolor}[2]{#2}

\newcommand{\sect}[1]{section~\ref{#1}}
\newcommand{\se}[2][nothere]{
\color{blue}
\ifthenelse{\equal{#1}{nothere}}{\section{#2}
\lhead{\textit{\thesection~#2}}
}{\section[#1]{#2} \lhead{\textit{\thesection~#1}}}  
\color{black} 
}
\newcommand{\sest}[1]{
\color{blue} 
\section*{#1} \color{black}}
\renewcommand{\thesection}{\arabic{chapter}.\arabic{section}}
\newcommand{\sse}[2][nothere]{
\color{blue} 
\ifthenelse{\equal{#1}{nothere}}{\subsection{#2}}{\subsection[#1]{#2}} 
\color{black}}
\newcommand{\ssest}[1]{
\color{blue} \subsection*{#1} \color{black}}
\renewcommand{\thesubsection}{\arabic{chapter}.\arabic{section}.\arabic{subsection}}
\newcommand{\ssse}[1]{
\color{blue} \subsubsection{#1}
\color{black}}
\renewcommand{\thesubsubsection}{\arabic{chapter}.\arabic{section}.\arabic{subsection}.\arabic{subsubsection}}
\newcommand{\ba}{\color{deepred}  
\begin{eqnarray}}
\newcommand{\ea}{\end{eqnarray}  
\color{black}}
\newcommand{\m}[1]{\textcolor{deepred}{$#1$}}
\newcommand{\ca}[2][nothere]{ 
\color{blue} \small \bfseries
\ifthenelse{\equal{#1}{nothere}}{\caption{#2}}{\caption[#1]{#2}} 
\color{black}}
\newcommand{\fig}[1]{figure~\ref{#1}}
\newcommand{\re}[1]{(\ref{#1})}
\newcommand{\maintitle}[1]{\vskip 0.9cm \begin{center}
{\LARGE{\bf \textcolor{deepblue}{#1}}} \end{center} \vskip 0.4cm}
\newcommand{\name}[1]{\begin{center} \textcolor{deepred}{#1} \end{center}}
\newcommand{\paperrefs}[2]{\color{deepred}
\begin{flushright} #1 \\ #2 \end{flushright} \color{black}}
\newcommand{\rname}[1]{\newblock {\em #1,}}
\newcommand{\rauthor}[1]{#1,}
\newcommand{\journal}[4]{#1 \textbf{#2} (#3) #4}
\newcommand{\preprint}[1]{, preprint \texttt{#1}}
\newcommand{\xpreprint}[1]{ preprint \texttt{#1}}
\newcommand{\book}[3]{#1, #2 #3} 
\newcommand{\chap}[1]{\chapter{#1}}
\newcommand{\bi}{\begin{itemize} \setlength{\parskip}{0pt}}
\newcommand{\ei}{\end{itemize}}
\newcommand{\quot}[2]{\begin{quote}\sffamily \slshape
\begin{center}``#1''\end{center}
\upshape {\raggedleft ---#2\\}\end{quote}}
\newcommand{\thline}{\hline\hline}
\newcommand\phup{^{\phantom{1}}}
\renewcommand{\contentsname}{\textrm{Contents}}
\renewcommand{\listfigurename}{\textrm{List of Figures}}
\renewcommand{\listtablename}{\textrm{List of Tables}}

\newcommand{\la}{\lambda} 
\newcommand{\xl}{\frac{\xi}{\lambda}}
\newcommand{\xp}{\frac{\xi}{\pi}} 
\newcommand{\g}{\Gamma}
\newcommand{\ul}{\frac{\la u}{\pi}}
\newcommand{\hf}{\frac{1}{2}}
\newcommand{\hp}{\frac{\pi}{2}}
\newcommand{\T}{\theta}
\newcommand{\pl}{\frac{\pi}{\lambda}}
\newcommand{\st}[1]{|#1\rangle}
\newcommand{\w}{\omega}
\newcommand{\hft}{{\textstyle\frac{1}{2}}}
\newcommand{\ep}{\epsilon}
\newcommand{\nn}{\nonumber}

\newcommand{\qbar}{\overline{q}}
\newcommand{\e}{\mathrm{e}}
\newcommand{\rv}{r^{\vee}}
\newcommand{\hv}{h^{\vee}}

\newcommand{\pam}[1]{{[\bf\small #1]}}
\newcommand{\ped}[1]{{[\bf\small #1]}}

\renewcommand{\pmb}[1]{\mathbf{#1}}

\newtheorem{lemma}{Lemma}


\pagestyle{fancy}
\rhead{\bfseries \thepage}
\cfoot{}

\frontmatter

\begin{titlepage}
{\raggedleft \ttfamily hep-th/0111261\\}
\vspace*{0.33in}
\Large 
\bfseries
\begin{center}
\rmfamily
Integrable Quantum Field Theories, in the Bulk and with a Boundary
\end{center}

\vspace{0.2in}
\large
\begin{center}
\rmfamily
Peter Aake Mattsson
\end{center}

\mdseries
\normalsize
\begin{center}
\itshape
Department of Mathematical Sciences,
University of Durham,
Durham DH1 3LE,
England.

\upshape
\end{center}
\vspace{0.2in}
\begin{center}
\Large
\bfseries
Abstract
\mdseries
\normalsize
\thispagestyle{plain}

\begin{minipage}{5.5in}
\setlength{\parskip}{6pt}
\vspace{0.1in}
In this thesis, we consider the massive field theories in 1+1 dimensions
known as affine Toda quantum field theories. These have the special
property that they possess an infinite number of conserved quantities, a
feature which greatly simplifies their study, and makes extracting exact information about them a tractable problem. We consider these theories
both in the full space (the bulk) and in the half space bounded by an
impenetrable boundary at \m{x=0}. In particular, we consider
their fundamental objects: the scattering matrices in the bulk, and
the reflection factors at the boundary, both of which can be found in a closed form.

\setlength{\parindent}{\boxparindent}
In Chapter 1, we provide a general introduction to the topic before going
on, in Chapter 2, to consider the simplest ATFT---the sine-Gordon
model---with a boundary. We begin by studying the classical limit, 
finding quite a clear
picture of the boundary structure we can expect in the
quantum case, which is
introduced in Chapter 3. We obtain the bound-state structure for all
integrable boundary conditions, as well as the corresponding reflection
factors. This structure turns out to be much richer than had
hitherto been imagined.

We then consider more general ATFTs in the bulk. The
sine-Gordon model is based on $a_{1}^{(1)}$, but there is an ATFT for any 
semi-simple Lie algebra. This underlying structure is known to show up
in their S-matrices, but the path back to the parameters in the
Lagrangian is still unclear. We investigate this, our main result 
being the discovery of a ``generalised bootstrap'' equation which 
explicitly encodes the Lie algebra into the S-matrix. This leads to a 
number of new S-matrix identities, as well as a generalisation of the 
idea that the conserved charges of the theory form an eigenvector of
the Cartan matrix.

Finally, our results are summarised in Chapter 5, and possible
directions for further study are highlighted.
\end{minipage}
\end{center}

\thispagestyle{empty}
\lhead{\textit{Contents}}
\end{titlepage}
\setlength{\parskip}{0pt}

\setlength{\parskip}{6pt}
\sest{Note added}
\thispagestyle{empty}
The work which follows was presented as a thesis for the degree of Doctor of
Philosophy at the University of Durham in September 2000. Since then,
Bajnok {\em et al.}~\cite{BPT1,BPT2,BPT3}
have also studied the problem of the boundary sine-Gordon model
with arbitrary integrable boundary conditions. Their results provide
independent confirmation of many of those reported here. We are also
grateful to Gabor Tak\'{a}cs
and Zoli Bajnok, in particular, for
pointing out a number of typos in~\cite{Mattsson2}. Since they apply equally
to the content of this thesis, we record them here.

\begin{itemize}
\item
We should perhaps have made it clear that the lemmas presented in Chapter 3
rested on an implicit assumption
that, when the the incoming particle decays into two other particles,
neither is the outgoing particle. Examples such as Figure
3.7 are therefore technically exceptions to the lemmas as stated.
These cases were not discussed explicitly in the
text as they are only relevant when the rapidity of the incoming particle is
independent of the boundary parameter.

\item
In equation (3.24) [equation (18) in~\cite{Mattsson2}], the two $l/2\lambda$ terms should read $l/\lambda$. 
\item
Just above Table 3.1 [Table 1 in~\cite{Mattsson2}], \m{b^{1}_{m,1}} should read \m{a^{1;m}_{1}}.
\item
In Table 3.7 [Table 7 in~\cite{Mattsson2}], the order of the pole at \m{\nu_{0}} is not correct; it
should be \m{k}.
\end{itemize}
The hep-th version of~\cite{Mattsson2} has been updated to incorporate the
last three amendments.
\pagebreak
\setcounter{page}{0}
\setlength{\parskip}{5pt}
\tableofcontents
\setlength{\parskip}{6pt}
\pagebreak
\sest{Preface}
\addcontentsline{toc}{chapter}{Preface}
\thispagestyle{plain}
\vspace{-6pt}
This thesis is the result of work carried out in the Department of
Mathematical Sciences at the University of Durham, between October 1996
and September 1999, under the supervision of Dr. Patrick Dorey. No part of
it has been previously submitted for any degree, either in this or any
other university.

No claim to originality is made for the review in Chapter 1, nor for the
introduction to the sine-Gordon model at the beginning of Chapter
2 or the introduction to ATFTs in Chapter 4. The work on the classical
sine-Gordon theory in Chapter 2 is new and due solely to the author. The
study of the quantum model with Dirichlet boundary conditions in 
Chapter 3 was carried out jointly with Patrick Dorey, and has been 
published as \cite{Mattsson2}. The extension to arbitrary boundary 
conditions is again new. 

The idea for the work on bulk ATFTs in Chapter 4 emerged from
conversations with Patrick Dorey (notably the generalised bootstrap) and with
Roberto Tateo (the idea of comparing the generalised bootstrap before
and after interchanging indices as a method of generating S-matrix 
identities). The implementation of these, however, and the motivation behind the bootstrap, is due to the author. Much of 
this has already been published \cite{Mattsson}, but the discussion 
presented here was lacking.

Some of the results presented here have also been obtained by other
groups, working independently. In particular, the ``generalised
bootstrap'' identity was found by Fring, Korff and Schultz \cite{Fring},
while some of the S-matrix identities were found in an ad-hoc fashion by
Khastgir \cite{Khastgir}. In addition, work on the boundary 
sine-Gordon
model with Neumann boundary conditions has been carried out by Brett
Gibson.

\vspace*{12pt}
\noindent {\bf \rmfamily \itshape Acknowledgements:}
\addcontentsline{toc}{chapter}{Acknowledgements}
I would like to thank Patrick Dorey for all his help, as well as Roberto
Tateo, Andreas Fring, Matthias Pillin, Subir
Ghoshal, Gustav Delius, Aldo Delfino and Brett Gibson for interesting
discussions. Financial support was provided by the EPSRC (who
funded a Research Studentship) and by a TMR grant of the European
Commission (reference ERBFMRXCT960012), both of whom I acknowledge and
thank.

Finally, thanks are undoubtedly due to Tuki, Adelene and my parents, for valiantly trying---against all odds---to keep me sane,
keeping my talent for procrastination in check, and so allowing this
thesis to be submitted \emph{before} the Last Trump, rather than just
afterwards as was once feared. I am also deeply grateful to
Mr Lee, my erstwhile English teacher, who
helped me to believe in myself, and to realise that maybe scientists
can write too!
(He also asked for my first book to be dedicated to him, without 
realising the consequences\ldots) 

\mainmatter

\chap{Integrable Quantum Field Theory in Two Dimensions}
\label{chap:intro}

\quot{I got really fascinated by these 1+1 models \ldots and how
mysteriously they jump out at you and work and you don't know
why. I am trying to understand all this better.}{Richard Feynman}

\renewcommand{\thepage}{\arabic{page}}
 

\se{Introduction}
\quot{`The time has come,' the walrus said, `to talk of many
things. Of ships and shoes and sealing-wax, and cabbages and
kings.'}{Lewis Carroll}
Why study a theory in \emph{two} dimensions, when the real world 
has at least four? The difficulty, at least at present, is that
realistic four-dimensional field theories are incredibly complicated,
even before the further dimensions proposed by superstring theories
are considered. Perturbative solutions
can be found, but exact, non-perturbative, results are few. In a
number of cases, sufficiently accurate perturbative answers are
enough, but many physical phenomena cannot be properly understood this
way. Uncovering exact solutions to quantum field theories is
considered by many to be one of the great remaining challenges to
particle theorists.

In view of this, it is perhaps logical to think of taking a step
backwards, to consider a simpler model which still exhibits some of
the same features. Understanding the issues involved a few at a time in
this way presents a more manageable problem, like climbing a
mountain in stages rather than hoping to stride straight to the top. 
A theory with two dimensions, one of space and one of time, is
a useful starting point.

The main focus of this thesis will be the scattering matrices
(S-matrices) of these theories, which describe how the final
state of the system is related to the initial state.
In general this is a messy object to deal with, since there
are potentially matrices for the evolution of any number of particles into
any other number. However, the theories considered here are
``simplified'' in one further respect in that they are 
\emph{integrable\/} theories. This has three main effects:
\begin{itemize}
\item there is no net particle production, so the number of
particles in the initial and final states must always be the
same;
\item the outgoing particles must have the same masses
and velocities as the incoming ones; 
\item a general S-matrix, for the
scattering of \m{n}
particles, can always be factorised into a product of two-particle
S-matrices. 
\end{itemize}
In essence, this means that, once the S-matrix for the
scattering of two particles into two particles has been calculated,
everything else follows, making a characterisation of the theory in
terms of these matrices a more attractive and tractable proposition.

The method for obtaining explicit expressions for these S-matrices is
in fact surprisingly straightforward (if not necessarily simple to put
into practice). It invokes four consistency requirements which, between
them, provide strong constraints on the S-matrix. This method, first 
formulated in the 1960s, was initially
developed to help explain the strong nuclear force (see
e.g. \cite{Chew, Eden}). In the 1970s, it was discovered that these
axioms could also be applied to some two dimensional quantum field
theories, and proved in certain cases to be powerful enough to
completely determine the S-matrix up to an overall factor. 

A theory is integrable if it has an infinite number of symmetries;
the particular theories we will be studying, the affine Toda field
theories (ATFTs), acquire these through being based on an
infinite-dimensional Lie algebra. We will study these both in the full
2-dimensional space, and in a half space (or half line) defined by
introducing an impenetrable boundary at \m{x=0}. As well as the usual
S-matrices, this requires the introduction of boundary factors to
describe particles scattering against the boundary. The particles can
either simply reflect from this ``wall'' or bind to it, and we will be
concerned with the bound state structure this introduces.

The situation with a boundary is the less well-understood of the two,
so we shall study only the simplest ATFT, the sine-Gordon model. Even
for this case, only the ground and first few excited states have been
explored.  We present a complete description of these matrices, for
any wall which leaves the resultant theory integrable.

Without a boundary, the picture is much clearer, and S-matrices have
been found for all real-coupling ATFTs. However, despite the manifest Lie
algebraic
structure of the theory, its path from the Lagrangian to the S-matrix
is still not precisely known, and remains an open problem. We will,
however,
present a convenient way of encoding the Lie algebra into the matrix,
with the aim of making the task a little easier. This process will
also throw up a number of new relationships between elements of the
S-matrix.

Apart from their interest in connection with higher-dimensional
theories, two-dimensional integrable models have an increasing number
of applications in their own right. They are, for example, useful in
studying impurity problems in an interacting 1D
electron gas~\cite{Kane} or edge excitations in fractional
quantum Hall states~\cite{Wen,Fendley}. (A recent review
can be found in \cite{Saleur}.)     

In the following sections all this will be put on a more formal
basis, paving the way for the discussion of the ATFTs which
will occupy our attention for the remainder of the thesis.       

\se{Exact S-matrices}

\quot{The three rules of the Librarians of Time and Space are: 1) Silence; 2)
Books must be returned no later than the date last shown; and \\3) Do not
interfere with the nature of causality.}{Terry Pratchett, Guards! Guards!}

Much of the discussion in this section is based on~\cite{Zsquare,
pedrev, GhoshZam}. Before proceeding to specifics, a gentle
introduction to two-dimensional field theory is perhaps appropriate. 

Let us begin by considering a general Euclidean field theory with one 
space and one time dimension \m{(x^{1},x^{2})=(x,t)} defined (in the 
Lagrangian approach) by the classical action
\begin{equation}
\mathcal{A}=\int_{-\infty}^{+\infty}dx\int_{-\infty}^{+\infty}dt
\, a(\varphi, \partial_{\mu}\varphi),
\label{eq:genfield}
\end{equation}
where \m{\varphi(x,t)} is some set of fundamental fields and the
action density \m{a(\varphi,\partial_{\mu}\varphi)} is a local function
of these fields and the derivatives \m{\partial_{\mu}\varphi=\partial
\varphi/\partial x^{\mu}} with \m{\mu=1,2}. For simplicity we shall also use
light-cone coordinates, so that, in place of \m{(p^{0},p^{1})} for the 
two-momentum, we will take \m{(p,\overline{p})=(p^{0}+p^{1},p^{0}-p^{1})}.

We will be considering the
particles only on mass-shell (i.e. real, rather than virtual, particles),
which means that their two-momenta satisfy the mass-shell condition
\begin{equation}
p_{a}\overline{p}_{a}=(p_{a}^{0})^{2}-(p_{a}^{1})^{2}=m_{a}^{2},
\, (a=1,2,\ldots,n).
\end{equation}
The two momenta can be conveniently parametrised in terms of their
rapidity \m{\T}:
\begin{equation}
 (p_{a}^{0},p_{a}^{1})=(m_{a}e^{\T_{a}},m_{a}e^{-\T_{a}}).
\label{eq:prapid}
\end{equation}

Suppose our theory contains \m{n} different types of particle \m{A_{a},
a=1,2,\ldots,n}, with masses \m{m_{a}}. The asymptotic particle states
are generated by the ``particle creation operators'' \m{A_{a}(\T)}:
\begin{equation}
\st{A_{a_{1}}(\T_{1})A_{a_{2}}(\T_{2})\cdots A_{a_{n}}(\T_{n})}=
A_{a_{1}}(\T_{1})A_{a_{2}}(\T_{2})\cdots A_{a_{n}}(\T_{n})\st{0}.
\label{eq:crnops}
\end{equation}
Looking into the far past, we shall call the state an \emph{in} state
if there are no further interactions as \m{t \rightarrow
-\infty}. This means that the fastest particle must be on the left,
and the slowest on the right, with all the others in order in-between, i.e.
\m{\T_{1} > \T_{2} > \cdots > \T_{n}}. Similarly, if there are no
further interactions as \m{t \rightarrow \infty}, the state will be
called an \emph{out} state, and the rapidities must be in the reverse
order.

The S-matrix can now be introduced as a mapping between the
\emph{in}-state basis and the \emph{out}-state basis. It is useful to
consider the \m{A_{x}(\T)}s as non-commuting symbols, giving them an existence
outside the ket vectors, so that we can write the above state simply
as \m{A_{a_{1}}(\T_{1})A_{a_{2}}(\T_{2})\ldots\linebreak
A_{a_{n}}(\T_{n})}. Considering an \m{m}-particle \emph{in}-state, we have
\ba
\lefteqn{A_{a_{1}}(\T_{1})A_{a_{2}}(\T_{2})\ldots A_{a_{n}}(\T_{n}) =}
\nonumber \\
&& \sum_{m=1}^{\infty}\sum_{\T_{1}'<\ldots<\T_{m}'}S_{a_{1}\ldots
a_{n}}^{b_{1}\dots b_{m}}(\T_{1}\ldots\T_{n};\T_{1}'\ldots \T_{m}')
A_{b_{1}}(\T_{1}')A_{b_{2}}(\T_{2}')\ldots A_{b_{m}}(\T_{m}')
\ea
where a sum on \m{b_{1}\ldots b_{n}} is
implied, and the sum on the \m{\T_{i}'} will, in general, turn out to
involve a number of integrals. The rapidities will also be constrained
by momentum conservation.

For a general theory, we can proceed no further, and introducing the
S-matrix would appear only to have complicated matters. However, for an
\emph{integrable\/} theory, the whole situation becomes dramatically
simpler. The name derives from the classical formulation of such
theories, which can be cast as partial differential equations; these
were said to be integrable if it was possible to find an explicit
solution. It was found that a solution was only possible if there were
an infinite number of symmetries constraining the behaviour of the
equation, and preventing it from becoming chaotic. The same applies
here: possessing so many symmetries constrains the S-matrix
sufficiently to allow an exact solution to be found.

Energy-momentum is always a conserved quantity, and its operator,
\m{\textbf P}, is said to have (Lorentz) spin 1 as it transforms under a
Lorentz boost \m{L_{\alpha}: \T \rightarrow \T'=\T+\alpha} as
\m{{\textbf P} \rightarrow
{\textbf P'}=e^{\alpha}{\textbf P}}. This means that a boost of
\m{2i\pi}---a
complete rotation---has no effect. On a one-particle state, the action of
\m{{\textbf P}=(P,\overline{P})} would be
\begin{equation}
P\st{A_{a}(\T)}=m_{a}e^{\T}\st{A_{a}(\T)},\quad
\overline{P}\st{A_{a}(\T)}=m_{a}e^{-\T}\st{A_{a}(\T)}\,.
\end{equation}
In general, there can also be other
conserved quantities, \m{\textbf Q_{s}}, which
transform in higher representations of the 1+1-dimensional Lorentz
group as \m{{\textbf Q_{s}} \rightarrow {\textbf Q_{s}'}=e^{\alpha s}
{\textbf Q_{s}}} and have spin
\m{s} since they rotate \m{s} times under a boost of \m{2i\pi}. This time,
the effect of \m{{\textbf Q_{s}}=(Q_{s},Q_{-s})} on a one-particle state is
\begin{equation}
Q_{s}\st{A_{a}(\T)}=q_{a}^{(s)}e^{s\T}\st{A_{a}(\T)}, \quad
Q_{-s}\st{A_{a}(\T)}=q_{a}^{(s)}e^{-s\T}\st{A_{a}(\T)}\,.
\end{equation}

In an integrable theory, there are an infinite number of these
conserved quantities (or ``charges''). It might, at first, appear that
such theories are quite improbable. In the theories to be considered later, 
these symmetries are due to an
underlying group structure which happens to be infinite-dimensional.

We will concentrate on local
conserved charges, which are those whose operators are integrals of strictly
local densities, meaning that their action on multi-particle states is
additive:
\begin{equation}
Q_{s}\st{A_{a_{1}}(\T_{1})\ldots A_{a_{n}}(\T_{n})}=
(q_{a_{1}}^{(s)}e^{s\T_{1}}+\ldots + q_{a_{n}}^{(s)}e^{s\T_{n}})
\st{A_{a_{1}}(\T_{1})\ldots A_{a_{n}}(\T_{n})}.
\label{eq:qact}
\end{equation}

Just as, above, momentum conservation
constrained the sum over the rapidities, so all the other conserved
quantities provide additional constraints, leaving an infinite number
of equations to be solved, of the form
\begin{equation}
q_{a_{1}}^{(s)}e^{s\T_{1}}+\ldots+q_{a_{n}}^{(s)}e^{s\T_{n}}=
q_{b_{1}}^{(s)}e^{s\T_{1}'}+\ldots+q_{b_{m}}^{(s)}e^{s\T_{m}'}.
\end{equation}
Since these must all hold for all possible sets of \emph{in}-momenta, the
only possible solution is the trivial one, i.e. \m{n=m}, and
\m{\T_{i}=\T_{i}'}, \m{q_{a_{i}}^{(s)}=q_{b_{i}}^{(s)}} for all \m{i}.

This establishes the fact that there can be no particle production in an 
integrable theory, and that the sets of incoming and outgoing momenta must be 
equal, thus reducing the workload involved in dealing with the S-matrix
just to the \m{n \rightarrow n} cases. There is, however, one further
property of integrable theories which makes them even easier to deal with:
\emph{factorisation}. This states that, for any \m{n \rightarrow n}
S-matrix, the
trajectories of the particles involved can
be shifted forwards or backwards in space so as to split the vertex
into a product of \m{\hf n(n-1)} two-particle vertices, as shown in
\fig{fig:yangbax}.

The origin of this property can be seen in the fact that, while the momentum
operator, for example, will act on a given state simply by shifting
the position of all the particles by a fixed amount, higher-spin
operators will, in general, change the positions by an amount depending
on the initial momentum of the particle. This argument was first
proposed by Shankar and Witten~\cite{Shankar} and elaborated by Parke
\cite{Parke}. In rough form it goes as follows.

The first step is to note that, since we are dealing with a local,
causal field theory, the particles in any process are sufficiently
well separated (at least most of the time) to be considered
individually, and it is reasonable to consider the effect of the
conserved charges particle by particle. If we consider a
single-particle state, with position approximately \m{x_{1}} and
spatial momentum approximately \m{p_{1}}, the position space
wavefunction will be
\begin{equation}
\psi(x) \propto \int_{-\infty}^{+\infty}dp
\, e^{-a^{2}(p-p_{1})^{2}}e^{ip(x-x_{1})}. 
\end{equation}
For simplicity, rather than considering a general spin-\m{s} operator,
we will now try acting on this with \m{P_{s}}, the spin \m{s}
operator which acts as \m{(P)^{s}}, i.e. as \m{s} copies of the
spatial part of the two-momentum operator. Applying \m{e^{-i\alpha
P_{s}}} to the above wavefunction gives
\begin{equation}
\overline{\psi(x)} \propto \int_{-\infty}^{+\infty}dp
e^{-a^{2}(p-p_{1})^{2}}
e^{ip(x-x_{1})}e^{-i\alpha p^{s}}.
\end{equation}
Since most of the value of the integral is due to the region around
\m{p \approx p_{1}}, we can Taylor expand the extra factor in powers of
\m{(p-p_{1})} to find new values for the position and momentum. For a
general momentum-dependent phase factor \m{e^{-i\phi(p)}}, this leaves
the momentum unchanged but shifts the position by
\m{\overline{x_{1}}=x_{1}+ \phi'(p_{1})}. Here, this gives a position
shift of \m{s\alpha p^{s-1}_{a}}. For momentum itself, this is just
\m{\alpha} but, for higher spins, the shift must depend on the initial
momentum. This, as Parke showed, is a general property of higher-spin 
operators.

Applying a suitable operator, \m{Q_{s}}, near an \m{n \rightarrow n}
vertex thus
separates the particles and splits up the vertex into \m{\hf n(n-1)}
\m{2 \rightarrow 2} vertices. However, since the operator is related
to a conserved charge, the amplitude for both processes must be the
same, giving us factorisability. In addition, applying \m{Q_{-s}}
rather than \m{Q_{s}} causes a mirror-image split, leading to 
\fig{fig:yangbax}.

This relies, of course, on the fact that, after splitting up the
vertex, the particle trajectories must still cross \emph{somewhere},
because we are restricted to only two dimensions. In higher numbers of
dimensions, it is quite easy to imagine splitting up such a vertex so
that the particles never meet at all, leading to a trivial
S-matrix. In fact, Coleman and
Mandula~\cite{ColeMand} proved the so-called Coleman-Mandula theorem,
showing that, for any theory with more than one space dimension and a 
conserved charge with spin 2 or more, the S-matrix must be trivial.

All this shows that an integrable theory in 1+1 dimensions is rather 
special, in that it has:
\begin{itemize}
\item no particle production;
\item equality of the sets of initial and final momenta;
\item factorisability of the \m{n \rightarrow n} S-matrix into a
product of \m{2 \rightarrow 2} S-matrices.
\end{itemize}

With these results, it is clear that the \m{2 \rightarrow 2} S-matrix
is the fundamental object of the theory, and that, once it has been
found, the full S-matrix is only a step away. The \m{2 \rightarrow 2}
process is just
\begin{equation}
A_{a_{1}}(\T_{1})A_{a_{2}}(\T_{2})=S^{b_{1}b_{2}}_{a_{1}a_{2}}(\T_{1} -
\T_{2})A_{b_{2}}(\T_{2})A_{b_{1}}(\T_{1}),
\label{eq:smatrix}
\end{equation}
and is shown graphically in \fig{fig:smatrix}. (In this, and in all
subsequent diagrams, time is taken to run up the page, and space from
left to right.) Note that momentum conservation demands
\m{m_{a_{1}}=m_{b_{1}}} and \m{m_{a_{2}}=m_{b_{2}}}, so that \m{a_{1}
\neq b_{1}} or \m{a_{2} \neq b_{2}} are only possible if there is a
degenerate mass spectrum.

\begin{figure}
\begin{center}
\parbox{2.0in}{
\unitlength 1.00mm
\linethickness{0.4pt}
\begin{picture}(47.33,55.00)
\color{green}
\put(46.11,49.89){\line(-1,-1){41.00}}
\put(6.33,49.89){\line(1,-1){41.00}}
\color{deepred}
\put(6.22,51.22){\makebox(0,0)[cc]{$a_{1}$}}
\put(46.11,51.33){\makebox(0,0)[cc]{$a_{2}$}}
\put(5.00,6.78){\makebox(0,0)[cc]{$b_{2}$}}
\put(47.33,6.56){\makebox(0,0)[cc]{$b_{1}$}}
\qbezier(22.33,33.89)(26.11,36.00)(30.33,34.11)
\put(26.33,36.56){\makebox(0,0)[cc]{$\theta$}}
\put(-5,25){\vector(0,1){10}}
\put(-5,25){\vector(1,0){10}}
\put(7,25){\makebox(0,0)[cc]{$x$}}
\put(-5,37){\makebox(0,0)[cc]{$t$}}
\color{black}
\end{picture}
} \ \m{=S_{a_{1}a_{2}}^{b_{1}b_{2}}(\T)}
\ca{S-matrix}
\label{fig:smatrix}
\end{center}
\end{figure}

The two-particle S-matrix has \m{n^{4}} elements, but these are not
all independent, and are in fact strongly constrained. Firstly, it is
generally assumed that parity, charge conjugation and time reversal
(P, C and T) symmetries hold. These impose the conditions
\begin{equation}
S^{kl}_{ij}(\T)=S^{lk}_{ji}(\T)=S^{\overline{k}\overline{l}}_{
\overline{\imath} \overline{\jmath}}(\T)=S^{ji}_{lk}(\T).
\end{equation}
In addition, they must satisfy four general axioms: the Yang-Baxter
equation; a unitarity condition; analyticity and crossing symmetry; and the
bootstrap condition. These are powerful demands; using just the first
three allows the S-matrix to be pinned down up to the so-called ``CDD
ambiguity'':
\begin{equation}
S^{kl}_{ij}(\T) \rightarrow S^{kl}_{ij}(\T)\Phi(\T),
\end{equation}
where the ``CDD factor'' satisfies
\ba
\Phi(\T)=\Phi(i\pi - \T)\,, && \Phi(\T)\Phi(-\T)=1\,,
\ea
but is otherwise arbitrary. This can often be further restricted by
the bootstrap.

\sse{Yang-Baxter (or ``factorisation'') equation}
The requirement of factorisability, and, in particular, the ability of
operators associated to
conserved charges to shift trajectories around, is only consistent if
\fig{fig:yangbax} is true. This gives rise to the condition
\begin{figure}
\begin{center}
\phantom{Humm}
\parbox{2.5in}{
\unitlength 1.00mm
\linethickness{0.4pt}
\begin{picture}(47.44,55.00)
\color{green}
\put(6.33,49.89){\line(1,-1){41.00}}
\put(46.11,49.89){\line(-1,-1){41.00}}
\put(15.44,49.89){\line(0,-1){41.11}}
\color{deepred}
\put(6.22,51.89){\makebox(0,0)[cc]{$a_{1}$}}
\put(15.44,51.56){\makebox(0,0)[cc]{$a_{2}$}}
\put(46.11,51.22){\makebox(0,0)[cc]{$a_{3}$}}
\put(5.00,7.44){\makebox(0,0)[cc]{$b_{3}$}}
\put(15.44,7.11){\makebox(0,0)[cc]{$b_{2}$}}
\put(47.44,6.78){\makebox(0,0)[cc]{$b_{1}$}}
\qbezier(15.56,24.00)(17.89,24.56)(18.89,22.89)
\put(17.78,26.00){\makebox(0,0)[cc]{$\theta'$}}
\qbezier(11.67,44.56)(12.89,46.67)(15.44,45.67)
\put(12.67,47.67){\makebox(0,0)[cc]{$\theta$}}
\color{black}
\end{picture}
} \ \LARGE = \normalsize \
\parbox{2.5in}{
\unitlength 1.00mm
\linethickness{0.4pt}
\begin{picture}(57.33,55.00)(-10,0)
\color{green}
\put(46.11,49.89){\line(-1,-1){41.00}}
\put(6.33,49.89){\line(1,-1){41.00}}
\put(37.00,49.89){\line(0,-1){41.11}}
\color{deepred}
\put(37.00,51.56){\makebox(0,0)[cc]{$a_{2}$}}
\put(37.00,7.11){\makebox(0,0)[cc]{$b_{2}$}}
\qbezier(36.89,24.00)(34.56,24.56)(33.56,22.89)
\put(34.67,26.00){\makebox(0,0)[cc]{$\theta'$}}
\qbezier(40.78,44.56)(39.56,46.67)(37.00,45.67)
\put(39.78,47.67){\makebox(0,0)[cc]{$\theta$}}
\put(6.22,51.22){\makebox(0,0)[cc]{$a_{1}$}}
\put(46.11,51.33){\makebox(0,0)[cc]{$a_{3}$}}
\put(5.00,6.78){\makebox(0,0)[cc]{$b_{3}$}}
\put(47.33,6.56){\makebox(0,0)[cc]{$b_{1}$}}
\color{black}
\end{picture}
}
\ca{Yang-Baxter equation}
\label{fig:yangbax}
\end{center}
\end{figure}
\begin{equation}
S_{a_{1}a_{2}}^{c_{1}c_{2}}(\T)S^{b_{1}c_{3}}_{c_{1}a_{3}}(\T+\T')
S_{c_{2}c_{3}}^{b_{2}b_{3}}(\T')=
S^{c_{2}c_{3}}_{a_{2}a_{3}}(\T')S^{c_{1}b_{3}}_{a_{1}c_{3}}(\T+\T')
S_{c_{1}c_{2}}^{b_{1}b_{2}}(\T).
\label{eq:yangbax}
\end{equation}

Formally, this is an associativity condition on the algebra of the
\m{A_{i}(\T)}s: moving from an \emph{in}-state \m{A_{a_{1}}(\T_{1})
A_{a_{2}}(\T_{2})\cdots A_{a_{n}}(\T_{n})} to an
\emph{out}-state by a series of pair transpositions, the result is
independent of their order if and only if the Yang-Baxter equation 
holds. For three particles, there are only two ways of doing this
(shown in \fig{fig:yangbax}). For the left-hand diagram, we find
\begin{eqnarray}
A_{a_{1}}(\T_{1})A_{a_{2}}(\T_{2})A_{a_{3}}(\T_{3})&=&
[S_{a_{1}a_{2}}^{c_{1}c_{2}}(\T)A_{c_{2}}(\T_{2})A_{c_{1}}(\T_{1})] 
A_{a_{3}}(\T_{3}) \nn \\
&=&S_{a_{1}a_{2}}^{c_{1}c_{2}}(\T)A_{c_{2}}(\T_{2})[S_{c_{1}a_{3}}^{b_{1}
c_{3}}(\T+\T')A_{c_{3}}(\T_{3})A_{b_{1}}(\T_{1})]
\\
&=& S_{a_{1}a_{2}}^{c_{1}c_{2}}(\T)S_{c_{1}a_{3}}^{b_{1}c_{3}}(\T+\T')
[S_{c_{2}c_{3}}^{b_{2}b_{3}}(\T')A_{b_{3}}(\T_{3})A_{b_{2}}(\T_{2})]A_{b_{1}}(\T_{1})\,, \nn
\end{eqnarray}
where we have set \m{\T=\T_{2}-\T_{1}} and \m{\T'=\T_{3}-\T_{2}}.
Doing the same for the right-hand diagram yields a relation between
the same \emph{in}- and \emph{out}-states with a different product of
S-matrices. Since these equations should be equivalent, the products
of S-matrices can be equated, to give
\re{eq:yangbax}. That no further conditions should arise in
considering larger numbers of particles can be seen through the fact
that the Yang-Baxter equation allows any trajectory to be moved past any
given vertex (by considering just the local area around the
vertex). Thus, with repeated applications, trajectories can be moved
arbitrarily, showing that all possible factorisations are
equivalent. 

That the Yang-Baxter equation is an associativity condition can most
easily be seen when we have a non-degenerate mass spectrum. In this case,
we can define operators \m{O_{ab}} which transpose the symbols
\m{A_{a}} and
\m{A_{b}}, and add a suitable S-matrix factor. The Yang-Baxter equation
then becomes just
\begin{equation}
O_{12}(O_{13}O_{23})=O_{23}(O_{13}O_{12}),
\end{equation}
which is indeed an associativity condition. The Yang-Baxter equation is
the extension of this to a degenerate spectrum.


\sse{Unitarity, analyticity and crossing symmetry}
The origin of these demands can best be seen by switching to
Mandelstam variables,
\ba
s=(p_{1}+p_{2})^{2}, & t=(p_{1}-p_{3})^{2}, & u=(p_{1}-p_{4})^{2}
\ea
with \m{s+t+u=\sum_{i=1}^{4}m_{i}^{2}}. Here,
\m{p_{1}} and \m{p_{2}} are the momenta of the incoming
particles, with \m{p_{3}} and \m{p_{4}} those of the outgoers. Only
one of these is independent, so we shall make the standard choice and
consider \m{s}. Making use of \re{eq:prapid}, this can be re-written
as
\begin{equation}
s=m_{i}^{2}+m_{j}^{2}+2m_{i}m_{j}\cosh (\T_{1}-\T_{2}).
\end{equation}

For a real, physical process, all rapidities are real, so \m{s} must
be real and satisfy \m{s \geq (m_{i}+m_{j})^{2}}. However, it is usual
to assume that the S-matrix \m{S(s)} is an analytic
function\footnote{It has been suggested \cite{Eden} that this is
connected to the causality of the theory.}, and can
so be continued into the complex plane to be single-valued, at least
after suitable cuts have been made. As it turns out, this can be
achieved with two cuts, as shown in \fig{fig:cuts}.

\begin{figure}
\unitlength 2.50mm
\linethickness{0.4pt}
\begin{center}
\begin{picture}(50.76,12.94)(0,25)
\color{blue}
\put(16.32,30.00){\circle*{0.80}}
\put(35.10,30.00){\circle*{0.80}}
\put(16.32,30.4){\line(-1,0){15.8}}
\put(16.32,29.56){\line(-1,0){15.8}}
\put(35.10,30.4){\line(1,0){15.5}}
\put(35.10,29.56){\line(1,0){15.5}}
\color{green}
\multiput(18.82,30.00)(2,0){8}{\makebox(0,0)[cc]{$\times$}}
\color{deepred}
\put(16.32,28.00){\makebox(0,0)[cc]{$(m_{i}-m_{j})^{2}$}}
\put(35.10,28.00){\makebox(0,0)[cc]{$(m_{i}+m_{j})^{2}$}}
\put(5.10,31.67){\makebox(0,0)[cc]{$D$}}
\put(4.98,28.22){\makebox(0,0)[cc]{$A$}}
\put(45.87,31.67){\makebox(0,0)[cc]{$C$}}
\put(45.87,28.22){\makebox(0,0)[cc]{$B$}}
\put(3.76,25.89){\vector(-3,4){2.25}}
\put(4.43,25.00){\makebox(0,0)[cc]{Cut}}
\put(47.87,25.89){\vector(3,4){2.25}}
\put(46.54,25.00){\makebox(0,0)[cc]{Cut}}
\put(40.54,35.56){\vector(0,-1){4.56}}
\put(40.54,36.67){\makebox(0,0)[cc]{Physical values}}
\color{black}
\end{picture}
\end{center}
\ca[The complex $s$-plane]{The complex $\pmb{s}$-plane}
\label{fig:cuts}
\end{figure}

The cut plane is the physical sheet of the Riemann surface for \m{S};
continuing through one of the cuts leads to one of the other,
unphysical, sheets. Making the cuts in this way, \m{S} is
single-valued, meromorphic and real-analytic\footnote{\m{S} takes
complex-conjugate values at complex conjugate points.}. Note also that
\m{S(s)} is real on the axis between the cuts, i.e. for
\m{(m_{i}-m_{j})^{2} \leq s \leq (m_{i}+m_{j})^{2}}.

Unitarity demands that \m{S(s)S^{\dagger}(s)=1} for physical values of
\m{s} (just above the right-hand cut). This is a matrix equation, so
there is an implicit sum over a complete set of asymptotic states
living between \m{S} and \m{S^{\dagger}}. Generally, as \m{s}
increases, states involving more and more particles become available,
bringing the \m{2\rightarrow n} \m{S}-matrix into play, for
\m{n=3,4,\ldots} Here, however, there is no particle production, so
this cannot happen and we are left with
\begin{equation}
S^{kl}_{ij}(s^{+})[S^{nm}_{kl}(s^{+})]^{*}=\delta_{i}^{n}\delta_{l}^{m}
\end{equation}
for all physical \m{s^{+}}, with \m{*} denoting the complex
conjugate. Considering \m{s^{+}} as \m{s+i\epsilon} (\m{\epsilon
\rightarrow 0}) to place it just above the cut, real analyticity
allows this to be re-written as
\begin{equation}
S^{kl}_{ij}(s^{+})S^{nm}_{kl}(s^{-})=\delta_{i}^{n}\delta_{l}^{m},
\end{equation}
with \m{s^{-}=s-i\epsilon}, just below the cut. (We have skipped many
of the details in the interests of simplicity. For a more rigorous
explanation, see \cite{Zsquare} or \cite{pedrev}.)

The other important constraint comes from the fundamentally
relativistic property of crossing. If the interaction is assumed to
take place at \m{t=0}, ``crossing'' one of the participating particles
involves inverting its path in time, so that incoming particles become
outgoing and vice versa. In general, if one of the incoming
particles to an interaction is crossed to become outgoing while one of
the outgoers is simultaneously crossed to become incoming, the
amplitude for another physical process is obtained.

In our case, this amounts to saying that we can look at
\fig{fig:smatrix} from the side, with the forward momentum taken as
\m{t} rather than \m{s}. Normally, \m{t=(p_{1}-p_{3})^{2}} but, here,
\m{p_{2}=p_{3}}, so it can be written as
\begin{equation}
t=(p_{1}-p_{2})^{2}=2p_{1}^{2}+2p_{2}^{2}-(p_{1}+p_{2})^{2}=2m_{i}^{2}
+ 2m_{j}^{2}-s\,.
\label{eq:stot}
\end{equation}
The amplitude for this process can be found by analytically continuing
from the original amplitude to a region where \m{t} is physical,
i.e. \m{t} is real and \m{t \geq (m_{i}+m_{j})^{2}}. From
\re{eq:stot}, this corresponds to \m{s \leq
(m_{i}-m_{j})^{2}}. Physical amplitudes come from approaching this
from above in \m{t} and hence from below in \m{s}; a suitable path for
continuation is shown in \fig{fig:crosscont}. As a result, we have
\begin{figure}
\unitlength 2.50mm
\linethickness{0.4pt}
\begin{center}
\begin{picture}(50.76,10.44)(0,25)
\color{blue}
\put(16.32,30.00){\circle*{0.80}}
\put(35.10,30.00){\circle*{0.80}}
\put(16.32,30.40){\line(-1,0){15.80}}
\put(16.32,29.56){\line(-1,0){15.80}}
\put(35.10,30.40){\line(1,0){15.5}}
\put(35.10,29.56){\line(1,0){15.5}}
\color{green}
\multiput(18.82,30.00)(2,0){8}{\makebox(0,0)[cc]{$\times$}}
\color{deepred}
\put(16.32,28.00){\makebox(0,0)[cc]{$(m_{i}-m_{j})^{2}$}}
\put(35.10,28.00){\makebox(0,0)[cc]{$(m_{i}+m_{j})^{2}$}}
\put(5.10,31.67){\makebox(0,0)[cc]{$D$}}
\put(4.98,28.22){\makebox(0,0)[cc]{$A$}}
\put(45.87,31.67){\makebox(0,0)[cc]{$C$}}
\put(45.87,28.22){\makebox(0,0)[cc]{$B$}}
\qbezier(41.32,30.78)(31.32,37.44)(25.87,29.89)
\qbezier(25.87,29.89)(20.98,21.00)(10.98,27.44)
\put(10.98,27.44){\vector(-3,2){1.89}}
\color{black}
\end{picture}
\end{center}
\ca{Crossing}
\label{fig:crosscont}
\end{figure} 
\begin{equation}
S^{kl}_{ij}(s^{+})=S^{k\overline{j}}_{i\overline{l}}(2m_{i}^{2}+2m_{j}^{2}-
s^{+})\, .
\end{equation} 

This picture becomes substantially simpler if we shift back to the
rapidity difference \m{\T} through the transformation
\ba
\T&=& \cosh^{-1}\left( \frac{s-m_{i}^{2}-m_{j}^{2}}{2m_{i}m_{j}} \right)\\
 &=& \log \left[ \frac{1}{2m_{i}m_{j}}\left( s - m_{i}^{2}-m_{j}^{2} +
\sqrt{\{s-(m_{i}+m_{j})^{2}\}\{s-(m_{i}-m_{j})^{2}\}}\right)\right]\,. \nn
\ea 
This maps the physical sheet to the ``physical strip'' \m{0 \leq
\mathrm{Im\ } \T \leq \pi}, with the unphysical sheets being mapped
onto the unphysical strips \m{n\pi \leq \mathrm{Im\ }\T \leq
(n+1)\T}. Also, the two branch points go to \m{0} and \m{i\pi}, with
the cuts opening up as shown in \fig{fig:thetacuts}.
\begin{figure}
\unitlength 1.25mm
\linethickness{0.4pt}
\begin{center}
\color{blue}
\begin{picture}(100.21,40.56)(0,11)
\put(50,11.67){\vector(0,1){38.89}}
\put(0.20,20.89){\line(1,0){100.11}}
\put(0.20,39.78){\line(1,0){100.00}}
\color{green}
\multiput(50,23.7)(0,4.5){4}{\makebox(0,0)[cc]{$\times$}}
\color{deepred}
\put(50,20.89){\circle*{0.63}}
\put(50,39.78){\circle*{0.63}}
\put(51.5,18.92){\makebox(0,0)[cc]{0}}
\put(52,41.86){\makebox(0,0)[cc]{$i\pi$}}
\put(6.64,41.86){\makebox(0,0)[cc]{$A$}}
\put(6.64,18.92){\makebox(0,0)[cc]{$B$}}
\put(94.42,41.86){\makebox(0,0)[cc]{$D$}}
\put(94.42,18.92){\makebox(0,0)[cc]{$C$}}
\put(75.52,30.39){\makebox(0,0)[cc]{Physical strip}}
\put(75.52,43.86){\makebox(0,0)[cc]{(Unphysical)}}
\put(75.52,16.92){\makebox(0,0)[cc]{(Unphysical)}}
\color{black}
\end{picture}
\end{center}
\ca[The $\T$ plane]{The $\pmb{\T}$ plane}
\label{fig:thetacuts}
\end{figure}

Analytically continuing to the entire plane, and re-writing in terms
of \m{\T}, the demands of analyticity and crossing symmetry become
\begin{equation}
S_{a_{1}a_{2}}^{c_{1}c_{2}}(\T)S_{c_{1}c_{2}}^{b_{1}b_{2}}(-\T)=
\delta_{a_{1}}^{b_{1}}\delta_{a_{2}}^{b_{2}}
\label{eq:analy}
\end{equation}
and
\begin{equation}
S^{b_{1}b_{2}}_{a_{1}a_{2}}(\T)=S_{a_{2}\overline{b}_{1}}^{b_{2}
\overline{a}_{1}}(i\pi-\T)
\end{equation}
respectively. (Note that, for physical \m{\theta},
\m{[S(\T)]^{*}=S(-\T)}.) These results can be combined into the 
``cross-unitarity equation''
\begin{equation}
S_{a_{1}c_{2}}^{c_{1}b_{2}}(i\pi-\T)S_{a_{2}c_{1}}^{c_{2}b_{1}}(i\pi +
\T)=\delta_{a_{1}}^{b_{1}}\delta_{a_{2}}^{b_{2}}\,.
\end{equation}

This unitarity result can also be understood in terms of the algebra
of the \m{A}
symbols. Since we are assuming that the S-matrix is analytic, and so can 
be defined for all complex \m{\T}, it seems reasonable to demand
that \re{eq:smatrix} still makes sense if we
interchange \m{\T_{1}} and \m{\T_{2}}. The equation now relates
\emph{in}-states to \emph{out}-states, rather than the other way round. If
the original equation is then applied to what is now an \m{in}-state on the
rhs, we find
\begin{equation}
A_{a_{1}}(\T_{1})A_{a_{2}}(\T_{2})=
\sum_{b_{1},b_{2}} S_{a_{1}a_{2}}^{b_{1}b_{2}}(\T_{1}-\T_{2})
S_{b_{1}b_{2}}^{c_{1}c_{2}}(\T_{2}-\T_{1})A_{c_{1}}(\T_{1})A_{c_{2}}(\T_{2}),
\end{equation}
which relates an \emph{out}-state to a sum of other
\emph{out}-states. However, the \emph{out}-states form an
asymptotically complete \emph{basis} (as do the \emph{in}-states) and
so cannot be broken down, leading
us to identify the states on each side of the equation and thus yielding
\re{eq:analy}.

If the time and space dimensions could be treated on an equal footing (e.g.
by working in Euclidean rather than Minkowski space) the crossing symmetry
result would have become \m{S^{b_{1}b_{2}}_{a_{1}a_{2}}(\T)=
S_{a_{2}\overline{b}_{1}}^{b_{2} \overline{a}_{1}}(\pi-\T)}, making it clear
that it amounted to allowing \fig{fig:smatrix} just to be rotated on the 
page. In Minkowski space, this is still true; rotating the diagram is just
not as trivial an operation.

\sse{The Bootstrap Principle}
It is normally assumed that at least some of the simple poles on the physical
strip indicate the presence of bound states, either in the forward (\m{s}) or
crossed (\m{t}) channel, as shown in \fig{fig:bound}. Note that this
is consistent with there being no particle production provided such
poles do not appear for physical values of \m{\T}. In fact, poles
corresponding to bound states only appear for purely imaginary \m{\T},
with resonance states possible at complex \m{\T}. Note also that
simple poles do not \emph{need} to correspond to bound states, a fact
that will become important later and will be discussed in Section
\ref{se:colethun}.

\begin{figure}
\begin{center}
\parbox{2.6in}{
\begin{center}
\unitlength 1.00mm
\linethickness{0.4pt}
\begin{picture}(42.11,58.00)
\color{blue}
\put(4.56,7.56){\vector(1,1){18.22}}
\put(41.89,7.56){\vector(-1,1){18.22}}
\put(23.33,26.44){\vector(0,1){10.11}}
\put(23.00,37.56){\vector(-1,1){18.22}}
\put(23.78,37.67){\vector(1,1){18.22}}
\qbezier(15.00,18.44)(23.00,13.00)(31.00,18.44)
\put(23.00,20.00){\makebox(0,0)[cc]{$U_{ij}^{k}$}}
\color{green}
\put(23.33,26.33){\circle*{1}}
\put(23.33,37.22){\circle*{1}}
\color{deepred}
\put(29.44,31.5){\makebox(0,0)[cc]{$A_{k}(\T_{3})$}}
\put(41.89,4.37){\makebox(0,0)[cc]{$A_{j}(\T_{2})$}}
\put(4.56,4.37){\makebox(0,0)[cc]{$A_{i}(\T_{1})$}}
\put(4.78,58.50){\makebox(0,0)[cc]{$A_{l}(\T_{2})$}}
\put(42.11,58.50){\makebox(0,0)[cc]{$A_{m}(\T_{1})$}}
\color{black}
\end{picture}
\end{center}
}
\parbox{2.6in}{
\begin{center}
\unitlength 1.00mm
\linethickness{0.4pt}
\begin{picture}(49.45,58.00)
\color{blue}
\put(47.56,13.01){\vector(-1,1){18.22}}
\put(17.45,32.23){\vector(-1,1){18.22}}
\put(-0.78,13.27){\vector(1,1){18.22}}
\put(17.99,31.79){\vector(1,0){10.11}}
\put(29.22,32.08){\vector(1,1){18.22}}
\color{green}
\put(28.79,31.79){\circle*{1}}
\put(17.90,31.79){\circle*{1}}
\color{deepred}
\put(23.34,34.40){\makebox(0,0)[cc]{$A_{k}(\T_{3})$}}
\put(49.45,52.56){\makebox(0,0)[cc]{$A_{j}(\T_{2})$}}
\put(49.45,10.01){\makebox(0,0)[cc]{$A_{i}(\T_{1})$}}
\put(-2.55,10.01){\makebox(0,0)[cc]{$A_{l}(\T_{2})$}}
\put(-2.55,52.56){\makebox(0,0)[cc]{$A_{m}(\T_{1})$}}
\color{black}
\end{picture}
\end{center}
}
\end{center}
\ca{Bound state formation}
\label{fig:bound}
\end{figure}

There are various reasons why this is taken to be true, such as:
\begin{itemize}
\item in quantum mechanics, if there is a pole in the S-matrix for
scattering a particle off a potential\footnote{Such poles are always
simple, though this is not necessarily the case in field
theory.} then the wavefunction for the particle bound to the potential
can be \emph{constructed};
\item tree-level Feynman diagrams.
\end{itemize}
In many other ways, however, it has to be taken as an axiom, without a
rigorous basis.

The ``fusing angle'' for \m{i j \rightarrow k} is denoted as \m{U_{ij}^{k}}
(as shown in \fig{fig:bound})
and indicates that \m{S_{ij}^{i'j'}} will have a simple pole at
\m{iU_{ij}^{k}} for the forward (\m{s}-channel) process, and
\m{\pi-iU_{ij}^{k}} for the crossed (\m{t}-channel) version. The intermediate
particle, \m{k}, is on-shell and so survives for a macroscopic length of time.
The ``bootstrap principle'' (or ``nuclear democracy'') then states that \m{k}
should be expected to be one of the other asymptotic one-particle states of
the model.

This has proved to be immensely useful in discovering the full structure of
models once at least the fundamental particles---those from which all
other particles can be built up as bound states---are known. By looking at
the interactions of all known particles, adding any new states that show up
as bound states of the known ones, and repeating the process until
everything can be accounted for, all the particles in the theory can
be found. That is not to say that the problem becomes trivial---the
fundamental particles must still be discovered by other means---but it
is simplified greatly.

Of course, it is not enough just to discover the presence of new states; we
need to know their properties as well, such as their mass and, in particular,
their S-matrices with other particles. Indeed, it is only through poles in
these S-matrices that further new bound states can appear.

For the forward-channel process, as particle \m{k} is on-shell,
\m{s=m_{k}^{2}}, so we have
\begin{equation}
m_{k}^{2}=m_{i}^{2}+m_{j}^{2}+2m_{i}m_{j}\cos U^{k}_{ij}.
\end{equation}
This is a well-known trigonometric formula, and implies that
\m{U^{k}_{ij}} can be represented as the exterior angle of a triangle of
sides \m{m_{i}, m_{j}} and \m{m_{k}}, as shown in \fig{fig:masstri}. This
also shows that the three fusing angles satisfy
\begin{equation}
U^{k}_{ij}+U^{i}_{jk}+U^{j}_{ki}=2\pi,
\end{equation}
as might be expected from looking at \fig{fig:bound}.

\begin{figure}
\begin{center}
\unitlength 1.50mm
\linethickness{0.4pt}
\begin{picture}(49.33,29.44)(0,20)
\color{blue}
\put(2.56,20.22){\line(3,5){17.53}}
\put(20.09,49.44){\line(1,-1){29.24}}
\put(59.33,20.20){\line(-1,0){56.67}}
\color{deepred}
\qbezier(5.89,25.78)(11.11,24.56)(10.67,20.22)
\qbezier(16.89,44.11)(22.00,41.78)(24.89,44.67)
\qbezier(43.67,25.78)(39.67,24.33)(40.67,20.22)
\qbezier(43.67,25.78)(52,28)(55.33,20.20)
\put(24.67,18.44){\makebox(0,0)[cc]{$m_{i}$}}
\put(36.00,36.78){\makebox(0,0)[cc]{$m_{j}$}}
\put(9.89,37.00){\makebox(0,0)[cc]{$m_{k}$}}
\put(7.29,22.56){\makebox(0,0)[cc]{$\overline{U}^{j}_{ik}$}}
\put(20.94,45.67){\makebox(0,0)[cc]{$\overline{U}^{i}_{jk}$}}
\put(43.54,22.67){\makebox(0,0)[cc]{$\overline{U}^{k}_{ij}$}}
\put(50.54,22.67){\makebox(0,0)[cc]{$U^{k}_{ij}$}}
\color{black}
\end{picture}
\end{center}
\ca{The mass triangle}
\label{fig:masstri}
\end{figure}

In addition, extending the dictates of factorisation (i.e. having to allow
trajectories to be moved past a vertex) to the case where a bound state is
formed yields \fig{fig:bulkboot}, and the corresponding ``bootstrap equation''
\begin{equation}
f^{c}_{a_{1}a_{2}}S^{b_{1}b_{2}}_{ca_{3}}(\T)= f^{b_{1}}_{c_{1}c_{2}}
S^{c_{1}b_{2}}_{a_{1}c_{3}}(\T+i\overline{u}^{\overline{a}_{2}}_{a_{1}
\overline{c}}) S^{c_{2}c_{3}}_{a_{2}a_{3}}(\T -
i\overline{u}^{\overline{a}_{1}}_{a_{2}\overline{c}}),
\label{eq:bulkboot}
\end{equation}
where \m{f^{c}_{ab}} is the ``three-particle coupling'', as shown in
\fig{fig:tpc}. At the pole where a bound state is formed, the S-matrix can
be considered as a pair of such couplings, giving
\begin{equation}
S_{ij}^{i'j'}(\T) \approx i\frac{f^{k}_{ij}f^{i'j'}_{k}}{\T-iu_{ij}^{k}}
\end{equation}

\begin{figure}
\parbox{2.5in}{
\unitlength 1.00mm
\linethickness{0.4pt}
\begin{picture}(47.33,55.00)
\color{green}
\put(46.11,49.89){\line(-1,-1){41.00}}
\put(47.33,8.78){\line(-1,1){27.11}}
\put(20.33,35.78){\line(-1,4){3.50}}
\put(20.33,35.78){\line(-4,1){15.44}}
\color{deepred}
\put(46.11,51.33){\makebox(0,0)[cc]{$a_{3}$}}
\put(5.00,6.78){\makebox(0,0)[cc]{$b_{2}$}}
\put(47.33,6.56){\makebox(0,0)[cc]{$b_{1}$}}
\put(16.89,50.78){\makebox(0,0)[cc]{$a_{2}$}}
\put(3.89,39.56){\makebox(0,0)[cc]{$a_{1}$}}
\put(21.89,32.11){\makebox(0,0)[cc]{$c$}}
\qbezier(22.89,33.22)(26.11,35.22)(29.33,33.22)
\put(26.11,35.89){\makebox(0,0)[cc]{$\theta$}}
\color{black}
\end{picture}
} \ \LARGE = \normalsize \
\parbox{2.5in}{
\unitlength 1.00mm
\linethickness{0.4pt}
\begin{picture}(47.44,55.00)
\color{green}
\put(46.11,49.89){\line(-1,-1){41.00}}
\put(47.44,8.78){\line(-1,1){14.22}}
\put(33.22,22.89){\line(-1,4){6.72}}
\put(33.22,22.89){\line(-4,1){26.89}}
\color{deepred}
\put(46.11,51.33){\makebox(0,0)[cc]{$a_{3}$}}
\put(5.00,6.78){\makebox(0,0)[cc]{$b_{2}$}}
\put(47.33,6.56){\makebox(0,0)[cc]{$b_{1}$}}
\put(26.56,51.00){\makebox(0,0)[cc]{$a_{2}$}}
\put(5.00,29.56){\makebox(0,0)[cc]{$a_{1}$}}
\qbezier(17.56,26.78)(20.11,28.89)(24.00,27.78)
\put(18.56,32.22){\makebox(0,0)[cc]
{$\theta+i\overline{u}^{\overline{a}_{2}}_{a_{1}\overline{c}}$}}
\put(27.22,28.78){\makebox(0,0)[cc]{$c$}}
\qbezier(29.22,39.22)(33.11,40.67)(34.78,38.56)
\put(43.44,38.22){\makebox(0,0)[cc]
{$\theta-i\overline{u}^{\overline{a}_{1}}_{a_{2}\overline{c}}$}}
\color{black}
\end{picture}
}
\ca{Bootstrap equation}
\label{fig:bulkboot}
\end{figure}

\begin{figure}
\begin{center}
\parbox{2in}{
\unitlength 1.00mm
\linethickness{0.4pt}
\begin{picture}(44.11,45.00)(0,10)
\color{blue}
\put(7.78,49.00){\line(1,-1){18.11}}
\put(44.00,49.00){\line(-1,-1){18.11}}
\put(25.89,30.89){\line(0,-1){20.67}}
\color{green}
\put(25.89,30.89){\circle*{1}}
\color{deepred}
\put(7.67,50.89){\makebox(0,0)[cc]{$a_{1}$}}
\put(44.11,50.67){\makebox(0,0)[cc]{$a_{2}$}}
\put(25.89,8.56){\makebox(0,0)[cc]{$c$}}
\color{black}
\end{picture}
} \ \m{=f^{c}_{a_{1}a_{2}}}
\end{center}
\ca{Three-particle coupling}
\label{fig:tpc}
\end{figure}

This is a great help to the aspiring state-hunter, as treating all the
relevant S-matrix elements at the point where the new bound state is
expected to be formed as a set of simultaneous equations for the \m{f}s
allows them to be found and substituted into \re{eq:bulkboot} to give the
S-matrices involving the new state.

Another useful relation comes from equating the action of one of
the conserved charges, \m{Q_{s}} on the state before 
fusing---\m{\st{A_{i}(\T_{1})
A_{j}(\T_{2})}}---and after---\m{\st{A_{\overline{k}}(\T_{3})}}. 
The action is given by \re{eq:qact} and leads to the ``conserved
charge bootstrap''
\begin{equation}
q_{\overline{k}}^{(s)}=q_{i}^{(s)}e^{is
\overline{U}^{j}_{ki}}+q_{j}^{(s)}e^{-is\overline{U}^{i}_{kj}}.
\label{eq:conschb}
\end{equation}

It is interesting to note, as was pointed out in \cite{BCDS}, that if
we take the logarithmic derivative of the S-matrix,
\begin{equation}
\varphi_{ab}(\T)=-i\frac{d}{d\T}\ln S_{ab}(\T),
\end{equation}
expanded according to
\begin{equation}
\varphi_{ab}(\T)=-\sum_{k=1}^{\infty}\varphi_{ab}^{(k)}e^{-k|\T|},
\end{equation}
and insert it into the logarithmic derivative of the bootstrap equation, we
recover
\begin{equation}
\varphi_{cd}^{(s)}=\varphi_{ad}^{(s)}e^{-is\overline{u}^{\overline{b}}_
{a\overline{c}}}+\varphi_{db}^{(s)}e^{is\overline{u}^{\overline{a}}_
{b\overline{c}}},
\end{equation}
showing that the rows and columns of \m{\varphi^{(s)}} provide solutions for
the conserved charge bootstrap \re{eq:conschb}.

\sse{The Coleman-Thun mechanism}
\label{se:colethun}
If all poles were simple, and inevitably corresponded to the creation of a
bound state, as in quantum mechanics, the story would now be complete. 
However, this is not the case; not only do some theories give rise to
double, triple, or higher order poles, but not all simple poles have a
natural interpretation in terms of bound states.

The solution to this problem was discovered by Coleman and Thun in 1978 
\cite{CT}, in terms of anomalous threshold singularities. For a given Feynman
diagram, if the external momenta are such that one or more of the internal 
propagators are simultaneously on-shell (i.e. can be considered as real, 
rather than virtual, particles) then it turns out that the loop integrals 
give rise to a singularity in the amplitude. The bound states considered 
above are simple examples of this, with one propagator (the bound state 
particle) on-shell.

In three or more dimensions, all the singularities which do not correspond 
to bound states show up as branch points, but in 1+1-dimensions, they can 
appear as poles. The practical upshot of this is that such poles should be 
considered as being due not to the tree-level diagrams we have been looking
at so far, but to more complicated diagrams which are nonetheless composed 
entirely of on-shell particles, such as the one shown in \fig{fig:onshell}.
This diagram, if it was possible, would be expected to produce a double pole 
in the appropriate S-matrix element.

\begin{figure}
\unitlength 1.00mm
\linethickness{0.4pt}
\begin{center}
\begin{picture}(50.67,39.50)(0,10)
\color{blue}
\put(11.67,44.44){\line(1,0){29.22}}
\put(40.89,44.44){\line(2,1){9.67}}
\put(11.67,44.44){\line(-2,1){9.67}}
\put(40.89,15.22){\line(2,-1){9.67}}
\put(11.67,15.22){\line(1,0){29.22}}
\put(11.67,15.22){\line(1,1){29.22}}
\put(40.89,15.22){\line(-1,1){29.22}}
\put(11.67,15.22){\line(-2,-1){9.67}}
\color{green}
\put(26.28,29.83){\circle*{1.33}}
\put(40.89,44.44){\circle*{1.33}}
\put(11.67,44.44){\circle*{1.33}}
\put(11.67,15.22){\circle*{1.33}}
\put(40.89,15.22){\circle*{1.33}}
\color{black}
\end{picture}
\end{center}
\ca{Example on-shell process}
\label{fig:onshell}
\end{figure}

A useful ``rule of thumb'' is that the
order of pole a diagram gives is equal to the number of ``degrees of
freedom'', e.g. the number of internal lengths in 
the diagram which can be independently adjusted without destroying it. For 
example, in the bound state diagram, the only internal length was the bound
state line, but this could be made as long or short as desired without 
problems. Similarly, in \fig{fig:onshell}, the upper or lower triangles can
be scaled independently.

The origin of this rule lies in the fact that, when the Feynman
integral of a diagram with \m{P} internal propagators and \m{L} loops
is calculated, it turns out to give a pole of order
\m{p=P-2L}. (Further details can be found in \cite{Eden}.) We now need
to apply Euler's well-known formula \m{\mathrm{vertices} -
\mathrm{edges} + \mathrm{faces}=1} for any closed diagram,
i.e. \m{V-P+L=1}, to get \m{p=2V-P-2}. Each of the \m{V} vertices is
of three-point type\footnote{By counting the
number of faces as the number of loops, we have implicitly taken the
points where two particles collide but do not bind \emph{not} to count
as vertices. This is different to the usual interpretation of Euler's
formula, but not inconsistent with it. By taking the diagram to exist in three
dimensions, a topological transformation can be applied to remove the
``extra'' vertices and faces.}, and each propagator is attached to two 
vertices, except for the four external ones (which are not counted in 
\m{P}), so \m{P=\frac{3V-4}{2}}. Thus, \m{p=L+1=\hft V=\frac{P+2}{3}}. 

The easiest way to proceed from here is to consider this purely as a 
problem of topology, and start with the diagram
without external legs (i.e. with 4 2-point vertices and \m{V-4}
3-point ones), then successively remove 2-point vertices and their attached
propagators. Since the position of these vertices is dictated by the 
other vertices present, this cannot change the number of degrees of 
freedom. Once this procedure has been exhausted, we can continue by 
removing the 1-point vertices (together with their attached propagator), 
at the cost of one degree of freedom per vertex. Proceeding in this way, 
we eventually end up with a diagram containing only 3- or 0-point 
vertices. A closed network of 3-point vertices can have had no 
propagators or vertices removed from it during the above process, and 
so, if present, must have existed as a disconnected set 
in the original diagram. Since this is not possible, and since such a 
network would, in any case, not permit momentum to be conserved at each 
vertex, we must in fact have only 0-point vertices, i.e. isolated 
points. We still have an arbitrary choice of origin to make, and so will 
choose to locate it at one of the vertices. Each of the 
remaining points can then be moved freely and independently, 
giving the diagram two degrees of freedom per vertex.

Removing \m{a} 2-point vertices and \m{b} 1-point ones leaves \m{V-a-b} 
vertices and \m{P-2a-b=0} propagators. Allowing for the \m{b} degrees of 
freedom which were lost along the way, this implies that the original 
diagram had \m{2(V-a-b-1)+b=2V-2a-b+2} degrees of freedom. Using the 
fact that there can be no propagators left, this is just \m{2V-P-2=p}. 
For later reference, note that this argument depends only on the fact 
that no initial vertex is of any \emph{more} than 3-point type, and not 
on the fact that \emph{all} vertices are of this type, as the first 
results do. This means that, although calculating the order of a diagram 
just by halving the number of vertices is probably the easiest approach 
in the bulk, using the number of degrees of freedom is a more generally 
applicable method. (Note, also, that it makes no reference to the 
integrability or otherwise of the theory.) 

Through this method, a pole of any order can be explained in terms of a 
sufficiently exotic on-shell (or Landau) diagram. The one remaining problem
is that the only such diagram which could ever explain a simple pole is the
formation of a bound state. If we are to argue that this does not always 
happen, we have to find a process to take its place.

Perhaps the most obvious way that the order of the diagram could be
reduced would be if it so happened that one of the ``internal'' 
S-matrix elements had a zero just at the right rapidity. However, even
if this does not happen, the order can still be lower than expected.
  
The explanation for this is that there is
not necessarily just one diagram which can be drawn to fit a given pair of
incoming particles. For example, in \fig{fig:onshell}, the theory might
allow a different set of particles to be used for the internal lines, e.g.
substituting anti-particle for particle on each line in the upper or lower
triangle, without disturbing the diagram. It is even possible that an 
entirely different diagram could be drawn to fit the same external lines.
In such a case, all possible
diagrams must be added together with appropriate relative weights. If a
cancellation occurs between the different diagrams, then the
overall order of the pole produced is lower than what would be expected for
any of the diagrams individually. For our example, if this sum came to
zero, then they would collectively contribute a simple, rather than a
double, pole. 

\se{Boundary field theory}

The theories we have been considering so far have lived on the ``full
line'' stretching to infinity in both directions. Many interesting new
features arise if we insert a ``wall'' at \m{x=0} to restrict the
world to the ``half line'' between zero and negative infinity. Far
away from the wall, particles behave in exactly the same way as before
but, when the approach the wall, two things can happen. Either they
will reflect off it, or they will bind to it, forming a boundary bound
state. The introduction of a wall is thus not just a simple matter of
geometry, and a boundary analogue of the S-matrix, termed the ``reflection
factor'' must also be introduced.

This idea was first introduced by Cherednik \cite{Cherednik}, though
it took 10 years or so for the topic to be put on a footing comparable
with the bulk theory. This was achieved by Ghoshal and Zamolodchikov
\cite{GhoshZam}, as well as Fring and K\"{o}berle \cite{FringK} and
Sasaki \cite{Sasaki}. A good review of the topic can be found in
\cite{Corriganb}.

In algebraic terms, Ghoshal and Zamolodchikov 
imagined the ground state of such a 
theory---\m{\st{0}_{B}}---as being formally created from the ground
state of the
bulk theory by a ``boundary creating operator'' \m{B}, creating an infinitely
heavy and impenetrable particle \m{B} sitting at \m{x=0}. Thus
\begin{equation}
\st{0}_{B}=B\st{0}\,.
\end{equation}
While this is a purely formal object, it makes analogy with the bulk theory
straightforward. Far from the wall, everything is exactly the same as for
the corresponding bulk theory, allowing the same set of asymptotic particle
states, so an \emph{in}-state of the boundary theory is just
\begin{equation}
A_{a_{1}}(\T_{1})A_{a_{2}}(\T_{2})\ldots A_{a_{N}}(\T_{N})\st{0}_{B}
= A_{a_{1}}(\T_{1})A_{a_{2}}(\T_{2})\ldots A_{a_{N}}(\T_{N})B\st{0}\,,
\end{equation}
with \m{\T_{1}>\T_{2}>\cdots>\T_{N}>0}.

By analogy with the bulk S-matrix, they then introduced a reflection factor to
interpolate between the \emph{in}- and \emph{out}-states through the
relations
\begin{equation}
A_{a}(\T)B=R^{b}_{a}(\T)A_{b}(-\T)B\,,
\label{eq:rmatrix}
\end{equation}
illustrated in \fig{fig:bounref}.

Following the previous discussion, we will consider the boundary
version of integrable theories. This means that the introduction of a
suitable ``wall'' will involve modifying the action
by adding a boundary potential term which will restrict the particles
to the half-line, but also leave the theory integrable, allowing us to
still have the useful features of factorisability and lack of particle
production. Importantly, this means that only the \m{1 \rightarrow 1}
reflection factor will need to be considered.

Assuming such a wall can be built, the logical next step is to search for
boundary analogues of the four conditions placed on the S-matrix above.
Of the three symmetries
enjoyed by the S-matrix---P, C and T---only time-reversal symmetry
inevitably
remains, demanding \m{R^{a}_{b}(\T)=R^{b}_{a}(\T)}. The presence of charge
conjugation symmetry is generally permitted by some choices of boundary
condition, but is not inevitable, as we shall see later. Finally, parity
symmetry must inevitably be broken by the introduction of a wall of any type.
The four S-matrix conditions, however, all have analogues for the boundary,
and are sufficient to specify the reflection factor up to a boundary 
CDD ambiguity which satisfies the same constraints as for the bulk.

\begin{figure}
\begin{center}
\parbox{2.0in}{
\unitlength 1.00mm
\linethickness{0.4pt}
\begin{picture}(48.00,55.00)
\color{green}
\put(45.89,30.00){\line(-3,2){34.22}}
\put(45.89,30.00){\line(-3,-2){34.22}}
\color{deepred}
\put(9.78,52.78){\makebox(0,0)[cc]{$a$}}
\put(9.78,7.11){\makebox(0,0)[cc]{$b$}}
\qbezier(40.22,33.78)(42.11,35.89)(45.89,35.33)
\put(42.56,37.22){\makebox(0,0)[cc]{$\theta$}}
\color{black}
\put(46.00,3.00){\rule{2.00\unitlength}{52.00\unitlength}}
\end{picture}
} \ \m{=R^{b}_{a}(\T)}
\ca{Boundary reflection factor}
\label{fig:bounref}
\end{center}
\end{figure}

\sse{Boundary Yang-Baxter equation}
The demands of factorisation again require that trajectories should be able
to be moved past boundary vertices, i.e. the points where particles interact
with the boundary. This is shown in \fig{fig:byangbax}, or algebraically as
\begin{figure}
\begin{center}
\parbox{2.5in}{
\unitlength 1.00mm
\linethickness{0.4pt}
\begin{picture}(48.00,55.00)
\color{green}
\put(35.89,53.44){\line(5,-6){10.09}}
\put(45.89,41.22){\line(-4,-5){30.22}}
\put(10.89,9.33){\line(5,2){35.00}}
\put(10.89,37.33){\line(5,-2){35.00}}
\color{deepred}
\put(34.33,52.00){\makebox(0,0)[cc]{$a_{2}$}}
\put(9.44,37.22){\makebox(0,0)[cc]{$a_{1}$}}
\put(9.44,9.22){\makebox(0,0)[cc]{$b_{1}$}}
\put(14.11,4.00){\makebox(0,0)[cc]{$b_{2}$}}
\qbezier(42.44,45.67)(43.56,47.67)(46.00,46.56)
\put(43.44,48.67){\makebox(0,0)[cc]{$\theta_{2}$}}
\qbezier(40.22,25.56)(42.22,28.44)(46.00,27.67)
\put(42.89,29.78){\makebox(0,0)[cc]{$\theta_{1}$}}
\color{black}
\put(46.00,3.00){\rule{2.00\unitlength}{52.00\unitlength}}
\end{picture}
} \ \LARGE = \normalsize \
\parbox{2.5in}{
\unitlength 1.00mm
\linethickness{0.4pt}
\begin{picture}(48.00,56.00)
\color{green}
\put(35.89,3.56){\line(5,6){10.09}}
\put(45.89,15.78){\line(-4,5){30.22}}
\put(10.89,47.67){\line(5,-2){35.00}}
\put(10.89,19.67){\line(5,2){35.00}}
\color{deepred}
\put(34.33,5.00){\makebox(0,0)[cc]{$b_{2}$}}
\put(9.44,19.78){\makebox(0,0)[cc]{$b_{1}$}}
\put(9.44,47.78){\makebox(0,0)[cc]{$a_{1}$}}
\put(14.78,52.44){\makebox(0,0)[cc]{$a_{2}$}}
\qbezier(40.11,36.00)(42.33,39.11)(45.89,37.67)
\put(42.44,39.67){\makebox(0,0)[cc]{$\theta_{1}$}}
\qbezier(41.78,21.00)(43.33,23.22)(45.89,21.89)
\put(43.33,24.00){\makebox(0,0)[cc]{$\theta_{2}$}}
\color{black}
\put(46.00,3.00){\rule{2.00\unitlength}{52.00\unitlength}}
\end{picture}
}
\ca{Boundary Yang-Baxter equation}
\label{fig:byangbax}
\end{center}
\end{figure}
\begin{multline}
R^{c_{2}}_{a_{2}}(\T_{2})S^{c_{1}d_{2}}_{a_{1}c_{2}}(\T_{1} + \T_{2})
R^{d_{1}}_{c_{1}}(\T_{1})S^{b_{2}b_{1}}_{d_{2}d_{1}}(\T_{1}-\T_{2})
= \\ S^{c_{1}c_{2}}_{a_{1}a_{2}}(\T_{1}-\T_{2})R^{d_{1}}_{c_{1}} (\T_{1})
S^{d_{2}b_{1}}_{c_{2}d_{1}}(\T_{1}+\T_{2}) R^{b_{2}}_{d_{2}}(\T_{2}).
\end{multline}

\sse{Boundary unitarity condition}
This is again a straightforward generalisation of the bulk requirement, and
results in the condition
\begin{equation}
R^{c}_{a}(\T)R^{b}_{c}(-\T)=\delta_{a}^{b}.
\end{equation}

Algebraically, this results from the demand that the reflection factor
should also be analytic, and
so \re{eq:rmatrix} should make sense for negative \m{\T}. The argument then
proceeds in the same way as for the S-matrix.

\sse{Boundary crossing symmetry condition}
This time, trying to find a boundary analogue is somewhat more tricky, and
in fact it turns out to be easier to find a ``boundary cross-unitarity''
condition
\begin{equation}
K^{ab}(\T)=S^{ab}_{a'b'}(2\T)K^{b'a'}(-\T),
\end{equation}
where
\begin{equation}
K^{ab}(\T)=R^{b}_{\overline{a}}\left( \frac{i\pi}{2} - \T \right).
\end{equation}
In terms of the reflection factor, this can also be written as
\begin{equation}
R^{a}_{b}(\T)=S^{\overline{b}\overline{a}}_{a'b'}(2\T)R^{a'}_{b'}(i\pi -
\T).
\end{equation}

\sse{Boundary bootstrap}
With the introduction of the boundary, there are now two types of bound
state to consider: bulk ``bound state'' particles, and ``boundary bound
states''. The second type arise due to an incoming particle binding to the
boundary, changing its state, as shown in \fig{fig:bbind}. For a particle
\m{a} changing the boundary from state \m{\alpha} to state \m{\beta}, we can
define a boundary fusing angle \m{u_{\alpha a}^{\beta}}, with a
corresponding pole in the reflection factor at \m{iu_{\alpha a}^{\beta}}.

\begin{figure}
\parbox{3in}{
\begin{center}
\parbox{2in}{
\unitlength 1.00mm
\linethickness{0.4pt}
\begin{picture}(48.00,55.00)
\put(46.00,3.00){\rule{2.00\unitlength}{52.00\unitlength}}
\color{blue}
\put(45.44,26.67){\line(-1,1){24.00}}
\color{red}
\multiput(45.44,2.89)(0,1){53}{\circle*{0.25}}
\color{deepred}
\qbezier(38.22,34.00)(40.44,37.44)(45.44,36.44)
\put(42.67,33.84){\makebox(0,0)[cc]{$u_{a\alpha}^{\beta}$}}
\put(19.78,52.11){\makebox(0,0)[cc]{$a$}}
\put(43.78,47.11){\makebox(0,0)[cc]{$\alpha$}}
\put(43.89,11.44){\makebox(0,0)[cc]{$\beta$}}
\color{black}
\end{picture}
} \ \m{=g^{\beta}_{a\alpha}}
\end{center}
\vspace*{-11pt}
\ca{Boundary bound state}
\label{fig:bbind}}
\parbox{3in}{
\begin{center}
\unitlength 1.00mm
\linethickness{0.4pt}
\begin{picture}(48.00,55.00)
\put(46.00,3.00){\rule{2.00\unitlength}{52.00\unitlength}}
\color{green}
\put(27.92,28.31){\line(-1,1){15.44}}
\put(27.81,28.31){\line(-1,-1){15.44}}
\color{blue}
\multiput(27.89,28.44)(1,0){5}{\circle*{0.25}}
\color{deepred}
\qbezier(23.47,32.53)(21.14,28.42)(23.69,24.19)
\put(15.37,28.31){\makebox(0,0)[cc]{$iu^{c}_{ab}$}}
\put(37.47,26.19){\makebox(0,0)[cc]{$c$}}
\put(11.00,44.00){\makebox(0,0)[cc]{$a$}}
\put(11.00,13.00){\makebox(0,0)[cc]{$b$}}
\put(51.00,28.44){\makebox(0,0)[cc]{$g^{c}$}}
\color{blue}
\multiput(33.00,28.44)(1,0){14}{\circle*{0.25}}
\color{black}
\end{picture}
\end{center}
\ca{Process involving a bulk and a boundary coupling}
\label{fig:bulkgc}}
\end{figure}

This also leads to the introduction of a set of boundary couplings 
\m{g^{c}}. Again, the reflection factor at a boundary fusing angle can
be considered as a pair of boundary couplings, giving
\begin{equation}
R(\T)^{a}_{b} \approx
\frac{i}{2}\frac{g^{\beta}_{a\alpha}g^{b\alpha}_{\beta}}{\T-iu_{\alpha
a}^{\beta}}. \end{equation}
Alternatively, if the particle \m{c} can be formed as the bound state of two
equal-mass particles in the bulk theory, we would expect the process
shown
in \fig{fig:bulkgc}, giving
\begin{equation}
K^{ab}(\T) \approx \frac{i}{2}\frac{f^{ab}_{c}g^{c}}{\T - iu_{ab}^{c}}.
\end{equation}

All this allows us to play a similar game to before to determine the boundary
spectrum. Assuming that all boundary states other than the lowest (vacuum)
state can be formed by the binding of a bulk particle to the boundary, and
that we can somehow construct reflection factors for the vacuum boundary
state for all the bulk particles, we can search their pole structures for
evidence of further boundary states. Constructing a new set of reflection
factors for these states, searching again, and repeating until all the poles
in all the reflection factors can be accounted for without introducing
further boundary states, we can hopefully obtain the entire spectrum. As
before, this relies on introducing no more states than are necessary to
complete the process, which might, in theory, mean some are
missed. However, that has so far never been found to happen in practice.

For the bulk bound states, factorisation demands that we be able to move the
boundary interaction past the bound state formation vertex as shown in
\fig{fig:boundboot1}, leading to
\begin{equation}
f^{ab}_{d}R^{d}_{c}(\T)=f^{b_{1}a_{1}}_{c}R^{a_{2}}_{a_{1}}(\T+
i\overline{u}^{b}_{ad})S^{b_{2}a}_{b_{1}a_{2}}(2\T+i\overline{u}^{b}_{ad}-
i\overline{u}^{a}_{bd})R^{b}_{b_{2}}(\T-i\overline{u}^{a}_{bd}).
\end{equation}

\begin{figure}[t]
\begin{center}
\parbox{2.5in}{
\unitlength 1.00mm
\linethickness{0.4pt}
\begin{picture}(48.00,55.00)
\color{blue}
\multiput(35.00,44.89)(-1,1){10}{\circle*{0.25}}
\color{green}
\put(35.11,44.89){\line(2,-1){10.78}}
\put(45.89,39.22){\line(-2,-1){31.33}}
\put(45.89,23.22){\line(-1,-2){9.61}}
\put(35.22,44.67){\line(1,-2){10.67}}
\color{deepred}
\put(12.22,22.33){\makebox(0,0)[cc]{$f_{1}$}}
\put(35.11,2.78){\makebox(0,0)[cc]{$f_{2}$}}
\put(33.61,42.61){\makebox(0,0)[cc]{$i_{2}$}}
\put(37.94,45.72){\makebox(0,0)[cc]{$i_{1}$}}
\put(41.11,28.67){\makebox(0,0)[cc]{$j_{2}$}}
\put(43.78,36.17){\makebox(0,0)[cc]{$j_{1}$}}
\qbezier(41.67,32.11)(43.78,33.89)(46.00,32.56)
\put(44.11,30.94){\makebox(0,0)[cc]{$\sigma$}}
\qbezier(39.78,42.44)(42.00,45.00)(46.00,44.44)
\put(43.78,42.56){\makebox(0,0)[cc]{$\alpha$}}
\qbezier(32.89,47.11)(38.89,50.78)(45.89,49.00)
\put(39.11,50.78){\makebox(0,0)[cc]{$u$}}
\qbezier(37.33,40.56)(39.56,40.33)(40.00,42.33)
\put(33.44,38.78){\vector(4,3){4.11}}
\put(31.67,37.67){\makebox(0,0)[cc]{$u_{n}$}}
\color{black}
\put(46.00,3.00){\rule{2.00\unitlength}{52.00\unitlength}}
\end{picture} } \ {\LARGE =} \ \parbox{2.5in}{
\unitlength 1.00mm
\linethickness{0.4pt}
\begin{picture}(48.00,55.00)
\put(46.00,3.00){\rule{2.00\unitlength}{52.00\unitlength}}
\color{green}
\put(16.56,11.89){\line(2,1){13.22}}
\put(29.78,18.50){\line(-1,-3){4.20}}
\color{blue}
\multiput(29.67,18.44)(1,1){17}{\circle*{0.25}}
\multiput(29.67,51.11)(1,-1){17}{\circle*{0.25}}
\color{deepred}
\qbezier(42.00,38.89)(43.56,40.56)(46.00,39.56)
\put(43.78,41.44){\makebox(0,0)[cc]{$u$}}
\qbezier(24.67,15.89)(25.22,13.56)(28.11,13.33)
\put(24.44,13.00){\makebox(0,0)[cc]{$u_{n}$}}
\put(14.33,10.89){\makebox(0,0)[cc]{$f_{2}$}}
\put(24.22,4.33){\makebox(0,0)[cc]{$f_{1}$}}
\color{black}
\end{picture} }
\end{center}
\ca{Boundary-bulk bootstrap}
\label{fig:boundboot1}
\end{figure}

A similar demand for the boundary bound states leads to
\fig{fig:boundboot2}, and the corresponding requirement
\begin{equation}
g^{\alpha a}_{\beta}R^{c}_{d\beta}(\T)=
g^{\alpha h}_{\beta}S^{dh}_{ge}(\T+iu_{\alpha h}^{\beta})R^{f}_{g\alpha}(\T)
S^{ef}_{ac}(\T-iu^{\alpha}_{\beta h})\,.
\end{equation}

\pagebreak
\begin{figure}
\begin{center}
\parbox{2.5in}{
\unitlength 1.00mm
\linethickness{0.4pt}
\begin{picture}(48.00,55.00)
\put(46.00,3.00){\rule{2.00\unitlength}{52.00\unitlength}}
\color{green}
\put(30.25,5.53){\line(1,1){15.56}}
\put(45.69,37.08){\line(-1,1){15.56}}
\put(17.89,6.22){\line(4,1){28.00}}
\put(17.89,20.22){\line(4,-1){28.00}}
\color{red}
\multiput(45.78,21.44)(0,1){16}{\circle*{0.25}}
\color{deepred}
\put(28.25,4.53){\makebox(0,0)[cc]{$a$}}
\put(15.89,5.22){\makebox(0,0)[cc]{$c$}}
\put(15.89,21.22){\makebox(0,0)[cc]{$d$}}
\put(28.13,53.64){\makebox(0,0)[cc]{$b$}}
\put(34.75,13.03){\makebox(0,0)[cc]{$e$}}
\put(40.89,9.22){\makebox(0,0)[cc]{$f$}}
\put(42.00,20.00){\makebox(0,0)[cc]{$h$}}
\put(51.00,6.00){\makebox(0,0)[cc]{$\st{\alpha}$}}
\put(51.00,52.00){\makebox(0,0)[cc]{$\st{\alpha}$}}
\put(51.00,29.00){\makebox(0,0)[cc]{$\st{\beta}$}}
\put(54.00,13.22){\makebox(0,0)[cc]{$R^{f}_{g\alpha}(\T)$}}
\color{black}
\end{picture} } \ \LARGE = \normalsize \
\parbox{2.5in}{
\unitlength 1.00mm
\linethickness{0.4pt}
\begin{picture}(48.00,55.00)
\put(46.00,3.00){\rule{2.00\unitlength}{52.00\unitlength}}
\color{green}
\put(30.25,5.53){\line(1,1){15.56}}
\put(45.69,37.08){\line(-1,1){15.56}}
\put(17.89,22.00){\line(4,1){28.00}}
\put(17.89,36.00){\line(4,-1){28.00}}
\color{red}
\multiput(45.78,21.44)(0,1){16}{\circle*{0.25}}
\color{deepred}
\put(28.25,4.53){\makebox(0,0)[cc]{$a$}}
\put(15.89,36.78){\makebox(0,0)[cc]{$d$}}
\put(15.89,21.22){\makebox(0,0)[cc]{$c$}}
\put(28.13,53.64){\makebox(0,0)[cc]{$b$}}
\put(51.00,6.00){\makebox(0,0)[cc]{$\st{\alpha}$}}
\put(51.00,52.00){\makebox(0,0)[cc]{$\st{\alpha}$}}
\put(51.00,23.00){\makebox(0,0)[cc]{$\st{\beta}$}}
\put(54.00,29.00){\makebox(0,0)[cc]{$R^{c}_{d\beta}(\T)$}}
\color{black}
\end{picture} }
\ca{Boundary-boundary bootstrap}
\label{fig:boundboot2}
\end{center}
\end{figure}

\begin{figure}
\begin{center}
\parbox{0.49\textwidth}{
\centering
\unitlength 1.00mm
\linethickness{0.4pt}
\begin{picture}(35.5,55.00)(12.5,0)
\put(46.00,3.00){\rule{2.00\unitlength}{52.00\unitlength}}
\color{green}
\put(27.92,28.31){\line(-1,1){15.44}}
\put(27.81,28.31){\line(-1,-1){15.44}}
\color{blue}
\put(27.89,28.44){\line(1,0){18}}
\color{black}
\end{picture} }
\parbox{0.49\textwidth}{ 
\centering
\unitlength 1.00mm
\linethickness{0.4pt}
\begin{picture}(38.00,55.00)(10.00,0)
\put(46.00,3.00){\rule{2.00\unitlength}{52.00\unitlength}}
\color{blue}
\put(46.00,29.00){\line(-2,1){20}}
\put(46.00,29.00){\line(-2,-1){20}}
\put(26.00,19.00){\line(0,1){20}}
\put(26.00,19.00){\line(-1,-1){15}}
\put(26.00,39.00){\line(-1,1){15}}
\color{black}
\end{picture} }
\ca{Some common boundary independent Coleman-Thun processes}
\label{fig:bcti}
\end{center}
\end{figure}

\sse{The boundary Coleman-Thun mechanism}
\label{sec:bct}
Though the discussion is essentially analogous to that of the bulk,
the Coleman-Thun mechanism becomes increasingly complicated with a 
boundary present \cite{DTW}. This is because, as well as the processes
which were possible in the bulk, a new set become possible involving
the boundary reflection factors. It is even possible to formulate
on-shell processes which involve cancellations between bulk
S-matrix elements and boundary reflection factors. One important result 
which does, however, remain true is that the na\"{\i}ve order of an 
on-shell diagram is equal to the number of degrees of freedom. This (or 
alternatively using \m{p=2V-P-2}) is perhaps the most useful way of 
proceeding, now that there will be a mixture of 3-point bulk vertices 
and 2-point boundary ones present. 

There are two types of process: ones which involve the boundary
vertices (``boundary dependent'') and those where the only
interaction between the particles and the boundary is to reflect
from it (``boundary independent''). Reflection
factors, in general, have a factor which is independent of any
boundary parameters present, but which nonetheless contains simple
poles. Without the Coleman-Thun mechanism, such poles would have no
explanation, since any pole which was due to the formation of a bound
state with the boundary would be expected to depend on the properties
of the boundary.

\pagebreak
Figure \ref{fig:bcti} shows two possible boundary independent
processes. In many models where two equal-mass particles can form a
bound state at relative rapidity \m{\alpha}, it would be expected that
the reflection factor would have a pole at \m{\hft(\pi-\alpha)},
explained by the left-hand diagram. The right-hand diagram shows a
more involved process,
which relies on having a suitable bulk vertex. The important point to
note here is that, to make the triangle close, the angle of incidence
on the boundary cannot depend on any of the boundary parameters. This
means that none of the boundary-dependent poles come into play, and so
there is always just a simple reflection from the boundary.

Some common boundary dependent processes are shown in \fig{fig:bct}. If an
incoming particle with rapidity \m{\delta} forms a boundary bound
state, then there will always be a pole in the reflection factor at
\m{\delta} for the same particle on the new state, explicable by the left-hand
diagram. The boundary initially emits the particle that helped to create it,
being reduced to the original state in the process. The incoming
particle then re-creates the new state. The other two diagrams simply rely on
there being suitable boundary and bulk vertices to make them close. They  
are na\"{\i}vely second order, but
could be reduced to first order if the boundary reflection factor had
a zero at the appropriate rapidity. 

The rightmost diagram is the most interesting, since it can be reduced
to first order either by a zero of the reflection factor, a zero of
the S-matrix element or (depending on the theory) cancellation between
diagrams for different arrangements of the internal loop. It is this
last, in particular, which shows how delicate the relationships
between the S-matrix and the reflection factors need to be to effect
the result.

Another point to make about boundary poles is that they can go from
describing a bound state to being due to a Coleman-Thun process by a
tuning of the boundary parameters. Often, at the point where this
happens, a process like \fig{fig:bctb} becomes possible. This is a modified
version of the right-hand diagram in \fig{fig:bcti}, where the
boundary parameter has been adjusted to make the particle reflect from
the boundary at a pole. As the parameters are tuned on through this
point, the diagram then collapses into a CT process such as the middle
diagram of \fig{fig:bct}. While there is no general proof, it appears
to be true for the sine-Gordon model at least that such a collision of 
boundary-independent and boundary-dependent processes must happen for a 
pole to cease to be due to a bound state.

\begin{figure}
\begin{center}
\parbox{0.30\textwidth}{
\centering
\unitlength 1.00mm
\linethickness{0.4pt}
\begin{picture}(22.00,55.00)(26.00,0)
\put(46.00,3.00){\rule{2.00\unitlength}{52.00\unitlength}}
\color{red}
\multiput(45.67,49.00)(0,1){6}{\circle*{0.25}}
\multiput(45.67,4.00)(0,1){6}{\circle*{0.25}}
\multiput(45.67,11.00)(0,2){19}{\circle*{0.25}}
\color{green}
\put(45.89,48.89){\line(-1,-2){20.06}}
\put(45.89,9.44){\line(-1,2){20.06}}
\color{black}
\end{picture}}
\parbox{0.34\textwidth}{
\unitlength 1.00mm
\linethickness{0.4pt}
\begin{picture}(48.00,55.00)(0,2)
\put(46.00,3.00){\rule{2.00\unitlength}{52.00\unitlength}}
\color{red}
\multiput(45.67,53.11)(0,1){2}{\circle*{0.25}}
\multiput(45.78,4.00)(0,1){2}{\circle*{0.25}}
\multiput(45.67,6.00)(0,2){24}{\circle*{0.25}}
\color{blue}
\put(37.33,13.44){\line(1,-1){8}}
\put(37.22,44.67){\line(1,1){8}}
\color{green}
\put(36.78,44.22){\line(3,-5){9.07}}
\put(36.78,14.00){\line(3,5){9.07}}
\put(36.78,14.00){\line(-2,-1){24.44}}
\put(36.78,44.22){\line(-2,1){24.44}}
\color{black}
\end{picture}}
\parbox{0.34\textwidth}{
\unitlength 1.00mm
\linethickness{0.4pt}
\begin{picture}(48.00,55.00)(0,2)
\color{blue}
\put(46.18,9.44){\line(-1,2){5}}
\put(46.18,49.67){\line(-1,-2){5}}
\color{red}
\multiput(45.67,49.11)(0,1){6}{\circle*{0.25}}
\multiput(45.78,4.00)(0,1){6}{\circle*{0.25}}
\multiput(45.67,10.00)(0,2){10}{\circle*{0.25}}
\multiput(45.67,32.00)(0,2){7}{\circle*{0.25}}
\color{green}
\put(40.96,20.00){\line(1,2){5.03}}
\put(40.96,19.89){\line(-5,3){28.11}}
\put(40.96,40.11){\line(1,-2){5.03}}
\put(40.96,40.22){\line(-5,-3){28.11}}
\color{black}
\put(46.00,3.00){\rule{2.00\unitlength}{52.00\unitlength}}
\end{picture}}
\ca{Some common boundary dependent Coleman-Thun processes}
\label{fig:bct}
\end{center}
\end{figure}

\begin{figure}[t]
\begin{center}
\parbox{3.5in}{
\begin{center}
\unitlength 1.00mm
\linethickness{0.4pt}
\begin{picture}(28.00,55.00)(20.00,0)
\put(46.00,3.00){\rule{2.00\unitlength}{52.00\unitlength}}
\put(31.89,42.89){\line(0,-1){28.22}}
\put(31.89,14.67){\line(5,3){14.00}}
\put(31.89,42.89){\line(5,-3){14.00}}
\put(31.89,42.89){\line(-4,5){9.60}}
\put(31.89,14.78){\line(-4,-5){9.33}}
\end{picture}
\ca{Coleman-Thun process possible only at special boundary parameter
values.}
\label{fig:bctb}\end{center}}
\end{center}
\end{figure}
\se{Summary}
This chapter has provided a brief overview of the world of 
1+1-dimensional integrable quantum field theory, and some of its most 
interesting features. The restrictions imposed by integrability 
make the axiomatic approach immensely powerful, allowing exact 
S-matrices to be found; this is the only arena where such results are 
possible at present, underlining its importance in uncovering 
non-perturbative results and pointing the way for tackling more realistic 
field theories.


\chap{Classical sine-Gordon Theory}
\quot{All these have never yet been seen---\\
But scientists who ought to know,\\
Assure us that they must be so...\\
Oh! let us never, never doubt\\
What nobody is sure about!}{Hilaire Belloc}


\se{Introduction}
\quot{First, establish a firm base.}{Sun Tzu}  
In the next chapter, we will study the effect of introducing a boundary 
into the sine-Gordon theory (which, as noted in the introduction, is the 
simplest ATFT). Before plunging ahead with the full quantum 
theory, however, it is worthwhile to take a look at the classical limit.
This 
exhibits essentially the same features as the quantum theory, but in a 
form that makes it much easier to gain a direct understanding of what is 
going on.

To take a step even further back, the first section discusses the
classical theory 
without a boundary, attempting to motivate the idea that it possesses 
an infinite number of conserved charges, and so is integrable, with all 
the simplifying features that entails. While not being a proof, it will 
offer a means of calculating as many conserved quantities as desired.
It will also help to show how ``special'' the
sine-Gordon theory really is: it is one of only two possible integrable 
field theories with a single scalar field.

Since it is not at all obvious that the introduction of a boundary 
should preserve many of these conserved quantities (let alone the 
infinite number required to maintain integrability) 
the restrictions integrability imposes on the possible boundary conditions
will 
then be examined, and the most general integrable boundary condition
found. 

To complete this chapter, and present a physical picture to take into 
the next, the first few classical boundary bound states will be 
constructed by the method of images. The idea---which 
is perhaps familiar from its use in electromagnetism---is that a given 
process on the half-line can be described by the theory on the full line 
with a set of ``image'' particles placed behind the boundary. This can
indeed be done, for any integrable boundary 
condition. Lastly---and with the benefit of hindsight---we will use
this to make some predictions for the full quantum theory, smoothing the 
path ahead.
     
\se{The bulk theory}
The classical action for the theory on the whole line is
\begin{equation}
\mathcal{A}_{SG}=\int_{-\infty}^{\infty}dx \int_{-\infty}^{\infty}dt~
\hft(\partial_{\mu}\varphi)^2
-\frac{m_0^{2}}{\beta^{2}}(\cos (\beta \varphi)-1)\,,
\label{eq:sgact}
\end{equation}
where \m{m_{0}} sets the mass scale and \m{\beta} is the coupling
constant. The particular form
of the potential term gives the theory its integrable properties, so let 
us, for the moment, consider a more general theory with a potential 
\m{-4V(\varphi)} so that we can investigate how special it really is. The
following argument
was first made by Ghoshal
and Zamolodchikov \cite{GhoshZam}; the form given here is taken from
\cite{Doreysete}.

To simplify the notation, it helps to use light-cone co-ordinates, 
defined through \m{\partial_{\pm}=\hft(\partial_{t} \pm \partial_{x})}. 
The equation of motion---\m{\partial \mathcal{A}=0}---then becomes 
\m{\partial_{+}\partial_{-}\varphi=-V'(\varphi)}. 

To construct conserved densities, imagine that there exist two 
quantities, \m{T} and \m{\Theta}, such that 
\m{\partial_{-}T=\partial_{+} \Theta}. Rewriting this using \m{x} and 
\m{t}, we find
\ba
\partial_{t}(T-\Theta)&=&\partial_{x}(T+\Theta), \\
\frac{\partial}{\partial t}\left[\int dx (T-\Theta)\right]&=&
[T+\Theta]^{\infty}_{-\infty}=0\,,
\label{eq:conservn}
\ea
showing that \m{\int dx (T-\Theta)} is conserved. The search now is 
for suitable quantities \m{T}; here, we will focus on polynomials in 
\m{\partial_{+}\varphi, \partial_{+}^{2}\varphi,\ldots}, and go 
order-by-order in the total number of \m{+}-derivatives. This number 
will be denoted as \m{s+1}, with \m{T_{s}} and 
\m{\Theta_{s}} standing for \m{T}s and \m{\Theta}s with \m{s+1} 
\m{+}-derivatives\footnote{Focusing simply on polynomial functions, 
these will turn out to be unique.}. The conserved charge will then be 
annotated as
\begin{equation}
Q_{s}=\int T_{s+1} - \Theta_{s-1} dx\,,
\end{equation}
where \m{s} can now be seen to stand for the spin of the charge.

Three other points are worth noting:
\begin{itemize}
\item total \m{\partial_{+}} derivatives can be dropped;
\item a polynomial in which no term has its highest derivative factor 
occurring linearly can never be a total \m{\partial_{+}} derivative;
\item for each \m{T_{s+1}}, there is a corresponding \m{T_{-s-1}}, 
obtained by interchanging \m{\partial_{+}} and \m{\partial_{-}} 
throughout.
\end{itemize}

Looking first at \m{s=1}, we find:
\ba
T_{2}&=&(\partial_{+}\varphi)^{2} \nn \\
\partial_{-}T_{2}&=&2(\partial_{+}\varphi)\partial_{+}\partial_{-}\varphi 
\nn \\
&=&-2(\partial_{+}\varphi)V(\varphi) \\
&=&\partial_{+}[-2V(\varphi)]\,, \nn
\ea
showing that \m{(\partial_{+}\varphi)^{2}} and \m{-2V(\varphi)} provide 
a suitable pair for any potential \m{V}. This, in fact, is not 
surprising, since \m{Q_{1}+Q_{-1}} is just energy, and 
\m{Q_{1}-Q_{-1}} is momentum, two quantities that are 
always conserved. 

There is no solution for \m{s=2}, and the first 
nontrivial result appears at \m{s=3}, where
\begin{equation}
T_{4}=\left(\frac{\beta}{2}\right)^{2}(\partial_{+}\varphi)^{4}
+(\partial_{+}^{2}\varphi)^{2}
\end{equation}
provides a solution for any real or imaginary \m{\beta}, but only if 
\m{V'''=\beta^{2}V'}. This has the solutions
\ba
\beta=0 &:& V=A+B(\varphi-\varphi_{0})^{2}, \\
\beta \neq 0 &:& V=A+B\e^{\beta\varphi}+C\e^{-\beta\varphi}\,,
\ea 
for any constants \m{A}, \m{B} and \m{C}. If \m{\beta=0}, this 
corresponds to either the massive or massless free field theory, 
depending on whether or not \m{B} is non-zero. For \m{\beta \neq 0}, we 
get the (massless) Liouville model if \m{B} or \m{C} is zero. Otherwise, 
it is the sinh-Gordon model if \m{\beta} is real, or sine-Gordon if it is 
imaginary.

If we were to proceed with this, we would find only one more model with 
any conserved charges above \m{s=1}, namely the Bullough-Dodd model, 
which appears at \m{s=5}. However, the sine-Gordon model would turn out 
to have conserved charges at all odd \m{s}, which is the crucial
point\footnote{In practice, the existence of an infinite number of
conserved charges was proved via the inverse scattering
method~\cite{ISM}.}. 
(In fact, since Parke's argument shows that any model with a conserved 
charge above \m{s=1} must have the properties needed to follow through 
the exact S-matrix approach, we already have all we need.) This, of
course, still does not answer the question as to \emph{why} this
should be true. A better understanding can be gained once the
sine-Gordon model is thought of as an ATFT, with an underlying Lie
algebra structure. It is this structure that endows it with the
symmetry that the charges flow from.   

\se{The theory on the half-line}
To introduce a boundary into the model, we must impose a boundary 
condition on the field, implemented through the addition of a boundary 
term to the action, i.e.
\begin{equation}
\mathcal{A}=\mathcal{A}_{\mathrm{bulk}} - 
\int_{-\infty}^{\infty}dt B(\varphi_{B})\,,
\end{equation}
where \m{\varphi_{B}(t)=\varphi(0,t)}, and so depends only on the value of
the field at the boundary. The term
\m{\mathcal{A}_{\mathrm{bulk}}} is
\begin{equation}
\mathcal{A}_{\mathrm{bulk}}=\int^{\infty}_{-\infty}dx\int_{-\infty}^{\infty}dt
~\hft(\partial_{\mu}\varphi)^{2}-4V(\varphi),
\end{equation}
and we are assuming that the bulk potential has been chosen so as to make
the bulk theory integrable. 

The equation of motion is the same as 
before (though restricted to apply only to the half-line) but the new 
term introduces the boundary condition
\begin{equation}
\partial_{x}\varphi|_{x=0}=-B'(\varphi_{B})\,.
\end{equation}

Clearly, not all the conservation laws from the bulk model can still 
apply now that we have introduced a boundary (momentum, for example) but
it still turns out to 
be possible to keep a (possibly infinite) subset. Working by analogy 
with the argument for the bulk, the problem arises because 
\re{eq:conservn} is modified to
\begin{equation}
\frac{\partial Q_{s}}{\partial t}=\frac{\partial}{\partial t}\left[
\int_{-\infty}^{0}(T_{s+1}-\Theta_{s-1})\right]= 
[T_{s+1}+\Theta_{s-1}]^{0}_{-\infty}=(T_{s+1}+\Theta_{s-1})|_{x=0}\,.
\end{equation}
The only way this can be saved is to demand that the rhs is a total 
\m{t}-derivative, allowing it to be incorporated into the lhs to give a 
new quantity which is conserved. (For the quantities found so far, the 
\m{T}s are \m{t}-derivatives, whereas the \m{\Theta}s are not.)

In general,
\ba
\frac{\partial}{\partial t}(Q_{s}\pm Q_{-s})&=&\frac{\partial}{\partial t} 
\int (T_{s+1} - \Theta_{s-1} \pm T_{-s-1} \mp \Theta_{-s+1}) dx \\
 &=&\int \partial_{t}[T_{s+1}-\Theta_{s-1} \pm T_{-s-1} \mp
\Theta_{-s+1}] dx \\
 &=& (T_{s+1} \mp T_{-s-1} \mp 
\Theta_{-s+1} - \Theta_{s-1} )|_{x=0}\,.
\ea
This explains why momentum---\m{Q_{1}-Q_{-1}}---is not conserved 
(the final line reading \m{(T_{2}+T_{-2}-2\Theta_{0})|_{x=0}}). For
energy,
on the other hand, we find
\ba
\frac{\partial}{\partial t}(Q_{1}+Q_{-1})&=&(T_{2}-T_{-2})|_{x=0} \nn \\
 &=& 
\partial_{t}\varphi\partial_{x}\varphi|_{x=0} \nn \\
&=&-\partial_{t}\varphi 
B'(\varphi_{B})|_{x=0} \\
&=& \frac{\partial}{\partial t}[-B(\varphi_{B})]\,, \nn
\ea
by making use of the boundary condition. This gives us
\begin{equation}
\frac{\partial}{\partial t}\left[\int_{-\infty}^{0} dx
(\hft(\partial_{t}\varphi)^{2}+\hft(\partial_{x}\varphi)^{2}+4V(\varphi)) 
dx +B(\varphi_{B})\right]=0\,,
\end{equation}
showing that energy is indeed still conserved on the half-line. The next 
natural step is to ask whether a \m{B} can be found that allows 
modified versions of all charges of the form \m{Q_{s}+Q_{-s}} to still 
be conserved. From above, this is true if 
\m{(T_{s+1}+\Theta_{s-1}-T_{-s-1}-\Theta_{-s+1})|_{x=0}} is a total 
\m{t}-derivative.

Imposing this restriction on the \m{s=3} charge of the bulk theory, we
find
\m{B'''=\left(\frac{\beta}{2}\right)^{2}B'}, whose most 
general solution is
\begin{equation}
B(\varphi_{B})=M\cos\frac{\beta}{2}(\varphi_{B}-\varphi_{0})\,,
\end{equation}
for some constants \m{M} and \m{\varphi_{0}}. The similarity of this
solution to the requirement on \m{V} in the bulk theory makes it
reasonable to imagine that, just as the bulk potential allowed
conserved charges for all higher odd \m{s}, this form for \m{B} should
too. This was finally proved for the classical theory through the
inverse scattering method~\cite{MacIntyre}, where it was found that
this is the most general integrable solution for \m{B}. It was also
found that all the charges discussed here---``even
parity'' charges from the bulk theory modified by a boundary 
term---survived with this boundary condition.

\begin{figure}
\begin{center}
\includegraphics{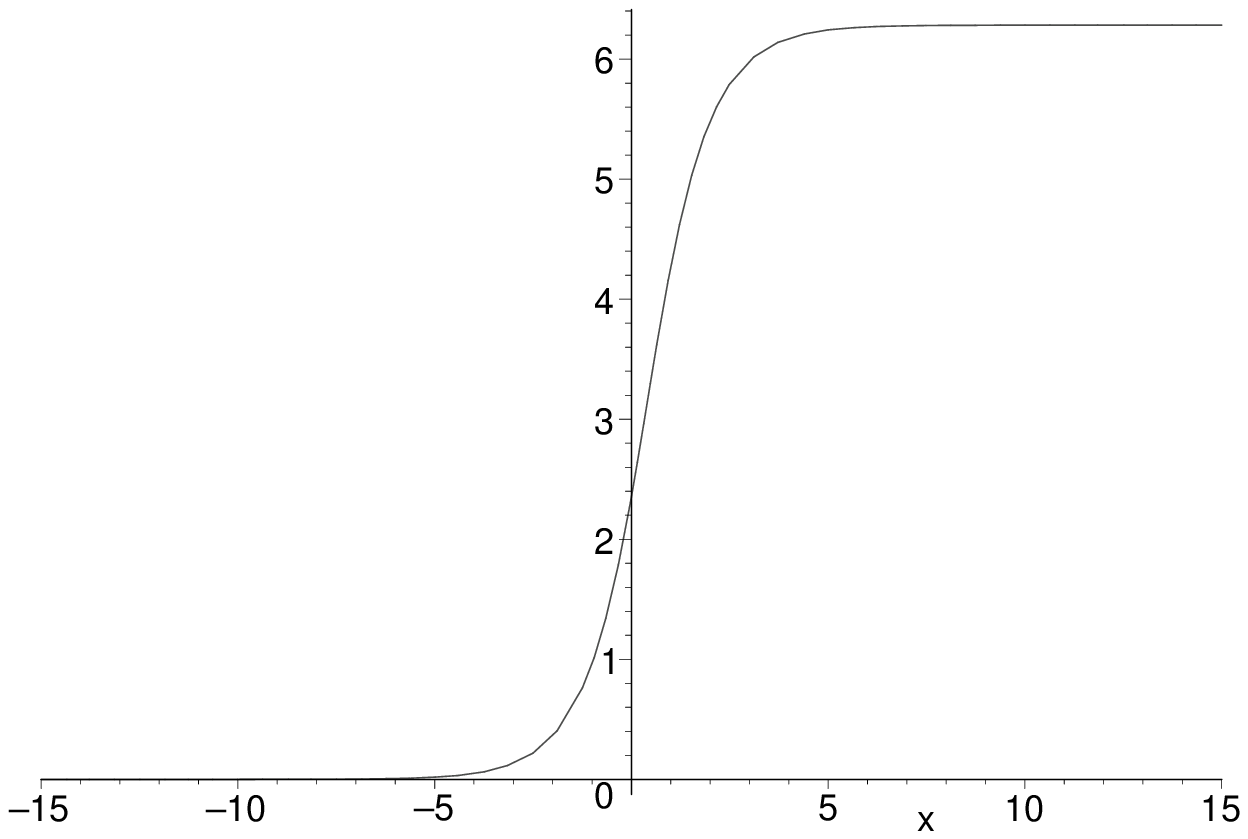} \\
\end{center}
\ca{Single soliton solution}
\label{fig:1soli}
\end{figure}

\se{Particle content}
Having established that both the bulk and boundary theories are
integrable, the next step is to find out what the theories actually
describe. Due to the periodic nature of the potential, the sine-Gordon
model is unusual in having an infinite number of vacua, at \m{\frac{2\pi
n}{\beta}} for any \m{n \in \mathbb{Z}}. The ``particles'' of the
theory therefore turn out not to be the usual localised humps in the
field, but rather a configuration that interpolates between two
neighbouring vacua, as shown in \fig{fig:1soli}.

This configuration has two useful properties. First, it is a
``soliton'', which means that it preserves its shape over time without
dissipation or decay. Secondly, because it interpolates \emph{between}
vacua, it cannot be destroyed as that would alter the value of the field
as \m{x \rightarrow \pm \infty}; for this reason, the theory is often
called ``topological''. The ``topological charge'' of a state is
defined as the difference (in units of \m{\frac{2\pi}{\beta}}) between
the value of the field at \m{-\infty} and \m{+\infty} and must be
conserved. A single soliton state has charge 1, while an
anti-soliton state (interpolating between the vacuum at \m{-\infty}
and the next lower one) has charge -1.

If two (anti-)solitons collide, they are simply transmitted through the
collision, without changing shape, and so can truly be considered as
particles (which, the theory being integrable, cannot be created or
destroyed). If their rapidities are allowed to be complex, however,
rather than purely real, the situation changes. If a soliton and an
anti-soliton are given
conjugate rapidities, a ``bound state'' appears. Due to the periodic
up-and-down motion of the field, these particles are known as
``breathers'' and
are categorised by the imaginary component of their rapidity, which
determines their period and mass. 

The soliton and anti-soliton both have the
same mass, which we shall call \m{m_{s}}, while the mass of
the breather formed by a soliton--anti-soliton pair at a relative
rapidity of \m{\T=iu} is \m{m=2m_{s}\cos \left(\frac{u}{2}\right)}.
That completes the particle spectrum of the bulk theory as, if two
breathers are persuaded to bind together, they simply form a third breather.

\se{Construction of boundary bound states}
Rather than try to analyse the boundary theory in a similar way to the
bulk, it is easier to use the method of images. The idea is to find a
particular configuration of the bulk theory where the value of the
field at \m{\varphi(0,t)} happens to obey one of the integrable
boundary conditions. The left-hand half line then provides a solution
to the boundary theory.

The vacuum state of the boundary theory simply requires that
\m{\varphi(0,t)=\varphi_{0}} for all time, whatever the value of
\m{M}, so a suitably-placed stationary soliton is all that is
required. By analogy with electromagnetism, we might then imagine that
the boundary state consisting of \m{n} particles with rapidities 
\m{\T_{1},\T_{2},\ldots,\T_{n}} corresponds to the bulk state with an
``image'' set of particles behind the boundary (with opposite
rapidities) and, again, a stationary soliton near the boundary. 

This problem was first tackled by Saleur, Skorik and Warner
\cite{SSW}, who found the 3-soliton solution. The
choice as to whether each particle was a soliton or anti-soliton and 
their relative initial positions selected which boundary condition was 
obeyed. In addition, in the Neumann limit the position of the 
stationary particle became infinite, reducing the result to a 
two-soliton solution.

The natural generalisation of this is to consider a \m{2n+1}-soliton 
solution (which reduces to a \m{2n}-soliton solution in the Neumann 
limit). This is easier than it might appear as, in the limit where the 
particles are well separated (i.e. \m{t \rightarrow \pm \infty}), the 
state of the field at the boundary is determined only by the central 
stationary soliton and the two moving solitons that are closest, 
allowing the 3-soliton solution to be re-used.

In addition, as SSW found in the 3-soliton case, the absolute positions 
of each pair of moving solitons are irrelevant in the solution of the
boundary 
condition; it is only the phase delay that is important. Using this 
fact, any pair of solitons in the \m{2n+1}-soliton solution can be moved 
off to infinity (provided their phase delay is preserved), reducing it 
to a \m{2n-1}-soliton solution. Using these two facts, it is perhaps 
beginning to become clear that the general solution can be built out of 
the 3-soliton solution with a little cunning.

\sse{Notation}
For the classical problem, it is convenient to re-scale the field and 
coupling constant to re-express the bulk sine-Gordon equation as
\begin{equation}
\phi_{tt}-\phi_{xx}=-\sin(\phi)\,,
\label{eq:classsg}
\end{equation}
where \m{\beta\varphi \equiv \phi}. On the half line, the most general 
boundary condition then becomes
\begin{equation}
\partial_{x}\phi|_{x=0}=M\sin \hft(\phi-\phi_{0})|_{x=0}\,.
\end{equation}

The classical multi-soliton solution for the whole line has been known 
for some time, and is generally expressed in terms of Hirota's 
\m{\tau}-functions \cite{Hirota} as
\begin{equation}
\phi(x,t)=4\arg(\tau) \equiv 
4\arctan\left(\frac{\Im(\tau)}{\Re(\tau)}\right) \,,
\end{equation}        
where the \m{\tau}-function for an \m{N}-soliton solution is
\begin{multline}
\tau(x,t) = \sum_{\mu_{j}=0,1}\e^{\frac{i\pi}{2}\left(\sum_{j=1}^{N} 
\epsilon_{j}\mu_{j}\right)}\exp\left[ - \sum_{j=1}^{N} \hft \mu_{j}
\left\{ \left(k_{j}+\frac{1}{k_{j}}\right)x + 
\left(k_{j}-\frac{1}{k_{j}}\right)t - a_{j}\right\}\right. \\
\left. + 2\sum_{1 \leq i < j \leq N} \mu_{i}\mu_{j} \ln \left( 
\frac{k_{i}-k_{j}}{k_{i}+k_{j}}\right)\right]\,.
\end{multline}
The parameters \m{k_{i}} are related to the soliton rapidities by 
\m{k_{i}=\e^{\T_{i}}}, so the solitons' velocities are given by
\begin{equation}
v_{i}=\left(\frac{k_{i}^{2}-1}{k_{i}^{2}+1}\right)\,.
\end{equation}
The \m{a_{i}} represent the initial 
positions of the solitons (but see below) while the \m{\epsilon_{i}} 
are +1(-1) for solitons (anti-solitons).

For the sake of simplicity, we shall number the particles in decreasing 
order of rapidity, so that particle 1 has the highest rapidity, particle 
2 has the next highest, and so on. This ensures the logarithm in the 
\m{\tau}-function is always real. Other orderings give the same
result---as they clearly must---but it is less transparent that the 
\m{\tau}-function is real.

\sse{The position problem}
Before going any further, a problem immediately arises with the 
interpretation of the \m{a_{i}} as the positions of the solitons. If 
this was truly the case, for example, a 3-soliton solution as \m{t 
\rightarrow \pm \infty} would reduce to a single-soliton solution with 
the same value of the position parameter. This, however, is not true. 
The one-soliton solution is just
\begin{equation}
\tau_{1}(x,t)=1 + i\epsilon_{1}\exp\left[ -x+\frac{a_{1}}{2}\right]\,,
\end{equation}
leading to
\begin{equation}
\phi_{1}(x,t)=4\arctan\left(\epsilon_{1}\exp\left[ -x+\frac{a_{1}}{2}
\right]\right)\,.
\label{eq:1soli}
\end{equation}

Taking the 3-soliton solution, note that, as \m{t \rightarrow \infty}, 
the soliton with positive rapidity will contribute a highly negative 
exponential whenever it appears in the sum, whereas the one with 
negative rapidity will contribute a correspondingly positive 
exponential. From this, it is clear that the two dominant 
terms will therefore be the ones where \m{\mu_{1}=0} and \m{\mu_{3}=1}.
Thus, 
as \m{t \rightarrow \infty},
\ba
\tau_{3}(x,t) &\approx&
\epsilon_{3}\exp \left[-\hf \left\{
\left(k_{3}+\frac{1}{k_{3}}\right)x+\left(k_{3}-\frac{1}{k_{3}}\right)t
-a'_{3}\right\}\right] \cdot \nonumber \\
&&\cdot \left(i-\epsilon_{2}\exp \left[-x+\frac{a'_{2}}{2}+2\ln\left(
\frac{1-k_{3}}{1+k_{3}}\right)\right]\right)\,, 
\ea
leading to
\begin{equation}
\phi_{3}(x,t) \approx 4\arctan\left(-\epsilon_{2}\exp\left[
x-\frac{a'_{2}}{2}
-2\ln\left(\frac{1-k_{3}}{1+k_{3}}\right)\right]\right)\,.
\end{equation}
If we now remember that \m{\tan \left(x\pm\hp\right) = -\tan (x)^{-1}}, 
this implies that \m{4\arctan x = y} can be re-written as
\m{4\arctan -\frac{1}{x} = y 
\pm2\pi}. Thus, we find
\begin{equation}
\phi_{3}(x,t) \pm 2\pi \approx 4\arctan\left(\epsilon_{1}
\exp\left[ -x+\frac{a'_{2}}{2}
+2\ln\left(\frac{1-k_{3}}{1+k_{3}}\right)\right]\right)\,.
\end{equation}
The \m{2\pi} on the lhs, which is equal to the spacing of the vacua, 
just represents the fact that the Hirota solution imposes \m{\phi=0} at 
\m{-\infty}, while the natural assumption here is that \m{\phi=\pm 
2\pi}, so that the leftmost soliton reduces the field to zero heading in 
towards the central particle. Thus, we do indeed have a single-soliton 
solution, but with
\begin{equation}
a_{1} = a'_{2} + 4\ln \left(\frac{1-k_{3}}{1+k_{3}}\right)\,.
\end{equation} 
Repeating this exercise with \m{t \rightarrow -\infty} instead gives the 
same result, but with \m{k_{3}} replaced by \m{\frac{1}{k_{1}}}. Since 
we would like the stationary soliton to solve the same boundary 
condition in both cases (as this condition stays unchanged for all
time), 
it is clear that we are forced to take \m{k_{3}=\frac{1}{k_{1}}}, as 
SSW did.

The reason for this is easy to see once it is realised that, to shift
the solution in time, all that is needed is to shift each position
parameter by velocity\m{\times}time, irrespective of any collisions
which may have happened in the interim. The effects of collisions are
thus built into the solution, and the parameters are only indirectly
related to particle positions at any given time. In what
follows, however, it will be easier to work in terms of ``actual''
parameters, and transform back to Hirota's parameters at the end. 

A more general analysis shows that the \m{2N+1}-soliton solution with
\m{N} pairs of solitons with opposite rapidities 
examined at \m{t \rightarrow -\infty}
reduces to 
\m{2N+1} single-soliton solutions as expected, but with each soliton 
position modified by a term involving all rapidities higher 
than its own. This means that the ``position'' parameters \m{a_{i}} only 
have a genuine interpretation as a position for the particle with 
the highest rapidity. As \m{t \rightarrow +\infty}, the opposite
situation arises, with the position parameters modified by all lower
rapidities. 
To be precise, let us take \m{x \rightarrow +\infty} and \m{t
\rightarrow -\infty} with \m{\frac{x}{t} 
\approx v_{i}}, to keep ourselves in the neighbourhood of particle 
\m{i}. This means that \m{i>N} (we are 
considering the particles with negative rapidities). Then, by the 
same reasoning as before, the two dominant terms will be those with 
\m{\mu_{j}=1, j<i} and \m{\mu_{j}=0, j>i}. This gives, for \m{i-1} even,
\begin{multline}
\tau_{2N+1}(x,t) \approx (-1)^{\frac{i-1}{2}}\prod_{j=1}^{i-1}
\ep_{j}\exp 
\left[-\sum_{j=1}^{i-1}\hf \left\{\left(k_{j}+\frac{1}{k_{j}}\right)x
+\left(k_{j}-\frac{1}{k_{j}}\right)t-a_{j}\right\} \right. \\
\left. + 2\sum_{1\leq i' < j 
\leq i-1} \ln 
\left(\frac{k_{i'}-k_{j}}{k_{i'}+k_{j}}\right)\right] \times \\
\left(1+
i\epsilon_{i}\exp \left[-\hf\left\{\left(
k_{i}+\frac{1}{k_{i}}\right)x+\left(k_{i}-\frac{1}{k_{i}}\right)t
-a_{i}
\right\} \right. \right. \\
\left. \left. + 2\sum_{1\leq j \leq i-1} \ln 
\left(\frac{k_{j}-k_{i}}{k_{i}+k_{j}}\right)\right]\right)\,,  
\end{multline}
leading to
\begin{multline}
\phi_{2N+1}(x,t) \approx 4\arctan\left(\epsilon_{i}\left[
-\hft\left\{\left(k_{i}+\frac{1}{k_{i}}\right)x+\left(k_{i}
-\frac{1}{k_{i}}\right)t-a_{i}\right\}
\right. \right.\\
\left. \left. +2\sum_{1\leq j \leq i-1} \ln 
\left(\frac{k_{j}-k_{i}}{k_{i}+k_{j}}\right)\right]\right)\,.
\end{multline}
Thus, compared with the appropriate single soliton solution,
\begin{equation}
a_{1}=a_{i}+4\sum_{1\leq j \leq i-1} \ln 
\left(\frac{k_{j}-k_{i}}{k_{i}+k_{j}}\right)\,.
\end{equation}
Note that, this time, \m{\phi \rightarrow 0} as \m{x \rightarrow 
\infty} is the natural situation, in agreement with the Hirota formula.
For \m{2N+1-i} odd, we need to use the same trick as for the 
three-soliton case, but finish up with the same formula. Finally, for 
\m{i \leq N} we need to take \m{x \rightarrow -\infty} but the result remains
true. Taking the other limit (as \m{t \rightarrow +\infty}), the
analogous result is
\begin{equation}
a_{1}=a_{i}+4\sum_{i+1\leq j \leq 2N+1} \ln 
\left(\frac{k_{i}-k_{j}}{k_{i}+k_{j}}\right)\,.
\end{equation} 
Note also that the only way to ensure the
boundary condition stays constant in time is to impose \m{k_{i}=1/k_{2N+2-i}}.

From this, we can calculate the phase delay between any pair of
particles with equal and opposite rapidities---\m{i} and
\m{2N+2-i}---in terms of their ``actual'' position parameters
\m{a_{n}'} as \m{t\rightarrow -\infty} as
\begin{equation}
a_{i}+a_{2N+2-i}=a'_{i}+a'_{2N+2-i}-4\sum_{j \neq i} \ln 
\left(\left|\frac{k_{j}-k_{i}}{k_{i}+k_{j}}\right|\right)\,,
\label{eq:actdelay}
\end{equation}
assuming \m{i\leq N}.

\sse{Solving the boundary condition}
As a warm up to the general solution, it is useful to consider the 
simplest possible solution, with only one stationary soliton. This 
corresponds to the ground state of the boundary model. In this case, all 
boundary conditions reduce to the demand that \m{\phi|_{x=0}=\Phi_{0}},
for some constant \m{\Phi_{0}} depending on the boundary conditions. (In
the Dirichlet case, \m{\Phi_{0}} becomes simply \m{\phi_{0}}.)  
Putting this into \re{eq:1soli}, we find
\begin{equation}
\phi_{1}(0,t)=4\arctan\left(\epsilon_{1}\e^{\frac{a_{1}}{2}}\right)
=\Phi_{0}\,,
\end{equation}
implying
\begin{equation}
a_{1}=2\ln\left(\epsilon_{1}\tan \frac{\Phi_{0}}{4}\right)\,.
\end{equation}

Taking advantage of the fact that the 3-soliton solution must tend to 
this near the origin as \m{t \rightarrow -\infty}, we can immediately
write down the 
position parameter of the stationary soliton in the 3-soliton solution as
\begin{equation}
a_{2}=2\ln\left(\epsilon_{1}\tan \frac{\Phi_{0}}{4}\right)
-4\ln \left(\frac{k_{1}-1}{k_{1}+1}\right)=
2\ln\left(\epsilon_{1}\tan 
\frac{\Phi_{0}}{4}\left(
\frac{k_{1}+1}{k_{1}-1}\right)^{2}\right)\,.
\end{equation}
Noting that, in terms of the rapidity variable, 
\m{(\frac{k_{1}-1}{k_{1}+1})^{2}=\tanh \left(\frac{\T}{2}\right)^{2}}, 
this agrees with the formula for \m{a_{3}}\footnote{Their \m{a_{3}} is 
twice the \m{a_{3}} which appears in the Hirota formula, accounting for 
their loss of the factor of 2 in front of the logarithm. Also, they
consider the left half-line rather then the right.} given by SSW 
in their appendix. They derive this specifically for the case of 
Dirichlet boundary conditions, but it can now be seen to have the same
form for \emph{all} boundary conditions. 

\sse{The general solution}
By extension of the above argument, the position parameter of the 
stationary soliton in the \m{2N+1}-soliton solution is
\begin{equation}
a_{N+1}=2\ln\left(\epsilon_{N+1}\tan 
\frac{\Phi_{0}}{4}\prod_{1\leq j \leq N} 
\tanh\left(\frac{\T_{j}}{-2}\right)^{2}\right)\,.
\end{equation}
As has already been mentioned, the general \m{2N+1}-soliton solution 
reduces to the 3-soliton solution (involving the 3 slowest solitons) as
\m{t \rightarrow -\infty}, so the phase delay for the slowest pair 
should be given by the SSW formula, which (with our conventions) is
\begin{equation}
a=2\ln\left\{-\ep_{1}\ep_{3}\tanh \left(\frac{\T}{2}\right)^{-2}
\tanh (\T)^{-2}\left[\frac{\tanh \hft(\T+i\eta) \tanh
\hft(\T-i\eta)}{\tanh \hft(\T+\zeta)\tanh 
\hft(\T-\zeta)}\right]^{\pm1}\right\}\,,
\label{eq:sswa}
\end{equation} 
where \m{\eta} and \m{\zeta} are the solutions of the simultaneous 
equations
\ba
M\cos (\hft\phi_{0})&=&2\cosh \zeta \cos \eta \nonumber \\
M\sin (\hft\phi_{0})&=&2\sinh \zeta \sin \eta\,.
\ea
The ambiguity in the sign of \re{eq:sswa} is simply a vagary of the 
solution method (due 
to the fact that the bulk vacua are \m{2\pi}-periodic, whereas the 
boundary vacua are only \m{4\pi}-periodic; working in terms of the bulk 
makes the stable and unstable possibilities appear together). We 
shall concentrate on the negative sign, which corresponds to the stable 
boundary value.

By virtue of \re{eq:actdelay}, this can be re-written using the
``actual'' position parameters instead, as
\begin{equation}
a'=2\ln\left\{-\ep_{1}\ep_{3}\left[\frac{\tanh
\hft(\T+i\eta) \tanh 
\hft(\T-i\eta)}{\tanh \hft(\T+\zeta)\tanh 
\hft(\T-\zeta)}\right]^{\pm1}\right\}\,.
\label{eq:acta}
\end{equation} 
Turning now to the faster particles, we need to use the fact that, for 
the slowest particles, only the phase delay is important. This means 
that we can take their actual positions off to \m{\pm \infty} without 
affecting the validity of the solution, and essentially reduce the 
problem to the \m{2N-1}-soliton case. Now, the 
\emph{next} slowest particles have gained the mantle of being the 
slowest, and so must have a phase delay of the same form. (Note that, in 
doing this, we have made the slowest particles collide with all the 
faster ones in turn on their way to infinity, changing their positions. 
The symmetry of the situation, however, ensures that the phase delays 
between the pairs of particles stay intact.)

Repeating this for all the particles shows that, for each pair, all that 
is relevant is the phase delay, and this always has the SSW
form. In terms of the position parameters, we then have
\begin{equation}
a^{i}=2\ln\left\{-\ep_{i}\ep_{2N+2-i}\prod_{j \neq i}
\tanh \left(\frac{\T_{i}-\T_{j}}{2}\right)^{-2}
\left[\frac{\tanh
\hft(\T+i\eta) \tanh
\hft(\T-i\eta)}{\tanh \hft(\T+\zeta)\tanh 
\hft(\T-\zeta)}\right]^{\pm1}\right\}\,,
\label{eq:sswai}
\end{equation} 
where \m{a^{i}=a_{i}+a_{2N+2-i}}.

This completes the solution, but for one point: through this argument 
we have shown that if the \m{2N+1}-soliton solution exists then it must 
have the given form, but we have \emph{not} shown that it actually 
exists. For that, we would need to substitute the results back into the 
Hirota formula to check---a cumbersome task, and one for which we lack 
the energy. In the meantime, we content ourselves with the observation 
that it seems a reasonable assumption, and bears up to all the numerical 
checks we have carried out.

\se{Boundary bound states}
\sse{Boundary breathers}
The natural progression from this is to consider extending the
rapidities to complex values. While this can be used give a solution
where breathers rather than solitons interact with the boundary, it
can also be used to construct ``boundary breathers'' or boundary bound
states. These solutions arise when the pair of particles which are
given complex conjugate rapidities consist of one in front of the
boundary and one behind.

Due to the requirement that their rapidities must also be equal and
opposite, this implies that they must be given purely imaginary
rapidities. Curiously, in the Dirichlet case (as SSW noted) the pair of
particles must also
consist of two solitons or two anti-solitons, not a particle and its
anti-particle as in the bulk.

Simply by continuing all rapidities to imaginary values, we can
generate a sequence of bound states through solutions with
successively greater numbers of solitons. The one subtlety is that
both members of each pair have to be given the same initial position
parameter for the solution to still obey the boundary condition. This
is a consequence of the way the solution was found: the
\m{\tau}-function was split into real and imaginary parts, assuming
all rapidities were real. Making some rapidities imaginary disturbs
this in general, but putting both members of each pair at the same
position allows the split used to remain valid. Since all other
solutions to the problem with real rapidities can be related to this
by a time translation, it is reasonable to assume that the same is
true for the imaginary case. The only difference is that, with
imaginary rapidities, there is no movement in real space.

As with the bulk breathers, the period of a boundary breather is given
through the imaginary part of the rapidity. For the 3-soliton
solution, with the moving pair given a rapidity of \m{\T=iu}, the
period is \m{2\pi/\sin(u)}. Now, however, for the breathers coming
from higher solutions, each pair has its own period. If there exists a
common period, whose length is an integer multiple of the periods of
all the pairs, then the motion is still periodic, but, in general, it
will now be aperiodic.

\begin{figure}
\begin{center}
\includegraphics{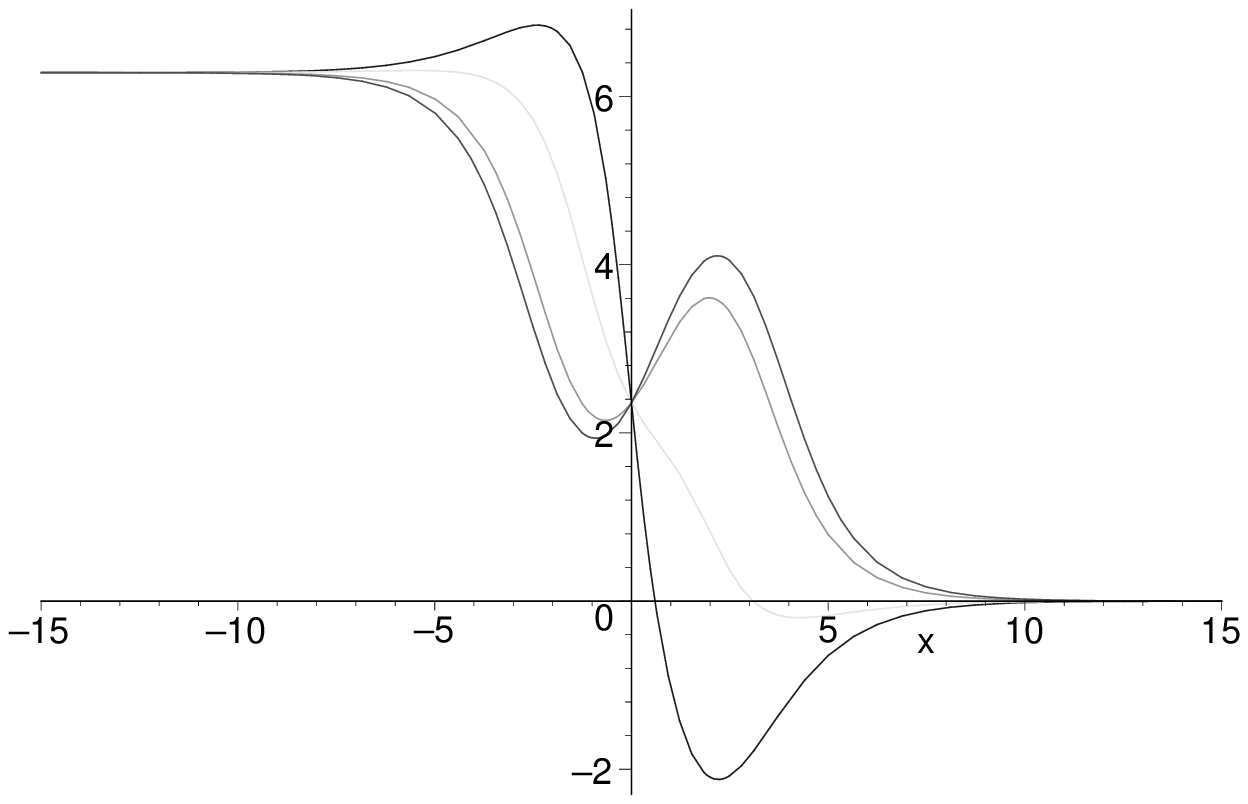} \\
\end{center}
\ca{3-soliton solution, period 10}
\label{fig:3soli}
\end{figure}

\begin{figure}
\begin{center}
\includegraphics{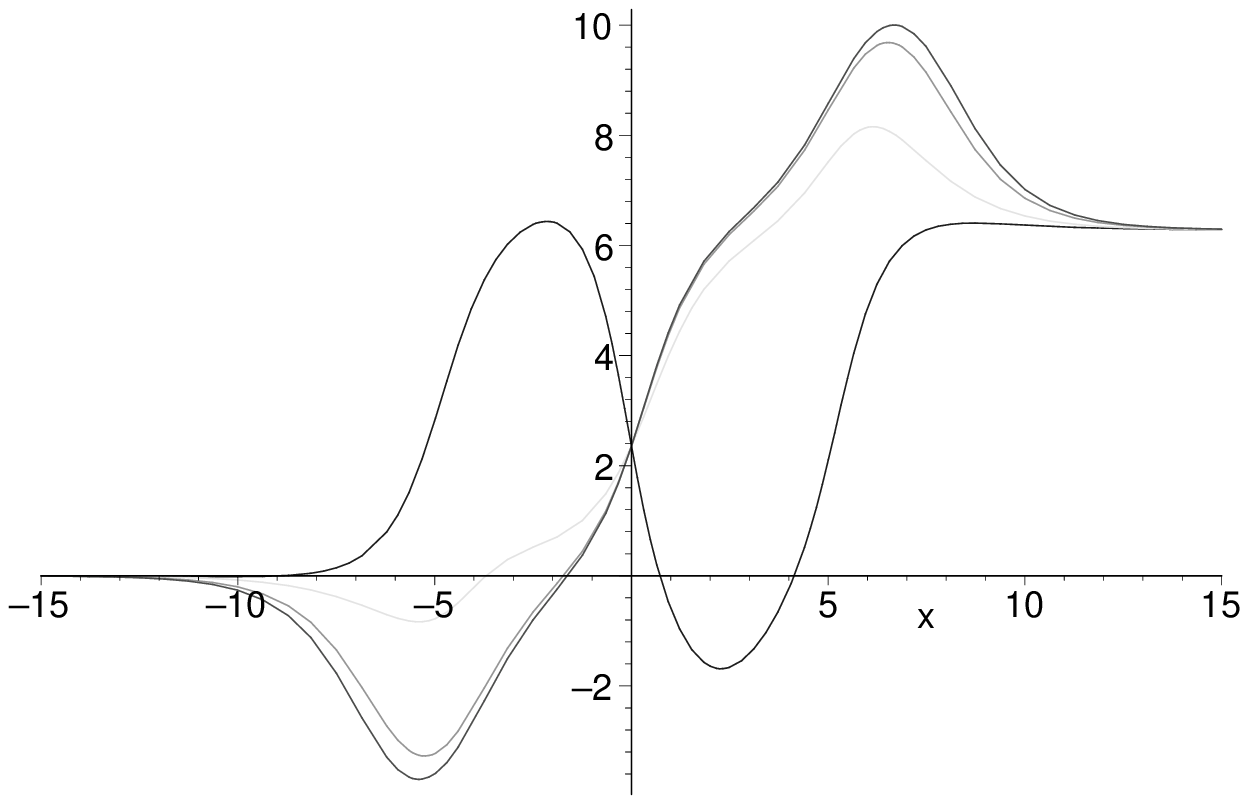} \\
\end{center}
\ca{5-soliton solution, periods 10 and 12}
\label{fig:5soli}
\end{figure}

\begin{figure}
\begin{center}
\includegraphics{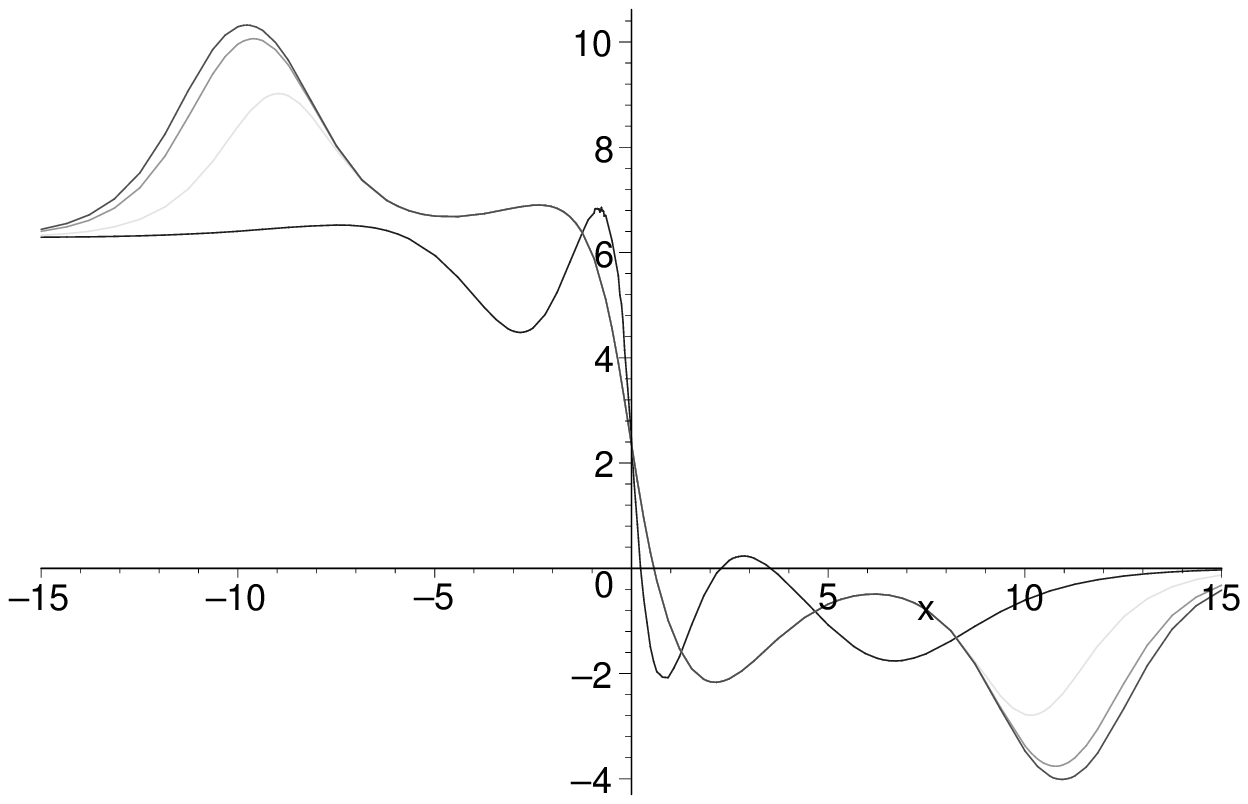} \\
\end{center}
\ca{7-soliton solution, periods 10, 12, and 14}
\label{fig:7soli}
\end{figure}

To demonstrate the form of the boundary breathers, we have chosen
periodic solutions by giving each pair an integer period. These are
shown in Figures \ref{fig:3soli}, \ref{fig:5soli}, and \ref{fig:7soli}
for the 3, 5, and 7 soliton solutions respectively.         

\sse{Another bound state}
The breathers mentioned above are not the only bound states in the
classical theory. The phase delay \re{eq:acta} becomes infinite at
\m{\T=\zeta}, which must be due to the formation of a stable bound
state. In the bulk theory, this could not happen (all bound states must
be formed at imaginary rapidities), so it is further evidence of the
changes wrought by the introduction of a boundary.

Considering the 3-soliton solution, we can imagine the particle with
negative rapidity as being taken off to infinity at this point,
leaving just a two-particle process. The remaining moving particle
sweeps past the boundary, shifting the stationary soliton on the way
past; this only appears as a bound state when we restrict ourselves to
the half line. Then, the incoming particle reaches the boundary and
disappears, leaving the boundary state changed.

The final state can be found by considering the limit of the
3-particle \m{\tau}-function where the position parameters of both
moving particles are taken to infinity. The result of this is that
\m{\phi_{0}} becomes \m{\phi'_{0}}, given by
\begin{equation}
\tan\left(\frac{\phi'_{0}}{4}\right)=\tan\left(\frac{\phi_{0}}{4}\right)
\tanh^{4}\left(\frac{\zeta}{2}\right)\,.   
\end{equation}

\se{Predictions}
The first prediction to take across to the discussion of the quantum
theory is that there should be a hierarchy of excited states. For
example, the states formed by binding a soliton to the boundary should
be analogous to the 3-soliton solution found above. After that,
further solitons should create the quantum versions of \m{5,7,\ldots}
soliton solutions. Furthermore, the introduction of a breather should
allow the formation of a state which could otherwise have been formed
by two successive solitonic particles.

\begin{figure}
\begin{center}
\includegraphics{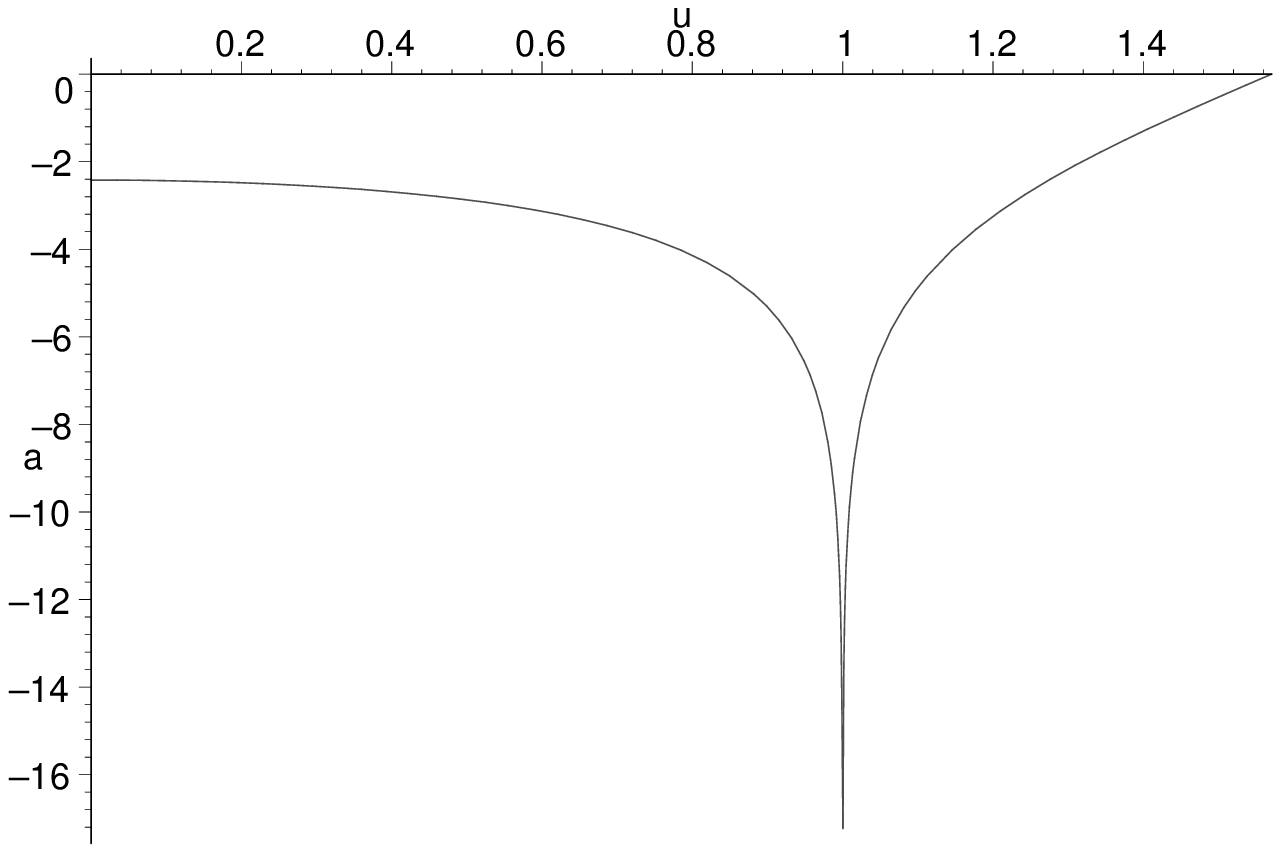} \\
\end{center}
\ca[Plot of $a'$ versus $u$ for $\eta=1.0$]{Plot of $\pmb{a'}$ versus $\pmb{u}$ for $\pmb{\eta=1.0}$}
\label{fig:aspectr}
\end{figure}

A final, and slightly more subtle point, is that the ``actual''
position parameter used for a given pair of particles (with imaginary
rapidity) is monotonically decreasing for \m{u < \eta}, as shown in
\fig{fig:aspectr}. This means that, in a given solution, the soliton
pair with the least rapidity will be positioned farthest from the
boundary. If we imagine that such a solution, translated into the
quantum regime, is built up with the soliton finally positioned
nearest the boundary interacting first, this means that particles must
interact in decreasing order of rapidity. 
     
\clearpage


\chap{Quantum Boundary sine-Gordon Theory}
\quot{The most exciting phrase to hear in science, the one that
heralds new discoveries, is not `Eureka!' but `That's
funny\ldots'}{Isaac Asimov}


\se{Introduction}
\quot{Mathematicians are a species of Frenchman: if you say
something to them they translate it into their own
language and presto! it is something entirely different.}{Goethe}
As we saw in the previous chapter, introducing a boundary into the
classical theory brings with it a number of new phenomena, and in
particular a new set of boundary
bound states. In this chapter, we will investigate these further in
the full quantum sine-Gordon model. 

A major complicating factor in this
work is the fact that even simple poles in the boundary reflection
factors should not necessarily be interpreted as being due to the
formation of bound states, since many have an interpretation through 
the Coleman-Thun mechanism. Indeed, this model provides a good arena
for demonstrating the range of possible explanations this mechanism
can throw up, in some cases involving bulk and boundary matrices
working together to induce a cancellation in the na\"{\i}ve order of a
diagram.   

The two main tasks, therefore, are to find suitable interpretations
where required, but also to find a method of proving that the
remaining poles are indeed associated with bound states. Two
elementary lemmas---which simply serve to impose momentum
conservation on boundary processes---will turn out to give us all
the ammunition we need for this, and should also be readily applicable 
to other models. 

The groundwork for the study of the boundary sine-Gordon model was
laid by Ghoshal and Zamolodchikov~\cite{GhoshZam}, before being taken
further by Ghoshal~\cite{Ghoshal} and Skorik and
Saleur~\cite{Skorik}. They provided the basic ground-state
reflection factors, and investigated the first few excited states; we
will take this forward to provide (hopefully) a full and rigorous
solution to the problem.

After reviewing these results in the first section, we will go on to a
detailed investigation of the Dirichlet boundary condition (where the
value of the field at the boundary is fixed for all time). This
displays most of the features of the general solution, and will allow
us to extend the results straightforwardly to all other integrable
boundary conditions.  

\se{Review of previous results}
\sse{The theory in the bulk}
As we discussed in the previous chapter, the classical sine-Gordon
model \re{eq:sgact} is integrable. This can be shown to be true at the
quantum level as well~\cite{Faddeev} and so
the exact quantum S-matrix can be found through the axiomatic
program. The essential difference between the classical and quantum
theories is that the breather particles, which could be formed
classically by a soliton--anti-soliton pair at any imaginary rapidity,
now become quantised. These states can now only be formed at relative
rapidities of \m{i\left(\pi - n\pi/2\la\right)},
\m{n=1,2,\ldots,<\la}, where
\begin{equation}
\la = \frac{8\pi}{\beta^{2}}-1\,,
\end{equation}  
and so will be labelled as \m{B_{n}}. Their mass is therefore
\m{m_{n}=2m_{s}\sin\left(n\pi/2\la\right)}.

If we denote the soliton S-matrix as
\m{S^{ab}_{cd}(\T)} for rapidity \m{\T}, with \m{a,b,c,d} taking the
value \m{+} (\m{-}) if the particle is a soliton (anti-soliton), the 
non-zero scattering amplitudes \cite{Zsquare} are 
\m{S^{++}_{++}(\T)=S^{--}_{--}(\T)=a(\T)}
(soliton-soliton or anti-soliton-anti-soliton scattering),
\m{S^{+-}_{+-}(\T)=S^{-+}_{-+}(\T)=b(\T)} (soliton-anti-soliton
transmission), and \m{S^{+-}_{-+}(\T)=S^{-+}_{+-}(\T)=c(\T)}
(soliton-anti-soliton reflection). Explicitly,
\ba
a(\T)&=&\sin [\la(\pi - u)]\rho(u)\,, \nonumber\\
b(\T)&=&\sin(\la u)\rho(u)\,, \\
c(\T)&=&\sin(\la \pi)\rho(u)\,, \nonumber
\ea
where \m{u=-i\T} and
\begin{equation}
\rho(u)=\frac{1}{\sin (\la (u-\pi))}\prod_{l=1}^{\infty}\left[\frac{\g
\left((2l-2) \la -
\ul\right) \g \left(1+2l\la -\ul \right)}{\g \left( (2l-1)\la - \ul
\right) \g \left(1+ (2l-1)\la - \ul \right)} / (u \rightarrow - u) \right].
\end{equation}
As pointed out in~\cite{Pillin},
this factor can also be written in terms of Barnes' 
diperiodic sine function \m{S_{2}(x|\omega_1,\omega_2)} \cite{Barnes,Jimbo}.
This is a meromorphic function parametrised by the pair of `quasiperiods'
\m{(\omega_1,\omega_2)}, with poles and zeroes at the following
points:
\ba
\hbox{poles}&:& x=n_1\,\omega_1+n_2\,\omega_2 ~~~~(n_1,n_2=1,2,\dots) \nn\\
\hbox{zeroes}&:& x=m_1\,\omega_1+m_2\,\omega_2~~~(m_1,m_2=0,-1,-2\dots)
\ea
In terms of this function,
\begin{equation}
\rho(u)=\frac{1}{\sin (\la
(u-\pi))}\frac{S_{2}\left(\pi-u\left|\pl,2\pi\right.
\right)S_{2}\left(u\left|\pl,2\pi
\right. \right)}{S_{2}\left(\pi+u\left| \pl,2\pi
\right. \right)S_{2}\left( -u \left| \pl,2\pi \right. \right)}\,.
\end{equation}

The amplitudes \m{b(\T)} and \m{c(\T)} have simple poles at \m{\T=i\left(\pi -
\frac{n\pi}{\la}\right)}, \linebreak \m{n=1,2,\ldots,<\la}, which can be
attributed to the creation of \m{B_{n}} in the forward channel. There
are also poles at \m{\T=\frac{i\pi n}{\la}} in \m{a(\T)} and \m{b(\T)}
corresponding to the same process in the cross channel. 
Since all poles that we will be discussing, both in the bulk and at the
boundary, occur at purely imaginary rapidities, from now on we will
use the variable \m{u=-i\T} and always work in terms of
purely imaginary rapidities. 

\sse{The theory with a boundary}
\label{sec:review}
Returning again to the previous chapter, the bulk theory
be restricted to the half-line \m{x
\in (-\infty,0]} while still preserving integrability by adding a
``boundary action'' term \cite{GhoshZam}
\begin{equation}
-\int_{-\infty}^{\infty}dt\, M \cos \left[\frac{\beta}{2}(\varphi\phup_B -
\varphi\phup_{0})\right],
\label{eq:boundact}
\end{equation}    
where \m{M} and \m{\varphi_{0}} are free parameters,
and $\varphi\phup_B(t)=\varphi(x,t)|_{x{=}0}$.

This does not conserve topological charge in general, so
four solitonic boundary reflection
factors need to be introduced, as well as a set of breather reflection
factors. The solitonic factors which we quote here were 
given in~\cite{GhoshZam}, while
breather factors can be found in~\cite{Ghoshal}.

\ssse{Solitonic ground state factors}
The reflection factors for the sine-Gordon solitons off the boundary
ground state will be denoted by
\m{P_{\pm}(u)} (a soliton or anti-soliton, incident on the boundary,
is reflected back unchanged) and \m{Q_{\pm}(u)} (a soliton is
reflected back as an anti-soliton, or vice versa). These are given by
\begin{equation}
\begin{array}{rcl}
P^{+}(u)&=&\cos (\xi+\la u) R(u) \\
P^{-}(u)&=&\cos (\xi - \la u) R(u) \\
Q^{\pm}(u)&=&\frac{k}{2}\sin (2\la u) R(u),
\end{array}
\label{eq:genboot}
\end{equation}
where
\begin{equation}
R(u)=R_{0}(u)R_{1}(u).
\end{equation}
The first factor---\m{R_{0}(u)}---is boundary-independent, and can be
written as
\begin{equation}
R_{0}(u)=\prod_{k=1}^{\infty}\left[\frac{\g \left( 1+\la(4k-4)
-\frac{2\la u}{\pi} \right)\g \left( 4\la k-\frac{2\la
u}{\pi}\right)}{\g \left( \la(4k-3)-\frac{2\la u}{\pi} \right) \g
\left( 1+ \la(4k-1)-\frac{2\la u}{\pi}\right)} /(u \rightarrow
-u)\right]\,.   
\label{eq:r0}
\end{equation}

The boundary-dependent term is \m{R_{1}(u)}, given by
\begin{equation}
R_{1}(u)=\frac{1}{\cos \xi}\sigma (\eta,u) \sigma (i\vartheta,u),
\end{equation}
where\footnote{Note that there is a
small error in Ghoshal and Zamolodchikov's formula (5.23) for
\m{\sigma}. This corrected version was supplied to Patrick Dorey and the
author
by Subir Ghoshal.}
\begin{equation}
\sigma (x,u) = \frac{\Pi\left(x,\hp-u\right)\Pi\left(-x,\hp-u\right)
\Pi \left(x,-\hp+u\right)\Pi\left(-x,-\hp+u\right)}{\Pi \left
( x,\hp\right) \Pi \left(x,-\hp\right) \Pi\left(-x,\hp\right) \Pi
\left( -x,-\hp \right)},
\end{equation}
and
\begin{equation}
\Pi(x,u)=\prod_{l=0}^{\infty}\frac{\g \left(\hf+\left(2l+\hf\right)\la
+ \frac{x}{\pi}-\frac{\la u}{\pi}\right)\g \left(\hf +
\left(2l+\frac{3}{2} \right) \la +\frac{x}{\pi}\right)}{\g \left( \hf+
\left( 2l+\frac{3}{2}\right)\la +\frac{x}{\pi} - \frac{\la
u}{\pi}\right) \g \left( \hf +\left(2l+\hf\right)\la +\frac{x}{\pi}\right)}.
\end{equation}

The parameters \m{\xi,\eta,\vartheta,} and \m{k} are real and arbitrary
apart from being constrained by
\begin{equation}
\begin{array}{rcl}
\cos (\eta)\cosh(\vartheta)&=&-\frac{1}{k}\cos \xi \\
\cos^{2}(\eta)+\cosh^{2}(\vartheta)&=&1+\frac{1}{k^{2}}.
\end{array}
\label{eq:paramconds}
\end{equation}
The relationship of these parameters---which arise in the course of 
finding the most general solution to the four requirements given in 
Chapter 1---to the ones which appear in the action was, for a long time, 
unknown. The problem has only recently been solved by Al. Zamolodchikov; 
further details can be found in Appendix~\ref{app:relation}. These 
formal parameters, however, are easier to work with in practice than the 
physical \m{\varphi_{0}} and \m{M}, and so we shall continue to use them.

The theory is invariant under 
\m{\varphi_0\to\varphi_0+\frac{2\pi}{\beta}}, and also
under the simultaneous transformations \m{\varphi_0 \rightarrow
-\varphi_0} and
\m{\mathrm{soliton\ } \rightarrow \mathrm{anti{-}soliton}}. 
Introducing the boundary breaks the degeneracy of the bulk vacua, and
selects the lower line in \fig{fig:vacua} as the lowest-energy state,
with the upper line as the first excited state. Continuing
\m{\varphi_{0}} through \m{\frac{\pi}{\beta}} thus simply interchanges
the r\^{o}les of these two states, and selects the upper one as the
ground state.

\begin{figure}
\begin{center}
\unitlength 1.00mm
\linethickness{0.4pt}
\begin{picture}(50.11,55.00)(0,5)
\put(46.00,3.00){\rule{2.00\unitlength}{52.00\unitlength}}
\color{green}
\qbezier(45.39,29.44)(41.28,23.33)(32.61,22.56)
\qbezier(45.39,29.56)(44.50,39.89)(32.61,40.89)
\put(32.61,40.89){\line(-1,0){27.17}}
\put(32.61,22.56){\line(-1,0){27.17}}
\color{blue}
\multiput(5.44,41.00)(1,0){41}{\circle*{0.25}}
\multiput(5.44,22.22)(1,0){41}{\circle*{0.25}}
\color{deepred}
\put(3.11,9.11){\vector(0,1){6.67}}
\put(3.11,9.11){\vector(1,0){7}}
\put(3.11,17.67){\makebox(0,0)[cc]{$\varphi$}}
\put(11.89,9.11){\makebox(0,0)[cc]{$x$}}
\put(50.11,22.22){\makebox(0,0)[cc]{$0$}}
\put(50.11,41.00){\makebox(0,0)[cc]{$\frac{2\pi}{\beta}$}}
\put(50.61,29.50){\makebox(0,0)[cc]{$\varphi_{0}$}}
\color{black}
\end{picture}
\end{center}
\ca{Vacuum structure}
\label{fig:vacua}
\end{figure}
     
In light of this, we are free to choose \m{\varphi_{0}}
to be in the interval \m{0<\varphi_{0}<\frac{\pi}{\beta}}.
Note also that the topological
charge of the ground state is no longer zero, as in the bulk model,
but
\begin{equation}
q=\frac{\beta}{2\pi}\int_{-\infty}^{0}dx\frac{\partial}{\partial
x}\varphi (x,t)=\frac{\beta}{2\pi}[\varphi (0,t)-\varphi
(-\infty,t)]=\frac{\beta \varphi_{0}}{2\pi}\,,
\end{equation}
with the charge of the first excited state being \m{1-\frac{\beta
\varphi_{0}}{2\pi}}. We will find---at least for the Dirichlet case---that 
all the boundary states have
one of these charges so, for convenience, we shall designate them
simply as 0 and 1 respectively.

\ssse{Breather ground state reflection factors}
For the breather sector, Ghoshal~\cite{Ghoshal} 
obtained the
relevant reflection factors---\m{R^{n}_{\st{0}}(u)} for breather \m{n}
and boundary ground state \m{\st{0}}---from the solitonic
reflection factors using the general boundary bootstrap equation
\cite{FringK,GhoshZam}
\begin{equation}
f^{n}_{i_{1}i_{2}}R^{i_{1}}_{j_{1}\st{x}}\left(u+\frac{u_{n}}{2}\right)
S_{j_{2}f_{1}}^{i_{2}j_{1}}(2u)R^{j_{2}}_{f_{2}\st{x}}\left(u-\frac{u_{n}}{2}
\right)=f^{n}_{f_{1}f_{2}}R^{n}_{\st{x}}(u),
\end{equation}
where \m{u_{n}=\pi - \frac{n\pi}{\la}}, and the \m{R^{a}_{b\st{x}}(u)} are
the solitonic reflection factors, such that \m{R^{+}_{-\st{x}}(u)} is the
factor for a soliton to be reflected back as an anti-soliton and so
on. The \m{f^{n}_{ab}} are the bulk vertices for the creation of
breather \m{n} from (anti-)solitons \m{a} and \m{b}. 
These obey \m{f^{n}_{+-}=(-1)^{n}
f^{n}_{-+}}. The bootstrap is illustrated in \fig{fig:brboot}. 

\begin{figure}
\parbox{2.5in}{
\unitlength 1.00mm
\linethickness{0.4pt}
\begin{picture}(48.00,55.00)
\color{blue}
\multiput(35.00,44.89)(-1,1){10}{\circle*{0.25}}
\color{green}
\put(35.11,44.89){\line(2,-1){10.78}}
\put(45.89,39.22){\line(-2,-1){31.33}}
\put(45.89,23.22){\line(-1,-2){9.61}}
\put(35.22,44.67){\line(1,-2){10.67}}
\color{deepred}
\put(12.22,22.33){\makebox(0,0)[cc]{$f_{1}$}}
\put(35.11,2.78){\makebox(0,0)[cc]{$f_{2}$}}
\put(33.61,42.61){\makebox(0,0)[cc]{$i_{2}$}}
\put(37.94,45.72){\makebox(0,0)[cc]{$i_{1}$}}
\put(41.11,28.67){\makebox(0,0)[cc]{$j_{2}$}}
\put(43.78,36.17){\makebox(0,0)[cc]{$j_{1}$}}
\qbezier(41.67,32.11)(43.78,33.89)(46.00,32.56)
\put(51.00,30.94){\vector(-1,0){7}}
\put(52.00,30.94){\makebox(0,0)[lc]{$u-\frac{u_{n}}{2}$}}
\qbezier(39.78,42.44)(42.00,45.00)(46.00,44.44)
\put(51.00,42.56){\vector(-1,0){7}}
\put(52.00,42.56){\makebox(0,0)[lc]{$u+\frac{u_{n}}{2}$}}
\qbezier(32.89,47.11)(38.89,50.78)(45.89,49.00)
\put(39.11,50.78){\makebox(0,0)[cc]{$u$}}
\qbezier(37.33,40.56)(39.56,40.33)(40.00,42.33)
\put(33.44,38.78){\vector(4,3){4.11}}
\put(31.67,37.67){\makebox(0,0)[cc]{$u_{n}$}}
\color{black}
\put(46.00,3.00){\rule{2.00\unitlength}{52.00\unitlength}}
\end{picture}
} \ \LARGE = \normalsize \
\parbox{2.5in}{
\unitlength 1.00mm
\linethickness{0.4pt}
\begin{picture}(48.00,55.00)
\put(46.00,3.00){\rule{2.00\unitlength}{52.00\unitlength}}
\color{green}
\put(16.56,11.89){\line(2,1){13.22}}
\put(29.78,18.50){\line(-1,-3){4.20}}
\color{blue}
\multiput(29.67,18.44)(1,1){17}{\circle*{0.25}}
\multiput(29.67,51.11)(1,-1){17}{\circle*{0.25}}
\color{deepred}
\qbezier(42.00,38.89)(43.56,40.56)(46.00,39.56)
\put(43.78,41.44){\makebox(0,0)[cc]{$u$}}
\qbezier(24.67,15.89)(25.22,13.56)(28.11,13.33)
\put(24.44,13.00){\makebox(0,0)[cc]{$u_{n}$}}
\put(14.33,10.89){\makebox(0,0)[cc]{$f_{2}$}}
\put(24.22,4.33){\makebox(0,0)[cc]{$f_{1}$}}
\color{black}
\end{picture}
}
\ca{Breather bootstrap}
\label{fig:brboot}
\end{figure}

\se{The boundary Coleman-Thun mechanism}
\label{sec:colethun}
To discover the boundary spectrum, the most natural approach is to
look for simple poles in the reflection factors, which might be
expected to be related to the formation of boundary bound states. 
As we have already mentioned in section \ref{se:colethun}, however, a 
complicating factor is the fact that not all simple poles correspond
to bound states, some having an interpretation as anomalous threshold
singularities.

This problem becomes especially serious once a boundary is involved,
due to the increased complexity of the on-shell diagrams which become
possible. This makes it hard
to be sure that any given pole really does correspond to a new boundary
bound state. In the bulk, a simple geometrical argument shows 
that poles in the S-matrix elements of
the lightest particle can never be explained by a Coleman-Thun
mechanism, and so must always be due to bound states~\cite{BCDS}. 
We wish to find analogous criteria for the boundary
situation. To this end, the following two lemmas turn out to be useful.
Suppose the incoming particle is of type \m{a}, and that its
reflection factor has a simple pole at \m{\T=iu}. 

\begin{lemma}
Let \m{\overline{U}_{a}=\min_{b,c} \left(\pi-U^{c}_{ab}\right)}. 
If \m{u<\overline{U}_{a}}, then the 
the pole at $iu$ cannot be explained by a Coleman-Thun mechanism, and so
must correspond to the binding of particle $a$ to the
boundary, either before or after crossing the outgoing particle.
\label{lemma:1}
\end{lemma}
{\bf Proof: }%
All processes must take the form shown in \fig{fig:lemma1} or the
crossed version shown in \fig{fig:lemma1c}.  Conservation of momentum
demands that all 
rescattering
must take place within the hatched region,
which is drawn 
from
the furthest point from the boundary where either
the incoming or outgoing particle undergoes any interaction.
If neither particle decays, we simply have a diagram of the form of
\fig{fig:sboun} or \fig{fig:uboun}. Otherwise, 
momentum conservation requires
that neither product of the particle which decays on the
boundary of the hatched region has a trajectory which takes it outside
that region. Fixing the notation by \fig{fig:decay} (with
angles \m{U^{b}_{ac}} and \m{U^{c}_{ab}} defined
correspondingly), this reduces to demanding \m{\pi -
U^{c}_{ab}\leq u \leq U^{b}_{ac}}.
If we introduce
\m{\overline{U}_{a}} then we must have \m{\overline{U}_{a} \leq u \leq
\pi - \overline{U}_{a}} (i.e. just \m{u \geq \overline{U}_{a}}, as
\m{u \leq \hp}). Thus, if \m{u < \overline{U}_{a}}, then the only
possible explanations for the pole are \fig{fig:sboun} and
\fig{fig:uboun}.

\begin{lemma}
If the boundary is in its ground state, then lemma~\ref{lemma:1} can
be strengthened, requiring that the incoming particle bind to the
boundary if \m{u} is outside the range
\m{\overline{U}_{a}<u<\hp-\overline{U}_{a}}. In addition, if
\m{\min_{b,c}U^{a}_{bc}>\hp}, the incoming particle must always bind
to the boundary.
\label{lemma:2}
\end{lemma}
{\bf Proof: }%
With the boundary in its ground state,
all rescattering must take
place in the area shown in \fig{fig:lemma2}. Reasoning as before but
demanding that both product particles be emitted into this more
restricted region, we find \m{\pi - U^{c}_{ab}\leq u \leq
U^{b}_{ac}-\frac{\pi}{2}}, or \m{\overline{U}_{a} \leq u
\leq \frac{\pi}{2}-\overline{U}_{a}}. 
In addition, both particles \m{b,c} must be 
emitted into an angle of \m{\hp}, so \m{U^{a}_{bc}<\hp} for at least
one pair of particles $b$, $c$.
If either of these conditions are violated,
then the incoming particle must bind to the boundary.

\begin{figure}
\begin{center}
\parbox{2.5in}{
\centering 
\unitlength 1.00mm
\linethickness{0.4pt}
\begin{picture}(42.00,55.00)(6,0)
\color{deepred}
\multiput(26.22,55.00)(0,-1){52}{\line(0,-1){0.5}}
\multiput(26.22,55.00)(0,-4){13}{\line(6,-1){19.78}}
\multiput(46.00,55.00)(0,-4){13}{\line(-6,-1){19.78}}
\color{green}
\put(6.11,5.33){\line(3,2){20.11}}
\put(6.11,52.56){\line(3,-2){20.11}}
\color{black}
\put(46.00,3.00){\rule{2.00\unitlength}{52.00\unitlength}}
\end{picture}
\ca{General process, with incoming particles uncrossed}
\label{fig:lemma1} } \
\parbox{2.5in}{ 
\centering
\unitlength 1.00mm
\linethickness{0.4pt}
\begin{picture}(42.00,55.00)(6,0)
\color{deepred}
\multiput(26.22,55.00)(0,-1){52}{\line(0,-1){0.5}}
\multiput(26.22,55.00)(0,-4){13}{\line(6,-1){19.78}}
\multiput(46.00,55.00)(0,-4){13}{\line(-6,-1){19.78}}
\color{green}
\put(26.22,18.74){\line(-3,2){20.11}}
\put(26.22,39.15){\line(-3,-2){20.11}}
\color{black}
\put(46.00,3.00){\rule{2.00\unitlength}{52.00\unitlength}}
\end{picture}
\ca{General process, with incoming particles crossed}
\label{fig:lemma1c} }
\end{center}
\end{figure}

\begin{figure}
\begin{center}
\parbox{2.5in}{ 
\centering
\unitlength 1.00mm
\linethickness{0.4pt}
\begin{picture}(42.00,55.00)(6,0)
\color{green}
\put(6.11,5.33){\line(3,2){20.11}}
\put(6.11,52.56){\line(3,-2){20.11}}
\color{deepred}
\put(26.22,18.67){\dashbox{0.5}(19.89,20.44)[cc]{}}
\multiput(26.22,38.61)(0,-4){5}{\line(6,-1){19.78}}
\multiput(46.00,38.61)(0,-4){5}{\line(-6,-1){19.78}}
\color{black}
\put(46.00,3.00){\rule{2.00\unitlength}{52.00\unitlength}}
\end{picture}
\ca{General process when boundary is in ground state}
\label{fig:lemma2} } \
\parbox{2.5in}{ 
\centering
\unitlength 1.00mm
\linethickness{0.4pt}
\begin{picture}(34.00,55.00)(9,0)
\color{green}
\put(27.92,28.31){\line(1,1){15.44}}
\put(27.81,28.31){\line(1,-1){15.44}}
\color{blue}
\put(27.89,28.44){\line(-1,0){19}}
\color{deepred}
\qbezier(32.31,32.53)(34.64,28.42)(32.09,24.19)
\put(40.41,28.31){\makebox(0,0)[cc]{$U^{a}_{bc}$}}
\put(18.31,26.19){\makebox(0,0)[cc]{$a$}}
\put(35.03,38.61){\makebox(0,0)[cc]{$b$}}
\put(35.03,17.59){\makebox(0,0)[cc]{$c$}}
\color{black}
\end{picture}
\ca{Decay process}
\label{fig:decay} 
}
\end{center}
\end{figure}

\vspace{3mm}

These two results, between them, will allow the spectrum of the 
boundary sine-Gordon model to be fixed
completely, provided it is assumed that no pole corresponds to the
creation of a boundary state if it has an alternative (Coleman-Thun)
explanation. 

For the problem under discussion, writing the 
rapidity bounds \m{\overline{U}_a}
as \m{\overline{U}_{+(-)}} for the soliton (anti-soliton)
and 
as \m{\overline{U}_{n}} for the \m{B_{n}}, we have
\ba
\overline{U}_{\pm} &=& \frac{\pi}{2}-\frac{n_{\mathrm{max}}\pi}{2\la}
\nonumber \\
\overline{U}_{n}&=&\frac{\pi}{2\la}~,~~~ n \neq n_{max} \\
\overline{U}_{n_{\mathrm{max}}}&=&\hp-\frac{n_{\mathrm{max}}\pi}{2\la}\,,
\nonumber 
\ea
where \m{B_{n_{max}}} is the highest-numbered breather present in the
model. To derive these results, note that a soliton (anti-soliton)
can only decay into an anti-soliton (soliton) and a breather (with
vertex \m{U^{\pm}_{\mp n}=\hp+\frac{n\pi}{2\la}}). A breather can
either decay into a soliton--anti-soliton pair
(\m{U^{n}_{+-}=\pi-\frac{n\pi}{\la}}) or a pair of breathers
(\m{U^{l}_{nm}=\pi - \frac{l\pi}{2\la}} with \m{n=m+l} or \m{m=n+l}, or 
\m{U^{l}_{nm}=\frac{\pi(n+m)}{2\la}} with \m{l=n+m}).

These restrictions can also be combined to produce a stronger version of
lemma 1 when the incoming particle is a soliton.
If \m{\overline{U}_{+}<u<\frac{\pi}{\la}}, decay within the hatched
region is only possible into
the topmost breather and an anti-soliton. One or other of these particles
will be heading away from the centre of the diagram. If the process in
uncrossed, as in
\fig{fig:lemma1}, the breather will be created heading towards
the centre of the diagram, the anti-soliton away (we are being
somewhat cavalier with the direction of time; this should be
considered as a purely geometric argument). The anti-soliton must itself
obey our lemmas; if in any further  decay before it reaches the boundary
one of the decay products is heading away from the boundary, then there
would be no way to close the diagram while conserving momentum at every
vertex. For a crossed process (\fig{fig:lemma1c}) the breather is
the outermost particle, and is again restricted in its decay by
our lemmas for the same reason.

The anti-soliton created by the uncrossed process
heads for the boundary with a rapidity less than
\m{\overline{U}_{-}} and so, by
lemma~\ref{lemma:1}, may not decay. By the same token, the breather of
the crossed process cannot decay either so, 
if the initial soliton is not to form a bound state, the only possible
alternative processes 
are \fig{fig:fb3} and \fig{fig:fb2}. If these are found not
to occur (for example, if the necessary boundary vertices are not
present) 
then the pole must correspond to a bound state for any \m{u<\frac{\pi}{\la}}.

\se{The Dirichlet case}
\sse{The soliton sector}
The Dirichlet case is exceptional in that topological charge \emph{is} 
conserved and so \m{Q_{\pm}=0}. The remaining factors can be rewritten as
\begin{multline} 
P^{\pm}(u)=R_{0}(u)
\prod_{l=1}^{\infty}\left[ \frac{\g \left(\hf + 2l\la \pm \xp+\ul
\right) \g \left( \hf + (2l-2)\la \mp \xp +\ul \right) }{\g \left( \hf
+ (2l-1)\la +\xp +\ul \right) \g \left( \hf + (2l-1)\la -\xp +\ul
\right)} / \right.\\ \left.(u \rightarrow -u) \right]\,, 
\end{multline} 
where \m{R_{0}(u)} is as before
and \m{\xi = \eta = \frac{4\pi \varphi_{0}}{\beta}}. Taking 
\m{\varphi_{0}} to lie in \m{0 < \varphi_{0} < \frac{\pi}{\beta}}, 
\m{\xi} is in the range
\begin{equation}
0 < \xi < \frac{\pi(\la +1)}{2}\,.
\end{equation}

These factors can again be written in terms of Barnes' multiperiodic
functions, as
\begin{equation}
P^{\pm}(u)=R_{0}(u)\frac{S_{2}\left(\frac{\pi}{2\la} \mp \xl+\pi+u \left|
\pl, 2\pi\right. \right)
S_{2}\left(\frac{\pi}{2\la} \mp \xl-u \left|
\frac{\pi}{\la},2\pi\right.\right)}
{S_{2}\left(\frac{\pi}{2\la} \mp \xl+\pi-u\left|\frac{\pi}{\la},2\pi\right.
\right)S_{2}\left(\frac{\pi}{2\la} \mp \xl+u\left|
\frac{\pi}{\la},2\pi\right. \right)},
\end{equation}
with
\begin{equation}
R_{0}(u)=\frac{S_{2}\left(\hp-u \left| \frac{\pi}{2\la},2\pi
\right. \right) S_{2}\left(\frac{\pi}{2\la}+u \left|
\frac{\pi}{2\la},2\pi \right. \right)}{S_{2}\left(\hp+u \left|
\frac{\pi}{2\la},2\pi \right. \right) S_{2}\left(\frac{\pi}{2\la}-u
\left| \frac{\pi}{2\la},2\pi \right. \right)}.
\end{equation}
\sse{The breather sector}
In the Dirichlet case, with topological
charge conserved, the bootstrap equation reduces to
\begin{equation}
f^{n}_{i_{1}i_{2}}P^{i_{1}}_{\st{x}}\left(u+\frac{u_{n}}{2}\right)S^{i_{2}
i_{1}}_{f_{2}f_{1}}(2u)P^{f_{2}}_{\st{x}}\left(u-\frac{u_{n}}{2}\right)=
f^{n}_{f_{1}f_{2}}R^{n}_{\st{x}}(u). 
\end{equation} 

Ghoshal found that, for the boundary ground state, the breather
reflection factors were
\begin{equation}
R^{n}_{\st{0}}(u)=R^{(n)}_{0}(u)R^{(n)}_{1}(u),
\label{gsrfl}
\end{equation}
where
\begin{equation} 
R_{0}^{(n)}(u)=\frac{\left(\hf\right)
\left(\frac{n}{2\la}+1\right)}{\left( \frac{n}{2\la} +
\frac{3}{2}\right)} \prod_{l=1}^{n-1}\frac{ \left( \frac{l}{2\la}
\right) \left( \frac{l}{2\la} +1 \right)}{ \left(
\frac{l}{2\la}+\frac{3}{2}\right)^{2}}, 
\end{equation} 
and 
\begin{equation}
R_{1}^{(n)}(u)=\prod_{l=\frac{1-n}{2}}^{\frac{n-1}{2}}\frac{\left(
\frac{\xi}{\la \pi}-\hf +\frac{l}{2\la}\right)}{\left( \frac{\xi}{\la
\pi} +\hf + \frac{l}{2\la}\right)}.  
\end{equation}
This makes use of the notation
\begin{equation}
(x)=\frac{\sinh \left( \frac{\T}{2}+\frac{i\pi x}{2}\right)}{
\sinh \left( \frac{\T}{2}-\frac{i\pi x}{2} \right)},
\end{equation}
which will also be helpful later.

\se{Initial pole analysis}
\label{sec:initpole}
\sse{Solitonic ground state factors}
The \m{R_{0}(u)} factor is insensitive to the boundary parameters, and so all
its poles should be explicable in terms of the bulk. The only 
poles are at \m{u=\frac{N\pi}{2\la}}, where
\m{N=1,2,3,\ldots}, with no zeroes. These can be explained by the
creation of a breather which is incident perpendicularly on the boundary,
as shown in \fig{fig:bbre}. Here, 
as in all subsequent diagrams, the time axis points up the page,
and the \m{x} axis points to the right. Solitons and anti-solitons
are drawn as solid lines, while breathers are drawn as dotted
lines.

\begin{figure}
\begin{center}
\parbox{0.39\textwidth}{
\centering
\unitlength 1.00mm
\linethickness{0.4pt}
\begin{picture}(35.5,55.00)(12.5,0)
\put(46.00,3.00){\rule{2.00\unitlength}{52.00\unitlength}}
\color{green}
\put(27.92,28.31){\line(-1,1){15.44}}
\put(27.81,28.31){\line(-1,-1){15.44}}
\color{blue}
\multiput(27.89,28.44)(1,0){5}{\circle*{0.25}}
\color{deepred}
\qbezier(23.47,32.53)(21.14,28.42)(23.69,24.19)
\put(15.37,28.31){\makebox(0,0)[cc]{$\pi-\frac{N\pi}{\lambda}$}}
\put(37.47,26.19){\makebox(0,0)[cc]{$B_{N}$}}
\color{blue}
\multiput(33.00,28.44)(1,0){14}{\circle*{0.25}}
\color{black}
\end{picture}
\ca[$\xi$-independent pole]{\phantom{poles}\newline
$\pmb{\xi}$-independent pole} 
\label{fig:bbre} }
\parbox{0.29\textwidth}{
\centering
\unitlength 1.00mm
\linethickness{0.4pt}
\begin{picture}(17.75,55.00)(30.25,0)
\put(46.00,3.00){\rule{2.00\unitlength}{52.00\unitlength}}
\color{green}
\put(30.25,5.53){\line(1,1){15.56}}
\put(45.69,37.08){\line(-1,1){15.56}}
\color{red}
\multiput(45.78,21.44)(0,1){16}{\circle*{0.25}}
\color{black}
\end{picture}
\ca[Bound state]{\phantom{Bound}\newline Bound state}
\label{fig:sboun} } \
\parbox{0.29\textwidth}{
\centering
\unitlength 1.00mm
\linethickness{0.4pt}
\begin{picture}(22.00,55.00)(26.00,0)
\put(46.00,3.00){\rule{2.00\unitlength}{52.00\unitlength}}
\color{red}
\multiput(45.67,49.00)(0,1){6}{\circle*{0.25}}
\multiput(45.67,4.00)(0,1){6}{\circle*{0.25}}
\multiput(45.67,11.00)(0,2){19}{\circle*{0.25}}
\color{green}
\put(45.89,48.89){\line(-1,-2){20.06}}
\put(45.89,9.44){\line(-1,2){20.06}}
\color{black}
\end{picture}
\ca[Crossed process]{\phantom{Crossed}\newline Crossed process}
\label{fig:uboun} }
\end{center}
\end{figure}

Turning now to \m{\xi}-dependent poles and zeroes, we find zeroes at
\begin{equation} 
u= -\xl +
\frac{(2n+1)\pi}{2\la},
\label{eq:basiczeroes}
\end{equation} 
where \m{n=0,1,2,\ldots}, for \m{P^{+}}, and at the same rapidities but
with \m{\xi \rightarrow -\xi} for \m{P^{-}}. There are also poles in
\m{P^{+}} only at \m{u=\nu_{n}}, with
\begin{equation}
\fbox{\,$\nu_{n}=\xl - \frac{(2n+1)\pi}{2\la_{\phantom{l}}}$\,}
\end{equation} 
A soliton can only decay into an anti-soliton and a breather, with a
rapidity difference between the two of \m{\hp+\frac{b\pi}{2\la}} for
breather \m{b}. Thus, by lemma~\ref{lemma:2}, all these poles must
correspond to bound states, as shown in \fig{fig:sboun}. For reasons
which will become clear 
in a moment, we
shall depart from the convention of \cite{Skorik} and, rather
than labelling the state corresponding to pole \m{\nu_{n}} as
\m{\beta_{n}}, will label it according to topological charge and \m{n}
as \m{\st{1;n}}. 

\sse{Solitonic excited state reflection factors}
Using the boundary bootstrap
equations given in~\cite{GhoshZam}---which come from considering
\fig{fig:bootstrap}---solitonic reflection
factors can be calculated for this first set of bound states. In our
case, these equations read
\begin{equation}
P^{b}_{\st{y}}(u)=
\sum_{c,d}P^{d}_{\st{x}}(u)S^{ab}_{cd}(u-\alpha_{ax}^{y})S^{dc}_{ba}
(u+\alpha_{ax}^{y}),
\end{equation}
where \m{a,b,c}, and \m{d} take the values \m{+} or \m{-} and
\m{\alpha_{ax}^{y}} is the (imaginary) rapidity of the pole at which
particle \m{a} binds to boundary state \m{\st{x}} to give state
\m{\st{y}}. 
The mass of state \m{\st{y}}---\m{m_{y}}---is given by
\begin{equation}
m_{y}=m_{x}+m_{s}\cos \alpha_{ax}^{y}\,.
\end{equation}

\begin{figure}
\begin{center}
\parbox{2.5in}{
\unitlength 1.00mm
\linethickness{0.4pt}
\begin{picture}(48.00,55.00)
\put(46.00,3.00){\rule{2.00\unitlength}{52.00\unitlength}}
\color{green}
\put(30.25,5.53){\line(1,1){15.56}}
\put(45.69,37.08){\line(-1,1){15.56}}
\put(17.89,6.22){\line(4,1){28.00}}
\put(17.89,20.22){\line(4,-1){28.00}}
\color{red}
\multiput(45.78,21.44)(0,1){16}{\circle*{0.25}}
\color{deepred}
\put(28.25,4.53){\makebox(0,0)[cc]{$a$}}
\put(15.89,5.22){\makebox(0,0)[cc]{$b$}}
\put(15.89,21.22){\makebox(0,0)[cc]{$b$}}
\put(28.13,53.64){\makebox(0,0)[cc]{$a$}}
\put(34.75,13.03){\makebox(0,0)[cc]{$c$}}
\put(40.89,9.22){\makebox(0,0)[cc]{$d$}}
\put(42.00,20.00){\makebox(0,0)[cc]{$a$}}
\put(51.00,6.00){\makebox(0,0)[cc]{$\st{x}$}}
\put(51.00,52.00){\makebox(0,0)[cc]{$\st{x}$}}
\put(51.00,29.00){\makebox(0,0)[cc]{$\st{y}$}}
\put(54.00,13.22){\makebox(0,0)[cc]{$P^{d}_{\st{x}}(u)$}}
\color{black}
\end{picture} } \ \LARGE = \normalsize \
\parbox{2.5in}{
\unitlength 1.00mm
\linethickness{0.4pt}
\begin{picture}(48.00,55.00)
\put(46.00,3.00){\rule{2.00\unitlength}{52.00\unitlength}}
\color{green}
\put(30.25,5.53){\line(1,1){15.56}}
\put(45.69,37.08){\line(-1,1){15.56}}
\put(17.89,22.00){\line(4,1){28.00}}
\put(17.89,36.00){\line(4,-1){28.00}}
\color{red}
\multiput(45.78,21.44)(0,1){16}{\circle*{0.25}}
\color{deepred}
\put(28.25,4.53){\makebox(0,0)[cc]{$a$}}
\put(15.89,36.78){\makebox(0,0)[cc]{$b$}}
\put(15.89,21.22){\makebox(0,0)[cc]{$b$}}
\put(28.13,53.64){\makebox(0,0)[cc]{$a$}}
\put(51.00,6.00){\makebox(0,0)[cc]{$\st{x}$}}
\put(51.00,52.00){\makebox(0,0)[cc]{$\st{x}$}}
\put(51.00,23.00){\makebox(0,0)[cc]{$\st{y}$}}
\put(54.00,29.00){\makebox(0,0)[cc]{$P^{b}_{\st{y}}(u)$}}
\color{black}
\end{picture} }
\ca{Boundary bound-state bootstrap}
\label{fig:bootstrap}
\end{center}
\end{figure}

Taking \m{x} to be the ground state \m{\st{0}} and \m{y} to be one
of the set of excited states \m{\st{1;n}}, this gives
\ba
P^{+}_{\st{1;n}}(u)&=&P^{+}_{\st{0}}(u)a(u-\nu_{n})a(u+\nu_{n}) \nn\\
P^{-}_{\st{1;n}}(u)&=&P^{-}_{\st{0}}(u)b(u-\nu_{n})b(u+\nu_{n})+ 
P^{+}_{\st{0}}(u)c(u-\nu_{n})c(u+\nu_{n}).
\label{eq:firstboot}
\ea 
Note that
\m{P^{\pm}_{\st{1;0}}(u)=\overline{P^{\mp}_{\st{0}}(u)}},
where \m{\overline{P^{\pm}(u)}} is \m{P^{\pm}(u)} under the transformation
\m{\xi \rightarrow \pi (\la +1) - \xi}.  The reason for
this is clear
if we look back at \fig{fig:vacua}; this transformation is equivalent
to reflecting the diagram in the horizontal axis, interchanging the
ground and first excited states. 

Perhaps the neatest way to write the new reflection factors is
\begin{equation}
P^{\pm}_{\st{1;n}}(u) = \overline{P^{\mp}(u)}a^{1}_{n}(u),
\label{eq:newreff}
\end{equation}
where
\begin{equation}
a^{1}_{n}(u)=\frac{a(u+\nu_{n})a(u-\nu_{n})}{a(u+\nu_{0})a(u-\nu_{0})}.
\end{equation}
The factor \m{a^{1}_{n}(u)} simplifies to
\begin{equation}
a^{1}_{n}(u)=\prod_{x=1}^{n}\frac{\left( \frac{\xi}{\la
\pi}+\frac{1}{2\la} -\frac{x}{\la} \right)\left( \frac{\xi}{\la \pi} -
\frac{1}{2\la}-\frac{x}{\la}\right)}{ \left( \frac{\xi}{\la
\pi}+\frac{1}{2\la} - \frac{x}{\la} +1\right) \left( \frac{\xi}{\la
\pi} - \frac{1}{2\la} - \frac{x}{\la} +1 \right)}.
\end{equation}

Looking at the pole structure, we find that the functions
\m{\overline{P^{\pm}(u)}} have common simple poles at \m{\nu_{0}} and
\m{\nu_{-N}} where \m{N=1,2,3,\ldots}\,. In addition,
\m{\overline{P^{+}(u)}} has simple poles at 
\m{u=w_{N'}}, where
\begin{equation}
\fbox{\,$w_{N'}=\pi - \xl -
\frac{\pi(2N'-1)}{2\la_{\phantom{l}}}=\overline{\nu_{N'}}$\,}
\end{equation}
and \m{N'=1,2,3,\ldots}, and simple zeroes at \m{-w_{N''}} for appropriate
values of \m{N''}. Finally, \m{a^{1}_{n}(u)} has simple poles at
\m{\nu_{0}} and \m{\nu_{n}}, and double poles at \m{\nu_{k}},
\m{k=1,2,\ldots,n-1}.

Before proceeding to a more rigorous discussion, we shall now digress
to give an outline of how the bootstrap might 
be expected to work. If one of these poles \emph{does} correspond to 
a new bound state, factorisability leads us to expect that moving the 
soliton and anti-soliton trajectories past each other (so that the 
anti-soliton is incident on the boundary first) should also create the 
same state. The most obvious explanation for this would 
be for the anti-soliton to bind to the boundary first, followed by the 
soliton, to form the state. From above, however, only solitons can bind 
to the ground state, so we must look further.

The \emph{next} most obvious way this could happen is via the soliton 
and anti-soliton forming a breather, either before or after the 
anti-soliton has reflected from the boundary. The poles required to 
allow the first process (of the form \m{\pi +\xl - 
\frac{\pi(2m+1)}{2\la}}) are not present, whereas those necessary 
for the second (of the form \m{w_{m}}) are. Our candidate process 
therefore becomes
\fig{fig:bresoli}, where the soliton and anti-soliton bind to form a 
breather, which then creates the state in one step. It is quite 
difficult to imagine any further alternatives, so let us---for the 
moment---take the existence of such a process as a necessary condition 
for a pole to be responsible for the formation of a boundary state.
\begin{figure}
\begin{center}
\parbox{2.5in}{
\unitlength 1.00mm
\linethickness{0.4pt}
\begin{center}
\begin{picture}(55.00,55.00)
\put(46.00,3.00){\rule{2.00\unitlength}{52.00\unitlength}}
\color{green}
\put(25.33,4.60){\line(1,1){20.56}}
\put(25.33,53.29){\line(1,-1){20.56}}
\put(25.33,49.11){\line(2,-1){20.56}}
\put(25.33,9.22){\line(2,1){20.56}}
\color{red}
\multiput(45.78,25.44)(0,1){8}{\circle*{0.25}}
\color{deepred}
\put(23.33,3.50){\makebox(0,0)[cc]{$a$}}
\put(23.33,54.29){\makebox(0,0)[cc]{$a$}}
\put(23.33,8.22){\makebox(0,0)[cc]{$b$}}
\put(23.33,50.11){\makebox(0,0)[cc]{$b$}}
\put(41.33,16.22){\makebox(0,0)[cc]{$c$}}
\put(41.33,43.29){\makebox(0,0)[cc]{$c$}}
\put(41.33,23.60){\makebox(0,0)[cc]{$d$}}
\put(41.33,34.00){\makebox(0,0)[cc]{$d$}}
\put(51.00,6.00){\makebox(0,0)[cc]{$\alpha$}}
\put(51.00,52.00){\makebox(0,0)[cc]{$\alpha$}}
\put(52.00,25.16){\makebox(0,0)[cc]{$g^{\beta}_{\gamma d}$}}
\put(52.00,19.50){\makebox(0,0)[cc]{$g^{\alpha}_{\beta c}$}}
\put(52.00,34.73){\makebox(0,0)[cc]{$x$}}
\put(52.00,39.83){\makebox(0,0)[cc]{$y$}}
\qbezier(46.00,35.73)(45.00,37.73)(43.00,35.73)
\qbezier(46.00,40.83)(45.00,42.83)(43.00,40.83)
\put(50.50,34.73){\vector(-1,0){6}}
\put(50.50,39.83){\vector(-1,0){6}}
\color{black}
\end{picture}
\end{center} } \ \LARGE = \normalsize \
\parbox{2.5in}{
\begin{center}
\unitlength 1.00mm
\linethickness{0.4pt}
\begin{picture}(48.00,55.11)(24.00,0)
\put(46.00,3.00){\rule{2.00\unitlength}{52.00\unitlength}}
\color{red}
\multiput(45.78,25.44)(0,1){8}{\circle*{0.25}}
\color{green}
\put(37.78,3.33){\line(1,1){8.22}}
\put(45.89,11.56){\line(-1,1){5.22}}
\put(25.33,9.22){\line(2,1){15.33}}
\put(37.78,55.11){\line(1,-1){8.22}}
\put(45.89,46.89){\line(-1,-1){5.22}}
\put(25.33,49.22){\line(2,-1){15.33}}
\color{blue}
\multiput(40.67,41.67)(1,-1.66){6}{\circle*{0.25}}
\multiput(40.67,16.78)(1,1.66){6}{\circle*{0.25}}
\color{deepred}
\put(35.78,2.33){\makebox(0,0)[cc]{$b$}}
\put(23.33,8.22){\makebox(0,0)[cc]{$a$}}
\put(23.33,50.22){\makebox(0,0)[cc]{$a$}}
\put(35.78,56.11){\makebox(0,0)[cc]{$b$}}
\put(42.00,23.00){\makebox(0,0)[cc]{$n$}}
\qbezier(46.00,8.55)(45.00,6.55)(43.00,8.55)
\put(50.50,9.55){\vector(-1,0){6}}
\put(52.00,9.55){\makebox(0,0)[cc]{$y$}}
\qbezier(46.00,21.74)(45.00,20.00)(43.50,21.24)
\put(50.50,22.74){\vector(-1,0){6}}
\put(56.00,22.74){\makebox(0,0)[cc]{$x - y$}}
\color{black}
\end{picture}
\end{center} }
\ca{States can be created either by breathers or solitons}
\label{fig:bresoli}
\end{center}
\end{figure}

The consequence of this is that the \m{w_{N}} poles are selected 
as the only possible candidates, and it appears that new bound states can 
only be formed by anti-solitons. Such states hence have charge 0 
(agreeing with the idea that they can also be formed from the ground 
state by the action of a breather). In addition, it is also clear that 
only those \m{w_{N}} such that \m{w_{N}<\nu_{n}} can be considered, as, 
otherwise, the breather version of the process would see the breather 
created heading away from the boundary, rather than towards it.

Designating such a new state as \m{\st{0;n,N}} and bootstrapping on it
leads to
\ba
P^{-}_{\st{0;n,N}}(u)&=&P^{-}_{\st{1;n}}(u)a(u-w_{N})a(u+w_{N}) \\
P^{+}_{\st{0;n,N}}(u)&=&P^{+}_{\st{1;n}}(u)b(u-w_{N})b(u+w_{N})+ 
P^{-}_{\st{1;n}}(u)c(u-w_{N})c(u+w_{N})\,. \nn
\ea 
Substituting in \re{eq:newreff} and taking advantage of the fact that 
\m{w_{N}=\overline{\nu_{N}}} (so \m{a(u\pm w_{N})=\overline{a(u\pm 
\nu_{N})}}), this becomes
\ba
P^{-}_{\st{0;n,N}}(u)&=&a_{n}^{1}(u)\overline{P^{+}_{\st{0}}(u)}
a(u-\overline{\nu_{N}})a(u+\overline{\nu_{N}}) \\
P^{+}_{\st{0;n,N}}(u)&=&a_{n}^{1}(\overline{P^{-}_{\st{0}}(u)}
b(u-\overline{\nu_{N}})b(u+\overline{\nu_{N}})+ 
\overline{P^{+}_{\st{0}}(u)}c(u-\overline{\nu_{N}})c(u+\overline{\nu_{N}}))\,,
\nn
\ea
which (apart from an extra factor of \m{a_{n}^{1}(u)}) is just the first 
bootstrap \re{eq:firstboot} under the transformation \m{\xi \rightarrow 
\pi(\la+1)-\xi} and with solitons and anti-solitons interchanged on the 
lhs. Thus, the pole structure follows naturally from the above. This can 
also be written as
\begin{equation}
P^{\pm}_{\st{0;n,N}}(u) = P^{\pm}_{\st{0}}(u)a_{n}^{1}(u)
\overline{a_{N}^{1}(u)}.
\end{equation}

Repeating the factorisation argument shows that now we should focus on 
\m{\nu_{n'}} poles such that \m{\nu_{n'}<w_{N}}. These are present now 
in the solitonic factor, though (due to the extra factor of 
\m{a_{n}^{1}(u)}) only for \m{n'>n}. However, since any such state obeys 
\m{\nu_{n}>w_{N}>\nu_{n'}} in any case, this restriction is not relevant.
The resultant state must now have charge 0.

A pattern is emerging, and it is not hard to see how the process 
would continue. Starting from the ground state, and taking the broadest 
guess (given our assumptions) for the spectrum, states can be formed by 
alternating solitons and anti-solitons, the solitons having rapidity 
\m{\nu_{n_{i}}} and the anti-solitons having rapidity \m{w_{N_{j}}} (for 
some sets \m{\underline{n}} and \m{\underline{N}}). 
An schematic pole structure is shown in \fig{fig:poles}, in terms of which 
the criterion
for a state to be in the spectrum should be that we begin with one of the
\m{\nu_{n}} and then, as we move along the index list, move down
the diagram, switching from side to side as we go. If we finish on a
\m{\nu_{m}} (indicating that the most recent particle to bind was a
soliton) the state has charge 1 while, if we finish on a \m{w_{m}}
(meaning an anti-soliton) the state has charge 0.
\begin{figure}
\begin{center}
\unitlength=1.00mm
\begin{picture}(82.35,53.75)(15.00,0.00)
\color{deepred}
\put(40.00,53.75){\line(0,-1){53.75}}
\put(20.00,53.75){\line(0,-1){53.75}}
\put(19.00,53.75){\line(1,0){2}}
\put(19.00,0.00){\line(1,0){2}}
\put(16.00,52.75){$\hp$}
\put(16.00,-1.00){$0$}
\put(16.00,25.87){$u$}
\multiput(30.00,47.00)(1,0){21}{\circle*{0.25}}
\put(52.00,47.00){\makebox(0,0)[cc]{$\xl$}}
\put(58.00,32.00){\makebox(0,0)[cl]{$\nu_{n}=\xl-\frac{\pi(2n+1)}{2\la}$}}
\put(58.00,23.00){\makebox(0,0)[cl]{$w_{m}=\pi-\xl-\frac{\pi(2m-1)}{2\la}$}}
\color{blue}
\multiput(40.00,42.00)(0,-10){5}{\line(1,0){5}}
\multiput(40.00,42.00)(0,-10){5}{\circle*{2}}
\put(46.00,41.00){$\nu_{0}$}
\put(46.00,31.00){$\nu_{1}$}
\put(46.00,21.00){$\nu_{2}$}
\put(46.00,11.00){$\nu_{3}$}
\put(46.00,1.00){$\nu_{4}$}
\color{red}
\multiput(40.00,49.00)(0,-10){5}{\line(-1,0){5}}
\multiput(40.00,49.00)(0,-10){5}{\circle*{2}}
\put(30.00,48.00){$w_{2}$}
\put(30.00,38.00){$w_{3}$}
\put(30.00,28.00){$w_{4}$}
\put(30.00,18.00){$w_{5}$}
\put(30.00,8.00){$w_{6}$}
\color{black}
\end{picture}
\ca[Location of poles.]{Location of poles. \mdseries \itshape (Note that, in this case, 
$w_{2}$ can never participate in bound state formation as it is above 
$\nu_{0}$.)}
\label{fig:poles} 
\end{center}
\end{figure}     

Annotating such a state by its topological charge, \m{c}, and the sets 
\m{\underline{n}} and \m{\underline{N}} as 
\m{\st{c;n_{1},N_{1},n_{2},\linebreak
N_{2},\ldots}} (noting
\m{\nu_{n_{1}}>w_{N_{1}} >\nu_{n_{2}} >w_{N_{2}}>\ldots}), the solitonic 
reflection factors should be
\begin{equation}
P^{\pm}_{\st{c;n_{1},N_{1},\ldots}}(u)=P^{\pm}_{(c)}(u)a^{1}_{n_{1}}(u)
\overline{a^{1}_{N_{1}}(u)}\ldots\,, 
\end{equation}
with \m{P^{\pm}_{0}(u)=P^{\pm}_{\st{0}}(u)} and \m{P^{\pm}_{1}(u)=
\overline{P^{\pm}_{\st{0}}(u)}}. From now on, however, it will be more
convenient to consider a single index list, and denote 
\m{\overline{a^{1}_{m}(u)}} as \m{a^{0}_{m}(u)}, giving
\begin{equation}
P^{\pm}_{\st{c;n_{1},n_{2},\ldots,n_{k}}}(u)=P^{\pm}_{(c)}(u)a^{1}_{n_{1}}(u)
a^{0}_{n_{2}}(u)a^{1}_{n_{3}}(u)\ldots a^{c}_{n_{k}}(u), 
\end{equation} 
where \m{k} is odd if \m{c} is 1 and \m{k} is even if \m{c} is 0.
We will call 
this a level \m{k} boundary bound state.
If we choose the ground state mass to be 
\m{m_s\sin^{2}\left(\frac{\xi-\hp}{2\la}\right)}, the mass of this
state is
\begin{align}\label{eq:mass}
m_{n_{1},n_{2},\ldots}&=
m_s\sin^{2}\left(\frac{\xi-\hp}{2\la}\right)+
\sum_{i \mathrm{\ odd}} m_s\cos(\nu\phup_{n_i})
+ \sum_{j
\mathrm{\ even}} m_s\cos(w\phup_{n_j})\\[4pt]
&=
m_s\sin^{2}\left(\frac{\xi-\hp}{2\la}\right)+
\sum_{i \mathrm{\ odd}} m_s\cos \left(
\xl - \frac{(2n_{i}+1)\pi}{2\la}\right) \\
& \quad - \sum_{j
\mathrm{\ even}} m_s\cos \left( \xl+
\frac{(2n_{j}-1)\pi}{2\la} \right)\,.
\end{align}
This choice is convenient in that, as \m{\xi} passes \m{\pi/\beta},
the masses of the ground and first excited states interchange, in line
with the idea that the states themselves swap at this point.
An important point to note is that, in deriving all this, we have simply 
been considering the soliton sector. However, we will see that allowing
breather processes as well does not give rise to any further states, 
merely additional ways to jump between states. The Dirichlet boundary 
condition is also special in that either the soliton or the anti-soliton 
can couple to a given boundary, but not both, as might be generically 
expected.

Although we have built up the states by applying the solitons and
anti-solitons in this alternating fashion, precisely how this happens in a
given situation
will of course depend on the impact
parameters of the incoming particles. 
In \fig{fig:bresoli} we already gave an example of the
complicated way in which a process may be rearranged as these impact
parameters vary, and the particular choices that we have adopted are
mainly motivated by a desire to assemble the full spectrum in the simplest
possible way.

\sse{Breather ground state reflection factors}
We now return to the pole analysis, and examine the breather ground state
reflection factors (\ref{gsrfl}).
Again, the factor \m{R^{n}_{0}} is boundary-independent, and so all 
its poles should have an explanation in terms of the bulk. There are 
(physical strip) poles at \m{\hp}, \m{\frac{l\pi}{2\la}}, 
\m{\hp-\frac{n\pi}{2\la}}, and double poles at \m{\hp-\frac{l\pi}{2\la}}, 
with \m{l=1,2,\ldots,n-1}. There are no zeroes.
The pole at \m{\hp} is simply due to the breather coupling
perpendicularly to the boundary, while the poles at 
\m{\frac{l\pi}{2\la}} are explained by \fig{fig:bretri}. Next, the pole at 
\m{\hp-\frac{n\pi}{2\la}} comes from a breather version of 
\fig{fig:bbre}, \m{B_{2n}} being formed. Finally, the double poles at 
\m{\hp - \frac{l\pi}{2\la}} are due to \fig{fig:bredoubletri}.

\begin{figure}
\begin{center}
\parbox{2.5in}{ 
\centering
\unitlength 1.00mm
\linethickness{0.4pt}
\begin{picture}(38.00,55.00)(10.00,0)
\put(46.00,3.00){\rule{2.00\unitlength}{52.00\unitlength}}
\color{blue}
\multiput(46.00,29.00)(-1,-0.5){21}{\circle*{0.25}}
\multiput(46.00,29.00)(-1,0.5){21}{\circle*{0.25}}
\multiput(26.00,19.00)(0,1){21}{\circle*{0.25}}
\multiput(26.00,19.00)(-1,-1){16}{\circle*{0.25}}
\multiput(26.00,39.00)(-1,1){16}{\circle*{0.25}}
\color{deepred}
\put(20.00,29.00){\makebox(0,0)[cc]{$n-l$}}
\put(36.00,21.00){\makebox(0,0)[cc]{$l$}}
\put(36.00,37.00){\makebox(0,0)[cc]{$l$}}
\put(18.50,8.50){\makebox(0,0)[cc]{$n$}}
\put(18.50,49.50){\makebox(0,0)[cc]{$n$}}
\qbezier(26.00,23.00)(29.00,24.00)(30.00,21.00)
\put(31.00,17.00){\vector(-1,1){4}}
\put(34.00,14.50){\makebox(0,0)[cc]{$\frac{n\pi}{2\la}$}}
\color{black}
\end{picture}
\ca[Breather triangle process]{Breather triangle \newline process}
\label{fig:bretri} } \
\parbox{2.5in}{ 
\centering
\unitlength 1.00mm
\linethickness{0.4pt}
\begin{picture}(37.00,55.00)(11.00,0)
\put(46.00,3.00){\rule{2.00\unitlength}{52.00\unitlength}}
\color{blue}
\multiput(31.00,44.00)(1,0){16}{\circle*{0.25}}
\multiput(31.00,14.00)(1,0){16}{\circle*{0.25}}
\multiput(31.00,44.00)(1,-1){16}{\circle*{0.25}}
\multiput(31.00,14.00)(1,1){16}{\circle*{0.25}}
\multiput(31.00,14.00)(-1,-0.5){21}{\circle*{0.25}}
\multiput(31.00,44.00)(-1,0.5){21}{\circle*{0.25}}
\color{deepred}
\put(38.5,47.00){\makebox(0,0)[cc]{$n-l$}}
\put(38.5,11.00){\makebox(0,0)[cc]{$n-l$}}
\put(38.5,33.5){\makebox(0,0)[cc]{$l$}}
\put(38.5,24.5){\makebox(0,0)[cc]{$l$}}
\put(21.00,52.00){\makebox(0,0)[cc]{$n$}}
\put(21.00,6.00){\makebox(0,0)[cc]{$n$}}
\qbezier(34.00,17.00)(36.50,16.50)(35.00,14.00)
\put(29.00,19.00){\vector(1,-1){4}}
\put(26.00,22.00){\makebox(0,0)[cc]{$\frac{n\pi}{2\la}$}}
\color{black}
\end{picture}
\ca[Breather double triangle process]{Breather double \newline triangle
process}
\label{fig:bredoubletri} }
\end{center}
\end{figure}

Moving on to the boundary-dependent part, there are 
poles at
\begin{equation}
u=\xl - \hp \pm \frac{l\pi}{2\la},
\end{equation}
and zeroes at
\ba
u&=&-\xl +\hp \pm \frac{l\pi}{2\la} \nonumber\\
u&=&\xl +\hp \pm \frac{l\pi}{2\la}\,,
\ea
where, for a breather \m{n}, \m{l=n-1, n-3,\ldots, l \geq 0}.

The set of poles can be re-written by noting
that, for breather \m{m}, there is a simple pole of the form
\m{\hf(\nu_{n}-w_{N})} for all \m{n,N \geq 0} and \m{n,N \in Z}
such that \m{m=n+N}. This ties in with the discussion in the previous 
section, since these are the rapidities predicted for the single-step 
process which is equivalent to a soliton binding at an angle of 
\m{\nu_{n}} followed by an anti-soliton at \m{w_{N}}.

\sse{Breather excited state reflection factors}
Following the discussion of the solitonic excited state
reflection factors, we can introduce corresponding breather reflection 
factors:
\begin{equation}
R^{m}_{\st{c;n_{1},n_{2},\ldots,n_{k}}}(u)=R^{m}_{(c)}(u)a^{1;m}_{n_{1}}(u)
a^{0;m}_{n_{2}}(u)a^{1;m}_{n_{3}}(u)\ldots a^{c;m}_{n_{k}}(u), 
\label{eq:breexc}
\end{equation} 
where
\m{R^{m}_{0}(u)=R^{m}_{\st{0}}(u)} and \m{R^{m}_{1}(u)= \overline{R^{m}_{
\st{0}}(u)}}. We have also defined 
\begin{equation}
a^{c;m}_{n}(u)=a_{n}^{c}\left(u+\frac{u_{m}}{2}\right)a_{n}^{c}\left(u
-\frac{u_{m}}{2}\right),
\label{eq:simpleacn}
\end{equation}
or
\begin{multline}
a^{1;m}_{n}(u)=\prod_{x=1}^{m}\left[\frac{\left( \frac{\xi}{\la
\pi}+\frac{1-2x-n}{2\la}+\hf\right)\left( \frac{\xi}{\la
\pi}-\frac{1+2x+n}{2\la}+\hf\right)}{ \left( \frac{\xi}{\la
\pi}+\frac{1-2x-n}{2\la}-\hf\right) \left( \frac{\xi}{\la \pi} -
\frac{1+2x+n}{2\la} - \hf\right)} \right. \times\\
\left.\frac{\left(\frac{\xi}{\la \pi}+
\frac{1-2x+n}{2\la} -\hf\right)\left( \frac{\xi}{\la \pi}
-\frac{1+2x-n}{2\la} -\hf \right)}{\left( \frac{\xi}{\la \pi}+
\frac{1-2x+n}{2\la} +\hf \right)\left( \frac{\xi}{\la \pi} - \frac{1+
2x-n}{2\la} + \hf \right)}\right]\,,
\label{eq:complexacn}
\end{multline}
with \m{a^{0;m}_{n}(u)=\overline{a^{1;m}_{n}(u)}}.

For \m{\overline{R^{m}_{\st{0}}(u)}}, there are poles at
\ba
u&=&\hp-\xl+\frac{\pi}{\la}\pm\frac{l\pi}{2\la} \nonumber \\
u&=&\hp+\xl-\frac{(l+2)\pi}{2\la}\,,
\ea
and zeroes at
\begin{equation}
u=\xl-\hp+\frac{(l-2)\pi}{2\la}\,.
\end{equation}

For the other factors, \m{a^{1;m}_{n}(u)} has
physical strip poles/zeroes at
\begin{equation}
\begin{array}{crl}
u = -\xl + \hp + \frac{p\pi}{2\la} & \mathrm{poles:\
} & p = 2n-m+2x \pm 1 \\
 & \mathrm{zeroes:\ } & p = - m +2x \pm 1 \\
u = \xl - \hp + \frac{p\pi}{2\la} & \mathrm{poles:\
} & p = m-2x\pm 1 \\
 & \mathrm{zeroes:\ } & p = -2n+m-2x\pm 1 \\
u = \xl + \hp + \frac{p\pi}{2\la} & \mathrm{poles:\
} & p = -2n+m-2x\pm 1 \\
 & \mathrm{zeroes:\ } & p = m-2x\pm 1
\end{array}
\end{equation}
while \m{a^{0;m}_{n}(u)} has them at
\begin{equation}
\begin{array}{crl}
u = -\xl + \frac{3\pi}{2} + \frac{p\pi}{2\la} & \mathrm{poles:\
} & p = -2N-m+2x\pm 1 \\
 & \mathrm{zeroes:\ } & - \\
u = -\xl + \hp + \frac{p\pi}{2\la} & \mathrm{poles:\
} & p = -m+2x \pm 1 \\
 & \mathrm{zeroes:\ } & p = -2N - m +2x \pm 1 \\
u = \xl - \hp + \frac{p\pi}{2\la} & \mathrm{poles:\
} & p = 2N + m-2x\pm 1 \\
 & \mathrm{zeroes:\ } & p = m-2x\pm 1 \\
u = \xl + \hp + \frac{p\pi}{2\la} & \mathrm{poles:\
} & p = m-2x\pm 1 \\
 & \mathrm{zeroes:\ } & p = 2N + m-2x\pm 1
\end{array}
\end{equation}
These poles will be further discussed 
in \sect{sec:general} below.

\se{An example}
\label{sec:example}
To get an idea of the full picture, and which processes are 
responsible for the remaining poles,
we will now look at one particular example in more detail. If we select
\m{\xi=1.6\pi} and \m{\la=2.5}, then we have the first two breathers in
the spectrum, with the solitonic poles taking the form \m{\nu_{n} =
\frac{\pi(2.2-2n)}{5}} and \m{w_{N}=\frac{\pi(2.8-2N)}{5}}. 
Thus, for this case, only the poles at \m{\nu_{0}, \nu_{1}} and
\m{w_{1}} are on the physical strip, and so \fig{fig:poles} is
simplified to \fig{fig:expoles}. This is
the simplest case which requires a broader spectrum than that
postulated in \cite{Skorik}. First, let us turn to the soliton
sector.
\begin{figure}
\begin{center}
\unitlength=1.00mm
\begin{picture}(32.35,53.75)(15.00,0.00)
\color{deepred}
\put(40.00,53.75){\line(0,-1){53.75}}
\put(20.00,53.75){\line(0,-1){53.75}}
\put(19.00,53.75){\line(1,0){2}}
\put(19.00,0.00){\line(1,0){2}}
\put(16.00,52.75){$\hp$}
\put(16.00,-1.00){$0$}
\put(16.00,25.87){$u$}
\color{blue}
\multiput(40.00,47.30)(0,-43){2}{\line(1,0){5}}
\multiput(40.00,47.30)(0,-43){2}{\circle*{2}}
\put(46.00,47.30){$\nu_{0}$}
\put(46.00,4.3){$\nu_{1}$}
\color{red}
\put(40.00,17.20){\line(-1,0){5}}
\put(40.00,17.20){\circle*{2}}
\put(30.00,17.20){$w_{1}$}
\color{black}
\end{picture}
\ca{Location of poles in the example}
\label{fig:expoles}
\end{center}
\end{figure}

\sse{Boundary ground state---soliton sector}
As argued above, the soliton can bind to the boundary at all
rapidities \m{\nu_{n}} which are in the physical strip, here just
comprising \m{\nu_{0}} and \m{\nu_{1}}.  This introduces the states
\m{\st{1;0}} and \m{\st{1;1}}.

\sse{Boundary ground state---breather sector}
The only breather poles are at \m{\xl-\hp+\frac{(m-1)\pi}{2\la}} for
breather \m{m}. In addition, breather \m{B_{2}} has a zero at
\m{-\xl+\hp+\frac{\pi}{2\la}}.

By lemma~\ref{lemma:1}, the pole for \m{B_{1}} must correspond to a
new bound state, the rapidity being less than
\m{\frac{\pi}{2\la}}. From \fig{fig:bresoli},
it is clear that \m{B_{1}} creates the
state which was labelled \m{\st{\delta_{0,1}}} in \cite{Skorik},
and which we have called \m{\st{0;0,1}}.

The pole for the second breather can be explained by
\fig{fig:brfake}, with the state \m{\st{1;0}} being formed. The
anti-soliton is reflected from the boundary at a rapidity of
\m{\xl-\pi+\frac{3\pi}{2\la}}---a zero of the \m{\st{1;0}} reflection
factor---reducing the diagram to first order through the boundary
Coleman-Thun mechanism.

\sse{First level excited states---soliton sector}
{}From before, \m{P^{+}_{\st{1;0}}} just has a simple pole at
\m{\nu_{0}}, which
can be explained by the crossed process in \fig{fig:uboun}, reducing
the boundary to the ground state.
For \m{P^{+}_{\st{1;1}}}, the pole
at \m{\nu_{1}} can be explained this way while, for the double pole at
\m{\nu_{0}}, \fig{fig:fb4} is required, the first breather being
formed while the boundary is reduced to the vacuum state.

For \m{P^{-}_{\st{1;n}}(u)}, we have the additional job of explaining
simple poles at \m{w_{N}}, for all \m{N} such that this pole is in the
physical strip. Here, this is only \m{w_{1}}. For \m{\st{1;0}}, this is
appropriate for the formation of \m{\st{0;0,1}} which, from the previous
section, must be present. For \m{\st{1;1}}, however, \fig{fig:fb2} is
invoked, the second breather being created, and the boundary reduced
to the vacuum state. The breather is incident on the boundary at an
angle of \m{\hf(w_{1}-\nu_{1})=\pi - \xl - \frac{\pi}{2\la}} which, looking
at the above breather reflection factors, is a zero, ensuring the
diagram is of the correct order.

\sse{First level excited states---breather sector}
The pole structure of
\m{{R_{\st{1;0}}^{m}}} 
can be found from \m{\overline{R_{\st{0}}^{m}}}, and is
\begin{equation}
\begin{array}{rrl}
B_{1}: & \mathrm{pole\ at\ } & \hp-\xl+\frac{\pi}{\la} \\
B_{2}: & \mathrm{poles\ at\ } & \hp - \xl+\frac{3\pi}{2\la},
\hp-\xl+\frac{\pi}{2\la}
\end{array}
\end{equation}

By
lemma~\ref{lemma:1}, the second pole for \m{B_{2}} must correspond to
a new bound state; by the previous arguments, this is
\m{\st{1;0,1,1}}. 
This state is not in the spectrum given in~\cite{Skorik}, 
but lemma~\ref{lemma:1} shows
that there is no way to avoid its introduction. Considerations such as
this will open the door to a much wider spectrum 
in the general case.

The \m{B_{1}} pole is suitable for the creation of \m{\st{1;1}}. The
first pole for \m{B_{2}} can be explained by \fig{fig:ob6}, with the
boundary being reduced to the ground state by emission of a soliton.

For \m{R_{\st{1;1}}^{m}}, the above poles are supplemented by
additional poles from
\m{b_{m,1}^{1}(u)} to give the poles shown in table \ref{tab:bre1poles}.
\begin{table}[ht]
\[
\begin{array}{c|c|c|c} \thline
 & -\xl+\hp+\frac{p\pi}{2\la} &
\xl-\hp+\frac{p\pi}{2\la} &
\xl+\hp+\frac{p\pi}{2\la} \\
\thline B_{1} &2 &0 &- \\
B_{2} &3^{2} &1 &-5 \\
\thline
\end{array}
\]
\vspace{-0.25in}
\ca[Breather pole structure for $\st{1;1}$.]{Breather pole structure for $\mathbf{\st{1;1}}$. \mdseries
\itshape Entries
are the values of
\m{p} for which there is a pole in the location given in the column
heading. The power of the entry
gives the order of the pole, so e.g. \m{3^{2}} indicates a double pole
when \m{p=3}. There are no physical strip zeroes for either breather.
}
\label{tab:bre1poles}
\end{table}

The pole at \m{\xl+\hp-\frac{5\pi}{2\la}}
can be explained by \fig{fig:ob4}, with the boundary being reduced to
the ground state by emission of a soliton. The pole at \m{\xl-\hp} for
\m{B_{1}} can be allocated to the creation of \m{\st{1;0,1,1}}, while
the pole at \m{\xl-\hp+\frac{\pi}{2\la}} for \m{B_{2}}
is due to \fig{fig:ob5}, where the boundary emits \m{B_{1}}, being reduced to
\m{\st{1;0}}. The pole at \m{-\xl+\hp+\frac{2\pi}{2\la}} for \m{B_{1}}
is responsible for
this reduction to \m{\st{1;0}}, while the double pole for \m{B_{2}}
comes from an all-breather version of \fig{fig:ob3}, the boundary
being reduced in the same way.

\sse{Second level excited states---soliton sector}

For \m{P^{-}_{\st{0;0,1}}(u)}, the only poles are simple,
at \m{\nu_{0}} and \m{w_{1}}. The pole at \m{w_{1}} can be explained
by \fig{fig:uboun} while, for \m{\nu_{0}}, we need \fig{fig:fb3}. The
second breather is emitted by the boundary, reducing it to the ground
state, while a soliton is incident on the boundary at a
rapidity \m{w_{1}}. For the ground state, this is neither a pole
nor a zero, but the diagram contains a solitonic loop which can either
be drawn to leave a soliton or an anti-soliton incident on the
boundary. Adding the contributions of these two diagrams gives an
additional zero.

For \m{P^{+}_{\st{0;0,1}}(u)}, we have additional poles at all \m{\nu},
i.e. a simple pole at \m{\nu_{1}}, with
\m{\nu_{0}} becoming a double pole.
By lemma~\ref{lemma:1}, \m{\nu_{1}} must correspond to
the creation of a new bound state, namely \m{\st{1;0,1,1}}, while, for
\m{\nu_{0}}, \fig{fig:fb2} should be considered. Again, the second
breather is created, the boundary is reduced to the ground state, and
the breather is incident on the boundary at a rapidity of
\m{\hf(\nu_{0}-w_{1})=\xi/\la-\pi/2}---a zero of the
reflection factor.

\sse{Second level excited states---breather sector}

For \m{\st{0;0,1}}, we have the pole structure given in table
\ref{tab:bre2poles}.
\begin{table}[ht]
\[
\begin{array}{c|c|c|c} \thline
 & -\xl+\frac{3\pi}{2} +\frac{p\pi}{2\la} &
-\xl+\hp+\frac{p\pi}{2\la} &
\xl-\hp+\frac{p\pi}{2\la} \\
\thline B_{1} &-2 &2 &0,2 \\
B_{2} &-3 &3 &1^{2} \\
\thline
\end{array}
\]
\vspace{-0.25in}
\ca{Breather pole structure for $\st{0;0,1}$.}
\label{tab:bre2poles}
\end{table}

The poles at \m{-\xl+\frac{3\pi}{2}+\frac{p\pi}{2\la}} are due to
\fig{fig:ob4}, while the poles in the second column are due to
\fig{fig:ob5}. For all these, the boundary is reduced to \m{\st{1;0}}.
The pole at \m{\xl-\hp+\frac{(m-1)\pi}{2\la}} for \m{B_{m}} (\m{m=2})
is due to \fig{fig:ob7}, while for \m{m=1} it is due to a breather
version of \fig{fig:uboun}. The pole at \m{\xl-\hp+\frac{2\pi}{2\la}}
for \m{B_{1}} is due to \fig{fig:ob6}.

\sse{Third level excited states---soliton sector}
The only third level excited state is \m{\st{1;0,1,1}}. For
\m{P^{+}_{\st{1;0,1,1}}}, there are simple poles at
\m{w_{1},\nu_{0}} and \m{\nu_{1}}. Again, the pole at \m{w_{1}} comes from 
the crossed process \fig{fig:uboun}. For \m{\nu_{1}}, \fig{fig:uboun}
suffices while, for \m{\nu_{0}}, \fig{fig:fb4} is required, the
boundary being reduced to \m{\st{0;0,1}} while the first breather is
incident on the boundary at \m{\hp-\xl+\frac{\pi}{\la}}, another zero.

\sse{Third level excited states---breather sector}
Here, the only possible boundary state is \m{\st{1;0,1,1}} and we find
the poles given in table \ref{tab:bre3poles}.
\begin{table}[ht]
\[
\begin{array}{c|c|c|c|c} \thline
 & -\xl+\frac{3\pi}{2} +\frac{p\pi}{2\la} &
-\xl+\hp+\frac{p\pi}{2\la} &
\xl-\hp+\frac{p\pi}{2\la} & \xl+\hp+\frac{p\pi}{2\la} \\
\thline B_{1} &-2 &2^{2},4 &0,2 &- \\
B_{2} &-3 &1,3^{3} &1^{2} &-5 \\
\thline
\end{array}
\]
\vspace{-0.25in}
\ca{Breather pole structure for $\st{1;0,1,1}$.}
\label{tab:bre3poles}
\end{table}

Comparing this with the structure given above for
\m{\st{1;1}}, it can easily be seen that, whenever the two both have a pole
at the same rapidity, essentially the same explanation can be used. For the
remaining poles, \m{-\xl+\frac{3\pi}{2}+\frac{p\pi}{2\la}} can be
explained by \fig{fig:ob5}, the boundary being reduced to \m{\st{1;0}},
while that at \m{-\xl+\hp+\frac{\pi}{2\la}} for \m{B_{2}} is due to
\fig{fig:uboun}, reducing the boundary to \m{\st{1;0}}, and that at
\m{\xl-\hp+\frac{2\pi}{2\la}} for \m{B_{1}} is due to an all-breather
version of \fig{fig:ob6}, again reducing the boundary to \m{\st{1;0}}.

\sse{Summary}
The above has shown that, by introducing only the states which are
required by lemmas 1 and 2, the complete pole structure can be explained.
Below, we shall find that this is a general feature. In addition, the
spectrum of states is broader than that introduced in~\cite{Skorik}
(containing, in addition to their states, \m{\st{1;0,1,1}}).
It should be noted that the mass of this extra state corresponds
to \m{m_{1,1}} of \cite{Skorik}, the mass of a boundary Bethe ansatz
(1,1)-string whose apparent
absence from the bootstrap spectrum was described in that paper as
``confusing''. 
This does at least show that the Bethe ansatz
results of \cite{Skorik} are not incompatible with
the bootstrap. However, in more general cases it turns out that the
bootstrap predicts yet further states, beyond those identified in the 
boundary Bethe ansatz calculations of \cite{Skorik}, so a full reconciliation
of the Bethe ansatz and bootstrap approaches remains an open problem.

\se{The general case}
\label{sec:general}
{}From the above, we might imagine that the boundary state
\m{\st{c;n_{1},n_{2},n_{3},\ldots,n_{m}}} exists iff \m{c} is \m{0} or
\m{1} and \m{n_{1},n_{2},n_{3},\ldots} are chosen such that
\m{\pi/2>\nu_{n_{1}}>w_{n_{2}}>\nu_{n_{3}}>\ldots>0}. This turns out to be 
correct, and will be proved in two stages. Firstly, we need to
show that all these states must be present, before going on to show
that, given this, all other poles can be explained without invoking
further boundary states.

\sse{The minimal spectrum}
The argument proceeds as follows: starting with the knowledge that
the vacuum state \m{\st{0}} and all appropriate states
\m{\st{1;n_{1}}} are in the spectrum, we use breather poles to
construct all the other postulated states.

These poles are of the form \m{\hf(w_{N}-\nu_{n})} for breather
\m{n+N} incident on a charge 0 state (or \m{\hf(\nu_{n}-w_{N})}
for a charge 1 state). If \m{\nu_{n}-w_{N}<\pl}, lemma~\ref{lemma:1}
shows that they must correspond either to the formation of a new
state, or the crossed process. From \fig{fig:bresoli}, this
corresponds either to adding indices \m{n} and \m{N} if they are
absent or---if they are already present---removing them. (Note that
any other option would give rise to a state with a mass outside the
scheme given by \re{eq:mass}, and therefore outside our postulated
spectrum.) The condition \m{\nu_{n}-w_{N}<\pl} is always satisfied
if \m{\nu_{n}>w_{N}} and \m{\nu_n} and \m{w_N}
are as close together as possible,
i.e. if \m{\st{0;n,N}} exists, but \m{\st{0;n,N-1}} does not.

The only subtlety in this argument arises when considering the topmost
breather. If \m{n+N=n_{max}}, lemma~\ref{lemma:1} on its own is not
strong enough to require the presence of the state we need, and we
must invoke the stronger version introduced at
the end of of section~\ref{sec:colethun}. This makes use the idea that there
must be a corresponding two-stage solitonic route to the same state, i.e. a
soliton with rapidity $\nu_{n}$ followed by an anti-soliton with
rapidity $w_{N}$. Considering these two processes instead, the
stronger lemma shows that both form bound states, as $\nu_{n}$ and
$w_{N}$ must be
the lowest poles of their type---and so have rapidity less than
$\frac{\pi}{\la}$---for $n+N$ to equal $n_{max}$. This shows that
the state exists, and hence that the breather pole is due to its formation.

Since the arguments for the two sectors are analogous, let us focus on
the charge 0 sector here. The challenge is to create any state
\m{\st{0;\underline{x}}}---for some set of indices
\m{\underline{x}=(n_{1},n_{2},\ldots,n_{2k})}---from the ground state
using just these poles. As a first step, consider creating
\m{\st{0;n_{1},n_{2}}}. If \m{\nu_{n_1}} and \m{w_{n_2}} 
are as close together as possible, we
simply make use of the pole at
\m{\hf(w_{n_{2}}-\nu_{n_{1}})}. Otherwise, introduce the set
\m{m_{1},m_{2},\ldots,m_{t}} such that
\m{\nu_{n_{1}}>w_{m_{1}}>\nu_{m_{2}}>w_{m_{3}}>\ldots
>\nu_{m_{t}}>w_{n_{2}}}, with each successive rapidity as close to the
previous one as possible.  Now, we can successively create
\m{\st{0;\underline{x},n_{1},m_{1}}}, then \m{\st{0;x,n_{1},m_{1},
m_{2}, m_{3}}}
and so on, up to \m{\st{0;\underline{x},n_{1},m_{1},m_{2},m_{3}, 
\ldots,m_{t},n_{2}}}.

By now invoking the crossed process, a
suitable breather can be used to removed the indices \m{m_{1},m_{2}},
followed by \m{m_{3},m_{4}} and so on, until all the \m{m} indices
have been removed to leave \m{\st{0;\underline{x},n_{1},n_{2}}}.

Repeating this procedure allows \m{\st{0;n_{1},n_{2},n_{3},n_{4}}} to
be created, and hence \m{\st{0;\underline{x}}}. Note that this
allows any state in our allowed spectrum to be created, but no others,
as the condition \m{\nu_{n_{1}}>w_{n_{2}}>\ldots} is imposed by the
existence of the necessary breather poles. Charge 1 states can be
created analogously by starting from a suitable state \m{\st{1;n_{1}}}.

One remaining point is to check that all the necessary breather
poles do indeed exist. However, starting from \re{eq:breexc}, they occur in
the \m{R^{n}_{(c)}(u)} factor, and it is straightforward to check that
they are never modified by the other \m{a} factors.

\sse{Reflection factors for the minimal spectrum}
\label{sec:refmin}

The boundary state can be changed by the solitonic
processes given in table \ref{tab:solproc}.
\begin{table}[ht]
\[
\begin{array}{c|c|c|c} \thline
\mathbf{Initial\ state} & \mathbf{Particle} & \mathbf{Rapidity} &
\mathbf{Final\ state} \\
\thline \st{0;n_{1},\ldots,n_{2k}} & \mathrm{Soliton} &
\nu_{n} & \st{1;n_{1},\ldots,n_{2k},n} \\
\st{1;n_{1},\ldots,n_{2k-1}} &
\mathrm{Anti-soliton} & w_{N} & \st{0;n_{1},\ldots, n_{2k-1},N} \\
\thline
\end{array}
\]
\vspace{-0.25in}
\ca{Solitonic processes which change the boundary state.}
\label{tab:solproc}
\end{table}

The breather sector is more complex, as indices can be added or
removed from any point in the list, and not just at the end, as for
solitons. In addition, processes exist which simply adjust the value
of one of the indices, rather than increasing the number of
indices. For breather \m{m}, these are given in table \ref{tab:breproc}.
\begin{table}[ht]
\[
\begin{array}{c|c|c} \thline
\mathbf{Initial\ state} & \mathbf{Rapidity} & \mathbf{Final\ state} \\
\thline \st{0/1;\ldots n_{2x},n_{2x+1} \ldots} & \hf(\nu_{n}-w_{N}),
n+N=m & \st{0/1;\ldots n_{2x},n,N,n_{2x+1} \ldots} \\
\st{0/1;\ldots n_{2x-1},n_{2x} \ldots} & \hf(w_{N}-\nu_{n}),
n+N=m & \st{0/1;\ldots n_{2x-1},N,n,n_{2x} \ldots} \\
\st{0/1;\ldots n_{2x} \ldots} &
\hf(\nu_{-n_{2x}}-w_{n_{2x}+m}) &
\st{0/1;\ldots n_{2x}+m \ldots} \\
\st{0/1;\ldots n_{2x-1} \ldots} &
\hf(w_{-n{2x-1}}-w_{n_{2x-1}+m}) &
\st{0/1;\ldots n_{2x-1}+m \ldots} \\
\thline
\end{array}
\]
\vspace{-0.25in}
\ca{Breather processes which change the boundary state.}
\label{tab:breproc}
\end{table}
This should be read as implying that any index can have its value
raised, and that a pair of indices can be inserted at any point in the
list, including before the first index and after the last (providing
the resultant state is allowed). Both these tables have been
derived on the basis that, whenever assuming that a pole corresponds to a
bound state leads to a state with the same mass and topological charge as 
one in our minimal spectrum, the assumption is taken to be correct. As
with our
earlier assumption (that, if a pole has another possible explanation, it is
not taken as forming a bound state), this is intuitively reasonable but not
necessarily rigorous. It does, however, lead to consistent results, and
there is no conflict between the two assumptions: we have been unable to
find any alternative explanation for any of the poles listed above. 

It is vital for what follows that, for all the above processes, there
is very little dependence on the existing boundary state. For the
solitons, the topological charge of the state and the value of the
last index in the list are all that matter. Any two states which have
the same topological charge and last index can undergo processes at
the same rapidities to add an index.
Similarly, for the breathers, provided either the relevant two
indices can be added at some point in the list to create an allowed
state, or that the index to be adjusted is present in the list, the
other characteristics of the state are irrelevant.

\goodbreak

\sse{Solitonic pole structure} 
\nobreak
This turns out to be relatively
straightforward. All poles are either of the form \m{\nu_{n}} or
\m{w_{N}}. Looking at a charge 0 state with \m{2k} indices, and
labelling this as 
\m{\underline{x}=(n_{1},n_{2},\ldots,n_{2k})}, we find the results shown
in table \ref{tab:p-0x} for \m{P^{-}_{\st{0;\underline{x}}}(u)}. These poles
come from the \m{a} factors so, for \m{P^{+}}, there is an additional
pole at all \m{\nu}.

\begin{table}[ht]
\[
\begin{array}{c|c|c} \thline
\mathbf{Pole} & \mathbf{Order} & \mathbf{Pole} \\
\thline w_{1}\ldots w_{n_{2}-1} &
2k & \nu_{1}\ldots \nu_{n_{1}-1} \\ w_{n_{2}} & 2k-1 & \nu_{n_{1}} \\
w_{n_{2}+1}\ldots w_{n_{4}-1} & 2k-2 & \nu_{n_{1}+1}\ldots \nu_{n_{3}-1} \\
w_{n_{4}} & 2k-3 & \nu_{n_{3}} \\ \vdots & \vdots & \vdots \\
w_{n_{2k-2}+1} \ldots w_{n_{2k}-1} & 2 & \nu_{n_{2k-3}+1} \ldots 
\nu_{n_{2k-1}-1}
\\ w_{n_{2k}} & 1 & \nu_{n_{2k-1}} \\
\thline
\end{array}
\]
\vspace{-0.25in}
\ca[Pole structure for
$P^{-}_{\st{0;x}}(u)$.]{Pole structure for
$\mathbf{P^{-}_{\st{0;\underline{x}}}(u)}$. \mdseries \itshape
An entry of, for example, \m{w_{1}\ldots w_{n_{2}-1}} indicates that
there is a pole of
order \m{2k} at \m{w_{1},w_{2},w_{3},\ldots,w_{n_{2}-1}}.}
\label{tab:p-0x}
\end{table}  

For the charge 1 states, the picture is very similar, and,
considering \m{P^{+}} first, we find the pattern given in
table~\ref{tab:solch1} for a state with \m{2k-1} indices. For \m{P^{-}} 
there are additional poles at
all \m{w}. (For the charge 0 case, there are poles at \m{w_{x}} for
\m{x \leq 0}, but none of these are in the physical strip.)

\begin{table}[ht]
\[
\begin{array}{c|c|c} \thline
\mathbf{Pole} & \mathbf{Order} & \mathbf{Pole} \\
\thline - & 1 & \nu_{0},\nu_{-1},\ldots \\
- & 2k &
\nu_{1}\ldots\nu_{n_{1}-1} \\ - & 2k-1 & \nu_{n_{1}} \\ w_{1} \ldots 
w_{n_{2}-1} &
2k-2 & \nu_{n_{1}+1} \ldots \nu_{n_{3}-1} \\ w_{n_{2}} & 2k-3 & \nu_{n_{3}}
\\ \vdots & \vdots & \vdots \\ w_{n_{2k-4}+1} \ldots w_{n_{2k-2}-1} & 2 &
\nu_{n_{2k-3}+1} \ldots \nu_{n_{2k-1}-1} \\ w_{n_{2k-2}} & 1 &
\nu_{n_{2k-1}} \\
\thline
\end{array}
\]
\vspace{-0.25in}
\ca[Pole structure for
$P^{+}_{\st{1;x}}(u)$.]{Pole structure for
$\mathbf{P^{+}_{\st{1;\underline{x}}}(u)}$.}
\label{tab:solch1}
\end{table}

An important point to note is that, comparing
\m{\st{0;n_{1},n_{2},\ldots,n_{2k-1},n_{2k}}} (a general
level \m{2k} state) 
with the level 2 state \m{\st{0;n_{2k-1},n_{2k}}}, we find no additional
poles, though the order of some poles has increased. In the example
above, all level 2 states were explained by diagrams where the
boundary was reduced
either to the vacuum by emission of a breather, or to a first level
excited state by emission of an anti-soliton.
The same processes turn out to be present for any level \m{2k} state
to be reduced to a level \m{2k-1} or \m{2k-2} state. Thus, we might
imagine explaining the poles in the level \m{2k} reflection factor via
similar processes to the ones which explained them in the level 2
factor. At times, however---as we shall see---parts of these
processes will need to be replaced by more complex subdiagrams to
allow for the fact that the boundary is
in a higher excited state, explaining the differences in the orders of
the poles. Considering the level 2 processes so far introduced as
``building blocks'', this can be considered as an iterative process: level 4
states can be explained by replacing parts of level 2 processes
with building blocks, while level 6 states can
be explained by similarly replacing parts of level 4 processes with
building blocks, and so on. A generic process of the type we will examine can
therefore be viewed as a cascade of building blocks, each appearing as
a subdiagram of the one before it.
    
A similar argument applies to level \m{2k+1} states and level 3 states,
drawing the same diagrams with all rapidities transformed via \m{\xi
\rightarrow \pi(\la+1)-\xi}. We will concentrate on the charge 0
sector below, and consider a generic level \m{2k} state.

For poles of the form \m{\nu_{n}}, consider \fig{fig:os1}. The
boundary decays to 
the state \m{\st{0;n_{1},n_{2},\ldots,\linebreak n_{2k-2}}} 
by emission of breather
\m{n_{2k}+n_{2k-1}} at a rapidity of
\m{\hf(\nu_{n_{2k-1}}-w_{n_{2k}})}. This then decays into
breather \m{n_{2k-1}-n} heading towards the boundary at a rapidity of
\m{\hf(w_{-n_{2k-1}}-\nu_{n})} and breather \m{n_{2k}+n} heading
away from the boundary at a rapidity of
\m{\nu_{n}-\left(\hp-\frac{(n_{2k}+n)\pi}{2\la}\right)}. This then
decays to give the outgoing particle and one heading towards the
boundary at a rapidity of \m{w_{n_{2k}}}. For \m{n<n_{2k-1}}, it is
straightforward to check that all these rapidities are within suitable
bounds.

This diagram is na\"{\i}vely third order. However, breather
\m{n_{2k-1}-n}, which is drawn as simply reflecting off the boundary,
in fact has a pole, meaning that the diagram should be treated as
schematic and the appropriate diagram from the next section inserted
instead. In addition, as noted in the discussion of the example, 
the soliton
loop contributes a zero for an incoming anti-soliton through the
Coleman-Thun mechanism. When this is taken into account, we obtain the
correct result.

For \m{\nu_{2k-1}}, the slightly simpler \fig{fig:fb3} suffices. The
remaining \m{\nu} poles are only present in the soliton reflection
factor, and can be explained by \fig{fig:fb2}, with the boundary
decaying by emitting an anti-soliton at \m{w_{n_{2k}}}, which then
interacts with the incoming soliton to give breather \m{n+n_{2k}},
heading towards the boundary at a rapidity of
\m{\hf(\nu_{n}-w_{n_{2k}})}. Looking ahead again, the interaction
of this breather with the boundary contributes the required zero. For
\m{\nu_{n}<w_{n_{2k}}}, this diagram fails, the breather being created
heading away from the boundary; this is the point when the pole is to
be considered as creating the bound state
\m{\st{1;n_{1},\ldots,n_{2k},n}}.

For the \m{w_{N}} poles, the story is very similar, this time being
based on \fig{fig:fb4} (requiring a 
suitable pole/zero
for \m{B_{N-n_{2k}}} on state
\m{\st{1;n_{1},\ldots,n_{2k-1}}} at
\m{\xl-\hp+\frac{\pi(N+n_{2k}-1)}{2\la}}) for \m{N<n_{2k}} and
\fig{fig:uboun} for \m{n_{2k}}. As noted above, all charge 1 state
poles can be explained by the same mechanisms, with the rapidities
transformed according to \m{\xi \rightarrow \pi(\la+1)-\xi}.

\sse{Breather pole structure}
This is considerably more complicated. However, with a bit of work it
turns out that, for breather \m{n} on the state
\m{\st{0;n_{1},n_{2},\ldots, n_{2k}}}, the pole
structure is as given in table \ref{tab:breps}.
\begin{table}[ht]
\[
\begin{array}{c|c|c} \thline
\mathbf{Pole} & \mathbf{Range} & \mathbf{Pole/zero\ order} \\
\thline \xl-\hp+\frac{\pi(n+2x-1)}{2\la} & n_{2q}<x<n_{2q+2}
 & 2(q'-q)+y \\
&n_{2q'}<n+x<n_{2q'+2}& \\
 & x<0, n_{2q-1}<|x|<n_{2q+1}  & 2(q'-q)+y+i
\\
&n_{2q'} < n- |x| < n_{2q'+2}& \\
-\xl+\hp+\frac{\pi(n+2x+1)}{2\la} & n_{2q-1}<x<n_{2q+1}
& 2(q'-q)+y \\
&n_{2q'-1}<n+x<n_{2q'+1}& \\
 & x<0, n_{2q}<|x|<n_{2q+2}, & 2(q'-q)-i+y \\
& n_{2q'-1}<n-|x|<n_{2q'+1} & \\
 & x<-n, n_{2q} < |x| < n_{2q+2}  &
2(q'-q) \\
&n_{2q'} < |x|-n < n_{2q'+2}& \\
\xl+\hp+\frac{\pi(n+2x-1)}{2\la} & \mathrm{As\ } \xl -
\hp+\frac{\pi(n+2x+1)}{2\la}  & \\
&\mathrm{\ with\
poles}\leftrightarrow \mathrm{zeroes}& \\
-\xl + \frac{3\pi}{2} + \frac{\pi(n+2x+1)}{2\la} & \mathrm{As\ } -\xl +
\hp + \frac{\pi(n+2x-1)}{2\la}  & \\
&\mathrm{\ with\ poles}\leftrightarrow
\mathrm{zeroes} & \\
\thline
\end{array}
\]
\vspace{-0.25in}
\ca[Breather pole structure for a generic charge 0 state.]{Breather pole structure for a generic charge 0 state. \mdseries \itshape
The variable
$x$ takes integer and half-integer values within the allowed ranges.
An entry in the third column represents a pole of that
order if it is positive, and a zero of appropriate order if it is
negative. (Thus an entry of +1 is a first-order pole, and an entry of
-1 is a first-order zero.) Also, for convenience, \m{y} is 1 if \m{x}
(or \m{|x|}) attains the lower limit, -1 if \m{n+x} (or \m{n-|x|})
attains the lower limit, and zero otherwise, while \m{i} is 1 if \m{x}
is integer, and 0 otherwise.}
\label{tab:breps}
\end{table}

In explaining all this, we can begin with the diagrams found previously.
For the first line, consider an all-breather version of
\fig{fig:brfake}, where the breather decay is chosen to produce breather
\m{n+x-n_{2q'}} on the left, which then binds to the boundary to raise
index \m{n_{2q'}} to \m{n+x}. In some cases, this is not possible, the
appropriate state not being in the spectrum, but, then, we can consider
an all-breather version of \fig{fig:ob1}, where the boundary decays so
as to remove the indices \m{n_{2q'}} and \m{n_{2q'+1}}, with the same
initial breather decay, and the additional breather reflecting from the
boundary contributes a zero. This diagram becomes
possible just as the other fails. In either case, the other breather
from the initial decay (which is drawn as simply reflecting from the
boundary), is breather \m{y=n_{2q'}-x} at rapidity \m{\xl-\hp +
\frac{\pi(y+2x-1)}{2\la}}. This has a pole of order 2 less than the
initial breather. If this order is less than or equal to zero, the
diagram stands as drawn while, otherwise, the simple reflection from the
boundary should be replaced by a repeat of this argument, iterating
until the result is less than or equal to zero. For the next line,
precisely the same argument can be used.

The next three lines can be explained by a similar argument, based on
either increasing the value of index \m{n_{2q'-1}} or removing indices
\m{n_{2q'-1}} and \m{n_{2q'}}.

For \m{\xl+\hp+\frac{\pi(n+2x-1)}{2\la}}, we invoke a similar
process. This time, however, the outer legs have rapidity
\m{\nu_{-(n+x)}} (where \m{-(n+x)} is actually a positive number if
the initial pole is to be in the physical strip), and so we need to
substitute in the explanation of soliton poles of this form from before,
leading, in simple cases, to \fig{fig:ob5}.

Finally, for \m{\frac{3\pi}{2}-\xl+\frac{\pi(n+2x+1)}{2\la}}, we begin with
\fig{fig:ob4}. This time, the reflection factor for the central
soliton always provides a zero, while
the outer soliton has rapidity \m{w_{n+x}}. If \m{n+x=n_{2k}}, the diagram is
as drawn while, otherwise, we need to replace the two outer anti-soliton
legs with the explanation of the appropriate pole in the anti-soliton
factor. The first iteration of this is shown in \fig{fig:ob1}.

\se{Number of states}
In this section, we examine how the size and content of the boundary
spectrum changes with variation in \m{\xi} and \m{\la}. Since any state can
be formed by a suitable sequence of solitons and anti-solitons, we will
focus on the solitonic sector.

The relevant poles, \m{\nu_{n}} and \m{w_{n}}, both have the same 
spacing---\m{\frac{\pi}{\la}}---but, interestingly, the range of \m{n}
for which
\m{\nu_{n}} is in the physical strip is independent of \m{\la}, while that
for the \m{w}-type poles is not. For \m{\nu_{n}}, \m{0 \leq n <
\frac{\xi}{\pi}-\hf} while, for \m{w_{n'}}, \m{\la-\frac{\xi}{\pi}-\hf < n' <
\frac{\la}{2}-\frac{\xi}{\pi}+\frac{3}{2}}. 
Designating the lowest-rapidity \m{\nu}-type pole
as \m{\nu_{n_{*}}}, there are \m{n_{*}+1} \m{\nu}-type poles, and either
\m{n_{*}} or \m{n_{*}+1} relevant \m{w}-type poles, depending on whether the
lowest-rapidity pole is \m{\nu}-type or \m{w}-type. (Note that any
\m{w}-type pole with a rapidity greater than \m{\nu_{0}} can never be
relevant in forming a bound state.)

Recall now that the criterion for a state
\m{\st{c;n_{1}, n_{2}, n_{3}, \ldots}} to be in the spectrum is that
\m{\nu_{n_{1}}>w_{n_{2}}>\nu_{n_{3}}>\ldots}, corresponding to
moving down \fig{fig:poles}, alternating from side to side. Since
movement must be strictly
downward, there are two cases to consider: when the \m{w} and \m{\nu} poles
occur at the same rapidities, and when they do not.

The first case is the simplest to deal with, as enumerating the states
in the spectrum becomes equivalent to calculating the number of ways of
making an ordered selection of an arbitrary number of objects from a set
of \m{n_{*}+1}. However, to simplify the rest of the argument, we shall
formulate it as a recursion relation.

We shall consider the situation where \m{w_{x}=\nu_{x}}
(realised when \m{\xi=\frac{\pi(\la+1)}{2}}). Clearly, all other cases
are similar, with the even indices uniformly increased by
\m{\frac{\xi}{\pi}-\frac{\la+1}{2}}\footnote{This is an integer when 
the two sets of poles occur at the same rapidities.} but with the
overall spectrum size unchanged.

Consider first a subset of the spectrum, with all indices less than,
say, \m{m}, leaving \m{m} poles to play with in each sector. Denote the
number of charge 0 and 1 states in this part of the spectrum as \m{c_{0}
(m)} and \m{c_{1}(m)} respectively. Now consider extending this to \m{m+1}
poles; all the states previously present are still there,
together with new states involving the extra index. For each sector,
a new state can be formed by taking an existing state in the opposite sector
and adding the new index, \m{m} (provided the
vacuum state is included in the list of charge 0 states to allow for the
possibility of forming \m{\st{1;m}}).

Overall, then, \m{c_{1}(m+1)=c_{1}(m)+c_{0}(m)=c_{0}(m+1)}. Solving
this gives \m{c_{0/1}(n_{*}+1)=2^{n_{*}}c_{1}(0)+2^{n_{*}}c_{0}(0)}. Without
allowing any poles, the spectrum consists of only the ground state, so
\m{c_{0}(0)=1} and \m{c_{1}(0)=0}. Thus \m{c_{0/1}(n_{*}+1)=2^{n_{*}}}, as
expected from the combinatoric approach.

Moving to the case where the \m{w} and \m{\nu} poles do not
coincide, the argument changes a little. Consider the case where
\m{w_{x}} lies between \m{\nu_{x}} and \m{\nu_{x+1}}, noting that, as
before, all other cases simply involve a uniform adjustment of the even
indices. Again, we can look at the subset with all indices less than
\m{m}, and compare it with that with all indices less than \m{m+1}. The
difference now is that we can potentially add two extra indices to an
existing state, one from each sector, since their rapidities no longer
coincide.

A new charge 1 state can only be formed by the addition of the index \m{m}
to an existing state, but a charge 0 state can either be formed by adding
\m{m} to an existing charge 1 state, or \m{m,m} to an existing charge 0
state. Thus, \m{c_{1}(m+1)=c_{0}(m)+c_{1}(m)}, but  \m{c_{0}(m+1)= 2c_{0}(m)
+ c_{1}(m)}. To solve these, it is useful to think of writing out the list
\m{c_{0}(0),c_{1}(1),c_{0}(1),c_{1}(2),c_{0}(2),\ldots} and note that the
relation for \m{c_{0}(m+1)} can be rewritten as \m{c_{0}(m+1)=c_{1}(m+1)
+c_{0}(m)}. These relations then demand that each element of the list is the
sum of the previous two. Since \m{c_{0}(0)=c_{1}(1)=1}, this is just a
Fibonacci sequence, and we can take advantage of the standard formula for
the \m{\mathrm{n}^{\mathrm{th}}} term of a Fibonacci sequence, \m{F(n)}:
\begin{equation}
F_{n}=\frac{\phi^{n}-(-\phi)^{-n}}{\sqrt{5}}
\end{equation}
where \m{\phi} is the so-called ``golden ratio''
\m{\phi=\frac{1+\sqrt{5}}{2}}. From this, \m{c_{0}(m)=F(2m+1)} and
\m{c_{1}(m)=F(2m)}.

\begin{figure}[t]
\begin{center}
\m{\frac{\xi}{\la+1}}
\parbox{5in}{\includegraphics{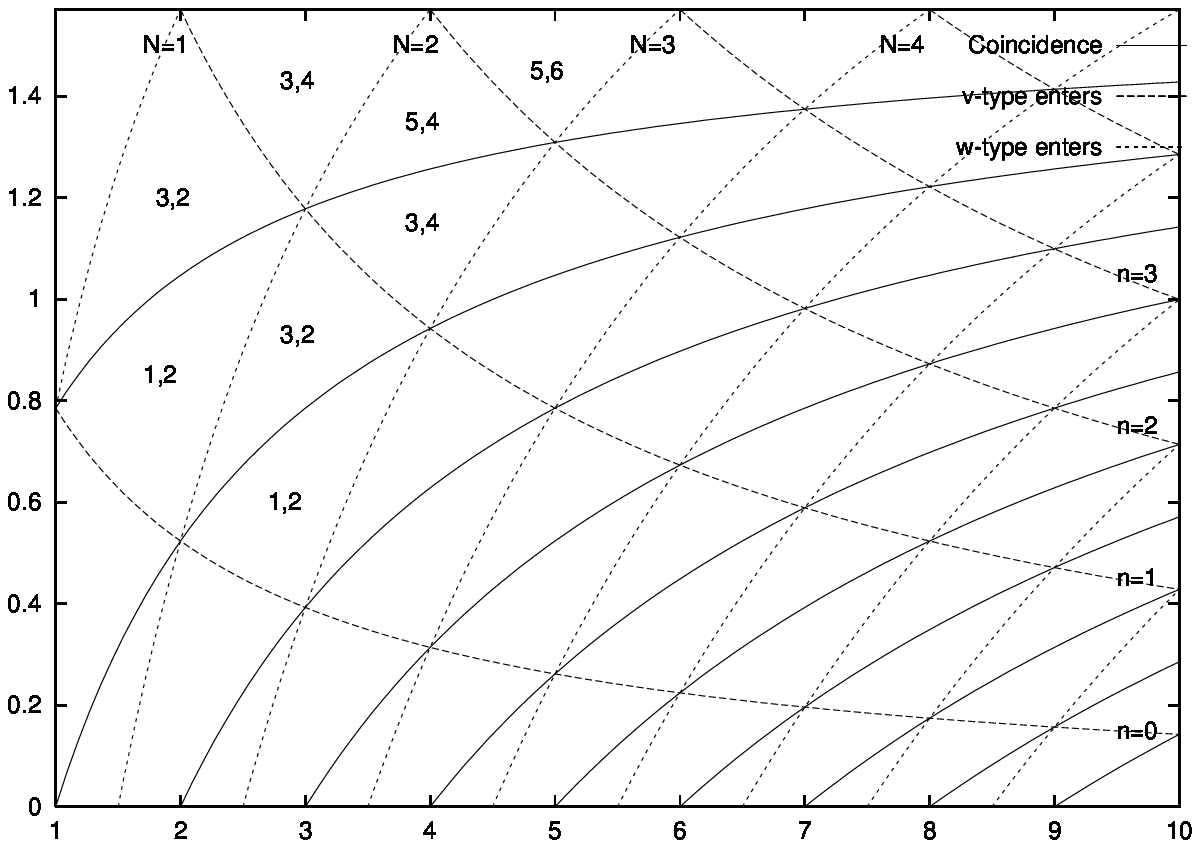}} \\
\m{\la}
\end{center}
\ca[Boundary bound state spectrum size.]{Boundary 
bound state spectrum size. \mdseries \itshape The number
of states 
present increases as
$\nu$-type and $w$-type poles enter the physical strip, but changes also
occur as the two sets of poles pass through coincidence: moving in the 
direction of increasing $\la$, the topmost relevant $w$-type pole
passes $\nu_{0}$
and ceases to be relevant, reducing the spectrum. (Notation $x,y$
implies $F(x)$ charge 0 states and $F(y)$ charge 1 states.)}
\label{fig:spectrum}
\end{figure}
One small complication is that, once \m{m=n_{*}}, \m{w_{n_{*}}} is not
necessarily in the physical strip. This means that, while the total number
of charge 1 states must be \m{c_{1}(n_{*}+1)}, the number of charge 0
states
will either be \m{c_{0}(n_{*}+1)} or \m{c_{0}(n_{*})} depending on whether
or not \m{w_{n_{*}}} is present. It is perhaps easiest to note that, with
\m{n'_{*}+1} relevant \m{w}-type poles, the number of charge 0 states is
\m{c_{0}(n'_{*}+1)}.

A plot of the spectrum size against \m{\la} and
\m{\frac{\xi}{\la+1}}\footnote{Since \m{\xi} lies between \m{0} and
\m{\frac{\pi(\la+1)}{2}}, it is more convenient to work with
\m{\frac{\xi}{\la+1}}, which lies between \m{0} and \m{\hp}.} is shown
in \fig{fig:spectrum}. Three sets of curves are shown: the points
where given \m{\nu} and \m{w} poles enter the physical strip, and the points
where the two sets of poles coincide. As drawn, a given \m{\nu} pole
will be in the spectrum above the appropriate line, and a given \m{w}
pole will be present to the right of its line. Note, however, that,
while \m{\nu} poles will never subsequently leave the spectrum, \m{w}
poles will; crossing a coincidence line to the right, the relevant \m{w}
pole with the highest rapidity passes \m{\nu_{0}} and ceases to be relevant,
reducing the number of relevant \m{w}-type poles by 1. The number of states
in each sector has been quoted in terms of Fibonacci numbers,
so that ``\m{x,y}'' implies a charge 0 sector of size \m{F_{x}} and a
charge 1 sector of size \m{F_{y}}. On the equality line, of course,
each sector has size \m{2^{n_{*}+1}}.

Finally, note that the top of the diagram represents \m{\xi=
\frac{\pi(\la+1)}{2}}, i.e. the coincidence case considered above, and the
region just below this represents the other case considered. Moving diagonally
down and to the right from there, the even indices receive successively
greater uniform additions but the spectrum size merely oscillates, as the
\m{\nu}-type and \m{w}-type poles take turns at having the lowest
rapidity.

\se{Other boundary conditions}
\label{sec:otherbcs}
Surprisingly, the extension of the Dirichlet results to encompass more
general boundary conditions does not require much more work. A hint as
to why can be gained from the fact that the general ground state
reflection factors can be rewritten in terms of those for Dirichlet
multiplied by terms which introduce no new poles at purely imaginary
rapidities. (They do, however, introduce poles at complex rapidities,
but these have an interpretation as resonances rather than bound
states and will be discussed separately.)

{}Despite the fact that the reflection factors appear to depend on
four parameters---\m{\xi},\m{k},\m{\eta} and \m{\vartheta}---it
is clear that essentially only two are independent, the other two
being determined by~\re{eq:paramconds}. If
we note that~\re{eq:paramconds} also implies
\begin{equation}
\sin(\eta)\sin(\vartheta)=-\frac{(-1)^{a}}{k}\sin \xi,
\end{equation}
with \m{a} either 0 or 1, we can re-write \m{\cos(\xi+\la u)} as
\ba
\cos(\xi+\la u)&=&\cos (\xi) \cos (\la u) - \sin (\xi)\sin (\la u)\\
 &=& -k\left[\cos (\eta)\cosh(\vartheta)\cos(\la u) -
(-1)^{a}\sin(\eta) \sinh(\vartheta)\sin(\la u)\right] \nn \\
 &=& -\frac{k}{2}\left[e^{\vartheta}\cos \left(\eta+(-1)^{a}\la
u\right) + e^{-\vartheta}\cos \left(\eta-(-1)^{a}\la u\right)\right] \nonumber.
\ea 

Denoting the reflection factors for the Dirichlet boundary condition
on the vacuum boundary state as \m{P^{\pm}_{D\st{0}}(u,\xi)},
\re{eq:genboot} can be re-written as
\begin{equation}
\begin{array}{rcl}
P^{\pm}(u)&=&R_{0}(u)\frac{\sigma(i\vartheta,u)}{2\cosh{\vartheta}}\left[
e^{\pm(-1)^{a}\vartheta}P^{+}_{D\st{0}}(u,\eta)+e^{\mp(-1)^{a}\vartheta}
P^{-}_{D\st{0}}\right] \\
Q^{\pm}(u)&=&-R_{0}(u)\frac{\sin(\la u)\sigma(i\vartheta,u)} 
{2\cosh(\vartheta)\cos (\eta)}\left[
P^{+}_{D\st{0}}(u,\eta)+P^{-}_{D\st{0}}(u,\eta)\right].
\end{array}
\end{equation}
Since the transformations \m{\vartheta \rightarrow -\vartheta},
\m{a \rightarrow a+1}, and \m{\eta \rightarrow -\eta} are all
equivalent to \m{\mathrm{soliton} \rightarrow \mathrm{anti-soliton}},
we shall set \m{a=0} and \m{\vartheta \geq 0} for simplicity. The
Dirichlet case corresponds to \m{\vartheta \rightarrow \infty}, in
which case \m{\eta \rightarrow \xi} and we recover the expected
factors.

In this form, it is clear that we will be able to re-use much of what
we have already found about the Dirichlet pole structure in the
general case. The one important difference is the factor of \m{\sigma
( i\vartheta,u)}. This only has poles at
complex \m{u}, however, and so will not contribute to the bound state
structure. We can thus ignore this factor for the present.

All the reflection factors have the same pole structure at purely imaginary
rapidities as
\m{P^{+}_{D\st{0}}}, though based on \m{\eta} rather than \m{\xi}. Arguing as
before, these must be responsible for the formation of a first set of
excited states. We will continue to use the notation \m{\nu_{n}} to
label these poles, on the understanding that it is more generally
defined as
\begin{equation}
\nu_{n}=\frac{\eta}{\la}-\frac{\pi(2n+1)}{2\la}\,.
\end{equation}

Unlike the Dirichlet case, however, where these poles appeared only in
one reflection factor, they now appear in all four. While
time-reversal symmetry argues that the poles in \m{Q^{+}} and
\m{Q^{-}} must form the same state, we must now deal with the
possibility that those in \m{P^{-}} and \m{P^{+}} potentially form
different states, degenerate in mass. This cannot be so, however,
since e.g. a soliton, incident on the boundary, cannot yet ``know''
whether it will ultimately be reflected back as a soliton or an
anti-soliton, meaning that the states formed by \m{Q^{+}} and
\m{P^{+}} must be the same. A similar argument holds for
anti-solitons, and so for \m{Q^{-}} and \m{P^{-}}. Since the states
formed by \m{Q^{+}} and \m{Q^{-}} must be the same, all four states
must, in fact, be the same state. This also means, for the solitonic
sector at least, that all states must be non-degenerate.

This degree of similarity with the Dirichlet case makes it a
reasonable guess that the entire structure should also be similar,
with reflection factors given by
\begin{equation}
\begin{array}{rcl}
P^{\pm}_{\st{x}}(u)&=&R_{0}(u)\frac{\sigma(i\vartheta,u)}{2\cosh{\vartheta}}
\left[ e^{\pm(-1)^{a}\vartheta}P^{+}_{D\st{x}}(u,\eta)+e^{\mp(-1)^{a}\vartheta}
P^{-}_{D\st{x}}\right] \\
Q^{\pm}_{\st{x}}(u)&=&-R_{0}(u)\frac{\sin(\la u)\sigma(i\vartheta,u)} 
{2\cosh(\vartheta)\cos (\eta_{c})}\left[
P^{+}_{D\st{x}}(u,\eta)+P^{-}_{D\st{x}}(u,\eta)\right]\,,
\end{array}
\label{eq:genform}
\end{equation} 
where \m{\eta_{0}=\eta} and \m{\eta_{1}=\pi(\la+1) -\eta}. The
breather factors, in turn, should be given by
\begin{equation}
R^{(n)}_{\st{x}}(u)=R^{(n)}_{0}(u)R^{(n)}_{1D\st{x}}(\eta,u)R^{(n)}_{1D\st{x}}
(i\vartheta u)\,,
\end{equation}
where \m{R^{(n)}_{1D\st{x}}(u)} is the boundary-dependent part of the
Dirichlet factor (i.e. without the \m{R^{(n)}_{0}(u)} term).

One difficulty that might be raised with this idea is that, since
topological charge is not in general conserved, the two charge sectors
might not translate into the general case. As we shall see in a
moment, the above reflection factors are correct as long as the bound
state poles at each level match the Dirichlet results. The 
argument given before for deciding whether or not a
pole is due to a bound state works just as well here, indicating
that this is indeed the case, so
the conclusion must be that there are still two sectors. In one sector
solitons bind with rapidities \m{\nu_{n}} and in the other they
bind at \m{w_{n}}, as is necessary for continuity with the Dirichlet
limit. The only difference is that the sector label now does not
correspond to topological charge; in fact, it does not appear to
correspond to anything other than the number of labels the state
carries. For this reason we shall now call them ``odd'' and ``even'',
rather than charge 1 or 0. The other difference is that, at all
stages, the poles appear in all four factors, allowing either a
soliton \emph{or} an anti-soliton to form a bound state. 

The formula \re{eq:genform} is most easily proven by induction. Since
we already know it is true for the ground state, all that remains is
to show that it is consistent with the bootstrap.

In its full glory, the boundary bootstrap equation reads
\begin{equation}
g^{\gamma}_{a\delta}R^{b'\beta}_{b\gamma}(\T)=g^{\beta}_{a_{2}\gamma}S^{b_{1}a_{1}}_{ba}
(\T-i\alpha^{\beta}_{a \delta}) R^{b_{2} \gamma}_{b_{1} \alpha} (\T)
S^{a_{2}b'}_{a_{1} b_{2}} (\T+ i\alpha_{a\delta}^{\beta}).
\end{equation}
Given that we are taking all boundary states to be non-degenerate, and
assuming that all states can be created by either a soliton or an
anti-soliton, we are free to take the incident particle to be whatever
we please. For convenience, then, we shall set \m{a=b}, leading to
\begin{equation}
g^{\beta}_{b\delta}R^{b'}_{b\st{\beta}}(\T)=g^{\beta}_{a_{2}\delta}S^{bb}_{bb}
(\T-i\alpha^{\beta}_{b \delta}) R^{b_{2}}_{b\st{\delta}} (\T)
S^{a_{2}b'}_{b b_{2}} (\T+ i\alpha_{b\delta}^{\beta}).
\end{equation}
(Making the other choice---\m{a\neq b}---can be shown to produce an
equivalent set of bootstrap equations, reinforcing the idea that all
reflection factors produce the same boundary state.)

The boundary couplings \m{g^{\beta}_{x\delta}} can be found from
\begin{equation}
\begin{array}{ccc}
P^{+}_{\st{\delta}} \approx
\frac{i}{2}\frac{g^{\beta}_{+\delta}g^{+\beta}_{\delta}}{\T-i\alpha_{\delta}^{\beta}},
&
P^{-}_{\st{\delta}} \approx
\frac{i}{2}\frac{g^{\beta}_{-\delta}g^{-\beta}_{\delta}}{\T-i\alpha_{\delta}^{\beta}},
&
Q^{\pm}_{\st{\delta}} \approx
\frac{i}{2}\frac{g^{\beta}_{\pm\delta}g^{\mp\beta}_{\delta}}{\T-i\alpha_{\delta}^{\beta}}.
\end{array}
\end{equation}

This means that, using our assumed form for the reflection factors,
\begin{equation}
g_{+\delta}^{\beta}=g_{-}(-1)^{n}e^{(-1)^{c}\vartheta}
\end{equation}
for pole \m{\nu_{n}} or \m{w_{n}} as applicable.

Overall, then, the bootstrap reads
\ba
P^{\pm}_{\st{\beta}}(u)&=&a(u-\alpha_{\delta}^{\beta})\left[
P^{\pm}_{\st{\delta}}(u)a(u+\alpha_{\delta}^{\beta})+(-1)^{n}e^{\mp(-1)^{c}\vartheta}Q^{\pm}_{\st{\delta}}(u)c(u+\alpha_{\delta}^{\beta})\right],
\nonumber \\
Q^{\pm}_{\st{\beta}}(u)&=&
a(u-\alpha_{\delta}^{\beta})b(u+\alpha_{\delta}^{\beta})Q^{\pm}_{\st{\delta}}
(u).
\nonumber 
\ea
  
Applying this to a state of our assumed form does indeed give (after
some cumbersome algebra) the requisite result. The other point which
remains is to show that, at each step, the spectrum is the same as
before. However, looking at the breather factors given above, it is
clear that their pole structure at imaginary rapidities is always the
same as for Dirichlet. The argument to determine the states which are
required
in the model depends exclusively on breather poles, and so must go
through precisely unchanged. The only danger is that the
remaining enumeration of the explanations for the other poles might
run into problems.

The solitonic factors have poles whenever either of the Dirichlet
factors do, the order being the higher of the two. Similarly, there
are zeroes whenever both Dirichlet factors have zeroes, the order
being the lower of the two. This turns out to mean that the
explanations used before still apply, with the difference that
the extra boundary vertices allow solitons and anti-solitons to
be interchanged within the diagrams in ways not possible in the
Dirichlet case. 

This allows a diagram which previously
explained a soliton pole to be re-used to explain an anti-soliton pole
at the same rapidity. In addition, the difference in the order of a pole
between the soliton and anti-soliton factors was due to loops which
allowed a cancellation between diagrams for one but not the
other (as in e.g. \fig{fig:fb3}). Altering the factors from their 
Dirichlet values destroys this delicate cancellation, raising the
order to the higher of the pair. With this borne in mind, the
discussion is completely analogous to that given previously, and so we
shall not repeat it here.

Finally, it is also worth noting that the general factors can still be
written in the form
\begin{equation}
P^{\pm}_{\st{c;n_{1},N_{1},\ldots}}(u)=P^{\pm}_{(c)}(u)a^{1}_{n_{1}}(u)
\overline{a^{1}_{N_{1}}(u)}\ldots\,, 
\label{eq:reltobs}
\end{equation}
with \m{P^{\pm}_{0}(u)=P^{\pm}_{\st{0}}(u)} and \m{P^{\pm}_{1}(u)=
\overline{P^{\pm}_{\st{0}}(u)}}. An analogous expression holds for the
\m{Q}s. 

\sse{Resonance states}
We now return to the extra factor of \m{\sigma(i\vartheta,u)}. This
provides extra complex poles, found from the imaginary poles we have
been discussing by replacing \m{\eta} with \m{i\vartheta}. Thus, the
most notable poles (and the ones we shall concentrate on) are at 
\m{u=\frac{i\vartheta}{\la}-\frac{\pi(2n+1)}{2\la}}. 

A feature of these poles is that they never fall into the physical
strip. Those which fall into the unphysical strip immediately below
the physical one (as the poles just given do) however, do have an
explanation as resonance states~\cite{Eden}. In bulk QFT, a
resonance state is an unstable bound state, and a similar idea applies
here. From the Breit-Wigner formula~\cite{BreitW}, we can find the
mass \m{M} and decay width \m{\g} of the state using the usual formulae 
with \m{M \rightarrow M+\frac{i\g}{2}}. For the bulk, this becomes
\begin{equation}
\left(M+\frac{i\g}{2}\right)^{2}=m_{1}+m_{2}+2m_{1}m_{2}\cosh
(\sigma-i\Theta)\,,
\end{equation}
for the binding of particles with masses \m{m_{1}} and \m{m_{2}}. In
our case, we find
\begin{equation}
M+\frac{i\g}{2}=m_{s}\cosh (\sigma-i\Theta)\,,
\end{equation}
or
\begin{align}
M&=m\cosh \sigma \cos \Theta \\ 
\g&= -2m_{s}\sinh \sigma \sin \Theta\,.
\end{align}
The lifetime of such a particle is \m{\tau \propto \frac{1}{\g}}; to
compare this with the discussion in the previous chapter, this can be 
converted into a phase delay by multiplying it by the real velocity 
\m{v=\tanh \sigma} to get \m{a \propto -2m_{s}\cosh \sigma
\sin \Theta)^{-1}}. For the poles
\m{\frac{i\vartheta}{\la}-\frac{\pi(2n+1)}{2\la}}, this then becomes
\begin{equation}
a \propto \left(2m_{s}\cosh \frac{\vartheta}{\la}\sin
\frac{\pi(2n+1)}{2\la}\right)^{-1}\,.
\end{equation}
In the classical limit \m{\beta \rightarrow 0}, taking
\m{m_{s}=\frac{8}{\beta^{2}}}, these become simply
\ba
\g &\propto& \frac{\beta^{2}(2n+1)\vartheta}{8\pi} \\
a &\propto& (2n+1)^{-1}\,.
\ea 
This means that, in this limit, the resonance states become stable,
though the phase delay remains finite. The poles collapse onto the
real axis, though at an infinite distance from the origin. In the
classical calculations of the previous chapter, however, due to the
re-scaling of the field, the poles collapse at a finite distance from
the origin with an infinite phase delay, as we have already found.

\se{Summary} 
\label{sec:conclusions}
\quot{No doubt aardvarks think that their offspring are beautiful too.}
{John Ellis}
We have found that the spectrum of boundary bound states
of the boundary sine-Gordon model can be
characterised in terms of two ``sectors''. With Dirichlet boundary 
conditions, these have topological charges
\m{\frac{\beta \varphi_{0}}{2\pi}} and \m{1-\frac{\beta
\varphi_{0}}{2\pi}} (which we labelled
as ``0'' and ``1'' respectively). Otherwise, if topological charge is 
not conserved, the sectors remain, but lose this interpretation. It is 
still useful to label them as ``0'' and ``1'', but this is best thought 
of as ``even'' and ``odd'', since they require even and odd numbers of 
solitonic particles for their creation.   

A boundary state
can be described in an index notation as
\m{\st{c;n_{1},n_{2},\ldots,n_{k}}} for sector \m{c}, with
\m{c=0} for
\m{k} even
and \m{c=1} for
\m{k} odd. For the Dirichlet case, such a state can be
created by a succession of alternating solitons and anti-solitons,
beginning with a soliton. With other boundary conditions, this 
requirement is eased, and any selection of solitonic particles becomes 
possible. To create a state in the odd sector, the necessary rapidities 
are of the form,
\begin{equation}
\nu_{n}=\frac{\eta}{\la}-\frac{\pi(2n+1)}{2\la}\,,
\end{equation}
while for the even sector they are
\begin{equation}
w_{m}=\pi-\frac{\eta}{\la}-\frac{\pi(2m-1)}{2\la}\,.
\end{equation}
These are interchanged by the transform \m{\eta
\rightarrow \pi(\la+1) - \eta}. Any such state can be formed, provided
the rapidities involved are monotonically decreasing,
i.e.\ \m{\nu_{n_{1}}>w_{n_{2}}>\nu_{n_{3}}>\ldots}, and its mass is
given by
\begin{align}
m_{n_{1},n_{2},\ldots}&=
m_s\sin^{2}\left(\frac{\eta-\hp}{2\la}\right)+
\sum_{i \mathrm{\ odd}} m_s\cos(\nu\phup_{n_i})
+ \sum_{j
\mathrm{\ even}} m_s\cos(w\phup_{n_j})\\
&=
m_s\sin^{2}\left(\frac{\eta-\hp}{2\la}\right)+
\sum_{i \mathrm{\ odd}} m_s\cos \left(
\frac{\eta}{\la} - \frac{(2n_{i}+1)\pi}{2\la}\right) \\
& \quad - \sum_{j
\mathrm{\ even}} m_s\cos \left( \frac{\eta}{\la}+
\frac{(2n_{j}-1)\pi}{2\la} \right)\,.\nn
\end{align}
This spectrum is considerably larger than that suggested
in \cite{Skorik}, 
though all the states introduced are required to satisfy our
lemmas. It is worth pointing out that a second part of the analysis
of \cite{Skorik} involved an examination of the (boundary) Bethe
ansatz for a lattice regularisation of the model.
Some of the masses which emerged in the course of that study---those of the
\m{(n,N)}-strings---were outwith the spectrum proposed
in \cite{Skorik}, but are now included as the masses of the states
\m{\st{1;0,n,N}}. It remains to be seen, however,
whether the other masses in our spectrum can be
recovered in the Bethe ansatz approach.

The number of states present in the spectrum clearly depends on the
boundary parameters, as illustrated in \fig{fig:spectrum}. It is
convenient to express this in terms of Fibonacci numbers, \m{F(x)}.
If
there are \m{n} \m{\nu}-type poles, and \m{m} relevant
\m{w}-type poles, there are, in general, \m{F(2n)} charge 1 states and
\m{F(2m+1)} charge 0 states. Explicitly, these are given by
\ba
n=\left[\frac{\eta}{\pi}-\hf\right]+1 &\mathrm{and} &
m=\left[\frac{\la}{2}-\frac{\eta}{\pi}+\hf\right] - \left[\la 
- \frac{\eta}{\pi}+\hf\right]\,,
\ea
where the square brackets denote the integer part of the number.
This changes when the two sets of poles
coincide, in which case there are \m{2^{n-1}} states in each sector.

Finally, we note that the general method used to
derive the spectrum, via the simple geometrical argument leading to
the two lemmas given in Section \ref{sec:colethun}, can be applied
equally well to any two-dimensional model. Using this to deduce the
existence of as many states as possible led---in our case---to the
full spectrum. In other cases, we may not be so fortunate, but using
it as a starting point should make the derivation of the full
spectrum a finite (though possibly lengthy) task.

\chap{Affine Toda Theory}
\label{chap:bulk}
\quot{It was here that the thaum, hitherto believed to be the smallest
possible particle of magic, was successfully demonstrated to  be made
up of `resons' (Lit.: `Thing-ies') or reality fragments. Currently
research indicates that each reson is itself made up of a combination
of at least five `flavours', known as `up', `down', `sideways', `sex
appeal' and `peppermint'.}{Terry Pratchett, Lords and Ladies}


\se{Introduction}
\quot{Once upon a time and a very good time it was.}{James Joyce}
The sine-Gordon model, which has occupied us for the previous two
chapters, is a member of the larger family of affine Toda field
theories (ATFTs), and it is to these that we will now turn our
attention. These theories are, in general, not as well understood as
the sine-Gordon model, even in the bulk.

ATFTs are also integrable, and rely on a Lie algebra structure built
into their Lagrangian to provide the necessary conserved
charges. A tantalising problem with them---and one which will provide the
basis for this work---is that the underlying structure shows up again
in their S-matrices, among other places, though it is not at all
clear how it arises. The difficulty is that the exact S-matrix
program, while it provides a good method for obtaining a result, is
totally disconnected from the original Lagrangian. For the boundary
sine-Gordon model, this caused problems in relating the parameters in
the reflection factors back to the parameters in the Lagrangian, while
here it hides the path of the Lie algebraic parameters into the
S-matrix.

Obtaining a better understanding of this is still an unsolved problem,
but we will find a neat method of constructing S-matrix elements
through rules based on the Lie algebra, generating a number of new
identities in the process.    

Before we plunge into the full quantum theory, a preliminary 
discussion of the classical version will serve to introduce much of the 
structure, as it did for the sine-Gordon model. In addition, building
an exact quantum S-matrix cannot begin 
without making some initial assumptions; looking at the classical theory 
will help us make a more educated guess. Classical theories are also 
more intuitively comprehensible, so the more that can be gleaned from 
them and transferred across to the full quantum case, the more 
tractable it becomes.   

We shall only attempt a relatively brief introduction to the topic
here, sufficient for our needs. For a more detailed review and further
references see e.g. \cite{Correview}.

\se{The Lagrangian}
Affine Toda field theory (ATFT) is a massive integrable 1+1--dimensional 
theory with a number---which we shall call \m{r}---of scalar fields 
\m{\phi^{a}}, and with a Lagrangian of the form
\begin{equation}
\mathcal{L}=\frac{1}{2}\partial^{\mu}\phi_{a}\partial_{\mu}\phi_{a}
-\frac{m^{2}}{\beta^{2}}\sum_{j=0}^{r}n_{j}\exp(\beta\alpha_{j}\cdot \phi)\,,
\label{eq:tlagr}
\end{equation}
where \m{m} determines the mass scale (though it does not equate to the 
mass of any individual particle) and \m{\beta} is a dimensionless 
coupling constant.  

The \m{\alpha_{j}} can, in principle, take any values, but it turns out 
\cite{Mikhailov} that the resulting theory is only integrable if, for 
\m{j=1 \ldots r}, they can be considered as the simple roots of a rank-\m{r}
semi-simple Lie algebra \m{g}. 

This is because leaving \m{\alpha_{0}} out of the sum gives a
conformal theory, known
as conformal Toda theory or just Toda theory. The possession of
conformal (or scale) invariance naturally gives such theories an
infinite number of symmetries, and hence any conformal field theory
must
be integrable. However, because this means that the theory cannot
depend on any fixed length scale, all the particles in it must be
massless (as, otherwise, the inverse of the mass would provide such a
scale). Including the extra root to form the ``affine'' theory can be 
considered as a perturbation which breaks the 
conformal invariance---and so provides its particles with
mass---while still retaining an infinite number of symmetries.
(Taking \m{\alpha_{0}} to be the affine root is purely a
conventional choice of labelling.) Interest in these theories was
initially stimulated by this connection to perturbed conformal field
theories \cite{Zamconf}, and the fact that, through the breaking of
the conformal symmetry, the particles acquired mass.

An important feature of such algebras
is that they can be conveniently classified \cite{Cartan} in terms of
a Cartan matrix---\m{C}---defined by
\begin{equation}
C_{ij}=\frac{2(\alpha_{i},\alpha_{j})}{(\alpha_{i},\alpha_{i})}\,,
\end{equation}
where \m{(\alpha_{i},\alpha_{j})} denotes an inner-product on the
roots \m{\alpha_{i}} and \m{\alpha_{j}}. This matrix encodes the
relationships between the simple roots, and is particularly simple in
that it is composed entirely of integer entries. The content of the
matrix is often described by a Dynkin \cite{Dynkin} diagram, where
each simple root is drawn as a ``spot'' and the spots corresponding to
roots \m{\alpha_{i}} and \m{\alpha_{j}} are connected by \m{n} lines
if \m{C_{ij}=n}. In the case where \m{C_{ij} \neq C_{ji}}, an arrow is
drawn on the lines pointing from the long root to the short root. 

In Cartan's classification, there are two infinite sets of untwisted
``simply-laced'' algebras (where all roots are of the same length)
known as \m{a_{r}^{(1)}} and \m{d_{r}^{(1)}}, with three exceptional
cases, \m{e_{6}^{(1)}, e_{7}^{(1)}} and \m{e_{8}^{(1)}}. There are also
``nonsimply-laced''
algebras (where one root has a different length to the others) divided
into two infinite sets, \m{b_{r}^{(1)}} and \m{c_{r}^{(1)}}, and two
exceptional
cases, \m{g_{2}^{(1)}} and \m{f_{4}^{(1)}}. A listing of their Cartan
matrices,
together with their Dynkin diagrams, can be found in Appendix
\ref{app:liedata}. A good introduction to the topic can be found in 
\cite{Carter}.    
  
While it appears on an equal footing with the other simple roots, and it can
be drawn as an extra spot on the Dynkin diagram to describe its inner
products with the other simple roots, \m{\alpha_{0}} is not itself
simple, in that it can be described as a linear combination of
the other simple roots:
\begin{equation}
\alpha_{0}=-\sum_{i=1}^{r}n_{i}\alpha_{i}\,.
\end{equation} 
The \m{n_{i}}---usually 
called marks or Kac labels---are
integers, chosen to make \m{\sum n_{i}\alpha_{i}=0}. (The value of 
\m{\alpha_{0}} is prescribed by demanding that this be true with 
\m{n_{0}=1}.) Two other useful pieces of notation are the Coxeter and 
dual Coxeter numbers, \m{h} and \m{\hv} defined by
\ba
h = 1+ \sum_{i=1}^{r}n_{i} & \mathrm{and} & \hv = 
1+\sum_{i=1}^{r}n_{i}^{\vee}\,,
\ea
where the co-marks, \m{n_{i}^{\vee}}, are related to the marks through 
\m{n_{i}^{\vee}=n_{i}\alpha_{i}^{2}/2}. The Coxeter and dual Coxeter 
numbers will arise frequently in the context of the periodicity of poles 
of the S-matrix or the number of distinct conserved charges.

Finally, the so-called ``incidence matrix'' \m{G} deserves a mention. 
This is just the negative of the Cartan matrix with all the diagonal 
terms set to zero, meaning that it simply encodes the relationships 
\emph{between} the roots.

Without wishing to delve too far into the development of the theory
(further details can be found in \cite{BCDS}), it contains \m{r} massive 
particles, which can be associated with spots on the Dynkin diagram of 
\m{g}. Their masses and couplings can be easily extracted from a 
perturbative expansion (in small \m{\beta}) of the the potential term of 
the Lagrangian \re{eq:tlagr}:
\ba
V(\phi)= \frac{m^{2}}{\beta^{2}}\sum_{j=0}^{r}n_{j}\exp(\beta\alpha_{j}\cdot 
\phi) &=& \frac{m^{2}}{\beta^{2}}\sum_{j=0}^{r}n_{j} + \frac{m^{2}}{2} 
\sum_{j=0}^{r}n_{j}\alpha_{j}^{a}\alpha_{j}^{b}\phi^{a}\phi^{b} 
\nonumber \\
&&+\frac{m^{2}\beta}{3!}\sum_{j=0}^{r}n_{j}\alpha_{j}^{a}\alpha_{j}^{b} 
\alpha_{j}^{c}\phi^{a}\phi^{b}\phi^{c}+\ldots
\ea
This allows us to read off a \m{\mathrm{(mass)}^{2}} matrix
\begin{equation}
(M^{2})^{ab}=m^{2}\sum_{j=0}^{r}n_{j}\alpha_{j}^{a}\alpha_{j}^{b}\,,
\end{equation}
and a set of three-point couplings
\begin{equation}
C^{abc}=\beta
m^{2}\sum_{j=0}^{r}n_{j}\alpha_{j}^{a}\alpha_{j}^{b}\alpha_{j}^{c}\,,
\label{eq:classc}
\end{equation}
as well as infinitely many higher couplings, at successively higher
orders in \m{\beta}.

If a basis of fields is chosen so as to make the bare propagator
diagonal, \m{(M^{2})} becomes diagonal also, allowing the classical
masses to be read off as eigenvalues. Finding such a basis, and
especially computing the three-point couplings in it, is too long a task
to be attempted here, but it can be done, and closed-form answers
obtained \cite{BCDS}. These results, together with other relevant Lie
algebraic data, can be found in Appendix \ref{app:liedata}.

\se{The Quantum Theory}
To find the S-matrix of the quantum theory through the bootstrap
approach, we need to begin with a suitable guess at one or more of its
elements. If, after working through the bootstrap, the result is
consistent---each three-point coupling must introduce poles in all
three relevant S-matrix elements---then the guess could be said to be
good. Otherwise, corrections need to be made until a consistent result
is achieved.

{}From the earlier classical results, we might guess that the same
couplings transfer across to the quantum case, and so predict a minimal pole
structure for the S-matrix. One potential problem with this approach is
that the classical case is the \m{\beta \rightarrow 0} limit of the
quantum theory so, as \m{\beta} moves away from zero, the mass ratios, 
and hence the pole positions, would be expected to change due to 
renormalisation. As luck would have it, moving away from
this limit in simply-laced cases does not change the position of the 
poles we have considered so far (one-loop calculations showing that the masses renormalise in
such a way as to leave their ratios unchanged). In an intuitive sense, 
the bootstrap equations determine the algebraic structure so precisely 
that any continuous change in the parameters (such as the coupling 
angles) disturbs the way the pieces fit together and destroys the 
solution. Thus the classical mass ratios remain even in the full quantum 
theory. For simplicity, we will go through this case in more detail, and 
just quote the results for the nonsimply-laced cases.

\sse{Simply-laced cases}
The next logical step is to
construct a putative S-matrix element with a suitable pole structure. 
A good ``building block'' for this is 
provided by
\begin{equation}
(x)=\frac{\sinh \hf\left(\theta+\frac{i\pi x}{h}\right)}{\sinh \hf
\left(\theta-\frac{i\pi x}{h}\right)}\,.
\end{equation}
As mentioned in Chapter \ref{chap:intro}, this automatically enforces
unitarity. It also has only one pole (at \m{\theta=\frac{i\pi x}{h}})
and one zero (at \m{\theta=-\frac{i\pi x}{h}}), making it easy to form a
suitable product. (In the nonsimply-laced cases, the poles are no 
longer always multiples of \m{\frac{i\pi}{h}}, so a different block is 
needed.) Crossing symmetry is enforced by demanding a suitable
pole structure, and---ATFTs being elastic scattering theories---we
need not worry about the Yang-Baxter relation, leaving just the
bootstrap to be satisfied. Building e.g. \m{S_{11}} in this way and
working through the bootstrap, we do indeed find that it is consistent.

While this turns out to encode the bound state poles correctly, there is
no mention of the coupling constant, so it is unlikely to be the 
complete story. Trying to introduce a dependence on the coupling constant
leads to the idea that the full S-matrix elements are the
elements found so far (usually termed ``minimal'' since they are also the
complete S-matrix elements of certain perturbed conformal field theories
known as minimal models) multiplied by a suitable factor. This factor 
is firstly determined by the fact
that the resultant S-matrix must still respect unitarity and crossing
symmetry, making it natural to also build it out of the \m{(x)} blocks. In
addition, all the necessary bound state poles are already encoded in the
minimal S-matrix, so the extra factor should not introduce any more physical
strip poles, at least for \m{\beta} small, though it must still respect the
bootstrap.

Finally---and this is the reason why an extra factor is not just an
aesthetic invention---at \m{\beta=0}, \re{eq:classc} shows that all 
the classical three-point couplings disappear, so the extra factor 
should provide zeros to cancel all the physical-strip poles in the 
minimal elements, tending to them as \m{\beta \rightarrow 0}. This means 
we might be tempted to build the full S-matrix out of blocks of the 
form \m{(x)/(x\pm B)}, where \m{B} is a
coupling-constant dependent constant. One final complication, however, is
that the sign of the residue at a pole determines whether it corresponds to
a forward or cross-channel process. This can be found to be correct for the
minimal elements, and must be kept so, determining the sign above.

Following this through motivates the introduction of an extended block
\begin{equation}
\{x\} = \frac{(x-1)(x+1)}{(x-1+B)(x+1-B)}\,,
\end{equation}
from which the S-matrices of all the simply-laced ATFTs can be built. 
The S-matrix elements are usually written in the form
\begin{equation}
S_{ab}(\beta)=\prod_{x=1}^{h}\{x\}^{m_{ab}(x)}\,,
\end{equation}
where the non-negative integers \m{m_{ab}(x)} denote the multiplicity of 
the block.

An interesting property of this block is that \m{\{x\}_{B}=\{x\}_{2-B}}, 
heralding a duality. Ensuring no extra physical poles for real \m{\beta}
means that \m{0 \leq B(\beta) \leq 2}, and we have constructed \m{B} to 
vanish at \m{\beta=0}, so we might imagine that \m{B \rightarrow 2} as 
\m{\beta \rightarrow \infty}. This would set up a strong-weak coupling 
duality, the theory becoming free in either limit. Determining the 
precise form of \m{B(\beta)} turns out to be difficult, but it is 
conjectured to be~\cite{Arinshtein,BCDS, CM, Christe}
\begin{equation}
B(\beta)=\frac{1}{2\pi}\frac{\beta^{2}}{1+\beta^{2}/4\pi}\,,
\end{equation}
implementing the duality as
\begin{equation}
B\left(\frac{4\pi}{\beta}\right)=2-B(\beta)\,.
\end{equation}

\sse{Nonsimply-laced cases}
For these, the S-matrix can be written in a product
form~\cite{Dorey2} as
\begin{equation}
S_{ab}(\T)=\prod_{x=1}^{h}\prod_{y=1}^{\rv\hv} \{x,y\}^{m_{ab}(x,y)}\,,
\end{equation}
where the \m{\{x,y\}} are of the form
\begin{equation}
\{x,y\}=\frac{\langle x-1,y-1 \rangle \langle x+1,y+1 \rangle}{
\langle x-1,y+1 \rangle \langle x+1,y-1 \rangle}\,,
\end{equation}
with
\begin{equation}
\langle x,y \rangle = \langle \frac{(2-B)x}{2h}+\frac{By}{2\rv\hv}
\rangle\,,
\end{equation}
and
\begin{equation}
\langle x \rangle = \frac{\sinh \left(\hf\left(\T+i\pi x \right)\right)}{
\sinh \left(\hf \left( \T - i\pi x \right)\right)}\,.
\end{equation}
The \m{m_{ab}(x,y)}s are again non-negative integers, serving to encode 
the Lie algebraic structure of the model. This time, \m{B(\beta)} is 
conjectured \cite{Dorey2} to be given by
\begin{equation}
B(\beta)=\frac{2\beta^{2}}{\beta^{2}+\frac{4\pi h}{\hv}}\,.
\end{equation}
A difference now, however, is that, while there is still a strong-weak 
coupling duality present, it relates the strong coupling r\'{e}gime 
of one theory to the weak coupling r\'{e}gime of a different theory. For 
example
\begin{equation}
B_{c_{n}^{(1)}}\left(\frac{4\pi}{\beta}\right)=2-B_{d_{n+1}^{(2)}}(\beta)\,.
\end{equation} 
For this reason, the algebras \m{c_{n}^{(1)}} and \m{d_{n+1}^{(2)}} are 
termed a ``dual pair''. The simply-laced algebras (and \m{a_{2n}^{(2)}}) 
are self dual, and all the other algebras fall naturally into the dual 
pairs \m{(b_{n}^{(1)}}, \m{a_{2n-1}^{(2)})}, \m{(c_{n}^{(1)}}, 
\m{d_{n+1}^{(2)})},
\m{(g_{2}^{(1)}}, \m{d_{4}^{(3)})}, and \m{(f_{4}^{(1)},e_{6}^{(2)})}. The 
S-matrices for each member of a dual pair are the same except for the 
interchange of \m{h} and \m{\hv}. In light of this, we will
concentrate on the untwisted algebras from now on, and drop the
superscripts.

\se{Lie algebra structure}
S-matrices were first found through this approach for the simply-laced 
cases~\cite{Arinshtein, BCDS, CM, Christe}, and later for 
nonsimply-laced cases~\cite{del, CDS}. (The results are summarised in
Appendix 
\ref{app:liedata}.) This was all accomplished on a case-by-case basis,
and, although there were many hints of the underlying Lie algebra in the 
results, it was not clear how that had been transferred across from the 
Lagrangian. This is frustrating as, apart from the general demands of 
unitarity, analyticity and crossing
symmetry, the ATFT S-matrix is principally shaped by the Lie algebra. 

These results were put on a uniform Lie algebraic basis for the
simply-laced cases by Dorey~\cite{Dorey}. He
considered the Weyl
reflection \m{w_{i}} corresponding to the simple root \m{\alpha_{i}},
defined by
\begin{equation}
w_{i}(x)=x-\frac{2}{\alpha_{i}^{2}}(\alpha_{i},x)\alpha_{i}\,.
\end{equation}
From this, he set \m{w=w_{1}w_{2}\ldots w_{r}} to be a Coxeter element,
with \m{\langle w \rangle} the subgroup of the Weyl group generated by
\m{w}, and defined roots \m{\phi_{i}} by
\begin{equation}
\phi_{i}=w_{r}w_{r-1}\ldots w_{i+1}(\alpha_{i})\,.
\end{equation}
The other crucial ingredient was a two-colouring of the spots on the
Dynkin diagram, where each spot was labelled as either ``black'' or
``white'' such that no two adjacent spots had the same colour. Then,
the integers \m{m_{ab}(x)} turned out to be just
\begin{equation}
m_{ab}(2p+1+u_{ab})=(\la_{a},w^{-p}\phi_{b})\,,
\end{equation}
where \m{\la_{a}} is the fundamental weight corresponding to root \m{a},
and \m{u_{ab}=(c(a)-c(b))/2}, with \m{c(a)=\pm 1} encoding the colour of
the roots. In addition, if we define
\m{\g_{i}} as the orbit of \m{\phi_{a}} under \m{\langle w
\rangle}, then \m{C^{ijk} \neq 0} iff there exists
\m{\alpha_{(i)} \in \g_{i}}, \m{\alpha_{(j)} \in \g_{j}}, and
\m{\alpha_{(k)} \in \g_{k}} such that
\m{\alpha_{(i)}+\alpha_{(j)}+\alpha_{(k)}=0}. 

These results were initially found by observation. However,
Freeman~\cite{Freeman} showed how to diagonalise the mass matrix in a
Lie algebraic way, allowing them to be re-derived more rigorously. 

Similar results were later found for the
nonsimply-laced algebras by Oota~\cite{Oota} through a deformation of the 
Coxeter element. Oota also produced an
integral formula for the S-matrix, explicitly built from the Cartan
matrix, which we shall discuss below. This formula was later
reproduced by Fring, Korff and Schultz~\cite{Fring}, while a similar
result was conjectured by Frenkel and Reshetikhin~\cite{Frenkel} in the
course of a general study of \m{\mathcal{W}}-algebras.

The starting point for our discussion will be the processes shown in
\fig{fig:lieproc}. When two identical initial particles have a
relative rapidity of \m{2\T_{h}+t_{a}\T_{H}}, these comprise all the
possible diagrams which can result. The interesting point about them,
however, is that it turns out that the particles present in the middle
of each diagram are always those adjacent to the initial particle on
the Dynkin diagram. All such particles are present, but no others.

For the simply-laced cases, only the first three diagrams are
relevant, applying to the cases where the initial particle has one,
two and three adjacent particles respectively. Note that all the
intermediate particles are parallel
and have zero rapidity in the centre of mass reference frame.

For the nonsimply-laced
cases, the situation is complicated slightly when the adjacent
particles are associated with roots which are longer than that of
the initial particle. The fourth and fifth diagrams describe this for the case
where there is only one adjacent particle, and the Cartan matrix entry
is 2 or 3 respectively. For more than one adjacent particle, the
relevant vertical line must be replaced with this more complex
pattern, as shown in the last diagram.

The precise makeup of these diagrams is as follows. The first diagram speaks
for itself, while, in the next (for two adjacent particles), the unspecified
particle is always the lightest in the theory. The case with three adjacent
particles only occurs for \m{d_{n}} and \m{e_{6-8}}, for which the particles
are given in table \ref{tab:3adj}.

The next two only occur for \m{c_{n}} and \m{g_{2}}, with the particles as
shown.
The last diagram occurs in \m{b_{n}} and \m{f_{4}}, with the particles
given
in table \ref{tab:2adj}.

\begin{table}[ht]
\[
\begin{array}{c|c|c|c|c|c|c} \thline
\mathrm{\bf Theory} & \mathbf{a} & \mathbf{b} & \mathbf{c} & \mathbf{d} &
\mathbf{e} & \mathbf{f} \\ \thline
d_{n} & n-2 & n-3 & n-1 & 1\phantom{-n} & n\phantom{-1} & n\phantom{-1} \\
e_{6} & 4 & 3 & 2 & 1 & 6 & 5 \\
e_{7} & 7 & 5 & 3 & 2 & 1 & 6 \\
e_{8} & 8 & 6 & 4 & 2 & 1 & 7 \\
\thline
\end{array}
\]
\vspace{-0.25in}
\ca{Diagrams for the cases with three adjacent particles}
\label{tab:3adj}
\end{table}

\begin{table}[ht]
\[
\begin{array}{c|c|c|c|c|c} \thline
\mathrm{\bf Theory} & \mathbf{a} & \mathbf{b} & \mathbf{c} & \mathbf{d} 
& \mathbf{e} \\ \thline
b_{n} & n-1 & n-2 & 1\phantom{-n} & n\phantom{-1} & n-1 \\
f_{4} & 3 & 1 & 1 & 3 & 4 \\
\thline
\end{array}
\]
\vspace{-0.25in}
\ca{Diagrams for nonsimply-laced cases with two adjacent particles}
\label{tab:2adj}
\end{table}

\begin{figure}
\parbox{0.49\textwidth}{
\unitlength 1.00mm
\linethickness{0.4pt}
\begin{picture}(40.78,50.44)
\put(12.78,50.44){\line(1,-1){14.00}}
\put(40.78,50.44){\line(-1,-1){14.00}}
\put(26.78,36.44){\line(0,-1){17.89}}
\put(12.78,4.56){\line(1,1){14.00}}
\put(40.78,4.56){\line(-1,1){14.00}}
\end{picture}}
\parbox{0.49\textwidth}{
\unitlength 1.00mm
\linethickness{0.4pt}
\begin{picture}(43.11,50.11)
\put(18.89,36.44){\line(0,-1){17.89}}
\put(18.89,36.44){\line(3,1){15.78}}
\put(19.00,18.44){\line(3,-1){15.78}}
\put(34.67,41.67){\line(0,-1){28.44}}
\put(34.67,13.22){\line(1,-1){8.44}}
\put(18.89,18.44){\line(-1,-1){13.56}}
\put(34.67,41.67){\line(1,1){8.44}}
\put(18.89,36.44){\line(-1,1){13.56}}
\end{picture}}\\
\parbox{0.49\textwidth}{
\unitlength 1.00mm
\linethickness{0.4pt}
\begin{picture}(47.89,62.11)
\put(25.00,34.89){\line(0,-1){17.11}}
\put(25.00,34.78){\line(2,1){11.11}}
\put(25.00,34.78){\line(-4,3){11.22}}
\put(25.00,17.89){\line(2,-1){11.11}}
\put(25.00,17.89){\line(-4,-3){11.22}}
\put(36.11,40.33){\line(0,-1){28.00}}
\put(13.78,43.22){\line(0,-1){33.78}}
\put(36.11,12.33){\line(1,-1){11.78}}
\put(13.78,9.44){\line(-1,-1){8.89}}
\put(36.11,40.33){\line(1,1){11.78}}
\put(13.78,43.22){\line(-1,1){8.89}}
\put(10.78,4.44){\makebox(0,0)[cc]{$a$}}
\put(41.67,4.56){\makebox(0,0)[cc]{$a$}}
\put(11.89,26.11){\makebox(0,0)[cc]{$b$}}
\put(23.56,26.22){\makebox(0,0)[cc]{$c$}}
\put(37.44,26.33){\makebox(0,0)[cc]{$f$}}
\put(10.78,48.11){\makebox(0,0)[cc]{$a$}}
\put(41.67,48.22){\makebox(0,0)[cc]{$a$}}
\put(20.44,40.00){\makebox(0,0)[cc]{$d$}}
\put(30.89,39.11){\makebox(0,0)[cc]{$e$}}
\put(20.44,13.00){\makebox(0,0)[cc]{$d$}}
\put(30.78,13.56){\makebox(0,0)[cc]{$e$}}
\end{picture}}
\parbox{0.49\textwidth}{
\unitlength 1.00mm
\linethickness{0.4pt}
\begin{picture}(47.89,58.78)
\put(14.00,41.11){\line(5,-6){23.52}}
\put(14.00,41.11){\line(1,0){23.56}}
\put(14.00,12.89){\line(5,6){23.52}}
\put(14.00,12.89){\line(1,0){23.56}}
\put(37.67,12.89){\line(4,-3){10.22}}
\put(13.89,12.89){\line(-4,-3){10.22}}
\put(37.67,41.11){\line(4,3){10.22}}
\put(13.89,41.11){\line(-4,3){10.22}}
\put(7.67,10.56){\makebox(0,0)[cc]{$n$}}
\put(44.11,10.78){\makebox(0,0)[cc]{$n$}}
\put(25.89,11.33){\makebox(0,0)[cc]{$1$}}
\put(25.89,42.78){\makebox(0,0)[cc]{$1$}}
\put(36.67,21.22){\makebox(0,0)[cc]{$n-1$}}
\put(15.67,21.33){\makebox(0,0)[cc]{$n-1$}}
\put(15.56,33.33){\makebox(0,0)[cc]{$n-1$}}
\put(36.67,33.56){\makebox(0,0)[cc]{$n-1$}}
\put(44.00,43.11){\makebox(0,0)[cc]{$n$}}
\put(7.78,43.22){\makebox(0,0)[cc]{$n$}}
\end{picture}}\\
\parbox{0.49\textwidth}{
\unitlength 1.00mm
\linethickness{0.4pt}
\begin{picture}(48.67,61.33)
\put(9.99,45.91){\line(5,-6){31.52}}
\put(9.99,8.09){\line(5,6){31.52}}
\put(9.89,46.00){\line(5,-2){15.89}}
\put(41.67,46.00){\line(-5,-2){15.89}}
\put(9.89,8.00){\line(5,2){15.89}}
\put(41.67,8.00){\line(-5,2){15.89}}
\put(25.78,39.78){\line(0,-1){25.44}}
\put(41.56,46.00){\line(4,3){7.11}}
\put(10.00,46.00){\line(-4,3){7.11}}
\put(41.56,8.00){\line(4,-3){7.11}}
\put(10.00,8.00){\line(-4,-3){7.11}}
\put(5.56,6.78){\makebox(0,0)[cc]{$2$}}
\put(46.22,6.89){\makebox(0,0)[cc]{$2$}}
\put(18.67,9.78){\makebox(0,0)[cc]{$1$}}
\put(32.00,10.00){\makebox(0,0)[cc]{$1$}}
\put(27.00,19.78){\makebox(0,0)[cc]{$1$}}
\put(27.00,34.67){\makebox(0,0)[cc]{$1$}}
\put(17.22,19.67){\makebox(0,0)[cc]{$1$}}
\put(33.89,19.78){\makebox(0,0)[cc]{$1$}}
\put(17.11,35.11){\makebox(0,0)[cc]{$1$}}
\put(34.00,34.78){\makebox(0,0)[cc]{$1$}}
\put(19.56,43.56){\makebox(0,0)[cc]{$1$}}
\put(32.56,43.67){\makebox(0,0)[cc]{$1$}}
\put(46.22,47.56){\makebox(0,0)[cc]{$2$}}
\put(5.44,47.44){\makebox(0,0)[cc]{$2$}}
\end{picture}}
\parbox{0.49\textwidth}{
\unitlength 1.00mm
\linethickness{0.4pt}
\begin{picture}(50.67,60.89)
\put(19.30,37.94){\line(5,-6){18.43}}
\put(19.30,37.94){\line(1,0){18.46}}
\put(19.30,15.83){\line(5,6){18.43}}
\put(19.30,15.83){\line(1,0){18.46}}
\put(19.22,15.83){\line(-4,-3){8.01}}
\put(19.22,37.94){\line(-4,3){8.01}}
\put(11.11,44.00){\line(0,-1){34.22}}
\put(11.11,9.78){\line(-1,-1){6.78}}
\put(37.78,15.78){\line(1,-1){12.89}}
\put(11.11,44.00){\line(-1,1){6.78}}
\put(37.78,38.00){\line(1,1){12.89}}
\put(9.00,5.33){\makebox(0,0)[cc]{$a$}}
\put(45.67,5.44){\makebox(0,0)[cc]{$a$}}
\put(9.22,26.78){\makebox(0,0)[cc]{$b$}}
\put(16.33,12.00){\makebox(0,0)[cc]{$c$}}
\put(16.33,41.78){\makebox(0,0)[cc]{$c$}}
\put(8.89,48.11){\makebox(0,0)[cc]{$a$}}
\put(45.67,48.00){\makebox(0,0)[cc]{$a$}}
\put(28.56,14.11){\makebox(0,0)[cc]{$d$}}
\put(28.56,39.33){\makebox(0,0)[cc]{$d$}}
\put(34.33,22.44){\makebox(0,0)[cc]{$e$}}
\put(22.78,22.56){\makebox(0,0)[cc]{$e$}}
\put(22.78,31.44){\makebox(0,0)[cc]{$e$}}
\put(34.33,31.56){\makebox(0,0)[cc]{$e$}}
\end{picture}}
\ca{Processes which impose Lie algebra structure on the S-matrix}
\label{fig:lieproc}
\end{figure}

These results come from a case-by-case analysis; it should be
possible to derive them from the Lie algebraic rule given above,
but, for the moment, we have not attempted to do this. However, if we take
them
as axiomatic of how to encode the Lie algebra into the S-matrix, there
are many consequences.

\se{The consequences}
The first important consequence of this result is that, as with the
bootstrap relations, another particle can be introduced, on a trajectory
which either crosses the two incoming particles before they interact, or
afterwards. This is shown in \fig{fig:conseq}. Due to factorisation, the
amplitudes for these two processes should be the same, giving rise to what
might be called the ``generalized bootstrap''
\begin{multline}
S_{ij}(\T+\T_{h}+t_{i}\T_{H})S_{ij}(\T-\T_{h}-t_{i}\T_{H})= \\ \e^{-2i\pi
\Theta(\T)G_{ij}}
\prod_{l=1}^{r}\prod_{n=1}^{G_{il}}S_{jl}(\T+(2n-1-G_{il})\T_{H})\,.
\label{eq:grtv}
\end{multline}
For conciseness, we have defined \m{\T_{h}=\frac{i\pi(2-B)}{2h}} and
\m{\T_{H}=\frac{i\pi B}{2\rv\hv}}, with the integer \m{\rv} being the 
maximum number of edges connecting any two
vertices of the Dynkin diagram\footnote{This is 1 for the \m{a,d} and 
\m{e} series, 2 for \m{f_{4}} and 3 for \m{g_{2}}.}. 
The integers \m{t_{i}} are defined by \m{t_{i}=\rv\frac{(\alpha_{i}, 
\alpha_{i})}{2}}, where the squared length
of the short roots is normalised to 2\footnote{Thus \m{t_{i}=1} for short
roots and \m{t_{i}=\rv} for long roots.}.

This formula was first discovered for simply-laced cases by Ravanini, Tateo
and Valleriani~\cite{Ravanini}, and was independently derived for
the nonsimply-laced algebras in unpublished work by the author and
P. Dorey~\cite{Mattsson3} (see also \cite{Mattsson}) and by Fring, Korff
and Schultz~\cite{Fring}.

\begin{figure}
\parbox{2.5in}{
\unitlength 1.00mm
\linethickness{0.4pt}
\begin{picture}(45.56,50.11)
\put(30.67,36.44){\line(0,-1){17.89}}
\put(30.67,36.44){\line(-3,1){15.78}}
\put(30.56,18.44){\line(-3,-1){15.78}}
\put(14.89,41.67){\line(0,-1){28.44}}
\put(14.89,13.22){\line(-1,-1){8.44}}
\put(30.67,18.44){\line(1,-1){13.56}}
\put(14.89,41.67){\line(-1,1){8.44}}
\put(30.67,36.44){\line(1,1){13.56}}
\put(4.67,6.44){\line(3,1){40.89}}
\end{picture}
} \ \LARGE = \normalsize \
\parbox{2.5in}{
\unitlength 1.00mm
\linethickness{0.4pt}
\begin{picture}(45.56,50.11)
\put(30.67,36.44){\line(0,-1){17.89}}
\put(30.67,36.44){\line(-3,1){15.78}}
\put(30.56,18.44){\line(-3,-1){15.78}}
\put(14.89,41.67){\line(0,-1){28.44}}
\put(14.89,13.22){\line(-1,-1){8.44}}
\put(30.67,18.44){\line(1,-1){13.56}}
\put(14.89,41.67){\line(-1,1){8.44}}
\put(30.67,36.44){\line(1,1){13.56}}
\put(4.67,20.22){\line(3,1){40.89}}
\end{picture}
}
\ca{The generalised bootstrap}
\label{fig:conseq}
\end{figure}

A subtlety is the exponential factor on the rhs, involving
the step function, \m{\Theta}, defined by
\begin{equation}
\Theta(x)=\lim_{\epsilon \rightarrow 0}\left[\frac{1}{2}+\frac{1}{\pi}
\arctan \frac{x}{\epsilon}\right]=\left\{
\begin{array}{l}
0\mathrm{\ if\ }x<0, \\
\frac{1}{2}\mathrm{\ if\ }x=0, \\
1\mathrm{\ if\ }x>0.
\end{array} \right.
\end{equation}
Due to the periodicity of the exponential, this term has no effect unless
\m{\T=0}, and accounts for the fact that, at this point, the additional
particle cannot really be said to cross either the incoming particles or the
intermediate particles. In applications such as the thermodynamic Bethe
ansatz, it is important that the formula nonetheless continue to make sense
at \m{\T=0}, so the extra term is introduced to keep the equation correct. A
more detailed discussion and derivation can be found in Appendix
\ref{app:details}.

Another form of this result was used by Oota~\cite{Oota} in his
derivation of an integral formula for the S-matrix. In Appendix
\ref{app:Ootasp}, we show that it can be re-stated as
\begin{equation}
m^{q}_{ab}(x+1)q^{-t_{a}}+m^{q}_{ab}(x-1)q^{t_{a}}=\sum_{c}m^{q}_{bc}(x)
[G_{ca}]_{q}\,,
\label{eq:chaprecursion}
\end{equation}
where
\begin{equation}
m^{q}_{ab}(x)=\sum_{y\in \mathbb{Z}}m_{ab}(x,y)q^{y}\,,
\end{equation}
and the standard notation \m{[n]_{q}=(q^{n}-q^{-n})/(q-q^{-1})} has
been used.

This restatement as a recursion relation makes it clear that,
with the input of \m{m^{q}_{ab}(0)} and \m{m^{q}_{ab}(1)}, all other
\m{m_{ab}(x,y)} follow.

These two inputs turn out to be
\ba
m^{q}(0)=0 && m^{q}(1)=q^{t_{a}}[t_{a}]_{q}\delta_{ab}\,.
\ea
The first of these,
\m{m^{q}(0)=0}, is trivial for simply-laced cases (as
\m{\{0\}}=1). Otherwise, it amounts to requiring that poles which are
on the physical strip at one value of the coupling stay there for all
values, which is necessary on physical grounds. (The existence of the 
three-point couplings is not dependent on the coupling, and hence
processes which are possible at one value of the coupling must be
possible for \emph{all} values.) If \m{\{0,y\}} was present in the
S-matrix, for example, it would lead to a pole at
\m{\T=(y-1)\T_{H}-\T_{h}=\frac{i\pi
B(y-1)}{2\rv\hv}-\frac{i\pi(2-B)}{2h}}. For sufficiently small \m{B},
this becomes negative.

The other condition, 
\m{m^{q}(1)=q^{t_{a}}[t_{a}]_{q}\delta_{ab}}, implies that
the part of the S-matrix coming from these blocks is just
\begin{equation}
\frac{\langle 2,2t_{a} \rangle}{\langle 2,0 \rangle \langle
0,2t_{i} \rangle}\,.
\end{equation}
The only pole is thus at \m{2\T_{h}+2t_{a}\T_{H}}, which is precisely
that required for our special processes. Thus, this is just the
statement that these processes should exist when the two incoming
particles are identical, but not otherwise. 

In sum, postulating the existence of these special processes, together
with basic physical requirements, serve to completely fix the minimal
S-matrix.
These processes, in other words, seem to completely encode all the
Lie algebraic information contained in the S-matrix.

The significance of \re{eq:grtv} can further be seen if we take its
logarithmic derivative, and use the fact \cite{BCDS} that we can
identify the resulting elements with conserved charges. This gives
\begin{equation}
\sum_{l=1}^{r}[G_{il}]_{\qbar(i\pi s)}q_{s}^{l}=2\cos\left[\pi s
\left(\frac{2-B}{2h}+\frac{Bt_{i}}{2\rv\hv}\right)\right]q_{s}^{i}\,.
\label{eq:eigen}
\end{equation}
The \m{i}th component of the conserved charge with spin \m{s} is
denoted by \m{q^{i}_{s}}. The forward-backward shifts on the rhs of
\re{eq:grtv} have been absorbed into the deformation of \m{G}, where
we have defined \m{\qbar(t)} (and \m{q(t)}, to be used later) as 
\ba
q(t)=\exp \left(\frac{(2-B)t}{2h}\right)& \mathrm{and} & \qbar(t)=\exp
\left(\frac{Bt}{2\rv\hv}\right)\,.
\ea

In simply-laced cases, since \m{[n]_{q}=n} for
\m{n=0,1} (as all entries of the incidence matrix are in these cases), and
we have all \m{t_{i}=1} and \m{h=\rv\hv}, this reduces to the 
eigenvector equation
\begin{equation}
\sum_{l=1}^{r}G_{il}q_{s}^{l}=2\cos\left(\frac{\pi 
s}{h}\right)q_{s}^{i}\,,
\label{eq:eigensl}
\end{equation}
a well-known but curious result~\cite{Klassen}. For 
nonsimply-laced cases, however, note that the \m{t_{i}} in the cos
term prevents this from being a proper eigenvalue equation.

\sse{An integral formula}
As well as the product form for the S-matrix elements introduced above, 
Oota also found an integral
form, which explicitly builds in the dependence on the Cartan
matrix. The proof of this exploits \re{eq:chaprecursion}. Since, in 
consequence, it
relies on little other than the Lie algebraic structure in the
particle couplings, it is perhaps not surprising that 
Frenkel and Reshetikhin \cite{Frenkel} also conjectured a very similar
result in their more general study of \m{\mathcal{W}}-algebras.

Further details can be found in 
Appendix \re{app:details}, but, for reference, the formula is
\begin{multline}\label{eq:Oota}
S_{ab}(\T)=(-1)^{\delta_{ab}}\exp\left(4\int_{-\infty}^{\infty}\frac{dk}{k}
\e^{ik\T}\left\{\sin k\T_{h}\cdot\sin k\T_{H}\cdot 
M_{ab}(q(\pi k),\qbar(\pi k)) \right. \right. \\ 
\left. \left. +\frac{\delta_{ab}}{4}\right\}\right),
\end{multline}

The matrix \m{M} introduces the 
dependence on the Cartan matrix and is defined by
\begin{equation}
M_{ij}(q,\qbar)=\left([K]_{q\qbar}\right)_{ij}^{-1}[t_{j}]_{\qbar}\,,
\end{equation}
where \m{K} is the ``deformed Cartan matrix'', given by
\begin{equation}
[K_{ij}]_{q\qbar}=(q\qbar^{t_{i}}+q^{-1}\qbar^{-t_{i}})\delta_{ij}
-[G_{ij}]_{\qbar}\,.
\label{eq:kdef}
\end{equation}
In the limit \m{q \rightarrow 1} and \m{\qbar \rightarrow 1}, 
this recovers the standard Cartan matrix. In some sense this can be
understood as a quantum deformation, since taking the classical limit (\m{B 
\rightarrow 0}) enforces \m{\qbar \rightarrow 1}. In the simply-laced 
cases, at least, this reduces the deformed Cartan matrix just to the
ordinary Cartan matrix with an additional factor proportional to the 
identity matrix. We should also note that our ``eigenvector'' result can
be neatly restated using this, as
as
\begin{equation}
\sum_{l=1}^{r}[K_{il}]_{q(i\pi s)\qbar(i\pi s)}q_{s}^{l}=0\,.
\end{equation}

To understand what \m{M} represents, think that, for the simply-laced 
cases (where all the \m{t_{i}} are 1), it is just the inverse deformed 
Cartan matrix. The consequences of the extra factor, which modifies it 
from this, will be seen later. 

The formula given by Frenkel and
Reshetikhin~\cite{Frenkel} is similar to \re{eq:Oota}, but without the 
factor of
\m{(-1)^{\delta_{ab}}\exp \left(\int_{-\infty}^{\infty}\frac{dk}{k}
\e^{ik\T}\delta_{ab}\right)}.  For real \m{\T}, this is 1 except when 
\m{\T = 0}, in which case it becomes -1 for \m{a=b}. Including the
factor or not
thus amounts to selecting the value of \m{S_{aa}(0)}; with the factor,
\m{S_{ab}(0)=1}, but without it \m{S_{ab}(0)=(-1)^{\delta_{ab}}}. This 
second is the value taken by the product form S-matrix, and so will be the
version we adopt here.  

With a little more work, this can be put into an slightly simpler form. If
we first define a new matrix $\phi$ as
\begin{equation}
\phi_{ab}=-i\frac{d}{d\T}\log S_{ab}(\T),
\end{equation}
we find
\begin{equation}
\phi_{ab}=\int_{-\infty}^{\infty}dk e^{ik\T}\left\{4\sin k\T_{h}\cdot
\sin k\T_{H} \cdot M_{ab}(q(\pi k),\qbar(\pi k)) +\delta_{ab}\right\}\,.
\end{equation}
Defining its Fourier transform $\tilde{\phi}$ as
\begin{equation}
\tilde{\phi}_{ab}(k)=\frac{1}{2\pi}\int_{-\infty}^{\infty}d\T
\phi_{ab}(\T)e^{-ik\T}
\end{equation}
then leads to
\begin{equation}
\tilde{\phi}_{ab}(k)=-2\pi(q(\pi k)-q(\pi k)^{-1})(\qbar(\pi k)-\qbar(\pi
k)^{-1})M_{ab}(q(\pi k),\qbar(\pi k)) + 2\pi\delta_{ab}\,.
\end{equation}
(Note that using Frenkel and Reshetikhin's form would have removed the
final \m{\delta_{ab}} term.)

\sse{A formula for the conserved charges}
An interesting consequence of the integral formalism---and our reason for 
introducing it here---is that it can be used to find a formula for the 
conserved charges of the theory, by taking the logarithmic derivative of 
\re{eq:Oota} and again identifying it with the conserved charges. Doing 
this, and noting that the resulting integral can 
be re-expressed as a contour integral over the upper half-plane, the 
problem is reduced to finding the poles of the expression. The only 
poles are in the matrix \m{M}, so, before we can continue, 
we must find a formula for this. The easiest route to the information 
we need is to compute \m{\tilde{\phi}_{ab}(k)} for the product form
and compare with the above.

The first step---calculating \m{\phi_{ab}}---is straightforward, and
yields
\begin{multline}
\phi_{ab}=-\frac{i}{2}\sum_{x=1}^{h}\sum_{y=1}^{\rv\hv}m_{ab}(x,y)\left[
\sum_{s(x,y) \in S_{1}}\{\tanh(\hft(\T+s(x,y)))\}^{-1}
\right. \\
\left. - \sum_{s'(x,y) \in S_{2}}\{\tanh(\hft(\T+
s'(x,y)))\}^{-1}\right]\,,
\end{multline}
where
\ba
S_{1}&=&\{(x-1)\T_{h}+(y-1)\T_{H},(x+1)\T_{h}+(y+1)\T_{H},
(1-x)\T_{h}-(y+1)\T_{H}, \nonumber \\
&&(1-x)\T_{h}-(y+1)\T_{H},-(x+1)\T_{h}+(1-y)\T_{H}\}\,, \\ 
S_{2}&=&\{(1-x)\T_{h}+(1-y)\T_{H}, -(x+1)\T_{h}-(y+1)\T_{H},
(x-1)\T_{h}+(y+1)\T_{H}, \nonumber \\
&&(x-1)\T_{h}+(y+1)\T_{H}, (x+1)\T_{h}+(y-1)\T_{H}\}\,.
\ea

The Fourier transform of these terms is given in Appendix \ref{sec-ft}
as
\begin{equation}
\int_{-\infty}^{\infty}\left(\tanh\left(\frac{\T}{2}+a
\right)^{-1}\right)e^{-ik\T}d\T
=-4\pi ie^{-ik(2a+in\pi)}\frac{\cosh(\pi k)}{\sinh(2\pi
k)}\,,
\end{equation}
where care must be taken to choose \m{n} such that there are no poles
between the real axis and the line \m{2a+in\pi}. Working this through 
finally gives
\begin{multline}
\tilde{\phi}_{ab}(k)=2\pi\delta_{ab}-2\pi\sum_{x=1}^{h}\sum_{y=1}^{\rv\hv}m_{ab}(x,y) (q(\pi k)-q(pi k)^{-1})(\qbar(\pi k)-\qbar(\pi k)^{-1})\times \\
\frac{q^{x}
\qbar^{y}-q^{-x}\qbar^{-y}}{1-q^{2h}\qbar^{2\rv\hv}} \,,
\end{multline}
and hence
\begin{equation}
M_{ab}(q,\qbar)=\sum_{x=1}^{h}\sum_{y=1}^{\rv\hv}m_{ab}(x,y)\frac{q^{x}
\qbar^{y}-q^{-x}\qbar^{-y}}{1-q^{2h}\qbar^{2\rv\hv}}\,.
\label{eq:mform}
\end{equation}

This shows that the only poles present are at \m{k=im}, \m{m} being any
integer, so the result is that we can re-express the integral in the form
of a Fourier expansion, and thus read off a relation between
\m{\varphi_{ab}^{(s)}} and \m{M} as
\begin{equation}
\varphi_{ab}^{(s)}=2 \sin \pi s \cdot \sinh s\T_{h} \cdot \sinh s\T_{H}
\cdot M(q(i\pi s),\qbar(i\pi s)).
\end{equation}
Of course, to find an expression in \m{q_{s}^{a}q_{s}^{b}}, we need to
include a scaling factor. Noting that
\m{\sum_{i=1}^{r}q_{s_{i}}^{a}q_{s_{i}}^{b}=\delta_{ab}}, where \m{s_{i}} is
the \m{i}th component of a rank-\m{r} algebra, we could use
\m{q_{s}^{a}q_{s}^{b}=\varphi_{ab}^{(s)}/\sum_{i=1}^{r}\varphi_{11}^{(s_{i})}}.

Combining this with the expression for \m{M}, we get
\begin{equation}
\varphi_{ab}^{(s)}=2 \sinh s\T_{h} \cdot \sinh s\T_{H} \cdot
\sum_{x=1}^{h}\sum_{y=1}^{\rv\hv}m_{ab}(x,y)\sin \left(
\frac{s\pi}{2}\left[ \frac{(2-B)x}{h}+\frac{By}{\rv\hv}\right]\right).
\end{equation}
{}From this, it is straightforward to see that the matrix
\m{\varphi_{ab}^{(s)}} is non-zero for generic \m{B} by simple case-by-case
analysis. (This is different from this minimal case where, as noted by
Klassen and Melzer~\cite{Klassen}, we can get a zero matrix for
\m{s=\frac{h}{2}} in simply-laced cases, even if that exponent is present.)
Had there been cases where \m{\varphi^{(s)}} was zero for some \m{s}, then
taking the logarithmic derivative of an S-matrix identity would sometimes
have resulted in a trivial conserved charge identity. As it is, however,
we can always derive a non-trivial conserved charge identity from an
S-matrix identity and vice versa.
\pagebreak
\sse{Multi-linear Identities}
\label{sec:multi}
\quot{Life must be understood backwards; but \ldots it must be lived
forwards.}{S\"{o}ren Kierkegaard}

The RTV result and its generalisation allow us to perform a simple trick
and generate a large number of S-matrix identities. Interchanging \m{i} and
\m{j} in~\re{eq:grtv} does not change the lhs if \m{t_{i}=t_{j}} - the two
roots are the same length - due to the symmetry of the S-matrix, so we can
equate the rhs before and after interchanging to get
\begin{equation}
\prod_{l=1}^{r}\prod_{n=1}^{G_{il}}S_{jl}(\T+(2n-1-G_{il})\T_{H})=
\prod_{l'=1}^{r}\prod_{n'=1}^{G_{jl'}}S_{il'}(\T+(2n'-1-G_{jl'})\T_{H}).
\label{eq:multi}
\end{equation}
(Note that the presence or absence of an exponential factor does not
affect this, as \m{t_{i}=t_{j}} ensures \m{G_{ij}=G_{ji}}.) If \m{i} and
\m{j} are
such that the corresponding rows of the incidence matrix consist of
entries no greater than 1, this reduces to
\begin{equation}
\prod_{l=1}^{r}S_{il}(\T)^{G_{lj}}=\prod_{l'=1}^{r}S_{jl'}(\T)^{G_{l'i}},
\end{equation}
and we can obtain identities for products of S-matrix elements, all
evaluated at the same rapidity. The existence of such identities was 
first discovered by Khastgir \cite{Khastgir}, though without such a
systematic method for describing them. In addition, we also have 
identities in which not all rapidities are equal.

To generalise the connection between S-matrix product identities and
conserved charge sum rules to this case, we can take logarithmic
derivatives to find that if
\begin{equation}
\prod_{a,b \in \{i,j\}}S_{ab}(\T+if^{1}_{ab})=\prod_{a',b' \in \{i',j'\}}
S_{a'b'}(\T+if^{2}_{a'b'}),
\end{equation}
for some sets \m{\{i,j\}} and \m{\{i',j'\}} then
\begin{equation}
\sum_{a,b \in \{i,j\}}\e^{-if_{ab}^{1}s}q_{s}^{a}q_{s}^{b}
=\sum_{a',b' \in \{i',j'\}} \e^{-if_{a'b'}^{2}s}q_{s}^{a'}q_{s}^{b'}.
\end{equation}
Applying this to~\re{eq:multi} gives
\begin{equation}
\sum_{l=1}^{r}[G_{il}]_{\qbar(i\pi s)}q_{s}^{l}q_{s}^{j}=\sum_{l'=1}^{r}
[G_{jl'}]_{\qbar(i\pi s)}q_{s}^{l'}q_{s}^{i},
\label{eq:ccid}
\end{equation}
where it should be noted that the sums over \m{n} and \m{n'}
in~\re{eq:multi} have been absorbed by the introduction of the
\m{[G_{ab}]_{\qbar(\pi s)}} notation.

To give a simple example of this result, in the \m{b_{r}^{(1)}}
algebra we have, for \m{1<i<r-1}
\begin{equation}
S_{(r-1)(i-1)}(\T)S_{(r-1)(i+1)}(\T)=S_{i(r-2)}(\T)S_{ir}(\T+\T_{H})
S_{ir}(\T-\T_{H}),
\end{equation}
and
\begin{equation}
q_{s}^{r-1}q_{s}^{i-1}+q_{s}^{r-1}q_{s}^{i+1}=q_{s}^{i}q_{s}^{r-2}+\frac{1}{2}
\cos \frac{B\pi s}{2\rv\hv} \cdot q_{s}^{i}q_{s}^{r},
\end{equation}
with (through the duality transformation \m{B \rightarrow 2-B})
corresponding identities for \m{a_{2r-1}^{(2)}}.

It is still an open question as to whether we have found all such 
identities, or merely a subset, but there is good reason to believe that 
these represent all that can be found. From the multi-linear
identities \re{eq:multi} come all possible identities involving shifts
only depending on \m{\T_{H}} while bringing the full machinery of the
generalised bootstrap into play ultimately allows the proof or
disproof of any identity.

In the first situation, case-by-case analysis shows that the first row
of all the S-matrices consists of linearly 
independent elements, as each has at least one pole which is not found 
in any of the others. If our identities provide a way to re-write all 
the other S-matrix elements in terms of this set, it can be 
used as a basis. Any other identity can then be proved or disproved by 
expanding it in the basis, and comparing terms.

In general, this idea works very well. The only 
difficulty arises for \m{d_{n}} due to the pair of degenerate 
particles. For \m{n} odd, the elements \m{S_{n(n-1)}(\T)} and 
\m{S_{aa}(\T)} (for \m{a=n-1,n}) cannot be separated, and the best that 
can be done is to say  
\begin{equation}
S_{aa}(\T)S_{n(n-1)}(\T)=\prod_{p=1\mathrm{\ step\ }2}^{n-2}S_{1p}(\T)\,.
\end{equation}
For \m{n} even, this separates into
\begin{eqnarray}
S_{n(n-1)}(\T)&=&\prod_{p=3\mathrm{\ step\ }4}^{n-x}S_{1p}(\T) \\
S_{aa}(\T)&=&\prod_{p=1\mathrm{\ step\ }4}^{n-4+x}S_{1p}(\T)\,,
\end{eqnarray}
(where $x=1$ for $n$ divisible by 4, and $x=3$ otherwise), but this 
cannot be done for \m{n} odd. However, in this case, e.g. 
\m{S_{n(n-1)}(\T)} becomes linearly independent of the first-row 
elements, and so can be added to the basis, allowing the argument to 
still be used.

For the more general situation, the generalised bootstrap (in
common with the usual bootstrap) allows 
the entire S-matrix to be built from an initial knowledge of one
element, usually \m{S_{11}}. 
This means that any other element can be written in terms of \m{S_{11}} 
(with various forward-backward shifts) by repeated use of the bootstrap. 
Inserting this into any identity to be proved then reduces it to a product of 
elements \m{S_{11}} with a variety of rapidities. If these are
linearly independent of each other (as seems reasonable) then simply
comparing terms would be sufficient to prove or disprove the identity.

Neither of these arguments is as rigorous as we would like, but they do 
hold out the reasonable possibility that the claim might be true. This 
would reinforce the idea that all the structure in the S-matrix is due to the 
underlying Lie algebra.

\se{Summary}
\quot{\ldots an ill-favoured thing, sir, but mine own \ldots}{William
Shakespeare}
The aim of this chapter was to find a concise way of encoding the Lie
algebraic information into the S-matrix of all ATFTs. This was
achieved by looking at the processes responsible for poles at
\m{2\T_{h}+2t_{i}\T_{H}} whenever the incoming particles were
identical. These could be explained by \fig{fig:lieproc}, where,
crucially, the intermediate particles consisted of those adjacent to
the initial particles on the Dynkin diagram on the algebra.

Sending in a third particle either before or after the interaction,
and using the principle of factorisation to equate the results led to
the ``generalised bootstrap''
\begin{multline}
S_{ij}(\T+\T_{h}+t_{i}\T_{H})S_{ij}(\T-\T_{h}-t_{i}\T_{H})= \\ \e^{-2i\pi
\Theta(\T)G_{ij}}
\prod_{l=1}^{r}\prod_{n=1}^{G_{il}}S_{jl}(\T+(2n-1-G_{il})\T_{H})\,.
\label{eq:grtv3}
\end{multline}
This, together with demanding the existence of these processes (and
their associated poles) completely fixes the minimal S-matrix. The
remaining question, however, is how the processes arise from the
initial Lagrangian formulation.

Taking the logarithmic derivative of \re{eq:grtv3} then leads to an
equation for the conserved charges of the theory, namely
\begin{equation}
\sum_{l=1}^{r}[G_{il}]_{\qbar(i\pi s)}q_{s}^{l}=2\cos\left[\pi s
\left(\frac{2-B}{2h}+\frac{Bt_{i}}{2\rv\hv}\right)\right]q_{s}^{i}\,,
\label{eq:eigen3}
\end{equation}
which reduces to a simple eigenvector equation in simply-laced
cases. These charges can also be written as
\begin{equation}
q_{s}^{a}q_{s}^{b} \propto 2 \sinh s\T_{h} \cdot \sinh s\T_{H} \cdot
\sum_{x=1}^{h}\sum_{y=1}^{\rv\hv}m_{ab}(x,y)\sin \left(
\frac{s\pi}{2}\left[ \frac{(2-B)x}{h}+\frac{By}{\rv\hv}\right]\right).
\end{equation}

Since the S-matrix is symmetric, the lhs of \re{eq:grtv3} is unchanged
by interchanging \m{i} and \m{j}, whereas the rhs is not, leading to
the identities
\begin{equation}
\prod_{l=1}^{r}\prod_{n=1}^{G_{il}}S_{jl}(\T+(2n-1-G_{il})\T_{H})=
\prod_{l'=1}^{r}\prod_{n'=1}^{G_{jl'}}S_{il'}(\T+(2n'-1-G_{jl'})\T_{H})\,,
\label{eq:multi2}
\end{equation}
which probably describe all identities with shifts only involving
\m{\T_{H}}, just as the generalised bootstrap contains enough
information to prove or disprove all possible identities.

 
\chap{Conclusions}

\quot{`Good morning,' said Deep Thought at last. 

`Er ... Good morning, O Deep Thought,' said Loonquawl nervously,
`do you have ... er, that is ...'

`An answer for you?' interrupted Deep Thought majestically. `Yes. I
have.'

`To Everything? To the great Question of Life, the Universe and
Everything?'
 
`Yes.'

`Though I don't think,' added Deep Thought, 
`that you're going to like it.' 

`Doesn't matter!' said Phouchg. `We must know it! Now!' 

`Alright,' said the computer and settled into silence again. The two
men fidgeted. The tension was unbearable. 

`You're really not going to like it,' observed Deep Thought. 

`Tell us!' 

`Alright,' said Deep Thought. `The Answer to the Great Question ...' 

`Yes ...!' 

`Of Life, the Universe and Everything ...' said Deep Thought. 

`Yes ...!' 

`Is ...' said Deep Thought, and paused. 

`Yes ...!!!...?' 

`Forty-two,' said Deep Thought, with infinite majesty and
calm.}{Douglas Adams, The Hitch
Hiker's Guide to the Galaxy}

\setlength{\parskip}{5pt}
\se{Introduction}
The aim of this study was to investigate the fundamental objects of 
ATFTs: the S-matrices of the theory in the bulk, and the reflection
factors of the theory with a boundary. 

For the bulk theory, the form of the 
S-matrices and the particle structure were already well-known; the 
intriguing question was how the Lie algebraic structure built into the 
Lagrangian manifested itself in the S-matrix. For the boundary theory, on 
the other hand, even for the simplest possible
ATFT---sine-Gordon---the 
reflection factors for all except the ground and lowest excited states of 
the theory were unknown, as was the boundary bound state structure. 

The focus for both pieces of
work could therefore be said to be their bootstrap structure: 
tying it in to the underlying Lie algebraic structure in the bulk; 
and finding a rigorous way to identify the bound states hidden in the
boundary reflection factors.  
\vspace*{-3pt}
\se{Bulk ATFTs}
\vspace*{-3pt}
This work was based on the observation that, for any ATFT, two 
identical particles (say \m{i}) colliding at a relative rapidity of 
\m{2\T_{h}+2t_{i}\T_{H}} results in the production of all the particles 
which are adjacent to it on the Dynkin diagram, and only 
those. Taking these processes as a starting point, a ``generalised
bootstrap'' was constructed, which explicitly related the
structure of the S-matrix elements to the Cartan matrix. By using
these equations, together with the requirement that no more couplings
than necessary be introduced, it was found that the complete minimal
S-matrix could be derived. 

The weak link is that these processes have been introduced
as axiomatic, rather than via a derivation from Dorey's Lie
algebraic coupling rule. An important open problem is whether our simple
tree-level argument will stand up to a perturbative verification to higher
loops in the Feynman diagrams. We can, however, gain some measure of
confidence 
from the fact that there is substantial evidence for the
validity of the S-matrix formulae, which are successfully reproduced.   

With this in place, it should then be possible to tie in all the other
results which have been found by observation on a more rigorous basis.
\vspace*{-3pt}
\section{Boundary sine-Gordon}
\vspace*{-3pt}
\lhead{\textit{5.3~Boundary sine-Gordon}}
The task here was more basic: a determination of the bound-state
structure and reflection factors for all integrable boundary
conditions. This was achieved, principally with the help of two rather
general lemmas which showed that poles at sufficiently small
rapidities could not correspond to anything other than a bound state
without violating momentum conservation. By taking the spectrum to
consist of just the states which were required to satisfy the lemmas,
we could then show that all the other poles had an explanation through
the Coleman-Thun mechanism. 

Since the lemmas are quite general, they
apply to all theories with a boundary, integrable or not. An
interesting open question is whether, as here, they are strong
enough to completely determine the spectrum, or merely provide a
starting point.

The natural way to continue the work would be to generalise it to
other ATFTs. It has been found \cite{BCDS} that, at the so-called
``reflectionless points'' (which occur at integer \m{\la}) the
full S-matrices for sine-Gordon become the minimal matrices for the
\m{d_{n}^{1}} theory. The soliton and anti-soliton correspond to the
two mass-degenerate particles, while the breathers correspond to all
the others. This might make the extension of the results found here to
\m{d_{n}^{1}} relatively straightforward. However, results for
\m{a_{n}^{1}} have already been found \cite{DG}, and indicate that the 
coupling plays a bigger r\^{o}le in the boundary spectrum than in the
bulk. Thus, while all the coupling information is contained in the
minimal S-matrix for the bulk, the story is probably not so simple
with a boundary. However, it might still provide a good starting
point. If this could be achieved, only the exceptional cases would remain
to complete the ADE series.

At this point, the position for the boundary theories would be analogous
to that for the bulk, in that the next logical step would be to put
everything on a manifestly Lie algebraic footing. While a unified
discussion of all boundary ATFTs is perhaps still some way off, it
should nonetheless be an attainable goal. The theories could then be
said to be under complete control, at least from this point of view.

\setlength{\parskip}{6pt}

\appendix
\renewcommand{\thesection}{\Alph{chapter}.\arabic{section}}
\renewcommand{\thesubsection}{\Alph{chapter}.\arabic{section}.\arabic{subsection}}

\chap{Boundary sine-Gordon Details}
\quot{This is a one line proof...if we start sufficiently far to the left.}{peter@cbmvax.cbm.commodore.com}

\se{Infinite products of gamma functions}
\label{app:convergence}
The products which arise in the course of this work are of the form
\begin{equation}
P(u)=\prod_{l=1}^{\infty}\left[\frac{\g (kl+a-x u)\g(kl+b-xu)}{\g
(kl+c-xu)\g (kl+d-xu)} / (u \rightarrow -u)\right]\,,
\label{eq:app1p}
\end{equation}
Rather than examine this product
directly, we take logs and use the standard formula
\begin{equation}
\ln \g (z) = z\ln (z) - z - \frac{1}{2}\ln (z) + \ln (\sqrt{2}) +
\frac{1}{12z} + O(z^{-3})
\end{equation}
Assuming that the sum over $l$ and the expansion in $z$ can be exchanged,
potential divergences arise from terms
of the form
\m{\sum_{l=1}^{\infty}\frac{a}{l^{n}}} with \m{a\neq 0} 
and \m{n < 2}. To begin with, we will consider
the terms arising from the block of four terms explicitly shown.

Firstly, there is a contribution of \m{\sum_{l=1}^{\infty}
a+b-c-d} from the \m{z} terms, which
can be set to zero by demanding \m{a+b=c+d}. For the \m{1/12z} terms,
the overall contribution from the
four terms is
\begin{equation}
\sum_{l=1}^{\infty}
\frac{1}{12}\left(\frac{a-c}{(kl+a-xu)(kl+c-xu)}
+\frac{b-d}{(kl+b-xu)(kl+d-xu)}\right)
\end{equation}
which
can be seen, for \m{a+b=c+d}, to be
of the form \m{1/l^{2}} and hence convergent. 

A similar argument applies to the \m{-\hf \ln (z)} terms, showing they
also provide a convergent contribution. This breaks down when
considering the \m{z\ln(z)} terms, however, and 
their contribution formally reduces to
\begin{equation}
\sum_{l=1}^{\infty}\Bigl( \frac{cd-ab}{kl} + 
O(l^{-2})\Bigr)~,
\end{equation}
which is divergent unless \m{a=c} or \m{b=c}, both of which are
trivial cases. However, repeating this argument on the other block (with
\m{u \rightarrow -u}) can be seen to yield the same result, allowing
the two divergent terms to cancel, and leaving a product which is
convergent overall.

For comparison with other results, it is useful to write $P(u)$ in other
ways. Firstly, it can be written in terms of Barnes' diperiodic sine
functions using the expansion as given in~\cite{Pillin}:
\begin{multline}
S_{2}(x|\w_{1},\w_{2})=\exp\left[\frac{(\w_{1}+\w_{2}-2x)\left(\gamma +
\log(2\pi) +
2\log\left(\frac{\w_{1}}{\w_{2}}\right)\right)}{2\w_{1}}\right] \times
\\
\frac{\g
\left( \frac{\w_{1}+\w_{2}-x}{\w_{1}}\right)}{\g
\left(\frac{x}{\w_{1}}\right)}
\prod_{n=1}^{\infty}\left[\frac{\g \left(\frac{\w_{1}+\w_{2}-x +
n\w_{2}} {\w_{1}}\right)}{\g\left( \frac{x+n\w_{2}}{\w_{1}}\right)}
\mathrm{e}^{-\frac{\w_{1}+\w_{2}-2x}{2n\w_{1}}}\left( \frac{n\w_{1}}
{\w_{2}} \right)^{-\frac{\w_{1}+\w_{2}-2x}{\w_{1}}} \right], 
\end{multline}
where \m{\gamma} denotes the Euler constant. For blocks
of the form we are interested in, this simplifies to
\ba
\lefteqn{\frac{S_{2}(x_{1}|\w_{1},\w_{2})S_{2}(x_{2}|\w_{1},\w_{2})}
{S_{2}(x_{3}|\w_{1},\w_{2})S_{2}(x_{4}|\w_{1},\w_{2})}=} \nonumber \\
&& \prod_{n=1}^{\infty} \left[\left\{\frac{\g
\left(\frac{n\w_{2}}{\w_{1}} + \frac{\w_{1}-\w_{2}}{2\w_{1}}-\frac{x'_{1}}{2\w_{1}}\right)\g
\left(\frac{n\w_{2}}{\w_{1}} + \frac{\w_{1}-\w_{2}}{2\w_{1}}-\frac{x'_{2}}{2\w_{1}}\right)}{
\g
\left(\frac{n\w_{2}}{\w_{1}} + \frac{\w_{1}-\w_{2}}{2\w_{1}}-\frac{x'_{3}}{2\w_{1}}\right)\g
\left(\frac{n\w_{2}}{\w_{1}} + \frac{\w_{1}-\w_{2}}{2\w_{1}}-\frac{x'_{4}}{2\w_{1}}\right)}
\right\}/(x'_{m}
\rightarrow -x'_{m})\right]\,,
\ea
(where \m{x'_{m}=x_{m}-\w_{1}-\w_{2}})
provided \m{x_{1}+x_{2}=x_{3}+x_{4}}. Comparing with \re{eq:app1p} we have
\begin{equation}
P(u)= \frac{S_{2}(\w_{1}(1-a+xu)|\w_{1},\w_{1}k)S_{2}(\w_{1} 
(1-b+xu)|\w_{1},\w_{1}k)}
{S_{2}(\w_{1}(1-c+xu)|\w_{1},\w_{1}k)S_{2}(\w_{1}(1-d+xu)|\w_{1},\w_{1}k)}\,,
\end{equation}
where \m{w_{1}} is arbitrary. In section~\ref{sec:review} we took 
\m{\w_{1}=x^{-1}} for simplicity. The identity
\begin{equation}
S_{2}(\w_{1}+\w_{2}-x|\w_{1},\w_{2})=\frac{1}{S_{2}(x|\w_{1},\w_{2})}
\end{equation}
was also used. 

These products can also be written in an integral form, through
\begin{equation}
\log \g (\zeta) = \int_{0}^{\infty} \frac{dx}{x}e^{-x}\left[\zeta-1 +
\frac{e^{-(\zeta-1)x}-1}{1-e^{-x}}\right], \mathrm{Re\ } \zeta>0.
\end{equation}
Since, for the expressions we consider, not all the
\m{\Gamma}-functions have arguments with
positive real part,
it is not possible to give a general formula for \m{P} solely in
these terms. Instead, we give expressions for the reflection
factors. To simplify matters, define
\ba
I^{1}(u)&=&\frac{2\la}{\pi}\int_{-\infty}^{+\infty}dx\cosh \left( \frac{2\la u
x}{\pi}\right)\left[ \frac{\sinh \left( \la -
\frac{2\xi}{\pi}\right)x}{2\sinh x \cosh \la x}\right] \nn \\
I^{2}(u)&=&
\frac{2\la}{\pi}\int_{-\infty}^{+\infty}dx\cosh \left( \frac{2\la u
x}{\pi}\right)\left[\frac{\sinh\left(\frac{2\xi}{\pi}
-2n_{*}-2\right)x}{\sinh x}\right] \\
I^{3}_{n}(u)&=&-\frac{2\la}{\pi}\int_{-\infty}^{+\infty}dx\cosh \left( 
\frac{2\la u x}{\pi}\right)\left[\frac{2\cosh x \sinh \left(\la +1
+2n-\frac{2\xi}{\pi}\right)x}{2\sinh x \cosh \la x}\right] \nn \\
I^{4}_{n}(u)&=&-\frac{2\la}{\pi}\int_{-\infty}^{+\infty}dx\cosh \left( 
\frac{2\la u x}{\pi}\right)\left[\frac{2\cosh x \sinh \left(\frac{2\xi}
{\pi}+2n-\la-1\right)x}{2\sinh x \cosh \la x}\right] \nn
\ea
(where \m{I^{3}_{n}(u)} and \m{I^{4}_{n}(u)} are related to each other 
through \m{\xi \rightarrow \pi(\la+1)-\xi}). The constant \m{n_{*}} is 
the number of \m{\nu}-type poles in the physical strip, which we
recall can be written as
\begin{equation}
n_{*}=\left[\frac{\xi}{\pi}-\hf\right]\,.
\end{equation}
 
The reflection factors
can then be written as
\ba
-\frac{d}{du}\log \left[\frac{P^{+}_{\st{c;\underline{x}}}(u)}
{R_{0}(u)}\right]&=& 
I^{1}(u)+cI^{2}(u)+\sum_{i \mathrm{\ odd}}I^{3}_{n_{i}}(u) +
\sum_{j \mathrm{\ even}}I^{4}_{n_{j}}(u) \\
-\frac{d}{du}\log \left[\frac{P^{-}_{\st{c;\underline{x}}}(u)}
{R_{0}(u)}\right]&=& 
I^{1}(u)-(1-c)I^{2}(u)+\sum_{i \mathrm{\ odd}}I^{3}_{n_{i}}(u) +
\sum_{j \mathrm{\ even}}I^{4}_{n_{j}}(u)\,, \nn
\ea
for topological charge \m{c} and \m{\underline{x}=(n_{1},n_{2},\ldots, 
n_{2k+c})}. These factors were given in~\cite{Skorik} for the first 
two levels of excited states (the extent of the spectrum they found); 
the above is simply a generalisation of this to the whole spectrum.

\se[Relation of \m{M} and \m{\varphi_{0}} to \m{\eta} and 
$\vartheta$]{Relation of \m{\pmb{M}} and \m{\pmb{\varphi_{0}}} to
\m{\pmb{\eta}} and \m{\pmb{\vartheta}}} 
\label{app:relation}
\lhead{\textit{\thesection~Relation of \m{M} and \m{\varphi_{0}} to
\m{\eta} and \m{\vartheta}}}
For the action defined as
\begin{equation}
\mathcal{A}_{SG}=\int_{-\infty}^{0}dx \int_{-\infty}^{\infty}dt~
\frac{1}{4\pi}(\partial_{\mu}\varphi)^2
+2\mu \cos (2\beta \varphi)+2\mu_{B}\int_{-\infty}^{\infty}dt \cos 
\beta(\varphi(0,t)-\varphi_{0})\,,
\end{equation}
Zamolodchikov \cite{Zamunpub} has claimed that
\begin{equation}
\cosh^{2}(\beta^{2}(\vartheta\pm i\eta))=\frac{\mu^{2}_{B}}{\mu}\sin 
(\pi\beta^{2})e^{\pm 2i\beta \varphi_{0}}\,,
\label{eq:zambe}
\end{equation}
where this should be read as two equations, one with the positive signs, 
and one with the negative. To match our conventions, we need to re-scale 
this according to \m{\varphi \rightarrow \sqrt{2\pi}\varphi}, 
\m{\varphi_{0} \rightarrow \sqrt{2\pi}\varphi_{0}} and \m{\beta 
\rightarrow \beta/2\sqrt{2\pi}}. Then we need to identify \m{\mu = 
m_{0}^{2}/2\beta^{2}} and \m{\mu_{B}=M/2}. This means that 
Zamolodchikov's formula becomes
\begin{equation}
\cosh^{2}\left(\frac{\beta^{2}}{8\pi}\left(\vartheta\pm i\eta\right)\right)
=\hf\left(\frac{M\beta}{m_{0}}\right)^{2}\sin 
\left(\frac{\beta^{2}}{8}\right)e^{\pm i\beta \varphi_{0}}\,.
\label{eq:zambe2}
\end{equation}

This result agrees with earlier results for special
cases~\cite{LeClair}. To get a better idea how the two sets of
parameters are related, it is
useful to deconstruct \re{eq:zambe} 
into equations for \m{M}, \m{\varphi_{0}}, \m{\eta} and \m{\vartheta} 
individually, giving each in terms of the other set of parameters. For
the first two, we get
\ba
\varphi_{0}&=&\frac{1}{i\beta}\ln \left[ \pm \frac{ \cosh \left( 
\frac{\beta^{2}}{8\pi}\left(\vartheta+i\eta\right)\right)}{\cosh \left(
\frac{\beta^{2}}{8\pi}\left(\vartheta-i\eta\right)\right)}\right] 
\nonumber \\
M^{2}&=&\pm \frac{2m_{0}^{2}}{\beta^{2}}\frac{ \cosh \left( 
\frac{\beta^{2}}{8\pi}\left(\vartheta+i\eta\right)\right)\cosh \left(
\frac{\beta^{2}}{8\pi}\left(\vartheta-i\eta\right)\right)}{\sin \left( 
\frac{\beta^{2}}{8}\right)}\,,
\ea
where the choices of sign must match, and are determined by requiring 
\m{M} to be real. (If we take \m{\eta} and \m{\vartheta} real, this means
we must take the positive sign.) For \m{\eta} and \m{\vartheta}, the task
is made much easier by introducing the change of variables given by
\begin{equation}
\begin{array}{rcl}
\cos \left(\frac{\beta^{2}\eta}{8\pi}\right)\cosh\left(\frac{\beta^{2}\vartheta}{8\pi}\right)&=&-k_{\beta}\cos \xi_{\beta} \\
\cos^{2}\left(\frac{\beta^{2}\eta}{8\pi}\right)+\cosh^{2}\left(\frac{\beta^{2}\vartheta}{8\pi}\right)&=&1+k_{\beta}^{2}\,,
\end{array}
\label{eq:betaparam}
\end{equation}
where, in the limit \m{\beta^{2} \rightarrow 8\pi}, \m{\xi_{\beta}
\rightarrow \xi} and \m{k_{\beta} \rightarrow k^{-1}}. In terms of
these, we find
\ba
\cos^{2} \xi_{\beta} &=& \hf\left(1 \pm
\cos(\beta\varphi_{0})\right) \\
k_{\beta}^{2}&=&\pm\left(\frac{M\beta}{m_{0}}\right)^{2}\sin\left(\frac{\beta^{2}}{8}\right)\,,
\ea
where the choice of sign is as above. The parameters \m{\eta} and
\m{\vartheta} are then determined by
\begin{equation}
\cosh\left(\frac{\beta^{2}\vartheta}{4\pi}\right) , \cos\left(\frac{\beta^{2}\eta}{4\pi}\right)=k_{\beta}^{2}\pm
\sqrt{k_{\beta}^{4}-2k_{\beta}^{2}\cos 2\xi_{\beta}+1}\,,
\label{eq:th-eta-form}
\end{equation}
where, in principle, either \m{\eta} or \m{\vartheta} can correspond
to either choice of sign, and the sign here is unconnected to the
earlier choice.

For sine-Gordon theory, \m{\beta} is taken 
to be a strictly real parameter. The boundary parameters, \m{M} and 
\m{\varphi_{0}}, must also be real to keep the boundary potential 
real\footnote{Allowing them to be complex (\m{M=M_{r}+iM_{i}} and
\m{\varphi_{0}=\varphi_{0r}+i\varphi_{0i}}) leads to the demand 
\m{M_{r}\sinh \frac{\beta\varphi_{0i}}{2}=M\cosh \frac{\beta 
\varphi_{0i}}{2}=0} if the potential is to be kept real. The only 
solution to this is \m{M_{i}=\varphi_{0i}=0}.}. 
The important point to note is that this means that the rhs of 
\re{eq:th-eta-form} is purely real, forcing
\m{\eta} and \m{\vartheta} to either be real or purely imaginary. In
addition, the choice with the negative sign has modulus
less than or equal to 1. This means that there is always a choice of
\m{\eta} and \m{\vartheta} where both are real. The symmetry between
\m{\eta} and \m{i\vartheta} makes the choice where both are purely
imaginary equivalent. The remaining two choices---where one is real,
the other imaginary---make the lhs's of \re{eq:zambe} real, while
the rhs's are complex conjugates of each other, and so are
untenable. Thus, we can take \m{\eta} and \m{\vartheta} real without
loss of generality.

In the Dirichlet limit, i.e. \m{M \rightarrow \infty}, we have 
\m{k_{\beta}^{2} \rightarrow \pm\infty} also, reducing
\re{eq:th-eta-form} to
\ba
\cos\left(\frac{\beta^{2}\eta}{4\pi}\right)&=&\cos 2\xi_{\beta}
\nonumber \\
\cosh
\left(\frac{\beta^{2}\vartheta}{4\pi}\right)&=&2k_{\beta}^{2}-\cos
2\xi_{\beta}\,.
\ea
This gives \m{\vartheta \rightarrow \pm\infty} and
\begin{equation}
\frac{\beta^{2}\eta}{4\pi}=n\pi \pm \beta\varphi_{0}\,,
\end{equation}
for any \m{n \in \mathbb{Z}}. All of these choices for \m{\eta}
correspond to the same physical reflection factors, so we can take
\m{\eta=\frac{4\pi\varphi_{0}}{\beta}}, recovering the result conjectured
in \cite{GhoshZam}.

\sse{Comparison with other results}
The sine-Gordon theory can be considered as the
continuation to imaginary coupling of the sinh-Gordon model. For this
model, an independent proposal for the relation between the
parameters in the lagrangian and the reflection factors was made by
Corrigan \cite{Corrigansinh1} and was futher studied in
\cite{Corrigansinh}. The field equation used there was
\begin{equation}
\partial_{t}^{2}\phi-\partial_{x}^{2}\phi
+\frac{\sqrt{8}m^{2}}{\beta}\sinh
(\sqrt{2}\beta \phi) = 0\,,
\end{equation}
with boundary condition
\begin{equation}
\partial_{x}\phi|_{0}=\frac{\sqrt{2}m}{\beta}\left(\epsilon_{0}\e^{-
\frac{\beta}{\sqrt{2}}\phi(0,t)}-\epsilon_{1}\e^{\frac{\beta}{\sqrt{2}}
\phi(0,t)}\right)\,,
\end{equation}
where the boundary parameters are \m{\epsilon_{0}} and
\m{\epsilon_{1}}.
The parameters in the reflection factors were then found to be
\begin{equation}
\frac{B\eta}{\pi} = (a_{0}+a_{1})(1-B/2)\quad \mathrm{and}\quad 
\frac{iB\vartheta}{\pi} = (a_{0}-a_{1})(1-B/2)\,,
\end{equation}
where \m{B} was related to the coupling constant by
\m{B=2\beta^{2}/(4\pi+\beta^{2})}, and \m{a_{0}}, \m{a_{1}} was given by
\begin{equation}
\epsilon_{0}=\cos \pi a_{0} \quad \epsilon_{1}=\cos \pi a_{1}\,.
\end{equation} 

Their conventions differ from ours in the bulk by the transformations
\m{\phi
\rightarrow \phi/2}, \m{\beta \rightarrow \sqrt{2}\beta}, 
and \m{m \rightarrow m_{0}/\sqrt{2\beta}}. Applying these to the boundary
condition gives
\begin{equation}
\partial_{x}\phi|_{0}=m_{0}/\beta^{3/2}\left(\epsilon_{0}\e^{-\frac{\beta}{2}\phi(0,t)}-\epsilon_{1}\e^{\frac{\beta}{2}\phi(0,t)}\right)\,.
\end{equation}
To finally turn this into a suitable form for comparison, we need the
trigonometric identity
\begin{equation}
a\e^{b}+c\e^{-b}=\sqrt{ac}\cosh(b+d)\quad\mathrm{if}\quad \cosh d =
\frac{a+c}{2\sqrt{ac}}\,.
\end{equation}
After a little algebra, and continuing \m{\beta} to \m{i\beta}, their 
results then become
\begin{eqnarray}
\cos\left(\frac{\beta^{2}\eta}{4\pi}\right)&=&\epsilon_{0}\epsilon_{1} -
\sqrt{1-\epsilon_{0}^{2}-\epsilon_{1}^{2}+\epsilon_{0}^{2}\epsilon_{1}^{2}}
\\
\cosh\left(\frac{\beta^{2}\vartheta}{4\pi}\right)&=&\epsilon_{0}\epsilon_{1}
+
\sqrt{1-\epsilon_{0}^{2}-\epsilon_{1}^{2}+\epsilon_{0}^{2}\epsilon_{1}^{2}}\,,
\end{eqnarray}
for the boundary condition
\begin{equation}
\partial_{x}\varphi|_{0}=\frac{2m_{0}\sqrt{\epsilon_{0}\epsilon_{1}}}{\beta^{3/2}}
\cos \left(\frac{\beta \varphi(0,t)}{2}-i\alpha\right)\,,
\end{equation}
where the value of \m{\alpha} is
\begin{equation}
\cosh \alpha =
\frac{\epsilon_{0}-\epsilon_{1}}{2\sqrt{\epsilon_{0}\epsilon_{1}}}\,.
\end{equation}

To match this boundary condition to ours, we need to identify
\begin{equation}
\epsilon_{0}\epsilon_{1}=\left(\frac{M\beta^{3/2}}{2m_{0}}\right)^{2}=
\frac{k_{\beta}^{2}\beta}{4\sin\left(\beta^{2}/8\right)}
\quad\mathrm{and}\quad i\alpha=\frac{\beta\varphi_{0}}{2}\,.
\end{equation}
More algebra then shows that this gives
\begin{eqnarray}
\cos\left(\frac{\beta^{2}\eta}{4\pi}\right)&=&\kappa_{\beta}^{2}-\sqrt{\kappa^{4}_{\beta}
-2\kappa_{\beta}^{2}\cos 2\xi_{\beta}+4\kappa_{\beta}^{2}+1} \\
\cosh
\left(\frac{\beta^{2}\varphi}{4\pi}\right)&=&\kappa_{\beta}^{2}
+\sqrt{\kappa_{\beta}^{4}-2\kappa_{\beta}^{2}\cos
2\xi_{\beta}+4\kappa_{\beta}^{2}+1}\,,
\end{eqnarray}
where
\begin{equation}
\kappa_{\beta}^{2}=\frac{k_{\beta}^{2}\beta}{4\sin (\beta^{2}/8)}\,.
\end{equation}
This is very similar to \re{eq:th-eta-form}, but it is not quite the same.
This does not necessarily mean that either is wrong; the differences could
simply be down to e.g. implicit choices of renormalisation scheme in the
derivation of the respective formulae. The resolution of this question is
still open.

\pagebreak

\se{On-shell diagrams}
In this appendix we collect together some of the on-shell diagrams
used in the main body of the thesis.
All boundaries are initially in the state
\m{\st{n_{1},n_{2},\ldots,n_{2k}}}, where \m{k} can be any integer, and
we have suppressed the topological charge index (which is
zero). Analogous processes for charge 1 states can be found by
applying the transformation \m{\xi \rightarrow \pi(\la+1)-\xi} to all
rapidities shown. 

In addition, where the boundary is shown decaying
through emission of a breather, only the process where this removes
the last two indices is given. Similar processes always exist to
remove any other adjacent pair of indices, or to simply modify an
index; see section \ref{sec:refmin} for the appropriate breather
boundary vertices.

\begin{figure}[!ht]
\parbox{0.49\textwidth}{
\unitlength 1.00mm
\linethickness{0.4pt}
\begin{picture}(48.00,60.00)(10,2)
\put(46.00,3.00){\rule{2.00\unitlength}{52.00\unitlength}}
\multiput(45.67,53.11)(0,1){2}{\circle*{0.25}}
\multiput(45.78,4.00)(0,1){2}{\circle*{0.25}}
\multiput(45.67,6.00)(0,2){24}{\circle*{0.25}}
\multiput(37.33,13.44)(1,-1){9}{\circle*{0.25}}
\multiput(37.22,44.67)(1,1){9}{\circle*{0.25}}
\put(36.78,44.22){\line(3,-5){9.07}}
\put(36.78,14.00){\line(3,5){9.07}}
\put(36.78,14.00){\line(-2,-1){24.44}}
\put(36.78,44.22){\line(-2,1){24.44}}
\multiput(36.78,11.00)(0,1){7}{\circle*{0.25}}
\qbezier(36.78,11.00)(34.00,10.00)(32.78,12.00)
\put(32.78,14){\makebox(0,0)[rc]{$w_{n_{2k}-n}$}}
\qbezier(42.22,49.67)(43.44,47.67)(45.56,48.33)
\qbezier(44.11,32.11)(44.56,33.44)(45.89,32.89)
\put(35.00,32.18){\makebox(0,0)[lc]{$w_{n_{2k}}$}}
\put(50.00,48.67){\makebox(0,0)[lc]{$\frac{\nu_{n_{2k-1}}-w_{n_{2k}}}{2}$}}
\put(50.00,30.00){\makebox(0,0)[lc]{$\st{n_{1},\ldots,n_{2k-2}}$}}
\put(40.00,8.00){\makebox(0,0)[cc]{$n$}}
\put(40.00,50.00){\makebox(0,0)[cc]{$n$}}
\end{picture}
\hspace*{-0.1\textwidth}
\ca{Incoming soliton, breather boundary decay}
\label{fig:fb1} } \
\parbox{0.49\textwidth}{
\unitlength 1.00mm
\linethickness{0.4pt}
\begin{picture}(48.00,60.00)(10,2)
\multiput(46.18,9.44)(-1,2){6}{\circle*{0.25}}
\multiput(46.18,49.67)(-1,-2){6}{\circle*{0.25}}
\multiput(45.67,49.11)(0,1){6}{\circle*{0.25}}
\multiput(45.78,4.00)(0,1){6}{\circle*{0.25}}
\multiput(45.67,10.00)(0,2){10}{\circle*{0.25}}
\multiput(45.67,32.00)(0,2){7}{\circle*{0.25}}
\put(40.96,20.00){\line(1,2){5.03}}
\put(40.96,19.89){\line(-5,3){28.11}}
\put(40.96,40.11){\line(1,-2){5.03}}
\put(40.96,40.22){\line(-5,-3){28.11}}
\multiput(23.96,24.99)(0,1){11}{\circle*{0.25}}
\qbezier(20.96,28.22)(21.74,26.56)(23.96,27.00)
\qbezier(44.05,45.78)(44.41,44.11)(46.52,44.67)
\put(20.96,29.89){\makebox(0,0)[rc]{$\nu_{n-n_{2k}}$}}
\put(50.00,23.18){\makebox(0,0)[lc]{$w_{n_{2k}}$}}
\put(50.00,45.67){\makebox(0,0)[lc]{$\frac{\nu_{n_{2k-1}}-w_{n_{2k}}}{2}$}}
\qbezier(43.33,24.56)(44.00,23.11)(46.00,23.78)
\qbezier(36.89,22.44)(39.11,24.67)(42.78,23.67)
\put(36.67,19.89){\vector(3,2){3.67}}
\put(30.89,18.78){\makebox(0,0)[cc]{$\pi-\frac{n\pi}{\la}$}}
\put(40.67,15.11){\makebox(0,0)[cc]{$n$}}
\put(41.11,45.22){\makebox(0,0)[cc]{$n$}}
\put(50.00,30.00){\makebox(0,0)[lc]{$\st{n_{1},\ldots,n_{2k-2}}$}}
\put(46.00,3.00){\rule{2.00\unitlength}{52.00\unitlength}}
\end{picture}
\ca[As \ref{fig:fb1} with incoming soliton crossed]{As \ref{fig:fb1} with incoming \newline soliton crossed}
\label{fig:fb3} }
\end{figure}

\begin{figure}[!ht]
\parbox{0.49\textwidth}{
\unitlength 1.00mm
\linethickness{0.4pt}
\begin{picture}(48.00,60.24)(10,2)
\put(46.00,3.00){\rule{2.00\unitlength}{52.00\unitlength}}
\multiput(45.67,52.11)(0,1){3}{\circle*{0.25}}
\multiput(45.78,4.00)(0,1){3}{\circle*{0.25}}
\multiput(45.78,8.00)(0,2){22}{\circle*{0.25}}
\put(46.00,5.67){\line(-1,1){7.78}}
\put(46.00,51.89){\line(-1,-1){7.78}}
\put(38.22,13.44){\line(-6,-1){24.78}}
\put(38.22,44.11){\line(-6,1){24.78}}
\multiput(38.22,13.44)(0.5,1){16}{\circle*{0.25}}
\multiput(38.22,44.11)(0.5,-1){16}{\circle*{0.25}}
\multiput(38.22,48.67)(0,-1){10}{\circle*{0.25}}
\qbezier(34.89,44.67)(35.00,47.56)(38.22,47.78)
\put(34.00,47.78){\makebox(0,0)[rc]{$\nu_{n-n_{2k}}$}}
\qbezier(42.44,9.22)(43.00,11.89)(45.89,11.78)
\qbezier(34.00,12.78)(36.33,9.78)(40.44,11.22)
\put(35.44,15.56){\vector(1,-4){1}}
\put(31.89,18.11){\makebox(0,0)[cc]{$\pi-\frac{n\pi}{\la}$}}
\put(40.00,22.78){\makebox(0,0)[cc]{$n$}}
\put(40.00,33.78){\makebox(0,0)[cc]{$n$}}
\qbezier(46.00,24.44)(45.00,21.44)(44.00,24.44)
\put(50.00,30.00){\makebox(0,0)[lc]{$\st{n_{1},\ldots,n_{2k-1}}$}}
\put(50.00,9.56){\makebox(0,0)[lc]{$w_{n_{2k}}$}}
\put(50.00,23.00){\makebox(0,0)[lc]{$\frac{\nu_{n}-w_{2n_{2k}}}{2}$}}
\end{picture}
\ca{Incoming soliton, soliton boundary decay}
\label{fig:fb2} } \
\parbox{0.49\textwidth}{
\unitlength 1.00mm
\linethickness{0.4pt}
\begin{picture}(48.00,60.00)(10,2)
\put(46.00,3.00){\rule{2.00\unitlength}{52.00\unitlength}}
\multiput(39.22,22.33)(1,1){7}{\circle*{0.25}}
\multiput(39.22,35.44)(1,-1){7}{\circle*{0.25}}
\put(4.33,18.00){\line(2,1){34.78}}
\put(39.22,22.44){\line(-2,1){34.78}}
\put(39.11,35.44){\line(1,2){6.61}}
\put(45.89,9.22){\line(-1,2){6.61}}
\qbezier(43.67,13.89)(44.11,15.78)(45.89,14.56)
\qbezier(20.89,26.22)(21.89,23.67)(26.11,24.67)
\qbezier(34.56,33.22)(35.67,38.56)(41.33,39.89)
\put(30.67,40.00){\makebox(0,0)[cc]{$\pi - \frac{n\pi}{\lambda}$}}
\put(20.89,27.22){\makebox(0,0)[rc]{$w_{n_{2k}-n}$}}
\multiput(26.11,22.33)(0,1){14}{\circle*{0.25}}
\qbezier(41.89,24.78)(43.00,22.22)(46.00,22.78)
\put(50.00,30.00){\makebox(0,0)[lc]{$\st{n_{1},\ldots,n_{2k-1}}$}}
\put(40.00,25.78){\makebox(0,0)[cc]{$n$}}
\put(40.00,31.78){\makebox(0,0)[cc]{$n$}}
\put(50.00,12.56){\makebox(0,0)[lc]{$w_{n_{2k}}$}}
\put(50.00,23.00){\makebox(0,0)[lc]{$\frac{\nu_{n}-w_{2n_{2k}}}{2}$}}
\multiput(45.67,49.00)(0,1){6}{\circle*{0.25}}
\multiput(45.67,4.00)(0,1){6}{\circle*{0.25}}
\multiput(45.67,11.00)(0,2){19}{\circle*{0.25}}
\end{picture}
\ca[As \ref{fig:fb2} with incoming soliton crossed]{As \ref{fig:fb2} with incoming \newline soliton crossed}
\label{fig:fb4} }
\end{figure}

\begin{figure}[!ht]
\parbox{0.49\textwidth}{
\unitlength 1.00mm
\linethickness{0.4pt}
\begin{picture}(48.00,60.00)(10,2)
\put(46.00,3.00){\rule{2.00\unitlength}{52.00\unitlength}}
\multiput(24.89,4.45)(1,1){11}{\circle*{0.25}}
\multiput(24.89,53.77)(1,-1){11}{\circle*{0.25}}
\put(34.78,43.78){\line(3,-1){11}}
\put(34.78,43.78){\line(3,-4){11}}
\put(34.78,14.44){\line(3,1){11}}
\put(34.78,14.44){\line(3,4){11}}
\multiput(45.78,18.22)(0,1){22}{\circle*{0.25}}
\qbezier(39.78,42.00)(39.67,40.22)(37.89,39.78)
\qbezier(42.44,16.89)(43.67,15.00)(45.89,15.22)
\put(28.88,39.78){\makebox(0,0)[cc]{$\pi-\frac{n\pi}{2\lambda}$}}
\put(50.00,15.78){\makebox(0,0)[lc]{$\nu_{m}$}}
\qbezier(42.33,24.44)(43.22,23.11)(46.00,23.44)
\put(50.00,23.44){\makebox(0,0)[lc]{$\nu_{m-n}$}}
\put(50.00,30.00){\makebox(0,0)[lc]{$\st{n_{1},\ldots,n_{2k},m}$}}
\put(28.78,10.89){\makebox(0,0)[cc]{$n$}}
\put(28.78,47.89){\makebox(0,0)[cc]{$n$}}
\multiput(34.78,11.44)(0,1){7}{\circle*{0.25}}
\qbezier(32.78,12.44)(33.00,10.00)(34.78,11.44)
\put(32.78,14.44){\makebox(0,0)[rc]{$\frac{\nu_{2m}-w_{n}}{2}$}}
\end{picture}
\ca{Incoming breather, soliton bound state}
\label{fig:brfake} } \
\parbox{0.49\textwidth}{
\unitlength 1.00mm
\linethickness{0.4pt}
\begin{picture}(48.00,60.00)(10,2)
\multiput(45.67,3.44)(0,1){52}{\circle*{0.25}}
\qbezier(41.44,9.22)(43.00,11.22)(45.89,10.44)
\qbezier(40.44,30.78)(42.67,33.00)(45.89,32.44)
\put(50.00,8.44){\makebox(0,0)[lc]{$w_{n_{2k}}$}}
\put(50.00,31.00){\makebox(0,0)[lc]{$w_{n_{2k}+n}+\pi$}}
\qbezier(21.89,24.67)(25.11,24.56)(24.44,21.78)
\put(23.89,18.44){\vector(-1,2){1.83}}
\put(23.89,17.11){\makebox(0,0)[cc]{$\pi-\frac{n\pi}{2\la}$}}
\put(9.78,15.89){\makebox(0,0)[cc]{$n$}}
\put(9.78,41.89){\makebox(0,0)[cc]{$n$}}
\put(50.00,25.00){\makebox(0,0)[lc]{$\st{n_{1},\ldots,n_{2k-1}}$}}
\multiput(16.78,16.22)(0,1){7}{\circle*{0.25}}
\qbezier(14.12,17.22)(15.00,16.22)(16.78,17.22)
\put(14.00,19.5){\makebox(0,0)[rc]{$\pi+\frac{w_{2n_{2k}}-\nu_{n}}{2}$}}
\put(45.89,3.67){\line(-5,6){29.07}}
\put(45.89,54.11){\line(-5,-6){29.07}}
\put(16.78,38.56){\line(3,-1){29.11}}
\put(16.78,19.22){\line(3,1){29.11}}
\multiput(16.78,38.56)(-1.33,1){13}{\circle*{0.25}}
\multiput(16.78,19.22)(-1.33,-1){13}{\circle*{0.25}}
\put(46.00,3.00){\rule{2.00\unitlength}{52.00\unitlength}}
\end{picture}
\ca[As \ref{fig:brfake} with outgoing soliton crossed]{As \ref{fig:brfake} with outgoing \newline soliton crossed}
\label{fig:ob3} }
\end{figure}
 
\begin{figure}[!ht]
\parbox{0.49\textwidth}{
\unitlength 1.00mm
\linethickness{0.4pt}
\begin{picture}(48.00,60.24)(10,0)
\put(46.00,3.00){\rule{2.00\unitlength}{52.00\unitlength}}
\multiput(45.67,52.11)(0,1){3}{\circle*{0.25}}
\multiput(45.78,4.00)(0,1){3}{\circle*{0.25}}
\multiput(45.78,8.00)(0,2){22}{\circle*{0.25}}
\put(46.00,5.67){\line(-1,1){7.78}}
\put(38.22,13.44){\line(1,2){7.67}}
\put(46.00,51.89){\line(-1,-1){7.78}}
\put(38.22,44.11){\line(1,-2){7.67}}
\multiput(38.22,13.56)(-1,-0.33){24}{\circle*{0.25}}
\multiput(38.22,44.11)(-1,0.33){24}{\circle*{0.25}}
\qbezier(42.44,9.22)(43.00,11.89)(45.89,11.78)
\put(50.00,9.56){\makebox(0,0)[lc]{$w_{n_{2k}}$}}
\qbezier(42.22,36.22)(43.33,37.78)(45.89,37.44)
\put(50.00,35.67){\makebox(0,0)[lc]{$-w_{n_{2k}+n}$}}
\qbezier(41.44,47.22)(43.33,43.89)(40.89,39.11)
\put(29.44,40.67){\makebox(0,0)[cc]{$\pi - \frac{n\pi}{\la}$}}
\put(36.44,41.67){\vector(4,1){4}}
\put(24.67,47.00){\makebox(0,0)[cc]{$n$}}
\put(24.56,10.11){\makebox(0,0)[cc]{$n$}}
\put(50.00,40.00){\makebox(0,0)[lc]{$\st{n_{1},\ldots,n_{2k-1}}$}}
\multiput(38.22,17.06)(0,-1){8}{\circle*{0.25}}
\qbezier(34.56,12.39)(35.45,10.28)(38.22,10.72)
\put(34.00,14.44){\makebox(0,0)[rc]{$\frac{\nu_{n}-w_{2n_{2k}}}{2}$}}
\end{picture}
\ca{Incoming breather, soliton boundary decay}
\label{fig:ob6} } \
\parbox{0.49\textwidth}{
\unitlength 1.00mm
\linethickness{0.4pt}
\begin{picture}(48.00,60.00)(10,0)
\put(46.00,3.00){\rule{2.00\unitlength}{52.00\unitlength}}
\put(33.22,35.89){\line(5,-3){12.78}}
\put(33.33,35.78){\line(3,4){12.67}}
\put(33.22,20.56){\line(5,3){12.78}}
\put(33.33,20.67){\line(3,-4){12.67}}
\multiput(33.44,20.56)(-1,0.33){30}{\circle*{0.25}}
\multiput(33.44,35.89)(-1,-0.33){30}{\circle*{0.25}}
\multiput(10.44,31.78)(0,-1){8}{\circle*{0.25}}
\qbezier(6.78,26.89)(7.67,24.78)(10.44,25.22)
\put(9.11,21.44){\makebox(0,0)[cc]{$\pi+\frac{w_{2n_{2k}}-\nu_{n}}{2}$}}
\qbezier(41.00,46.11)(43.22,44.11)(45.89,44.22)
\put(50.00,47.00){\makebox(0,0)[lc]{$w_{n_{2k}}$}}
\qbezier(40.11,24.78)(41.67,22.22)(45.89,22.67)
\put(50.00,24.67){\makebox(0,0)[lc]{$-w_{n_{2k}+n}$}}
\qbezier(37.22,23.00)(39.33,18.89)(36.56,16.44)
\put(30.78,17.11){\vector(2,1){5.33}}
\put(24.78,16.11){\makebox(0,0)[cc]{$\pi - \frac{n\pi}{\la}$}}
\put(21.22,33.44){\makebox(0,0)[cc]{$n$}}
\put(21.11,23.00){\makebox(0,0)[cc]{$n$}}
\put(50.00,30.00){\makebox(0,0)[lc]{$\st{n_{1},\ldots,n_{2k-1}}$}}
\end{picture}
\ca[As \ref{fig:ob6} with incoming breather crossed]{As \ref{fig:ob6} with incoming \newline breather crossed}
\label{fig:ob4} }

\end{figure}
\begin{figure}[!ht]
\parbox{0.49\textwidth}{
\unitlength 0.80mm
\linethickness{0.4pt}
\begin{picture}(48.00,75.5)(0.5,-7.5)
\put(46.00,-3.00){\rule{2.50\unitlength}{65.00\unitlength}}
\multiput(45.67,-2.00)(0,2){32}{\circle*{0.25}}
\multiput(18.44,58.94)(-2,1){7}{\circle*{0,25}}
\multiput(18.44,4.72)(-2,-1){7}{\circle*{0.25}}
\multiput(35.78,8.06)(1,-1){11}{\circle*{0.25}}
\multiput(40.44,54.94)(1,1){6}{\circle*{0.25}}
\put(45.89,31.84){\line(-1,1){27.34}}
\put(45.89,31.84){\line(-1,-1){27.34}}
\put(18.66,58.94){\line(5,-1){21.56}}
\put(40.22,54.86){\line(1,-3){5.60}}
\put(45.89,38.50){\line(-1,-3){10.14}}
\put(18.88,4.94){\line(5,1){16.66}}
\multiput(18.44,55.94)(0,1){7}{\circle*{0.25}}
\qbezier(18.44,61.94)(16.00,63.00)(14.44,60.94)
\put(14.44,58.94){\makebox(0,0)[cc]{$u$}}
\put(50.00,58.94){\makebox(0,0)[lc]{$u=\frac{w_{0}-\nu_{2n_{2k-1}+n}}{2}$}}
\qbezier(41.00,27.00)(43.00,24.33)(46.00,26.00)
\put(50.00,30.00){\makebox(0,0)[lc]{$-\nu_{n_{2k-1}+n}$}}
\qbezier(43.22,46.00)(43.89,47.78)(45.89,46.78)
\put(50.00,43.44){\makebox(0,0)[lc]{$w_{n_{2k}}$}}
\qbezier(23.89,9.78)(27.67,9.44)(26.22,6.33)
\put(19.89,10.44){\vector(4,-3){4.44}}
\put(13.00,12.11){\makebox(0,0)[cc]{$\pi-\frac{n\pi}{\la}$}}
\qbezier(30.56,7.22)(31.56,11.78)(37.56,13.67)
\put(31.56,4.56){\vector(1,2){3.17}}
\put(26.56,1.67){\makebox(0,0)[cc]{$\pi-\frac{l\pi}{\la}$}}
\qbezier(40.00,4.11)(42.22,7.89)(45.89,6.78)
\put(50.00,9.78){\makebox(0,0)[lc]{$\frac{\nu_{n_{2k-1}}-w_{n_{2k}}}{2}$}}
\put(37.56,2.22){\makebox(0,0)[cc]{$l$}}
\put(50.00,52.22){\makebox(0,0)[lc]{$l=n_{2k}+n_{2k-1}$}}
\put(11.67,3.00){\makebox(0,0)[cc]{$n$}}
\put(50.00,23.56){\makebox(0,0)[lc]{$\st{n_{1},\ldots,n_{2k-2}}$}}
\end{picture}
\ca{As \ref{fig:brfake}, outer legs replaced by \ref{fig:fb1}}
\label{fig:ob5} } \
\parbox{0.49\textwidth}{
\unitlength 1.00mm
\linethickness{0.4pt}
\begin{picture}(48.00,60.00)(10,0)
\put(46.00,3.00){\rule{2.00\unitlength}{52.00\unitlength}}
\multiput(45.67,3.44)(0,1){52}{\circle*{0.25}}
\put(31.22,41.77){\line(5,6){6.94}}
\put(45.89,17.77){\line(-1,-4){1.67}}
\put(38.00,49.78){\line(1,-4){8.00}}
\put(31.22,41.78){\line(4,-1){14.67}}
\put(25.56,32.78){\line(5,-6){18.52}}
\put(45.89,17.89){\line(-1,-4){1.81}}
\put(25.56,32.78){\line(4,1){20.44}}
\multiput(2.11,42.23)(1.25,-0.5){19}{\circle*{0.25}}
\multiput(31.11,41.67)(-1.25,-0.5){24}{\circle*{0.25}}
\multiput(44.00,10.56)(0.25,-1){7}{\circle*{0.25}}
\multiput(38.00,49.78)(0.25,1){5}{\circle*{0.25}}
\multiput(17.30,39.00)(0,-1){7}{\circle*{0.25}}
\qbezier(17.3,34.00)(16.00,33.00)(14.8,35.00)
\put(13.00,36.00){\makebox(0,0)[cc]{$u$}}
\put(8.00,42.00){\makebox(0,0)[cc]{$n$}}
\put(8.00,30.00){\makebox(0,0)[cc]{$n$}}
\put(50.00,51.94){\makebox(0,0)[lc]{$u=\frac{2\pi+\nu_{2n_{2k-1}+n}
-w_{0}}{2}$}}
\qbezier(44.33,9.22)(44.67,11.00)(46.00,10.00)
\put(50.00,8.00){\makebox(0,0)[lc]{$\frac{\nu_{n_{2k-1}}-w_{n_{2k}}}{2}$}}
\qbezier(44.00,26.33)(44.44,28.00)(45.89,27.00)
\put(50.00,23.78){\makebox(0,0)[lc]{$w_{n_{2k}}$}}
\qbezier(42.00,39.11)(42.78,41.89)(45.89,42.00)
\put(50.00,40.11){\makebox(0,0)[lc]{$-\nu_{n_{2k-1}}+n$}}
\qbezier(31.56,34.33)(32.33,30.44)(30.44,27.00)
\put(16.44,28.44){\makebox(0,0)[cc]{$\pi - \frac{n\pi}{2\la}$}}
\put(22.44,30.44){\vector(4,1){6}}
\qbezier(33.89,45.22)(36.56,43.11)(39.22,44.89)
\put(29.44,48.33){\vector(4,-1){7.22}}
\put(23.33,47.56){\makebox(0,0)[cc]{$\pi - \frac{n\pi}{2\la}$}}
\put(37.33,53.00){\makebox(0,0)[cc]{$l$}}
\put(43.44,6.22){\makebox(0,0)[cc]{$l$}}
\put(50.00,30.00){\makebox(0,0)[lc]{$\st{n_{1},\ldots,n_{2k-2}}$}}
\end{picture}
\ca[As \ref{fig:ob4}, outer legs replaced by \ref{fig:fb3}]{As \ref{fig:ob4}, outer legs \newline replaced by \ref{fig:fb3}}
\label{fig:ob1} }
\end{figure}

\clearpage

\begin{figure}[!ht]
\parbox{0.49\textwidth}{
\unitlength 1.00mm
\linethickness{0.4pt}
\begin{picture}(48.00,60.00)(10,0)
\multiput(45.67,3.44)(0,1){52}{\circle*{0.25}}
\put(34.22,30.11){\line(5,-6){4.44}}
\put(45.89,39.45){\line(-1,2){2.83}}
\put(38.78,24.89){\line(1,2){7.67}}
\put(34.22,30.11){\line(5,-2){11.67}}
\put(45.89,25.33){\line(-5,-2){27.78}}
\put(43.00,45.11){\line(-4,-5){24.71}}
\multiput(18.22,14.22)(-1.5,-1){8}{\circle*{0.25}}
\multiput(34.22,30.11)(-1.5,1){16}{\circle*{0.25}}
\multiput(43.00,45.23)(0.5,1.5){6}{\circle*{0.25}}
\multiput(38.78,24.67)(0.5,-1.5){14}{\circle*{0.25}}
\multiput(18.22,11.22)(0,1){7}{\circle*{0.25}}
\qbezier(18.22,12.22)(16.50,11.00)(15.22,12.22)
\put(16.5,10.00){\makebox(0,0)[cc]{$u$}}
\put(50.00,51.94){\makebox(0,0)[lc]{$u=\frac{2\pi+\nu_{2n_{2k-1}}
-w_{0}}{2}$}}
\put(50.00,45.94){\makebox(0,0)[lc]{$a=2\pi-\frac{\pi(n_{2k-1}
+n_{2k}+n)}{\la}$}}
\put(50.00,39.94){\makebox(0,0)[lc]{$l=n_{2k}+n_{2k-1}$}}
\qbezier(43.78,10.12)(45.00,11.67)(45.89,11.00)
\qbezier(43.56,24.44)(44.00,23.11)(45.89,23.00)
\qbezier(43.78,34.45)(44.56,33.23)(45.89,33.67)
\qbezier(39.44,26.44)(41.00,25.67)(41.66,27.11)
\put(50.00,9.67){\makebox(0,0)[lc]{$\frac{\nu_{n_{2k-1}}-w_{n_{2k}}}{2}$}}
\put(50.00,22.22){\makebox(0,0)[lc]{$\pi+\nu_{n_{2k-1}+n}$}}
\put(50.00,34.45){\makebox(0,0)[lc]{$w_{n_{2k}}$}}
\put(39.00,31.00){\makebox(0,0)[cc]{$a$}}
\put(39.33,30.11){\vector(1,-3){1.2}}
\qbezier(24.00,21.22)(27.56,20.56)(26.78,17.67)
\put(27.11,13.78){\vector(-2,3){3}}
\put(27.00,11.67){\makebox(0,0)[cc]{$\pi-\frac{n\pi}{\la}$}}
\put(12.33,12.56){\makebox(0,0)[cc]{$n$}}
\put(22.22,36.00){\makebox(0,0)[cc]{$n$}}
\put(43.00,8.00){\makebox(0,0)[cc]{$l$}}
\put(43.00,48.00){\makebox(0,0)[cc]{$l$}}
\put(50.00,29.00){\makebox(0,0)[lc]{$\st{n_{1},\ldots,n_{2k-2}}$}}
\put(46.00,3.00){\rule{2.00\unitlength}{52.00\unitlength}}
\end{picture}
\ca[As \ref{fig:ob3}, outer legs replaced by \ref{fig:fb3}]{As \ref{fig:ob3}, outer legs \newline replaced by \ref{fig:fb3}}
\label{fig:ob2} } \
\parbox{0.49\textwidth}{
\unitlength 1.00mm
\linethickness{0.4pt}
\begin{picture}(48.00,60.00)(10,0)
\multiput(45.67,4.00)(0,1){10}{\circle*{0.25}}
\multiput(45.67,15.00)(0,2){16}{\circle*{0.25}}
\multiput(45.67,46.00)(0,1){9}{\circle*{0.25}}
\multiput(32.22,43.22)(-2,1){8}{\circle*{0,25}}
\multiput(32.22,16.11)(-2,-1){8}{\circle*{0.25}}
\multiput(43.11,41.18)(0.5,-1.5){6}{\circle*{0.25}}
\multiput(45.89,33.00)(-0.5,-1.5){11}{\circle*{0.25}}
\put(45.89,29.67){\line(-1,1){13.67}}
\put(45.89,29.67){\line(-1,-1){13.67}}
\put(40.89,17.78){\line(1,-1){5}}
\put(43.22,41.22){\line(1,1){4}}
\put(32.33,43.33){\line(5,-1){10.78}}
\put(32.44,16.22){\line(5,1){8.33}}
\multiput(32.22,40.22)(0,1){7}{\circle*{0.25}}
\qbezier(32.22,45.22)(30.5,46.50)(29.22,44.72)
\put(31.00,47.00){\makebox(0,0)[cc]{$u$}}
\qbezier(41.44,17.22)(43.00,20.44)(46.00,20.44)
\put(50.00,17.67){\makebox(0,0)[lc]{$w_{n_{2k}}$}}
\qbezier(44.00,40.22)(44.89,41.11)(45.89,41.11)
\put(49.33,39.22){\vector(-1,0){5.11}}
\put(50.78,39.22){\makebox(0,0)[lc]{$\frac{\nu_{l}-w_{2n_{2k}}}{2}$}}
\qbezier(41.56,25.33)(42.44,23.78)(46.00,23.89)
\put(49.22,26.33){\vector(-1,0){5.44}}
\put(51.22,26.33){\makebox(0,0)[lc]{$-w_{n_{2k}+n-l}$}}
\qbezier(36.67,17.11)(39.22,15.44)(42.78,15.89)
\put(38.22,12.56){\vector(1,2){2.06}}
\put(34.33,9.11){\makebox(0,0)[cc]{$\pi-\frac{l\pi}{\la}$}}
\put(35.67,24.11){\vector(2,-1){4.44}}
\put(33.22,24.78){\makebox(0,0)[cc]{$l$}}
\qbezier(36.56,20.44)(39.22,20.11)(38.44,17.56)
\put(32.11,19.22){\vector(1,0){5.44}}
\put(25.89,20.00){\makebox(0,0)[cc]{$\pi-\frac{n\pi}{\la}$}}
\put(23.56,13.89){\makebox(0,0)[cc]{$n$}}
\put(23.56,44.89){\makebox(0,0)[cc]{$n$}}
\put(50.00,32.00){\makebox(0,0)[lc]{$\st{n_{1},\ldots,n_{2k-1}}$}}
\put(50.00,51.94){\makebox(0,0)[lc]{$u=\frac{\nu_{2l}-w_{2n_{2k}+n}}{2}$}}
\put(46.00,3.00){\rule{2.00\unitlength}{52.00\unitlength}}
\end{picture}
\ca[As \ref{fig:brfake}, outer legs replaced by \ref{fig:fb2}]{As \ref{fig:brfake}, outer legs \newline replaced by \ref{fig:fb2}}
\label{fig:ob7} }
\end{figure}

\begin{figure}[!ht]
\begin{center}
\parbox{0.8\textwidth}{
\begin{center}
\unitlength 1.00mm
\linethickness{0.4pt}
\begin{picture}(48.00,60.57)(10,0)
\put(46.00,3.00){\rule{2.00\unitlength}{52.00\unitlength}}
\multiput(45.67,4.00)(0,2){26}{\circle*{0.25}}
\put(32.92,10.23){\line(-5,1){25.26}}
\put(25.44,38.67){\line(-5,-1){15.00}}
\put(38.37,41.33){\line(-5,-1){25.26}}
\put(45.89,29.89){\line(-2,-3){13.04}}
\put(45.89,29.89){\line(-2,-3){12.07}}
\put(45.89,29.78){\line(-2,-3){10.15}}
\put(46.00,29.89){\line(-2,3){7.63}}
\multiput(34.67,6.56)(-0.5,1){5}{\circle*{0.25}}
\multiput(38.33,41.33)(0.5,0.83){7}{\circle*{0.25}}
\multiput(41.33,46.33)(0.25,1.5){8}{\circle*{0.25}}
\multiput(34.72,6.66)(0.25,-1.25){6}{\circle*{0.25}}
\multiput(45.89,34.78)(-0.5,-1.25){21}{\circle*{0.25}}
\multiput(45.89,34.89)(-0.5,-1.25){23}{\circle*{0.25}}
\multiput(41.33,46.33)(0.5,-1.25){10}{\circle*{0.25}}
\multiput(45.89,34.78)(-0.5,-1.25){18}{\circle*{0.25}}
\qbezier(32.89,40.22)(35.33,36.00)(40.56,38.00)
\put(29.11,35.44){\makebox(0,0)[cc]{$\pi-\frac{n\pi}{2\la}$}}
\qbezier(39.33,43.00)(41.00,41.00)(42.67,43.00)
\put(32.33,45.34){\makebox(0,0)[cc]{$\frac{n\pi}{2\la}$}}
\put(38.23,43.34){\makebox(0,0)[cc]{$n$}}
\put(36.89,51.78){\makebox(0,0)[cc]{$n+l$}}
\put(44.11,42.67){\makebox(0,0)[cc]{$l$}}
\bezier{12}(43.89,40.11)(44.67,41.33)(46.00,40.67)
\put(50.00,35.67){\makebox(0,0)[lc]{$\frac{w_{n_{2k}}-
\nu_{n_{2k-1}+n}}{2}$}}
\bezier{24}(42.78,25.22)(43.22,22.22)(45.89,23.22)
\put(50.00,24.33){\makebox(0,0)[lc]{$w_{n_{2k}}$}}
\multiput(36.11,-1.23)(0,1){8}{\circle*{0.25}}
\bezier{8}(35.22,4.00)(35.45,5.44)(36.11,4.66)
\put(50.00,3.55){\makebox(0,0)[lc]{$\frac{\nu_{n_{2k-1}}-w_{n_{2k}}}{2}$}}
\put(30.00,3.55){\makebox(0,0)[cc]{$n+l$}}
\put(36.00,45.33){\vector(4,-1){5.11}}
\put(50.00,30.00){\makebox(0,0)[lc]{$\st{n_{1},\ldots,n_{2k-2}}$}}
\put(50.00,51.94){\makebox(0,0)[lc]{$u=\nu_{n-n_{2k}}$}}
\end{picture}
\end{center}
\ca{As \ref{fig:fb3}, outer legs replaced by all-breather version}
\label{fig:os1} }
\end{center}
\end{figure}

\chap{Miscellaneous Proofs}
\quot{The trouble with facts is that there are so many of
them.}{Samuel McChord Crothers}
\label{app:details}

\quot{Basic research is what I am doing when I don't know what I am
doing.}{Wernher von Braun}

In this appendix, we present various proofs which are subsidiary to the 
main text, but serve either to fill out the bare bones of it, or provide 
cross-checks on the results presented.

\se{Oota's starting point}
\label{app:Ootasp}
Oota, in deriving his integral formula for the S-matrix, began by 
defining
\begin{equation}
m^{q}_{ab}(x)=\sum_{y\in \mathbb{Z}}m_{ab}(x,y)q^{y}\,,
\end{equation}
as well as the matrices
\ba
(D_{q})_{ab}=q^{t_{a}}\delta_{ab}\,, & 
(T_{q})_{ab}=[t_{a}]_{q}\delta_{ab}\,, & (I_{q})_{ab}=[G_{ab}]_{q}\,.
\ea
He then stated (after a case-by-case analysis) that the matrices 
\m{m^{q}(x)} satisfied
\ba
m^{q}(0)=0\,, & m^{q}(1)=D_{q}T_{q}\,,
\label{eq:m01}
\ea
as well as the recursion relation
\begin{equation}
D_{q}^{-1}m^{q}_{ab}(x+1)+D_{q}m^{q}_{ab}(x-1)=I_{q}m^{q}(x)\,.
\end{equation}

As we shall see, the recursion relation follows from the generalised 
bootstrap \re{eq:grtv}.
Examining first the recursion relation, note that it can be re-written as
\begin{equation}
m^{q}_{ab}(x+1)q^{-t_{a}}+m^{q}_{ab}(x-1)q^{t_{a}}=\sum_{c}[G_{ac}]_{q}
m^{q}_{cb}(x)\,.
\label{eq:recursion}
\end{equation}
Turning now to \re{eq:grtv}, we can use the product-form
notation to re-write the rhs as
\begin{equation}
\mathrm{rhs\ }=\prod_{c, G_{ac}\neq 0}\prod_{x=1}^{h}\prod_{y=1}^{r^{\vee}
h^{\vee}}\left\{
\begin{array}{l}
\{x,y\}^{m_{bc}(x,y)}, G_{ac}=1 \\
(\{x,y-1\}\{x,y+1\})^{m_{bc}(x,y)}, G_{ac}=2 \\
(\{x,y-2\}\{x,y\}\{x,y+2\})^{m_{bc}(x,y)}, G_{ac}=3
\end{array} \right. .
\end{equation}
This comes about because the forward-backward shift on the 
rhs has the effect of shifting \m{y}
forward and backward by 1 or 2, though it should be noted that this is not 
quite as straightforward as it seems and, for example, a forward shift 
on its own does \emph{not} have the effect of producing any neat shift in 
\m{y}. To put this another way,
\begin{equation}
\mathrm{rhs\ }=\prod_{c, G_{ca}\neq 0}\prod_{x=1}^{h}\prod_{y=1}^{r^{\vee}
h^{\vee}}\left\{
\begin{array}{l}
\{x,y\}^{m_{bc}(x,y)}, G_{ca}=1 \\
\{x,y\}^{m_{bc}(x,y-1)+m_{bc}(x,y+1)}, G_{ca}=2 \\
\{x,y\}^{m_{bc}(x,y-2)+m_{bc}(x,y)+m_{bc}(x,y+2)}, G_{ca}=3
\end{array} \right. .
\end{equation}

Looking at the lhs of \re{eq:grtv} and writing it the same way, we find
\begin{equation}
\mathrm{lhs\ }= \prod_{x=1}^{h}\prod_{y=1}^{r^{\vee}
h^{\vee}} \{x,y\}^{m_{ab}(x-1,y-t_{a})+m_{ab}(x+1,y+t_{a})}\,.
\end{equation}

Comparing block multiplicities, this reduces to
\begin{multline}
m_{ab}(x-1,y-t_{a})+m_{ab}(x+1,y+t_{a})= \\
\sum_{c}\left\{
\begin{array}{l}
m_{bc}(x,y), G_{ca}=1 \\
m_{bc}(x,y-1)+m_{bc}(x,y+1), G_{ca}=2 \\
m_{bc}(x,y-2)+m_{bc}(x,y)+m_{bc}(x,y+2), G_{ca}=3
\end{array}
\right. .
\end{multline}
Multiplying through by \m{q^{y}}, this can be rearranged to
\ba
\lefteqn{m_{ab}(x-1,y-t_{a})q^{y-t_{a}}q^{t_{a}}+m_{ab}(x+1,y+t_{a})q^{y+t_{a}} 
q^{-t_{a}}=} \nonumber \\
&& \sum_{c}\left\{
\begin{array}{l}
m_{bc}(x,y)q^{y}, G_{ca}=1 \\
m_{bc}(x,y-1)q^{y-1}q+m_{bc}(x,y+1)q^{y+1}q^{-1}, G_{ca}=2 \\
m_{bc}(x,y-2)q^{y-2}q^{2}+m_{bc}(x,y)q^{y}+m_{bc}(x,y+2)q^{y+2}q^{-2}, 
G_{ca}=3
\end{array}
\right. .
\ea
Summing both sides over all integer \m{y}, we can then re-write this in 
terms of the matrices \m{m^{q}(x)} as
\begin{equation}
m_{ab}^{q}(x-1)q^{t_{a}}+m_{ab}^{q}(x+1)q^{-t_{a}}=
\sum_{c}\left\{
\begin{array}{l}
m_{bc}^{q}(x), G_{cb}=1 \\
m_{bc}^{q}(x)(q+q^{-1}), G_{cb}=2 \\
m_{bc}^{q}(x)(q^{2}+1+q^{-2}), G_{cb}=3
\end{array}
\right. ,
\end{equation}
which, noting that \m{[n]_{q}=q^{n-1}+q^{n-3}+\ldots+q^{-(n-1)}}, is 
just \re{eq:recursion}.

\se[The generalised bootstrap at \m{\T=0}]{The generalised bootstrap at \m{\pmb{\T=0}}}
The generalised bootstrap is na\"{\i}vely
\begin{equation}
S_{ab}(\T+\T_{h}+t_{a}\T_{H})S_{ab}(\T-\T_{h}-t_{a}\T_{H})= 
\prod_{l=1}^{r}\prod_{n=1}^{G_{al}}S_{bl}(\T+(2n-1-G_{al})\T_{H})\,.
\label{eq:grtv-e}
\end{equation}
A subtlety arises when we consider the case $\T=0$, since we can
either consider the lhs as
\begin{equation}
\lim_{\T\rightarrow 0}\lim_{H\rightarrow \T_{h}+t_{a}\T_{H}}
S_{ab}(\T+H)S_{ab}(\T-H)
\end{equation}
or as
\begin{equation}
\lim_{H\rightarrow \T_{h}+t_{a}\T_{H}}\lim_{\T\rightarrow 0}
S_{ab}(\T+H)S_{ab}(\T-H)\,.
\end{equation}
For $\T \neq 0$, this distinction makes no difference, since we are not near
a pole of $S$, but at $\T=0$ we are potentially considering a pole,
and hence way the limit is taken is important. By leaving $\T$
arbitrary and fixing $H$ from the beginning, we have been
implicitly using the first form, but is perhaps more sensible to consider
$H$ as a shift from $S(\T)$---as would be the case in the Bethe ansatz
approach---in which case the second form would be more appropriate.

To see the difference between these two forms, we can consider one basic block
of $S$, $\langle x \rangle$ for some $x$. Shifted forward and back by
$H$, this becomes
\begin{equation}
\frac{\sinh \frac{1}{2}(\T+H+i\pi x)}{\sinh \frac{1}{2}(\T +H - i\pi x)} \cdot
\frac{\sinh \frac{1}{2}(\T-H+i\pi x)}{\sinh \frac{1}{2}(\T-H - i\pi x)}.
\end{equation}
For the sake of argument, we shall take $H$ positive. It is clear that as
long as $H \neq i\pi x$, all the arguments of the sinh functions are
non-zero (noting that we need not worry about periodicity as $x \leq 1$) and
thus the result is the same in either limit. However, at $H \rightarrow i\pi
x$, a discrepancy arises as, if we take this limit first, we find
\begin{equation}
\frac{\sinh \frac{1}{2}(\T+2H)}{\sinh \frac{1}{2}(\T -2H)}\cdot
\frac{\sinh \frac{1}{2}(\T)}{\sinh \frac{1}{2}(\T)}
\end{equation}
which reduces to -1 if we take the $\T$ limit as well. Taking the
limits in the other order, however, we get
\begin{equation}
\frac{\sinh \frac{1}{2}(H+i\pi x)}{\sinh \frac{1}{2}(H - i\pi x)} \cdot
\frac{\sinh \frac{1}{2}(-H+i\pi x)}{\sinh \frac{1}{2}(-H - i\pi x)}
\end{equation}
which reduces to 1 even before taking the $H$ limit.

For $\T=0$, then, if we want to take the $\T\rightarrow 0$ limit first, we
must modify \re{eq:grtv-e} by a factor of -1 for every ``problem''
S-matrix block, i.e. for every block of the form $\langle x,y \rangle
= \langle 1,t_{a}\rangle$ in
$S_{ab}$. Going to the larger block, $\{x,y\}$, this turns out to mean a
factor for every block $\{2,t_{a}\pm 1\}$.

The easiest way to go from here is to appeal to~\re{eq:recursion}, with
$x=1$, which gives
\begin{equation}
q^{-t_{a}}m_{ab}^{q}(2)=q^{t_{b}}[t_{b}]_{q}[G_{ba}]_{q}
\end{equation}
or
\begin{equation}
m_{ab}^{q}(2)=q^{t_{a}+t_{b}}(q^{t_{b}-1}+q^{t_{b}-3}+\cdots+q^{1-t_{b}})
[G_{ba}]_{q}.
\end{equation}
We are now looking for terms in $q^{t_{a}\pm1}$ in this expansion. If
$G_{ba}=1$, then the lowest term is $q^{t_{a}+1}$, meaning we need one minus
sign. If $G_{ba}=2$, we
introduce a factor of $(q+q^{-1})$, leaving us with terms like $q^{r_{a}}$
and $q^{r_{a}+2}$, but none of the right form. If $G_{ba}=3$, it is simplest
to note that we must therefore be looking at $G_{2}$, and that $r_{a}=3$,
$r_{b}=1$, showing $m_{ab}^{q}(2)=q^{8}+2q^{6}+3q^{4}+2q^{2}+1$, with us
searching for powers of $q^{2}$ or $q^{0}$. Thus, $G_{ab}=3$ leaves us
needing to introduce a net minus sign as well.

To summarise, we need to introduce a minus sign to one side
of \re{eq:grtv-e} for $G_{ab}$ odd\footnote{Note that $G_{ab}$ is
necessarily odd if $G_{ba}$ is, though the two need not be equal.} in 
the case $\T=0$; the term used
in \re{eq:grtv} is perhaps as good a way as any, and turns out to be
useful in further calculations.

\se{Check that generalised bootstrap follows from integral formula}

The most straightforward method (and the one we shall use) is to propose
an identity of the form
\begin{equation}
S_{ij}(\T+\T_{h}+t_{i}\T_{H})S_{ij}(\T-\T_{h}-t_{i}\T_{H})=\e^{y}
\prod_{l=1}^{r}\prod_{n=1}^{G_{il}}S_{jl}(\T+(2n-1-G_{il})\T_{H}),
\label{eq:grtvy}
\end{equation}
and aim to find $y$ by substituting in the integral formula for the
S-matrix.

Since equation~\re{eq:grtv} applies to the case where the \m{\T}
limit is taken first, we need a prescription for taking the other limit.
It turned out to be simplest to replace $\T_{h}$ and $\T_{H}$
in~\re{eq:grtvy} by $\T_{h}+i\epsilon$ and $\T_{H}+i\epsilon$, and take
the limit $\epsilon \rightarrow 0$ last. Substituting in
\re{eq:Oota} and simplifying, we find
\begin{multline}
\e^{y}= \lim_{\epsilon \rightarrow 0}\exp
\left(\sum_{l=1}^{r}\int_{-\infty}^{\infty}\frac{dk}{k}\e^{ik\T}
\left\{q(\pi
k)-q(-\pi k)\right\}\left\{\qbar(\pi k)^{t_{j}}-\qbar(-\pi k)^{t_{j}}
\right\}\cdot \right. \\
\left. [K_{il}]_{q'(\pi k)\qbar'(\pi k)}M_{lj}(q(\pi k),\qbar(\pi
k))\right)
\label{eq:ey}
\end{multline}
where $q'(t)=q(t)\e^{\frac{\epsilon}{\pi}}$ and
$\qbar'(t)=\qbar(t)\e^{\frac{\epsilon}{\pi}}$. 
Looking back to~\re{eq:mform}, we can see that
when the integrand in~\re{eq:ey} is expanded out,
all the terms are of the form
$t(x,\T)=\int_{-\infty}^{\infty}\frac{dk}{k}\e^{ik\T}\e^{x|k|}$, with
$x$ real, which is divergent if $x$ is positive. It is, however,
implicit in Oota's formulation that any terms which are na\"{\i}vely
divergent must be analytically continued. For $x$ negative, $t(x,\T)$
is just a standard
Fourier transform which has the result $2i \arctan-\frac{\T}{x}$. Thus
the analytic continuation $x \rightarrow -x$ to $x$ positive should just
introduce a minus sign, so each divergent term of this type with $x$
positive should be replaced by the same term with $x \rightarrow -x$ and
an additional minus sign. 

If $\T=0$, each term $t(x,0)$ becomes 0, unless $x=0$. If there
is no $t(0,\T)$ term, the rhs must therefore
reduce to 1. If $\T \neq 0$, the limit ordering does not matter, so we
can take the $\epsilon$ limit first and reduce
$\sum_{l=1}^{r}[K_{il}]_{q\qbar}M_{lj}(q(\pi k),\qbar(\pi k))$ to
$\delta_{ij}[t_{j}]_{\qbar(\pi k)}$. Each $t(x,\T)$ is then matched by a $t(-x,\T)$, so
the rhs again reduces to 1. The only way the rhs can come to
anything other than 1 overall for \emph{any} $\T$ is
if there are terms like $t(0,\T)$ present. 

For this to happen, we require
$[K_{il}]_{q'\qbar'}=(q',\qbar'$-independent part)$+$(terms in
$q',\qbar'$). From the definition~\re{eq:kdef}, and the fact that
$[n]_{q}$ can be expanded out as $q^{n-1}+q^{n-3}+\ldots+q^{-(n-1)}$
for $n$ integer, this reduces to requiring $G_{il}$ odd, in which case
the $q',\qbar'$-independent part is -1. We also need
$M_{lj}(q(\pi k),\qbar(\pi k))=q(\pi k)\qbar(\pi k)^{t_{j}}+$(terms in higher
powers of $q,\qbar$) for the same $l$. Expanding out~\re{eq:mform},
the lowest power of $q^{a}\qbar^{b}$ present is $q^{x}\qbar^{y}$, for
the smallest $x$ and $y$ such that $m_{lj}(x,y) \neq 0$. The second condition
can thus only be satisfied if the product-form S-matrix $S_{lj}(\T)$
contains a block $\{1,t_{j}\}$, and~\re{eq:m01} shows that
this occurs iff $l=j$. This should be compared with the discussion of
equation~\re{eq:grtv-e}, where the discrepancy between the two
possible limit prescriptions was caused by a pole from this block; we
are essentially approaching the same pole here.

Overall, then, we find that there is one block of the form $t(0,\T)$
if $G_{ij}$ is odd, but none otherwise. In this case, we find $y=0$
for $G_{ij}$ even and 
\begin{equation}
\e^{y}=\exp \left( -\int_{-\infty}^{\infty} \frac{dk}{k}
\e^{ik\T}\right), G_{ij} \mathrm{\ odd.}
\end{equation}
This is just the $x \rightarrow 0$ case of the previous Fourier
transform, so we find $y=-2i\pi \Theta(\T)$ for $G_{ij}$ odd or $y=0$
otherwise. This is equivalent to $y=-2i\pi \Theta(\T)G_{ij}$, showing
that we have indeed found a generalisation of the RTV formula.

To complete this section, we must discuss the exceptional case
$a_{2n}^{(2)}$. Being self-dual, the S-matrix for this theory cannot be
found from the above. Following Oota, however, we note that the necessary
prescription is to replace each reference to $\rv\hv$ by $\hv=h=2n+1$,
take all $t_{a}=1$, and replace the incidence matrix by the ``generalised
incidence matrix''~\cite{Klassen}, which is obtained from the incidence
matrix of $a_{n}^{(1)}$ by replacing the last zero on the the diagonal by
a one. Doing this, we obtain the correct integral S-matrix, and hence a
generalised RTV identity, for this case. 

\se{Fourier transforms}
\label{sec-ft}
Here we attempt to find
\begin{equation}
\tilde{\phi}(k)=\int_{-\infty}^{\infty}\left(\frac{\cosh\left(\frac{\T}{2}+a
\right)}{\sinh\left(\frac{\T}{2}+a\right)}\right)e^{ik\T}d\T
\end{equation}

To do this, we need to use the Convolution Theorem, which states
that, if $F(\alpha)$ and $G(\alpha)$ are the Fourier transforms of
$f(x)$ and $g(x)$ respectively, then
\begin{equation}
\frac{1}{2\pi}\int_{-\infty}^{\infty}F(\alpha)G(\alpha)e^{-i\alpha
x}d\alpha = \int_{-\infty}^{\infty}f(u)g(x-u)du.
\label{eq:ft1}
\end{equation}
This, together with the standard results
\begin{eqnarray}
f(x)=\frac{\cosh(ax)}{\sinh(bx)}, 0<a<b & \rightarrow &
F(\alpha)=\frac{i\pi \sinh\left(\frac{\pi \alpha}{b}\right)}{b
\left[\cosh\left(\frac{\pi \alpha}{b}\right)+\cos\left(\frac{\pi
a}{b}\right)\right]} \label{eq:std1}\\
f(x)=\delta(x) & \rightarrow & F(\alpha)=1
\end{eqnarray}
allow us, setting $a=\pi$ and $b=2\pi$ in \ref{eq:std1} to find
\begin{equation}
\frac{i}{4\pi}\int_{-\infty}^{\infty}\left(\frac{\sinh\left(\frac{\alpha}{2}
\right)}{\cosh\left(\frac{\alpha}{2}\right)}\right)e^{-i\alpha x}d\alpha=
\int_{-\infty}^{\infty}\frac{\cosh(\pi u)}{\sinh(2\pi u)}\delta(x-u)du
\label{eq:ft2}
\end{equation}
Returning now to \ref{eq:ft1}, if we make the change of variables $\T'=
\T+2a+in\pi$, where $n$ is an odd integer, we find it becomes
\begin{equation}
\tilde{\phi}(k)=e^{ik(2a+2in\pi)}\int_{-\infty}^{\infty}\left(\frac{\sinh
\left(\frac{\T'}{2}\right)}{\cosh\left(\frac{\T'}{2}\right)}\right)
e^{ik\T'}d\T'
\label{eq:ft3}
\end{equation}
where we have implicitly moved the contour of integration from the real axis
to a line $2a+in\pi$ above it. We can do this provided there are no poles of the
function between the real axis and this line, and, when we make use of this result,
we will pick the arbitrary constant $n$ to make sure this happens. If we
were to take $n$ such that there were $m$ simple poles in this region, we
would incur a correction term of $i2\pi m$, being a contour integral
of the function with the contour
going along the real axis to infinity, up to $2a+in\pi$, back along this line
to minus infinity and then back down to the real axis and back to the start.

We are now in a position to connect the above together, and find,
finally
\begin{equation}
\tilde{\phi}(k)=4\pi ie^{ik(2a+in\pi)}\frac{\cosh(\pi k)}{\sinh(2\pi
k)}
\end{equation}
(being careful over the sign, due to the discrepancy in the sign
of the exponential between \ref{eq:ft2} and \ref{eq:ft3}).

\pagebreak


\se{Dynkin diagrams}
\label{app:liedata}
Where there are roots of different lengths, the
filled in spots refer to short roots. 

\unitlength 0.80mm
\begin{picture}(140,110)


\multiput(25,100)(16,0){3}{\circle{4}}
\multiput(27,100)(16,0){2}{\line(1,0){12}}
\put(59,100){\line(1,0){6}}
\multiput(67,100)(4,0){3}{\circle{0.2}}
\put(77,100){\line(1,0){6}}
\multiput(85,100)(16,0){1}{\circle{4}}
\put(86.5,99){\line(1,0){13}}
\put(86.5,101){\line(1,0){13}}
\put(101,100){\circle*{4}}
\put(91,102.5){\line(2,-1){5}}
\put(91,97.5){\line(2,1){5}}
\put(0,95){\makebox(10,10){$b_{n}^{(1)}$}}
\put(25,95){\makebox(0,0)[cc]{$1$}}
\put(41,95){\makebox(0,0)[cc]{$2$}}
\put(57,95){\makebox(0,0)[cc]{$3$}}
\put(85,95){\makebox(0,0)[cc]{$n-1$}} 
\put(101,95){\makebox(0,0)[cc]{$n$}}


\multiput(25,70)(16,0){3}{\circle*{4}}
\multiput(27,70)(16,0){2}{\line(1,0){12}} 
\put(59,70){\line(1,0){6}}
\multiput(67,70)(4,0){3}{\circle{0.2}}
\put(77,70){\line(1,0){6}}
\multiput(85,70)(16,0){1}{\circle*{4}}
\put(86.5,69){\line(1,0){13}}
\put(86.5,71){\line(1,0){13}}
\put(101,70){\circle{4}}
\put(95,72.5){\line(-2,-1){5}}
\put(95,67.5){\line(-2,1){5}}
\put(0,65){\makebox(10,10){$c_{n}^{(1)}$}}
\put(25,65){\makebox(0,0)[cc]{$1$}}
\put(41,65){\makebox(0,0)[cc]{$2$}}   
\put(57,65){\makebox(0,0)[cc]{$3$}}
\put(85,65){\makebox(0,0)[cc]{$n-1$}} 
\put(101,65){\makebox(0,0)[cc]{$n$}}


\multiput(41,40)(16,0){2}{\circle{4}}
\multiput(43,40)(16,0){1}{\line(1,0){12}} 
\multiput(73,40)(16,0){2}{\circle*{4}}
\multiput(75,40)(16,0){1}{\line(1,0){12}}
\put(58.5,39){\line(1,0){13}}
\put(58.5,41){\line(1,0){13}}
\put(67,40){\line(-2,-1){5}}
\put(67,40){\line(-2,1){5}}
\put(0,35){\makebox(10,10){$f_{4}^{(1)}$}}
\put(41,35){\makebox(0,0)[cc]{$1$}}
\put(57,35){\makebox(0,0)[cc]{$3$}}  
\put(73,35){\makebox(0,0)[cc]{$4$}}
\put(89,35){\makebox(0,0)[cc]{$2$}}   


\put(57,10){\circle{4}}
\put(73,10){\circle*{4}}   
\put(58.5,11){\line(1,0){13}}
\put(58.5,9){\line(1,0){13}}
\put(59,10){\line(1,0){12}}
\put(63,12.5){\line(2,-1){5}}
\put(63,7.5){\line(2,1){5}}
\put(0,5){\makebox(10,10){$g_{2}^{(1)}$}}
\put(57,5){\makebox(0,0)[cc]{$2$}}
\put(73,5){\makebox(0,0)[cc]{$1$}}


\put(115,105){\line(1,0){5}}
\put(120,5){\line(0,1){100}}
\put(115,5){\line(1,0){5}} 
\put(122,50.0){\makebox(30,15){Nonsimply-}}
\put(122,45.0){\makebox(30,15){laced}}   

\end{picture}

\unitlength 0.90mm
\vspace*{10mm}
\begin{picture}(140,130)


\multiput(15,130)(16,0){3}{\circle{4}}
\multiput(17,130)(16,0){2}{\line(1,0){12}}
\put(49,130){\line(1,0){6}}
\multiput(57,130)(4,0){3}{\circle{0.2}}
\put(67,130){\line(1,0){6}}
\multiput(75,130)(16,0){2}{\circle{4}}
\put(77,130){\line(1,0){12}}
\put(0,125){\makebox(10,10){$a_{n}^{(1)}$}}
\put(15,125){\makebox(0,0)[cc]{$1$}}
\put(31,125){\makebox(0,0)[cc]{$2$}}
\put(47,125){\makebox(0,0)[cc]{$3$}}
\put(75,125){\makebox(0,0)[cc]{$n-1/\overline{2}$}}
\put(91,125){\makebox(0,0)[cc]{$n/\overline{1}$}}


\multiput(15,100)(16,0){2}{\circle{4}}
\multiput(17,100)(16,0){1}{\line(1,0){12}}
\put(33,100){\line(1,0){6}}
\multiput(41,100)(4,0){3}{\circle{0.2}}
\put(51,100){\line(1,0){6}}
\multiput(59,100)(16,0){2}{\circle{4}}
\put(61,100){\line(1,0){12}}
\put(76.5,101.5){\line(1,1){8}}
\put(76.5,98.5){\line(1,-1){8}}
\put(86,111){\circle{4}}
\put(86,90){\circle{4}}
\put(0,95){\makebox(10,10){$d_{n}^{(1)}$}}
\put(15,95){\makebox(0,0)[cc]{$1$}}
\put(31,95){\makebox(0,0)[cc]{$2$}}
\put(57,95){\makebox(0,0)[cc]{$n-3$}}
\put(73,95){\makebox(0,0)[cc]{$n-2$}}
\put(95,111){\makebox(0,0)[cc]{$n-1$}}
\put(91,90){\makebox(0,0)[cc]{$n$}}


\multiput(15,70)(16,0){5}{\circle{4}}
\multiput(17,70)(16,0){4}{\line(1,0){12}}
\put(47,72){\line(0,1){12}}
\put(47,86){\circle{4}}
\put(0,65){\makebox(10,10){$e_{6}^{(1)}$}}
\put(15,65){\makebox(0,0)[cc]{$6$}}
\put(31,65){\makebox(0,0)[cc]{$3$}}
\put(47,65){\makebox(0,0)[cc]{$4$}}
\put(63,65){\makebox(0,0)[cc]{$5$}}
\put(79,65){\makebox(0,0)[cc]{$1$}}
\put(42,86){\makebox(0,0)[cc]{$2$}}


\multiput(15,40)(16,0){6}{\circle{4}}
\multiput(17,40)(16,0){5}{\line(1,0){12}}
\put(47,42){\line(0,1){12}}
\put(47,56){\circle{4}}
\put(0,35){\makebox(10,10){$e_{7}^{(1)}$}}
\put(15,35){\makebox(0,0)[cc]{$2$}}
\put(31,35){\makebox(0,0)[cc]{$5$}}
\put(47,35){\makebox(0,0)[cc]{$7$}}
\put(63,35){\makebox(0,0)[cc]{$6$}}
\put(79,35){\makebox(0,0)[cc]{$4$}}
\put(95,35){\makebox(0,0)[cc]{$1$}}
\put(42,56){\makebox(0,0)[cc]{$3$}}


\multiput(15,10)(16,0){7}{\circle{4}}
\multiput(17,10)(16,0){6}{\line(1,0){12}}
\put(47,12){\line(0,1){12}}
\put(47,26){\circle{4}}
\put(0,5){\makebox(10,10){$e_{8}^{(1)}$}}
\put(15,5){\makebox(0,0)[cc]{$2$}}
\put(31,5){\makebox(0,0)[cc]{$6$}}
\put(47,5){\makebox(0,0)[cc]{$8$}}
\put(63,5){\makebox(0,0)[cc]{$7$}}
\put(79,5){\makebox(0,0)[cc]{$5$}}
\put(95,5){\makebox(0,0)[cc]{$3$}}
\put(111,5){\makebox(0,0)[cc]{$1$}}
\put(42,26){\makebox(0,0)[cc]{$4$}}


\put(115,135){\line(1,0){5}}
\put(120,135){\line(0,-1){130}}
\put(115,5){\line(1,0){5}}
\put(117,65){\makebox(30,15){Simply-}}
\put(117,60){\makebox(30,15){laced}}

\end{picture}

\se{Cartan matrices for simple Lie algebras}
\label{sec-cmats}

Here, we give explicitly the Cartan matrices for all the untwisted simple
Lie algebras, with the root ordering and normalisation we have used. 

\ssest{$a_{n}^{(1)}$}

\begin{displaymath}
\left(
\begin{array}{ccccccc}
2&-1&0&\cdots& & & \\
-1&2&-1&\cdots& & & \\
0&-1&2&\cdots& & & \\
\vdots &\vdots &\vdots &\ddots&\vdots &\vdots &\vdots \\
 & & &\cdots&2&-1&0 \\
 & & &\cdots&-1&2&-1 \\
 & & &\cdots&0&-1&2
\end{array}
\right)
\end{displaymath}

\ssest{$b_{n}^{(1)}$}

\begin{displaymath}
\left(
\begin{array}{ccccccc}
2&-1&0&\cdots& & & \\
-1&2&-1&\cdots& & & \\
0&-1&2&\cdots& & & \\
\vdots &\vdots &\vdots &\ddots&\vdots &\vdots &\vdots \\
 & & &\cdots&2&-1&0 \\
 & & &\cdots&-1&2&-1 \\
 & & &\cdots&0&-2&2
\end{array}
\right)
\end{displaymath}

\ssest{$c_{n}^{(1)}$}

\begin{displaymath}
\left(
\begin{array}{ccccccc}
2&-1&0&\cdots& & & \\
-1&2&-1&\cdots& & & \\
0&-1&2&\cdots& & & \\
\vdots &\vdots &\vdots &\ddots&\vdots &\vdots &\vdots \\
 & & &\cdots&2&-1&0 \\
 & & &\cdots&-1&2&-2 \\
 & & &\cdots&0&-1&2
\end{array}
\right)
\end{displaymath}

\ssest{$d_{n}^{(1)}$}

\begin{displaymath}
\left(
\begin{array}{cccccccc}
2&-1&0&\cdots& & & & \\
-1&2&-1&\cdots& & & & \\
0&-1&2&\cdots& & & & \\
\vdots&\vdots&\vdots&\ddots&\vdots&\vdots&\vdots&\vdots \\
 & & & &2&-1&0&0 \\
 & & & &-1&2&-1&-1 \\
 & & & &0&-1&2&0 \\
 & & & &0&-1&0&2
\end{array}
\right)
\end{displaymath}

\ssest{$e_{6}^{(1)}$}

\begin{displaymath}
\left(
\begin{array}{cccccc}
2&0&0&0&-1&0 \\
0&2&0&-1&0&0 \\
0&0&2&-1&0&-1 \\
0&-1&-1&2&-1&0 \\
-1&0&0&-1&2&0 \\
0&0&-1&0&0&2
\end{array}
\right)
\end{displaymath}

\ssest{$e_{7}^{(1)}$}

\begin{displaymath}
\left(
\begin{array}{ccccccc}
2&0&0&-1&0&0&0 \\
0&2&0&0&-1&0&0 \\
0&0&2&0&0&0&-1 \\
-1&0&0&2&0&-1&0 \\
0&-1&0&0&2&0&-1 \\
0&0&0&-1&0&2&-1 \\
0&0&-1&0&-1&-1&2
\end{array}
\right)
\end{displaymath}

\ssest{$e_{8}^{(1)}$}

\begin{displaymath}
\left(
\begin{array}{cccccccc}
2&0&-1&0&0&0&0&0 \\
0&2&0&0&0&-1&0&0 \\
-1&0&2&0&-1&0&0&0 \\
0&0&0&2&0&0&0&-1 \\
0&0&-1&0&2&0&-1&0 \\
0&-1&0&0&0&2&0&-1 \\
0&0&0&0&-1&0&2&-1 \\
0&0&0&-1&0&-1&-1&2
\end{array}
\right)
\end{displaymath}

\ssest{$f_{4}^{(1)}$}

\begin{displaymath}
\left(
\begin{array}{cccc}
2&0&-1&0 \\
0&2&0&-1 \\
-1&0&2&-2 \\
0&-1&-1&2
\end{array}
\right)
\end{displaymath}

\ssest{$g_{2}^{(1)}$}

\begin{displaymath}
\left(
\begin{array}{cc}
2&-3 \\
-1&2 \\
\end{array}
\right)
\end{displaymath}
\pagebreak
\se{S-matrices}
For the self-dual cases, the S-matrices were originally found in
\cite{Arinshtein,CM,BCDS}. The non-self-dual cases took a little longer
but were finally obtained in \cite{del,CDS,Dorey2}. We adopt the general
notation of \cite{Dorey2} and write the S-matrix as
\begin{equation}
S_{ab}(\T)=\prod_{x=1}^{h}\prod_{y=1}^{\rv\hv} \{x,y\}^{m_{ab}(x,y)}\,,
\end{equation}
where the \m{\{x,y\}} are of the form
\begin{equation}
\{x,y\}=\frac{\langle x-1,y-1 \rangle \langle x+1,y+1 \rangle}{
\langle x-1,y+1 \rangle \langle x+1,y-1 \rangle}\,,
\end{equation}
with
\begin{equation}
\langle x,y \rangle = \langle \frac{(2-B)x}{2h}+\frac{By}{2\rv\hv}
\rangle\,,
\end{equation}
and
\begin{equation}
\langle x \rangle = \frac{\sinh \left(\hf\left(\T+i\pi x \right)\right)}{
\sinh \left(\hf \left( \T - i\pi x \right)\right)}\,.
\end{equation}

For convenience, this notation can be extended to include
\begin{displaymath}
_{a}[x,y]_{b}=_{a}\{x,y\}_{b} \times \mathrm{crossing\ } = _{a}\{x,y\}_{b}
\times
\{h-x,r^{\vee}h^{\vee}-y\}_{b}
\end{displaymath}
with the subscript being omitted if it is equal to one and
\begin{eqnarray*}
\{x,y\}_{2}&=&\{x,y-1\}\{x,y+1\} \\
(x,y)_{3}&=&\{x,y-2\}\{x,y\}\{x,y+2\} \\
_{2}\{x,y\}_{2}&=&\{x-1,y\}_{2}\{x+1,y\}_{2} \\
 &=& \{x-1,y-1\}\{x-1,y+1\}\{x+1,y-1\}\{x+1,y+1\}.
\end{eqnarray*}

Whenever an entry appears to the power $n$ below, this means that
$m_{ab}(x,y)$ should be taken to be $n$ rather than 1 for that entry.

\ssest{$a_{n}^{(1)}\quad h=n+1 \mathrm{\ and\ } r^{\vee}h^{\vee}=n+1$}

\begin{equation*}
S_{ab}(\T)=\prod_{ p=|a-b|+1
\mathrm{\ step\ }2}^{a+b-1}\{p,p\}
\end{equation*}

\ssest{$b_{n}^{(1)} \quad h=2n \mathrm{\ and\ } r^{\vee}h^{\vee}=4n-2$}

\begin{eqnarray*}
S_{ab}(\T)&=&\prod_{p=|a-b|+1 \mathrm{\ step\
}2}^{a+b-1}[p,2p]_{2}\,,\quad
a,b < n \\
S_{an}(\T)&=&\prod_{p=|a-b|+1 \mathrm{\ step\ }2}^{a+b-1}[p,2p]\,,\quad
a<n \\
S_{nn}(\T)&=&\prod_{p=1-n \mathrm{\ step\ }2}^{n-1}\{n-p,2n-1-2p\}
\end{eqnarray*}

\ssest{$c_{n}^{(1)} \quad h=2n \mathrm{\ and\ } r^{\vee}h^{\vee}=2n+2$}

\begin{equation*}
S_{ab}=\prod_{p=|a-b|+1 \mathrm{\ step\ }2}^{a+b-1} [p,p]
\end{equation*}

\ssest{$d_{n}^{(1)} \quad h=2(n-1) \mathrm{\ and\ }
r^{\vee}h^{\vee}=2(n-1)$}

\begin{eqnarray*}
S_{ab}(\T)&=&\prod_{p=|a-b|+1 \mathrm{\ step\ }2}^{a+b-1}[p,p]\,,\quad a,b
< n-1 \\
S_{ab}(\T)&=&\prod_{p=n-2 \mathrm{\ step\ }2}^{n-2+b}\{p,p\}\,,\quad
b <n-1, a=n-1 \mathrm{\ or\ }n \\
S_{ab}(\T)&=&\prod_{p=1 \mathrm{\ step\ }4}^{2n-4+x}\{p,p\}\,,\quad a=b=n
\mathrm{\ or\ }a=b=n-1 \\
S_{n(n-1)}(\T)&=&\prod_{p=3 \mathrm{\ step\ }4}^{2n-4-x}\{p,p\}  
\end{eqnarray*}
(In the above, \m{x=1} for \m{n} even and \m{x=-1} for \m{n} odd.)

\ssest{$e_{6}^{(1)} \quad h=12 \mathrm{\ and\ } r^{\vee}h^{\vee}=12$}

(In this and the subsequent sections, \m{x} listed on its own should be
taken to mean \m{\{x\}}.)

\noindent\small
\begin{tabular}{c|c|c|c|c|c|c|c|c} \thline
$\mathbf{a}$&$\mathbf{b}$&{\bf
Block}&$\mathbf{a}$&$\mathbf{b}$&{\bf
Block}&$\mathbf{a}$&$\mathbf{b}$&{\bf Block} \\
\thline 1 & 1 &$ 1 , 7 $ & 2 & 3 &$ [3] , [5] $ & 3 & 6 &$ 2 , 6 , 8 $ \\
1 & 2 &$ [4] $ & 2 & 4 &$ [2] , [4] , 6^{2} $ & 4 & 4 &$ [1] , 
[3]^{2} , [5]^{3} $ \\
1 & 3 &$ 4 , 6 , 10 $ & 2 & 5 &$ [3] , [5] $ & 4 & 5 &$ [2] , 
[4]^{2} , 6^{2} $ \\
1 & 4 &$ [3] , [5] $ & 2 & 6 &$ [4] $ & 4 & 6 &$ [3] , [5] $ \\
1 & 5 &$ 2 , 6 , 8 $ & 3 & 3 &$ 1 , [3] , 5 , 7^{2} $ & 5 & 5 &$ 1 , 
[3] , 5 , 7^{2} $ \\
1 & 6 &$ 5 , 11 $ & 3 & 4 &$ [2] , [4]^{2} , 6^{2} $ & 5 & 6 &$ 4 , 
6 , 10 $ \\
2 & 2 &$ [1] , [5] $ & 3 & 5 &$ [3] , 5^{2} , 7 , 11 $ & 6 & 6 &$ 1 ,
7 $ \\ \thline
\end{tabular}
\normalsize

\ssest{$e_{7}^{(1)} \quad h=18 \mathrm{\ and\ } r^{\vee}h^{\vee}=18$}

$m_{ab}(x,y)=1$ for $(x,y)=(p,p)$ where

\noindent\small
\begin{tabular}{c|c|c|c|c|c|c|c|c} \thline
$\mathbf{a}$&$\mathbf{b}$&{\bf
Block}&$\mathbf{a}$&$\mathbf{b}$&{\bf
Block}&$\mathbf{a}$&$\mathbf{b}$&{\bf Block} \\ \thline
1 & 1 &$ [1] , 9 $ & 2 & 5 &$ [2] , [6] , [8] $ & 4 & 6 &$ [2] , [4] , [6] , [8]^{2} $ \\
1 & 2 &$ [6] $ & 2 & 6 &$ [4] , [6] , [8] $ & 4 & 7 &$ [3] , [5]^{2} , [7]^{2} , 9^{2} $ \\
1 & 3 &$ [5] , 9 $ & 2 & 7 &$ [3] , [5] , [7] , 9^{2} $ & 5 & 5 &$ [1] , [3] , [5] , [7]^{2} , 9^{2} $ \\
1 & 4 &$ [2] , [8] $ & 3 & 3 &$ [1] , [5] , [7] , 9 $ & 5 & 6 &$ [3] , [5]^{2} , [7]^{2} , 9^{2} $ \\
1 & 5 &$ [5] , [7] $ & 3 & 4 &$ [4] , [6] , [8] $ & 5 & 7 &$ [2] , [4]^{2} , [6]^{2} , [8]^{3} $ \\
1 & 6 &$ [3] , [7] , 9 $ & 3 & 5 &$ [3] , [5] , [7] , 9^{2} $ & 6 & 6 &$ [1] , [3] , [5]^{2} , [7]^{2} , 9^{3} $ \\
1 & 7 &$ [4] , [6] , [8] $ & 3 & 6 &$ [3] , [5] , [7]^{2} , 9 $ & 6 & 7 &$ [2] , [4]^{2} , [6]^{3} , [8]^{3} $ \\
2 & 2 &$ [1] , [7] $ & 3 & 7 &$ [2] , [4] , [6]^{2} , [8]^{2} $& 7 & 7
&$ [1] , [3]^{2} , [5]^{3} , [7]^{4} , 9^{4} $ \\
2 & 3 &$ [4] , [8] $ & 4 & 4 &$ [1] , [3] , [7] , 9^{2} $ & & & \\
2 & 4 &$ [5] , [7] $ & 4 & 5 &$ [4] , [6]^{2} , [8] $ & & & \\ \thline
\end{tabular}
\normalsize

\ssest{$e_{8}^{(1)} h=30 \mathrm{\ and\ } r^{\vee}h^{\vee}=30$}

$m_{ab}(x,y)=1$ for $(x,y)=(p,p)$ where

\noindent \small
\begin{tabular}{c|c|c|c|c|c} \thline
$\mathbf{a}$&$\mathbf{b}$&{\bf Block}&$\mathbf{a}$&$\mathbf{b}$&{\bf
Block} \\ \thline
1 & 1 &$ [1] , [11] $ & 3 & 6 &$ [5] , [7]^{2} , [9] , [11] , 
[13]^{2} , 15^{2} $ \\
1 & 2 &$ [7] , [13] $ & 3 & 7 &$ [3] , [5] , [7] , [9]^{2} , 
[11]^{2} , [13]^{2} , 15^{2} $ \\
1 & 3 &$ [2] , [10] , [12] $ & 3 & 8 &$ [4] , [6]^{2} , [8]^{2} , 
[10]^{2} , [12]^{2} , [14]^{3} $ \\
1 & 4 &$ [6] , [10] , [14] $ & 4 & 4 &$ [1] , [5] , [7] , [9] , 
[11]^{2} , [13] , 15^{2} $ \\
1 & 5 &$ [3] , [9] , [11] , [13] $ & 4 & 5 &$ [4] , [6] , [8]^{2} , 
[10] , [12]^{2} , [14]^{2} $ \\
1 & 6 &$ [6] , [8] , [12] , [14] $ & 4 & 6 &$ [3] , [5] , [7] , 
[9]^{2} , [11]^{2} , [13]^{2} , 15^{2} $ \\
1 & 7 &$ [4] , [8] , [10] , [12] , [14] $ & 4 & 7 &$ [3] , [5] , 
[7]^{2} , [9]^{2} , [11]^{2} , [13]^{3} ,
[15]^{2} $ \\
1 & 8 &$ [5] , [7] , [9] , [11] , [13] , 15^{2} $ & 4 & 8 &$ [2] , 
[4] , [6]^{2} , [8]^{2} , [10]^{3} , [12]^{3} ,
[14]^{3} $ \\
2 & 2 &$ [1] , [7] , [11] , [13] $ & 5 & 5 &$ [1] , [3] , [5] , [7] , 
[9]^{2} , [11]^{3} , [13]^{2} , 15^{2} $ \\
2 & 3 &$ [6] , [8] , [12] , [14] $ & 5 & 6 &$ [4] , [6]^{2} , [8]^{2}
, [10]^{2} , [12]^{2} , [14]^{3} $ \\
2 & 4 &$ [4] , [8] , [10] , [12] , [14] $ & 5 & 7 &$ [2] , [4] , 
[6]^{2} , [8]^{2} , [10]^{3} , [12]^{3} ,
[14]^{3} $ \\
2 & 5 &$ [5] , [7] , [9] , [11] , [13] , 15^{2} $ & 5 & 8 &$ [3] , 
[5]^{2} , [7]^{3} , [9]^{3} , [11]^{3} , [13]^{4} ,
[15]^{4} $ \\
2 & 6 &$ [2] , [6] , [8] , [10] , [12]^{2} , [14] $ & 6 & 6 &$ [1] , 
[3] , [5] , [7]^{2} , [9]^{2} , [11]^{3} , [13]^{3} ,
15^{2} $ \\
2 & 7 &$ [4] , [6] , [8] , [10]^{2} , [12] , [14]^{2} $ & 6 & 7 &
$ [3] , [5]^{2} , [7]^{2} , [9]^{3} , [11]^{3} , [13]^{3} ,
15^{4} $ \\
2 & 8 &$ [3] , [5] , [7] , [9]^{2} , [11]^{2} , [13]^{2} , 15^{2} $ & 
6 & 8 &$ [2] , [4]^{2} , [6]^{2} , [8]^{3} , [10]^{4} , [12]^{4} ,
[14]^{4} $ \\
3 & 3 &$ [1] , [3] , [9] , [11]^{2} , [13] $ & 7 & 7 &$ [1] , [3] , 
[5]^{2} , [7]^{3} , [9]^{3} , [11]^{4} , [13]^{4}, 
15^{4} $ \\
3 & 4 &$ [5] , [7] , [9] , [11] , [13] , 15^{2} $ & 7 & 8 &$ [2] , 
[4]^{2} , [6]^{3} , [8]^{4} , [10]^{4} , [12]^{5} ,
 [14]^{5} $ \\
3 & 5 &$ [2] , [4] , [8] , [10]^{2} , [12]^{2} , [14] $ & 8 & 8 &
$ [1] , [3]^{2} , [5]^{3} , [7]^{4} , [9]^{5} , [11]^{6} ,
[13]^{6} , 15^{6} $ \\ \thline
\end{tabular}
\normalsize

\ssest{$f_{4}^{(1)} \quad h=12 \mathrm{\ and\ } r^{\vee}h^{\vee}=18$}

\noindent \small
\begin{tabular}{c|c|c|c|c|c} \thline
$\mathbf{a}$&$\mathbf{b}$&{\bf Block}&$\mathbf{a}$&$\mathbf{b}$&{\bf
Block} \\ \thline
$1$&$1$&$[1,1],[5,7]$ & $2$&$3$&$[3,5]_{2},[5,7]_{2}$ \\
$1$&$2$&$[4,6]_{2}$ & $2$&$4$&$[2,4]_{2},[4,6]_{2},[6,8]_{2}$ \\
$1$&$3$&$[2,2],[4,6],\{6,9\}_{2}$ &
$3$&$3$&$[1,1],[3,4]_{2},[5,8]_{2},[5,7]$ \\
$1$&$4$&$[3,4]_{2}[5,8]_{2}$ & $3$&$4$&$[2,3]_{2},[4,5]_{2},[4,7]_{2},
[6,9]_{2}$ \\
$2$&$2$&$[1,2]_{2},[5,8]_{2}$ & $4$&$4$&$[1,2]_{2},[3,4]_{2},
(_{2}[4,6]_{2}),[5,8]_{2}^{2}$ \\ \thline
\end{tabular}
\normalsize

\ssest{$g_{2}^{(1)} \quad h=6 \mathrm{\ and\ } r^{\vee}h^{\vee}=12$}

\noindent \small
\begin{tabular}{c|c|c} \thline
$\mathbf{a}$&$\mathbf{b}$&{\bf Block} \\ \thline
$1$&$1$&$[1,1],\{3,6\}_{2}$ \\
$1$&$2$&$[2,4]_{3}$ \\
$2$&$2$&$[1,3]_{3},[3,5]_{3}$ \\ \thline
\end{tabular}
\normalsize


\addcontentsline{toc}{chapter}{Bibliography}
\lhead{\textit{Bibliography}}

\pagebreak

\sest{Epilogue}
\addcontentsline{toc}{chapter}{Epilogue}
\thispagestyle{plain}
\quot{The Road goes ever on and on,\\
Down from the door where it began.\\
Now far ahead the Road has gone,\\
And I must follow, if I can,\\
Pursuing it with eager feet,\\
Until it joins some larger way\\
Where many paths and errands meet.\\
And whither then? I cannot say.}{J.R.R. Tolkien,The Hobbit}


\begin{thebibliography}{99}
\bibitem[0]{}
\quot{`What is the use of a book',
thought Alice, `without pictures or conversations?'}{Lewis Carroll}

\bibitem{Mattsson2}
\rauthor{P. Mattsson and P. Dorey}
\rname{Boundary spectrum in the sine-Gordon model with Dirichlet
boundary conditions}
\journal{J. Phys.}{A33}{2000}{9065--9093}
\preprint{hep-th/0008071}

\bibitem{Mattsson}
\rauthor{P. Mattsson}
\rname{S-matrix identities in affine Toda field theories}
\journal{Phys. Lett.}{B468}{1999}{233--238}%
\preprint{hep-th/9908184}

\bibitem{Fring}
\rauthor{A. Fring, C. Korff, and B.J. Schulz}
\rname{On the universal representation of the scattering matrix of
affine Toda field theory}
\journal{Nucl. Phys.}{B567}{2000}{409--453}%
\preprint{hep-th/9907125}

\bibitem{Khastgir}
\rauthor{S.P. Khastgir}
\rname{S-matrices and bi-linear sum rules of conserved charges
in affine Toda field theories}
\journal{Phys. Lett.}{B451}{1999}{68--72}%
\preprint{hep-th/9805197}

\bibitem{Chew}
\rauthor{G.F. Chew}
\rname{The Analytic S-matrix}
\book{W.A. Benjamin Publ.}{New York}{1966}

\bibitem{Eden}
\rauthor{R.J. Eden, P.V. Landshoff, D.I. Olive and P.V. Polkinghorne}
\rname{The Analytic S-matrix}
\book{Cambridge University Press}{Cambridge}{1966}

\bibitem{Kane}
\rauthor{C. Kane and M. Fisher}
\rname{Transmission through barriers and resonant tunneling in an
interacting one-dimensional electron gas}
\journal{Phys. Rev.}{B46}{1992}{15233--15262}%

\bibitem{Wen}
\rauthor{X.G. Wen}
\rname{Chiral Luttinger liquid and the edge excitations in the
fractional quantum Hall states}
\journal{Phys. Rev.}{B41}{1990}{12838--12844}%

\bibitem{Fendley}
\rauthor{P. Fendley, A.W.W. Ludwig and H. Saleur}
\rname{Exact Conductance through Point Contacts in the $\nu=1/3$
Fractional Quantum Hall Effect}
\journal{Phys. Rev. Lett.}{74}{1995}{3005--3008}%
\preprint{cond-mat/9408068}

\bibitem{Saleur}
\rauthor{H. Saleur}
\rname{Lectures on non perturbative field theory and quantum
impurity problems}
in the proceedings of the 1998 Les Houches Summer School%
\preprint{cond-mat/9812110}\\
\rauthor{H. Saleur}
\rname{Lectures on non perturbative field theory and quantum
impurity problems. Part II}%
\xpreprint{cond-mat/0007309}

\bibitem{Zsquare}
\rauthor{A.B. Zamolodchikov and Al.B. Zamolodchikov}
\rname{Factorized S-matrices in two dimensions as the exact
solutions of certain relativistic quantum field theory models}
\journal{Ann. Phys.}{120}{1979}{253--291}%

\bibitem{pedrev}
\rauthor{P. Dorey}
\rname{Exact S-matrices}
in the proceedings of the 1996 E\"{o}tv\"{o}s Graduate School,
\xpreprint{hep-th/9810026}

\bibitem{GhoshZam}
\rauthor{S. Ghoshal and A. Zamolodchikov}
\rname{Boundary S matrix and boundary state in two-dimensional integrable
quantum field theory}
\journal{Int. J. Mod. Phys. }{A9}{1994}{3841--3885}%
\preprint{hep-th/9306002}

\bibitem{Shankar}
\rauthor{R. Shankar and E. Witten}
\journal{Phys. Rev.}{D17}{1978}{2134}

\bibitem{Parke}
\rauthor{S. Parke}
\rname{Absence of particle production and factorization of the
S-matrix in 1+1 dimensional models}
\journal{Nucl. Phys.}{B174}{1980}{166--182}

\bibitem{ColeMand}
\rauthor{S. Coleman and J. Mandula}
\rname{All possible symmetries of the S-matrix}
\journal{Phys. Rev.}{159}{1967}{1251--1256}

\bibitem{BCDS}
\rauthor{H.W. Braden, E. Corrigan, P.E. Dorey and R. Sasaki}
\rname{Affine Toda field theory and exact S-matrices}
\journal{Nucl. Phys.}{B338}{1990}{689--746}

\bibitem{CT}
\rauthor{S. Coleman and H.J. Thun}
\rname{On the Prosaic Origin of the Double Poles in the Sine-Gordon
S-matrix}
\journal{Comm. Math. Phys.}{61}{1978}{31--39}\\
\rauthor{H.W. Braden, E. Corrigan, P.E. Dorey and R. Sasaki}
\rname{Multiple poles and other features of Affine Toda Field Theory}
\journal{Nucl. Phys.}{B356}{1991}{469--498}%

\bibitem{Cherednik}
\rauthor{I.V. Cherednik}
\rname{Factorizing particles on a half line and root systems}
\journal{Theor. Math. Phys.}{61}{1984}{977--983}

\bibitem{FringK}
\rauthor{A. Fring and R. K\"{o}berle}
\rname{Factorized Scattering in the Presence of Reflecting Boundaries}
\journal{Nucl. Phys.}{B421}{1994}{159--172}%
\preprint{hep-th/9304141}

\bibitem{Sasaki}
\rauthor{R. Sasaki}
In \textit{``Interface Between Physics and Mathematics''} eds. W. Nahm
and J-M. Shen, World Scientific (1994) 201%
\preprint{hep-th/9311027}

\bibitem{Corriganb}
\rauthor{E. Corrigan}
\rname{Integrable field theory with boundary conditions}
Talks given at the workshop \textit{Frontiers in Quantum Field
Theory}, held in Urumqi, Xinjiang Uygur Autonomous Region, People's
Republic of China, 11--19 August, 1996%
\preprint{hep-th/9612138}

\bibitem{Doreysete}
\rauthor{P. Dorey}
Lectures given at the TMR Montpellier Summer School \textit{Recent
advances and applications of Conformal Field Theory} held in S\`{e}te,
France, 22--28 May 2000

\bibitem{ISM}
\rauthor{C.S. Gardner, J.M. Greene, M.D. Kruskal, R.M. Miura}
\rname{Method for solving the Korteweg-deVries equation}
\journal{Phys. Rev. Lett.}{19}{1967}{1095--1097}\\
\rauthor{V.E. Sakharov and A.B. Shabat}
\rname{Exact theory of two-dimensional self-focusing and
one-dimensional self-modulation of waves in nonlinear media}
\journal{Sov. Phys. JETP}{34}{1972}{62--69}\\
\rauthor{M.J. Ablowitz, D.J. Kaup, A.C. Newell and H. Segur}
\rname{Nonlinear-evolution equations of physical significance}
\journal{Phys. Rev. Lett.}{31}{1973}{125--127}\\
\rauthor{D.W. McLaughlin}
\rname{Four examples of the inverse method as a canonical
transformation}
\journal{J. Math. Phys.}{16}{1975}{96--99}

\bibitem{MacIntyre}
\rauthor{A. MacIntyre}
\rname{Integrable boundary conditions for classical sine-Gordon
theory}
\journal{J. Phys.}{A28}{1995}{1089--1100}%
\preprint{hep-th/9410026}

\bibitem{SSW}
\rauthor{H. Saleur, S. Skorik and N.P. Warner}
\rname{The boundary sine-Gordon theory: classical and semi-classical 
analysis}
\journal{Nucl. Phys.}{B441}{1995}{421--436}%
\preprint{hep-th/9408004}

\bibitem{Hirota}
\rauthor{R. Hirota}
\rname{Exact solutions of the Korteweg-de Vries equation for multiple
collisions of solitons}
\journal{Phys. Rev. Lett.}{27}{1972}{1192-1194}

\bibitem{Ghoshal}
\rauthor{S. Ghoshal}
\rname{Bound State Boundary S-Matrix of the sine-Gordon Model}
\journal{Int. J. Mod. Phys.}{A9}{1994}{4801--4810}%
\preprint{hep-th/9310188}

\bibitem{Skorik}
\rauthor{S. Skorik and H. Saleur}
\rname{Boundary bound states and boundary bootstrap in the sine-Gordon
model with Dirichlet boundary conditions}
\journal{J. Phys.}{A28}{1995}{6605--6622}%
\preprint{hep-th/9502011}

\bibitem{Faddeev}
\rauthor{L. Takhtadjian and L. Faddeev}
\rname{Essentially nonlinear one-dimensional model of classical field theory}
\journal{Theor. Math. Phys.}{21}{1974}{1046--57}\\
\rauthor{V. Korepin and L. Faddeev}
\rname{Quantization of solitons}
\journal{Theor. Math. Phys.}{25}{1975}{1039--49}

\bibitem{Pillin}
\rauthor{M. Pillin}
\rname{Exact two-particle matrix elements in S-matrix preserving
deformation of integrable QFTs}
\journal{Phys. Lett.}{B448}{1999}{227-233}%
\preprint{hep-th/9812106}

\bibitem{Barnes}
\rauthor{E.W. Barnes}
\rname{The Theory of the Double Gamma Function}
\journal{Phil. Trans. Roy. Soc.}{A196}{1901}{265--387}\\
\rauthor{E.W. Barnes}
\rname{On the Theory of the Double Gamma Function}
\journal{Trans. Cambridge Phil. Soc.}{19}{1904}{376--425}

\bibitem{Jimbo}
\rauthor{M. Jimbo and T. Miwa}
\rname{QKZ equation with $|q|=1$ and correlation functions of the XXZ model 
in the gapless regime}
\journal{J. Phys.}{A29}{1996}{2923--2958}%
\preprint{hep-th/9601135}

\bibitem{BreitW}
\rauthor{G. Breit and E.P. Wigner}
\journal{Phys. Rev.}{49}{1936}{519}\\
\rauthor{S. Weinberg}
\rname{The Quantum Theory of Fields}
\book{Cambridge University Press}{Cambridge}{1995}

\bibitem{Correview}
\rauthor{E. Corrigan}
\rname{Recent developments in affine Toda quantum field theory}
\newblock Invited lectures at the CRM-CAP Summer School `Particles and
Fields 94' Banff, Alberta, Canada%
\preprint{hep-th/9412213}

\bibitem{Mikhailov}
\rauthor{A.V. Mikhailov, M.A. Olshanetsky and A.M. Perelomov}
\rname{Two-dimensional generalized Toda lattice}
\journal{Comm. Math. Phys.}{79}{1981}{473--488}\\
\rauthor{D.I. Olive and N. Turok}
\rname{The symmetries of Dynkin diagrams and the reduction of Toda
field-equations}
\journal{Nucl. Phys.}{B215}{1983}{470--494} 

\bibitem{Cartan}
\rauthor{E. Cartan}
\rname{Oeuvres Compl\`{e}tes}
\book{Gauthier-Villars}{Paris}{1952}\\
\rauthor{W. Killing}
\rname{Die Zusammensetzung der stetigen endlichen
Transformationsgruppen, I--IV}
\journal{Math. Ann.}{31}{1888}{252--290},
\journal{Math. Ann.}{33}{1889}{1--48},
\journal{Math. Ann.}{34}{1889}{57--122},
\journal{Math. Ann.}{36}{1890}{161--189}

\bibitem{Dynkin}
\rauthor{E.B. Dynkin}
\rname{The structure of semi-simple Lie algebras}
\journal{Amer. Math. Soc. Transl.}{9}{1955}{328--469}

\bibitem{Carter}
\rauthor{R.W. Carter}
\rname{Simple Groups of Lie Type}
\book{Wiley-Interscience}{London}{1972}

\bibitem{Zamconf}
\rauthor{A.B. Zamolodchikov}
\rname{Integrals of motion in scaling 3-state Potts-model field-theory}
\journal{Int. J. Mod. Phys.}{A3}{1988}{743--750}\\
\rauthor{A.B. Zamolodchikov}
\rname{Integrals of motion and S-matrix of the (scaled) $T=T_{c}$
Ising-model with magnetic field}
\journal{Int. J. Mod. Phys.}{A4}{1989}{4235--4248}

\bibitem{Arinshtein}
\rauthor{A.E. Arinshtein, V.A. Fateev, and A.B. Zamolodchikov}
\rname{Quantum S-matrix of the (1+1)-dimensional Todd chain}
\journal{Phys. Lett.}{B87}{1979}{389--392}

\bibitem{CM}
\rauthor{P. Christe and G. Mussardo}
\rname{Elastic S-matrices in (1+1) dimensions and Toda field theories}
\journal{Int. J. Mod. Phys.}{A5}{1990}{4581--4627}

\bibitem{Adams}
\rauthor{D. Adams}
\rname{The hitch hiker's guide to the galaxy}
\book{Pan}{London}{1979}

\bibitem{Christe}
\rauthor{P. Christe and G. Mussardo}
\rname{Integrable systems away from criticality: the Toda field 
theory and S-matrix of the tricritical Ising model}
\journal{Nucl. Phys.}{B330}{1990}{465--487}

\bibitem{Dorey2}
\rauthor{P. Dorey}
\rname{A remark on the coupling dependence in affine Toda field theories}
\journal{Phys. Lett.}{B312}{1993}{291--298}%
\preprint{hep-th/9304149}

\bibitem{del}
\rauthor{G.W. Delius, M.T. Grisaru, and D. Zanon}
\rname{Exact S-matrices for Nonsimply-Laced Affine Toda Theories}
\journal{Nucl. Phys.}{B382}{1992}{365--408}%
\preprint{hep-th/9201067}

\bibitem{CDS}
\rauthor{E. Corrigan, P.E. Dorey and R. Sasaki}
\rname{On a Generalised Bootstrap Principle}
\journal{Nucl. Phys.}{B408}{1993}{579--599}%
\preprint{hep-th/9304065}

\bibitem{Dorey}
\rauthor{P. Dorey}
\rname{Root systems and purely elastic S-matrices}
\journal{Nucl. Phys.}{B358}{1991}{654--676}\\
\rauthor{P. Dorey}
\rname{Root systems and purely elastic S-matrices II}
\journal{Nucl. Phys.}{B374}{1992}{741--762}%
\preprint{hep-th/9110058}

\bibitem{Oota}
\rauthor{T. Oota}
\rname{q-deformed Coxeter element in non-simply-laced affine Toda
field theory}
\journal{Nucl. Phys.}{B504}{1997}{738--752}%
\preprint{hep-th/9706054}

\bibitem{Frenkel}
\rauthor{E. Frenkel and N. Reshetikhin}
\rname{Deformations of $\mathcal{W}$-algebras associated to simple
Lie algebras}
\journal{Comm. Math. Phys.}{197}{1998}{1--32}%
\preprint{q-alg/9708006}

\bibitem{Ravanini}
\rauthor{F. Ravanini, R. Tateo, and A. Valleriani}
\rname{Dynkin TBA's}
\journal{Int. J. Mod. Phys.}{A8}{1993}{1707--1727}%
\preprint{hep-th/9207040}

\bibitem{Mattsson3}
\rauthor{P. Dorey and P. Mattsson}
unpublished (1998).

\bibitem{Klassen}
\rauthor{T.R. Klassen and E. Melzer}
\rname{Purely elastic scattering theories and their ultraviolet limits}
\journal{Nucl. Phys.}{B338}{1990}{485--528}\\
\rauthor{T.R. Klassen and E. Melzer}
\rname{The thermodynamics of purely elastic scattering theories and   
conformal perturbation theory}
\journal{Nucl. Phys.}{B350}{1991}{635--689}

\bibitem{Zamunpub}
\rauthor{Al.B. Zamolodchikov}
unpublished private communication

\bibitem{Corrigansinh1}
\rauthor{E. Corrigan}
\rname{On duality and reflection factors for the sinh-Gordon model}
\journal{Int.J.Mod.Phys.}{A13}{1998}{2709--2722}%
\preprint{hep-th/9707235}

\bibitem{Corrigansinh}
\rauthor{E. Corrigan and G.W. Delius}
\rname{Boundary breathers in the sinh-Gordon model}
\journal{J. Phys.}{A32}{1999}{8601--8614}%
\preprint{hep-th/9909145}\\
\rauthor{A. Chenaghlou and E. Corrigan}
\rname{First order quantum corrections to the classical reflection
factor of the sinh-Gordon model \textup{(2000)}}%
\xpreprint{hep-th/0002065}\\
\rauthor{E. Corrigan and A. Taormina}
\rname{Reflection factors and a two-parameter family of boundary bound
states in the sinh-Gordon model}
\xpreprint{hep-th/0008237}

\bibitem{LeClair}
\rauthor{M. Ameduri, R. Konik and A. LeClair}
\rname{Boundary Sine-Gordon Interactions at the Free Fermion Point}
\journal{Phys. Lett.}{B354}{1995}{376--382}%
\preprint{hep-th/9503088}\\
\rauthor{Z-M. Sheng and H-B Gao}
\rname{On the sine-Gordon--Thirring equivalence in the presence of a
boundary}
\journal{Int. J. Mod. Phys.}{A11}{1996}{4089-4102}%
\preprint{hep-th/9512011}

\bibitem{DPTW}
\rauthor{P. Dorey, A. Pocklington, R. Tateo and G.M.T. Watts}
\rname{TBA and TCSA with boundaries and excited states}
\journal{Nucl. Phys.}{B525}{1998}{641--663}%
\preprint{hep-th/9712197}

\bibitem{DG}
\rauthor{G.W. Delius and G.M. Gandenberger}
\rname{Particle reflection amplitudes in $a^{(1)}_n$ Toda field theories}
\journal{Nucl. Phys.}{B554}{1999}{325--364}%
\preprint{hep-th/9904002}

\bibitem{Freeman}
\rauthor{M.D. Freeman}
\rname{On the mass spectrum of affine Toda field theory}
\journal{Phys. Lett.}{B261}{1991}{57--61}


\bibitem{DTW}
\rauthor{P. Dorey, R. Tateo and G.M.T. Watts}
\rname{Generalisations of the Coleman-Thun mechanism and boundary
reflection factors}
\journal{Phys. Lett.}{B448}{1999}{249--256}%
\preprint{hep-th/9810098}

\newpage
{\bf References added}

\bibitem{BPT1}
\rauthor{Z. Bajnok, L. Palla and G. Tak\'{a}cs}
\rname{Boundary states and finite size effects in sine-Gordon model with
Neumann boundary condition}
\xpreprint{hep-th/0106069}

\bibitem{BPT2}
\rauthor{Z. Bajnok, L. Palla, G. Tak\'{a}cs and G.Zs. T\'{o}th}
\rname{The spectrum of boundary states in sine-Gordon model with integrable
boundary conditions}
\xpreprint{hep-th/0106070}

\bibitem{BPT3}
\rauthor{Z. Bajnok, L. Palla and G. Tak\'{a}cs}
\rname{Spectrum and boundary energy in boundary sine-Gordon theory}
\xpreprint{hep-th/0108157}

\end{thebibliography}
\end{document}